%% file: ms.tex
\documentclass[10pt,twocolumn,letterpaper]{article}

\usepackage{cvpr}
\usepackage{times}
\usepackage{epsfig}
\usepackage{xcolor}
\usepackage{graphicx}
\usepackage{amsmath}
\usepackage{amssymb}
\usepackage{booktabs}
\usepackage[percent]{overpic}
\usepackage[font={footnotesize}]{caption}

\usepackage[pagebackref=true,breaklinks=true,letterpaper=true,colorlinks,bookmarks=false]{hyperref}

\cvprfinalcopy 

\def\cvprPaperID{7440} 

\ifcvprfinal\fi
\begin{document}

\title{OctSqueeze: Octree-Structured Entropy Model for LiDAR Compression}
\author{
    Lila Huang$^{1, 2}$ \quad Shenlong Wang$^{1, 3}$ \quad Kelvin Wong$^{1, 3}$ \quad Jerry Liu$^{1}$ \quad Raquel Urtasun$^{1, 3}$ \\
    $^{1}$Uber Advanced Technologies Group \quad $^{2}$University of Waterloo \quad $^{3}$ University of Toronto  \\
    {\tt\small \{lila.huang,slwang,kelvin.wong,jerryl,urtasun\}@uber.com}
}

\maketitle
\thispagestyle{empty}

\input{sections/abstract}
\input{sections/intro}
\input{sections/related}
\input{sections/method}
\input{sections/exp}
\input{sections/conc}

{\small
\bibliographystyle{ieee_fullname}
\bibliography{egbib}
}

\end{document}


\title{Supplementary Material -- OctSqueeze: Octree-Structured Entropy Model for LiDAR Compression}
\author{
    Lila Huang$^{1, 2}$ \quad Shenlong Wang$^{2, 3}$ \quad Kelvin Wong$^{2, 3}$ \quad Jerry Liu$^{2}$ \quad Raquel Urtasun$^{2, 3}$ \\
    $^{1}$University of Waterloo \quad $^{2}$Uber Advanced Technologies Group \quad ${^3}$University of Toronto \\
    {\tt\small \{lila.huang,slwang,kelvin.wong,jerryl,urtasun\}@uber.com}
}

\maketitle

\begin{abstract}
In this supplementary material, we describe additional experimental
results that further validate the efficacy of our proposed method.
We also benchmark the runtime of our proposed method and demonstrate its
ability to encode LiDAR point clouds in real-time.
Moreover, we exhibit an extensive array of qualitative results on NorthAmerica
and KITTI that compares our method against Draco in terms of
reconstruction quality and downstream task performance.
\end{abstract}

\section{Additional Ablation Studies}
We conduct a more thorough analysis on our entropy model to validate
our choice of model architecture and the feature set we use.
In Sec.~\ref{sec:saturation}, we show that our model performs best
when using $K=4$ levels of aggregation.
Then, in Sec.~\ref{sec:capacity}, we demonstrate that our model's performance
improvements arise as a result of our hierarchical feature aggregation scheme,
and not because of the increase in model capacity.
Finally, in Sec.~\ref{sec:exp_ablation}, we present an expanded ablation study
on our input feature set at the best level of aggregations; \ie, $ K = 4 $.
All experiments are conducted on the NorthAmerica evaluation set.

\subsection{Number of Aggregations} \label{sec:saturation}

\begin{table}[!htbp]
    \centering
    \resizebox{0.5\linewidth}{!}{%
    \begin{tabular}{c|c|c|c}
    \toprule
                    & \multicolumn{3}{c}{Bitrate} \\
    \# Aggregations & Depth = 12     & Depth = 14    & Depth = 16    \\
    \midrule
    0               & 3.48           & 8.91          & 14.97          \\
    1               & 3.39           & 8.78          & 14.84          \\
    2               & 3.31           & 8.59          & 14.64          \\
    3               & 3.25           & 8.47          & 14.51          \\
    4               & \textbf{3.17}  & \textbf{8.32} & \textbf{14.33} \\
    5               & 3.27           & 8.51          & 14.55          \\
    \bottomrule
    \end{tabular}%
    }
    \caption{Ablation study on the number of aggregations.}
    \label{tab:saturation}
\end{table}

Tab.~\ref{tab:saturation} extends Tab.~2 in the main paper with an
additional row entry for $K=5$ aggregations.
We found that $K=5$ aggregations performs worse than $K=4$ in terms of bitrate
reduction, suggesting that our choice of $K=4$ aggregations in the main paper
is best for our architecture.

\subsection{Aggregation of Parental Context Features}  \label{sec:capacity}
\begin{table}[!htbp]
    \centering
    \resizebox{0.55\linewidth}{!}{%
    \begin{tabular}{c|c|c|c|c}
    \toprule
                          &                 & \multicolumn{3}{c}{Bitrate} \\
    Parental Aggregations & K               & Depth = 12     & Depth = 14    & Depth = 16    \\
    \midrule
                          & 0               & 3.48           & 8.91          & 14.97          \\
    \midrule
                          & 4               & 3.47           & 8.92          & 14.98          \\
    \checkmark            & 4               & \textbf{3.17}  & \textbf{8.32} & \textbf{14.33} \\
    \bottomrule
    \end{tabular}%
    }
    \caption{
        We compare the performance of our entropy model with and without
        aggregating parental context features (bottom two rows).
        Both models have the same model capacity as one with $K=4$ aggregations.
        For completeness, we also show the performance of our model with $K=0$ aggregations.
    }
    \label{tab:stacks}
\end{table}

We also investigate whether a model that does not aggregate parental context features
can achieve similar bitrate reductions as our model with $ K = 4 $
aggregations, holding all else equal.
To perform this study, we trained an entropy model with the same architecture
as our model with $ K = 4 $ aggregations, except that each node takes in a copy
of its own context feature in the aggregation stage, rather than that of its parent.
Tab.~\ref{tab:stacks} shows our results.
Surprisingly, we found that not only did the model without parental aggregation
perform worse than the one with aggregation, it also performed only as well
as our original, smaller capacity model with $K=0$ aggregations!
This result suggests that adding more layers to the network alone does not
translate to performance gains.
Moreover, it validates our design of a tree-structured entropy model
that progressively incorporates parental information through aggregations.

\subsection{Input Context Features} \label{sec:exp_ablation}
\begin{table}[!htbp]
    \centering
    \resizebox{0.5\linewidth}{!}{%
    \begin{tabular}{cccc|c|c|c}
    \toprule
               &            &            &            & \multicolumn{3}{c}{Bitrate} \\
    L          & P          & O          & LL         & Depth = 12                      & Depth = 14                      & Depth = 16        \\
    \midrule
    \checkmark &            &            &            & 3.86                            & 9.79                            & 15.91             \\
    \checkmark & \checkmark &            &            & 3.44                            & 8.89                            & 14.94             \\
    \checkmark & \checkmark & \checkmark &            & 3.34                            & 8.72                            & 14.76             \\
    \checkmark & \checkmark & \checkmark & \checkmark & \textbf{3.17}                   & \textbf{8.32}                   & \textbf{14.33}    \\
    \bottomrule
    \end{tabular}%
    }
    \caption{
        Ablation study on input context features for our model with $K=4$ aggregations.
        L, P, O, and LL stand for the node's octree level,
        its parent occupancy symbol,
        its octant index,
        and its spatial location respectively.
    }
    \label{tab:ablation}
\end{table}

We conduct an ablation study on the input context features used by our entropy
model at $ K = 4 $ aggregations: the node's octree level, its parent occupancy
symbol, its octant index, and its spatial location.
Note that in Tab.~1 of the main paper, we presented an analogous ablation study
on a model with $ K = 0 $ aggregations.
In Tab.~\ref{tab:ablation}, we observe a similar decrease in bitrate as we
increase the number of input context features.
This result further corroborates our hypothesis that all four input
context features contribute to the predictive power of our entropy model.

\section{Additional Baselines}
We conduct experiments comparing our compression method with two additional
baselines.
In Sec.~\ref{sec:jpeg-range}, we experiment with a range view-based compression
method that leverages the popular JPEG2000 image codec.
Then in Sec.~\ref{sec:deep-voxel}, we present results from our experiments with
the voxel-based point cloud compression algorithm by Quach \etal \cite{quach2019}.

\subsection{JPEG Range Encoder}
\label{sec:jpeg-range}

\begin{figure*}[t]
    \vspace{-0.15in}
    \begin{center}
    \includegraphics[height=0.25\textwidth]{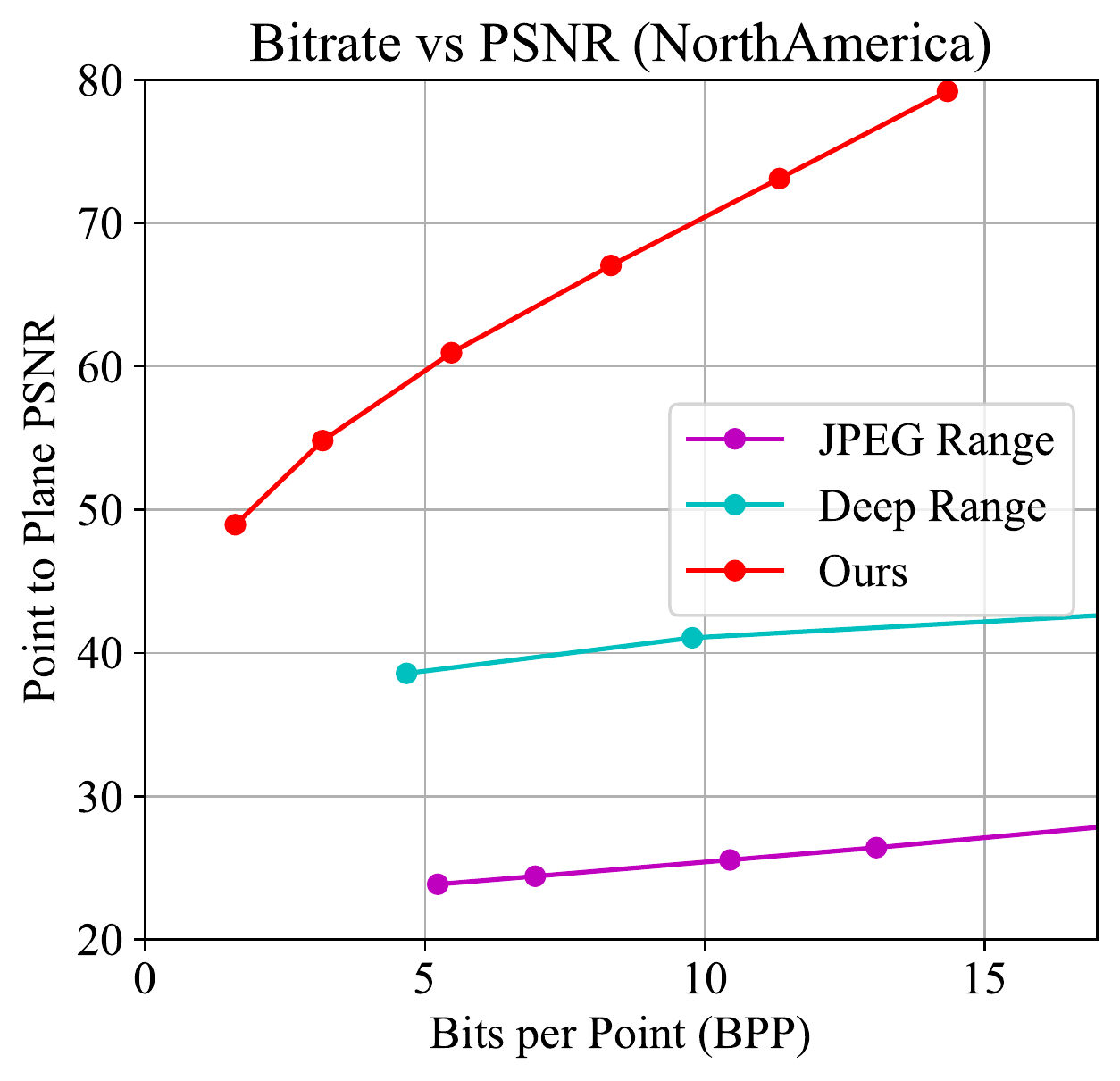}
    \includegraphics[height=0.25\textwidth]{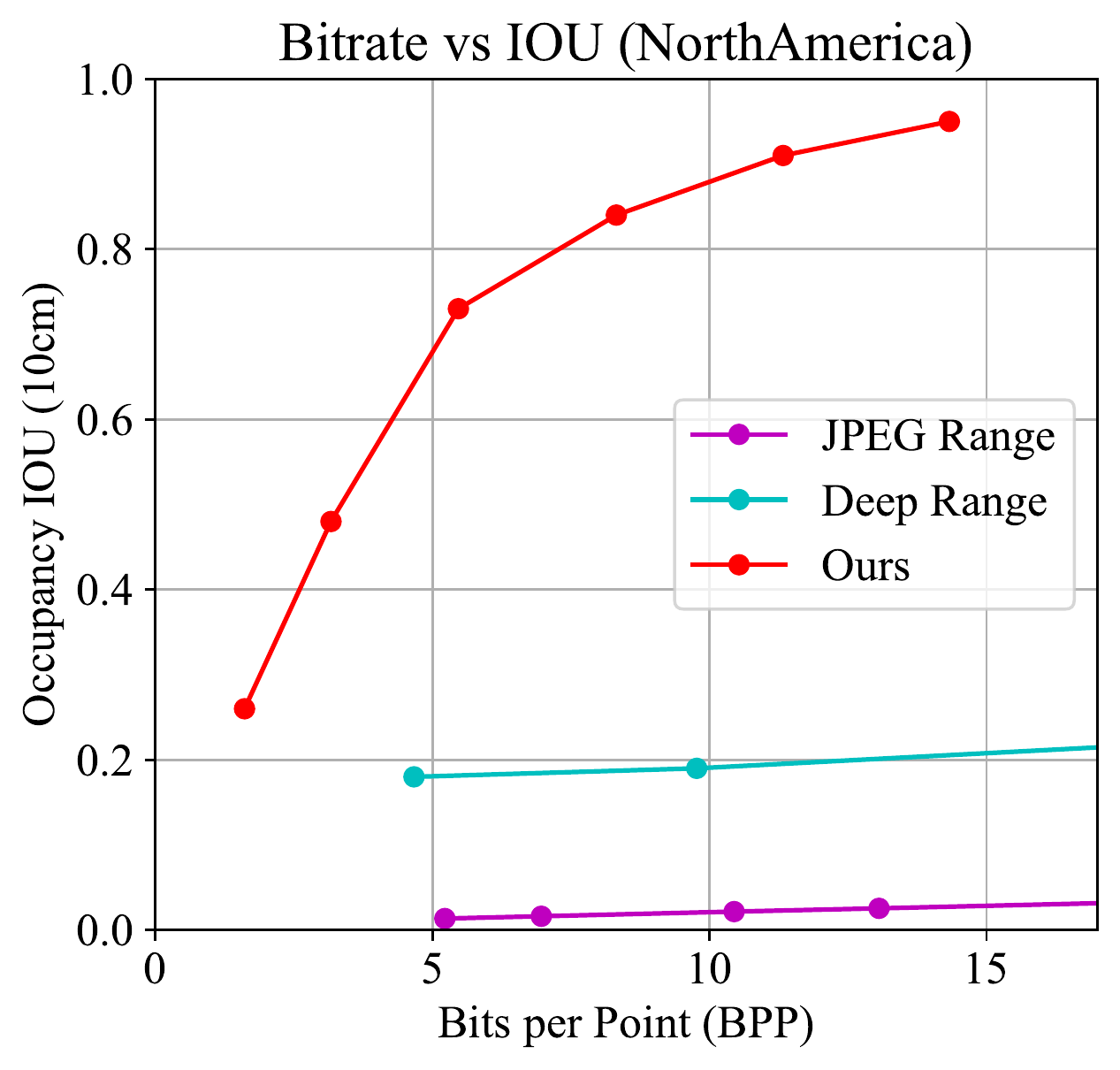}
    \includegraphics[height=0.25\textwidth]{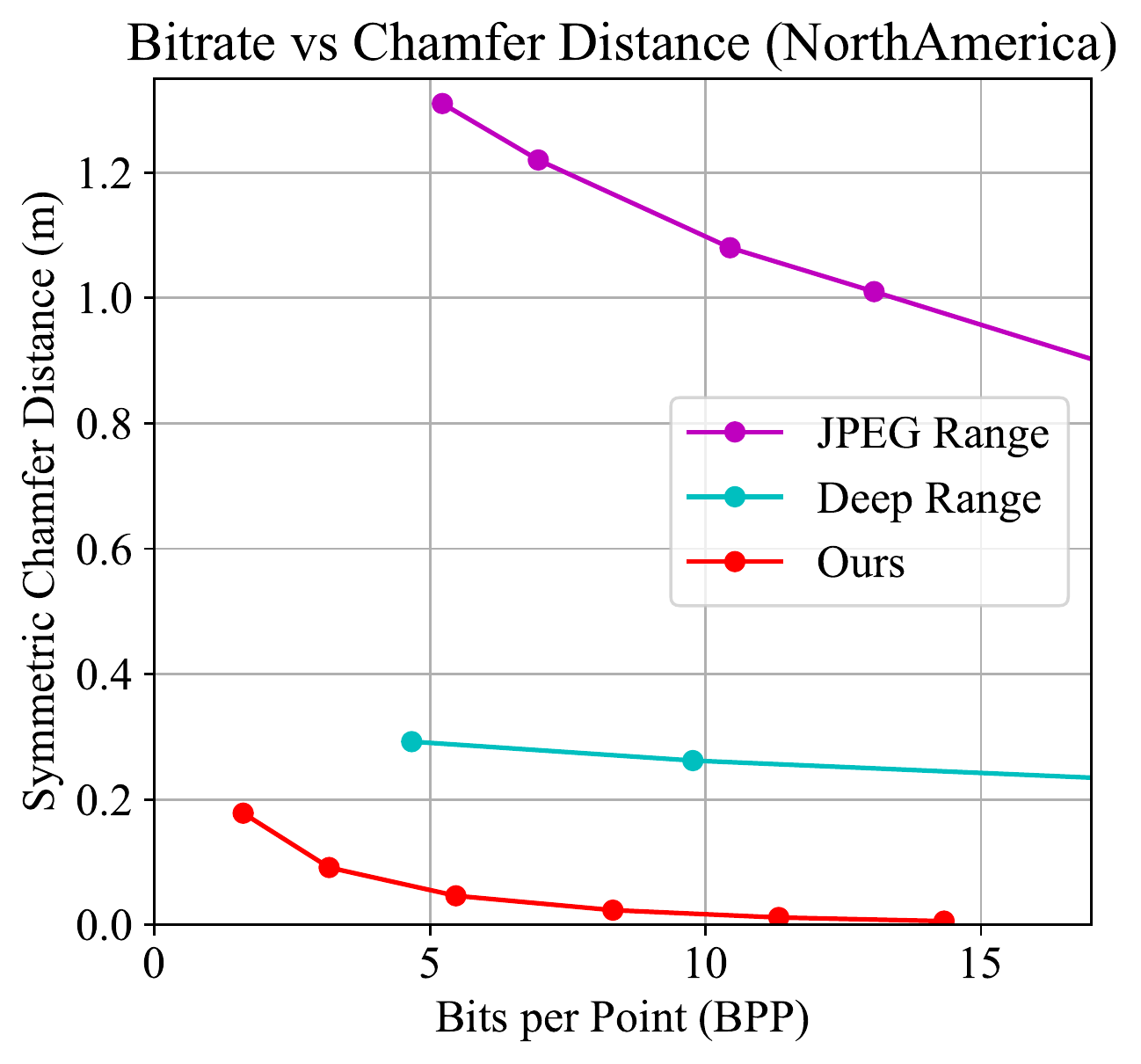}
    \end{center}
    \vspace{-0.25in}
    \begin{center}
    \includegraphics[height=0.25\textwidth]{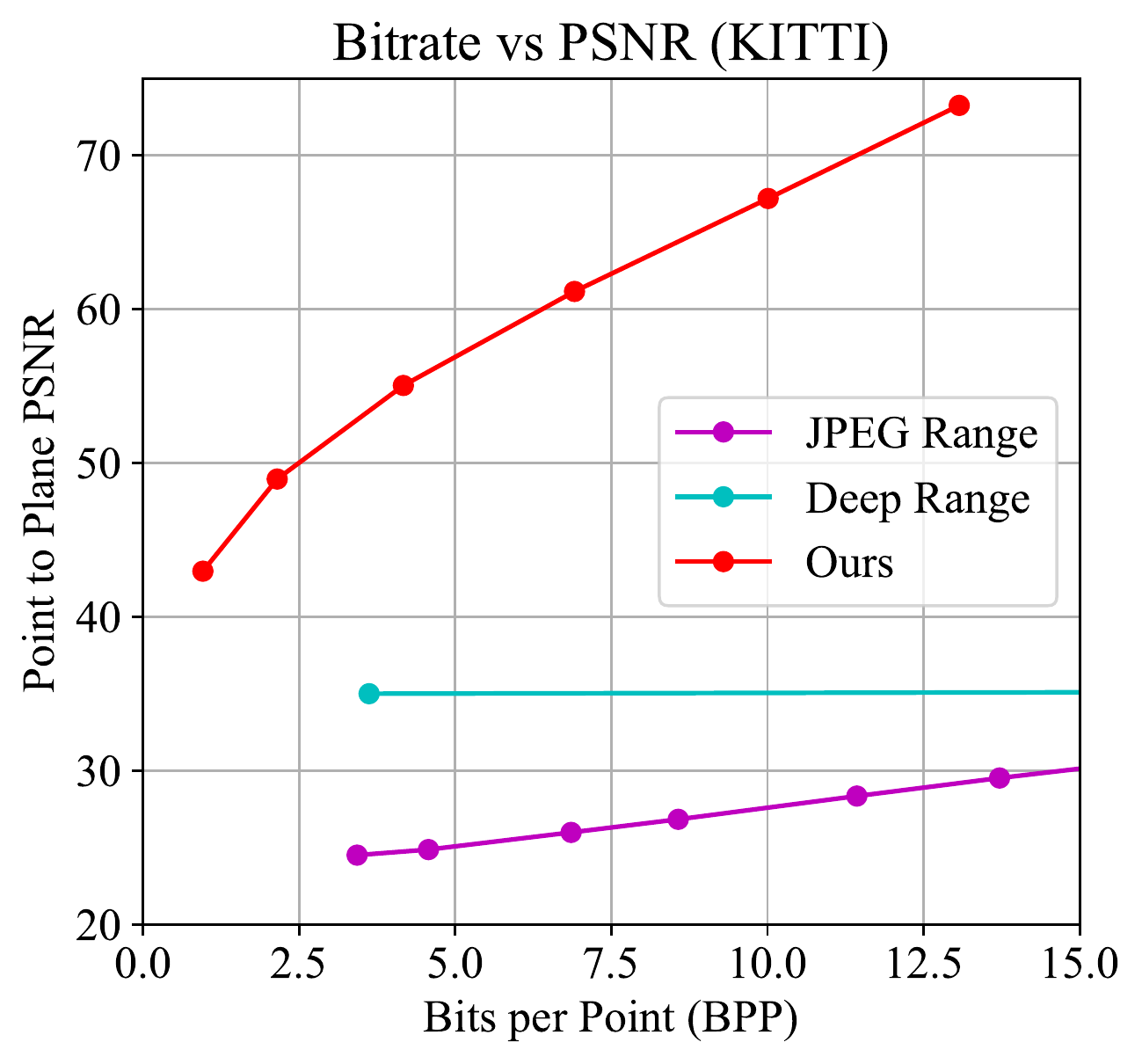}
    \includegraphics[height=0.25\textwidth]{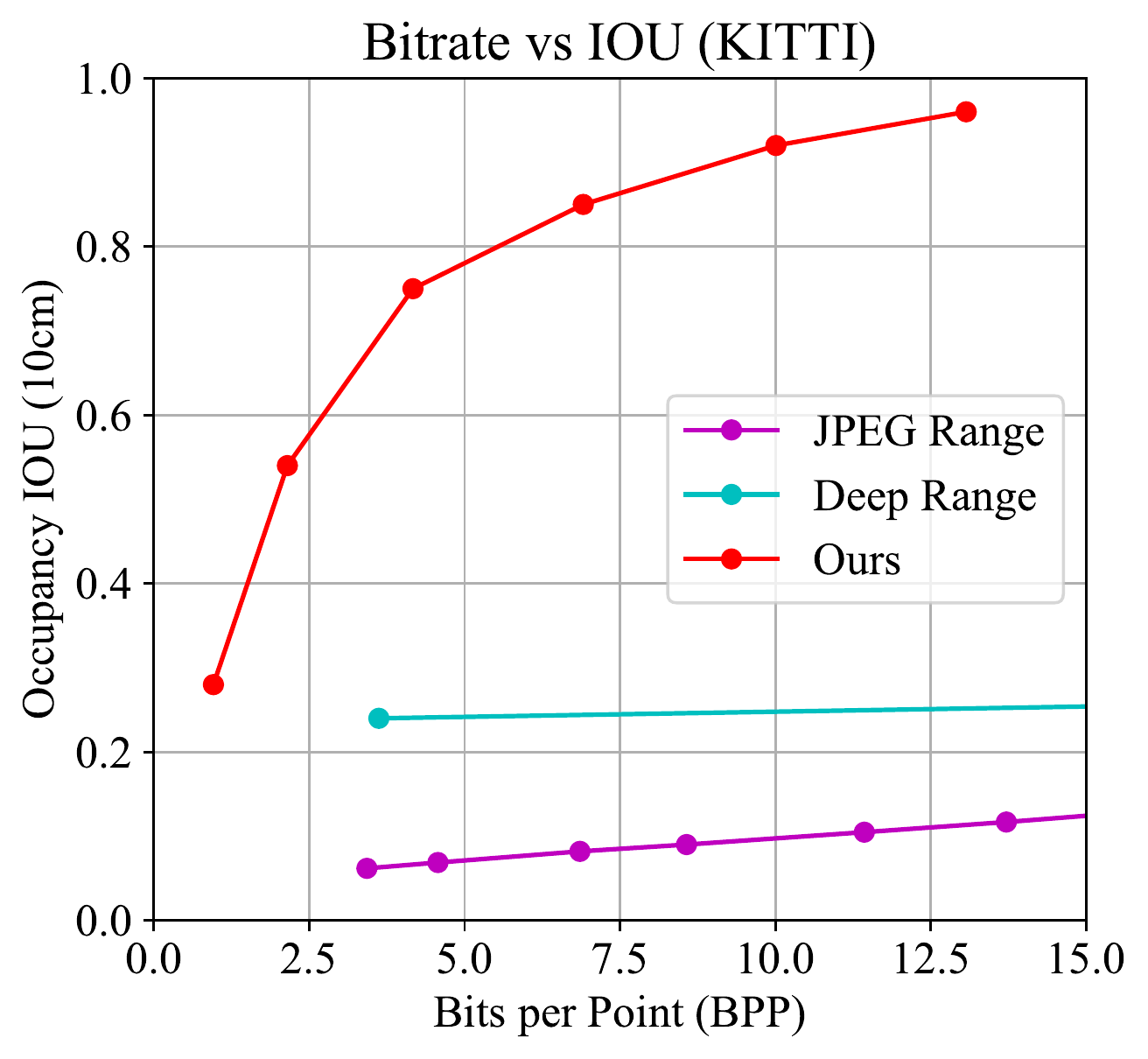}
    \includegraphics[height=0.25\textwidth]{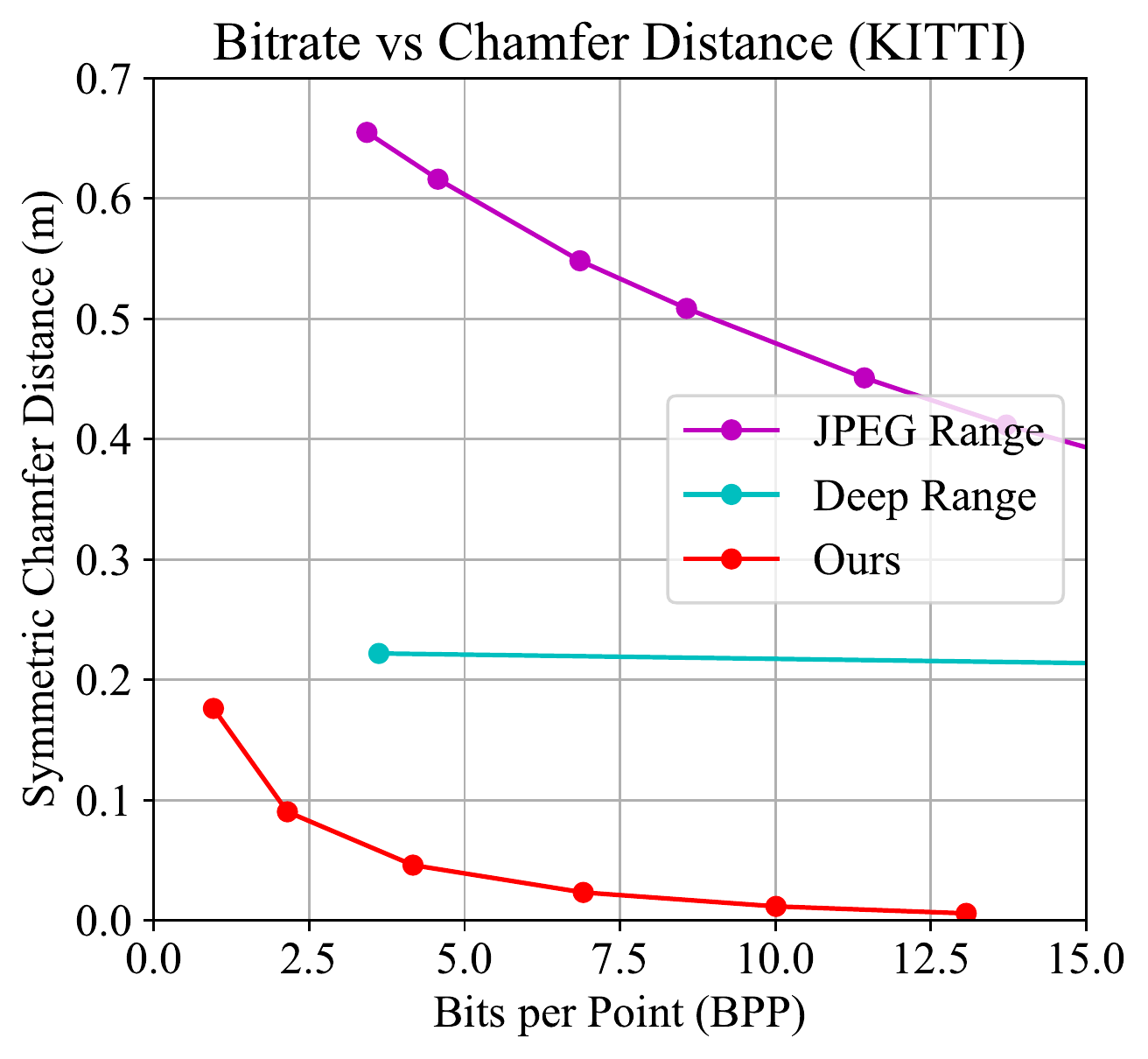}
    \end{center}
    \vspace{-0.2in}
    \caption{
        Quantitative results on NorthAmerica and KITTI.
        From left to right: point-to-plane PSNR, IOU, and Chamfer distance.
    }
    \label{fig:quant_range}
    \vspace{-0.15in}
\end{figure*}

We compare against two baselines that use a range image representation of the
input point cloud: Deep Range and JPEG Range.
Deep Range is the range view-based method discussed in Sec.~4.2 of the main paper.
In particular, given a LiDAR point cloud, we first construct a range image
by converting it from Euclidean coordinates to polar coordinates, and then
storing it as a 2.5D range image.
Deep Range then uses a Ball\'{e} hyperprior model~\cite{balle_varhyperprior}
to compress the 2.5D range image.
In contrast, JPEG Range uses the popular JPEG2000 image codec to compress
the 2.5D range image.

As shown in Fig.~\ref{fig:quant_range}, Deep Range outperforms JPEG Range
across all reconstruction quality metrics on both NorthAmerica and KITTI.
This is a testament to the performance of deep learning-based methods for
image compression.
Moreover, as we alluded to in Sec.~4.4 of the main paper, our approach
significantly outperforms both Deep Range and JPEG Range owing to its use
of an octree data structure to represent the LiDAR point cloud and
an octree-structured entropy model to compress it.

\subsection{Deep Voxel Encoder}
\label{sec:deep-voxel}
We additionally implemented the voxel-based point cloud compression algorithm by
Quach \etal \cite{quach2019}, consisting of a deep 3D convolutional autoencoder
architecture with a fully-factorized entropy model inspired from \cite{balle_varhyperprior}.
We trained and evaluated these models on the NorthAmerica LiDAR point cloud dataset,
voxelizing points at (0.25m, 0.25m, 1.0m) for length, width, and depth dimensions respectively.
Our best-performing reference model achieves a point-to-plane PSNR of 33.76 at a
bitrate of 26.81---this performs much worse than the tree-based methods of Draco
(PSNR: 48.47, bpp: 2.778) and our approach (PSNR: 48.95, bpp 1.61).
The underperformance of the voxel-based compression method indicates that a
dense voxel representation may not be the best fit for compressing LiDAR point
clouds due to the inherent sparsity and high frequency information in this data.

\section{Runtime}

\begin{table}[!htbp]
\centering
\begin{tabular}{c|cccc|c}
\toprule
           & \multicolumn{4}{c|}{Encoding (\emph{ms})} & Decoding (\emph{ms}) \\
Depth      & Octree & Network & Range Coding & Total   & Total \\
\midrule
11         & 21.10  & 13.85   & 0.80         &  35.75  & 92.96  \\
12         & 23.73  & 24.67   & 1.18         &  49.58  & 165.70 \\
13         & 25.51  & 39.82   & 2.19         &  67.52  & 299.41 \\
14         & 32.34  & 56.04   & 3.15         &  91.53  & 486.36 \\
15         & 34.85  & 65.17   & 3.55         & 103.57  & 698.98 \\
16         & 35.52  & 66.83   & 3.61         & 105.96  & 902.27 \\
\bottomrule
\end{tabular}
\caption{
Runtime of our model with $ K = 4 $ aggregations (in \emph{milliseconds}).
`Depth' is the maximum depth of the octree.
`Octree' is the time to build the octree;
`Network' the time to run our entropy model;
and `Range Coding' the time of range coding.
}
\label{tab:encoding-timing}
\end{table}

We benchmarked our approach on a workstation with an Intel Xeon E5-2687W
CPU and a Nvidia GeForce GTX 1080 GPU.
See Tab.~\ref{tab:encoding-timing} for the results.
In our experiments, octree building and range coding were implemented in C++,
and our entropy model was implemented in Python with PyTorch.
Our approach achieves end-to-end encoding in real-time, meaning that our
algorithm can be deployed in an online setting.
Moreover, we believe there are many opportunities to speed up our research
code significantly, especially in CPU/GPU I/O.

\section{Additional Qualitative Results}
\begin{figure*}[h]
\vspace{-0.15in}
\begin{center}
\begin{overpic}[width=0.32\textwidth]{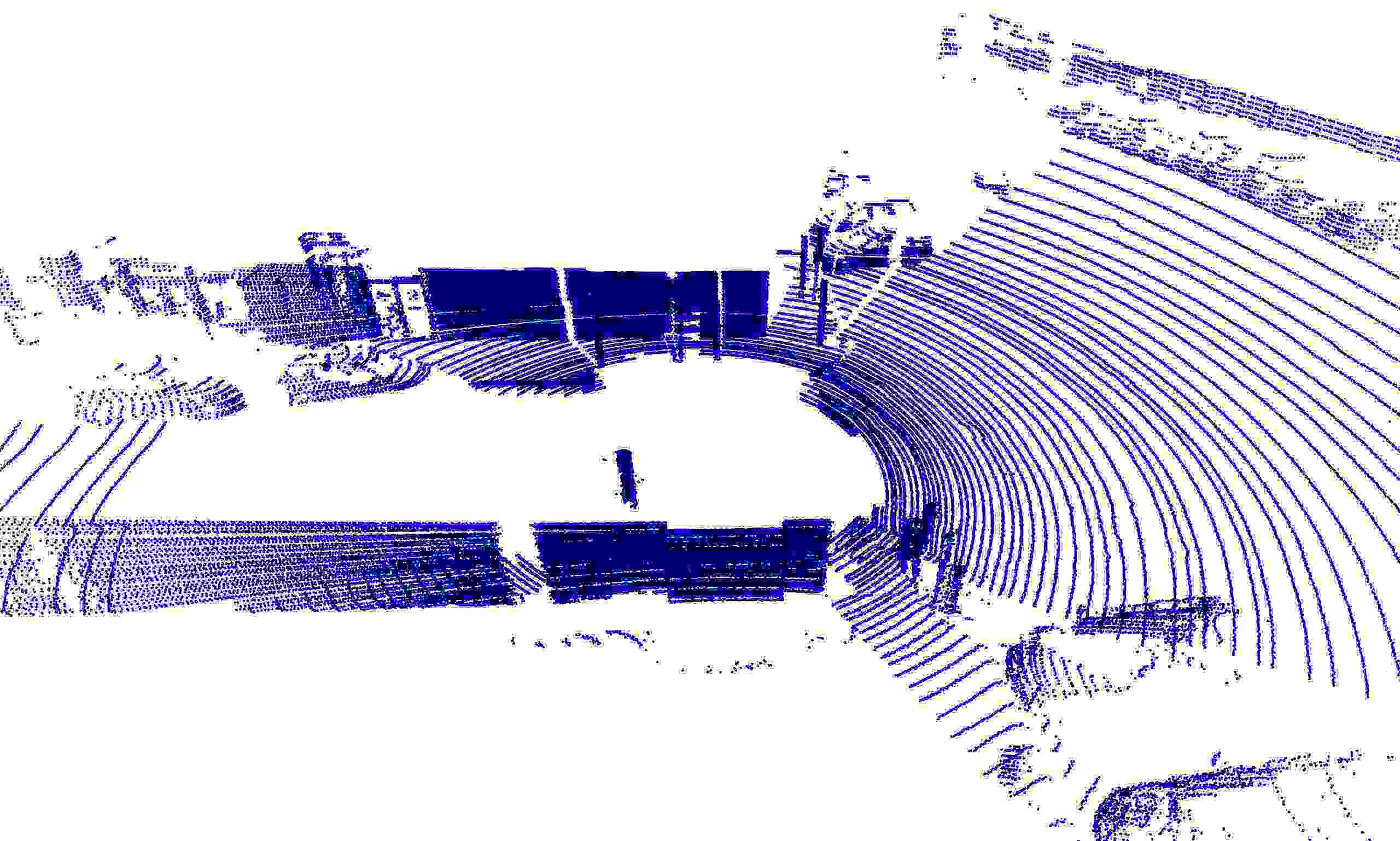}
    \put (0,55) {\colorbox{gray!30}{\scriptsize GT (NorthAmerica)}}
\end{overpic}
\begin{overpic}[width=0.32\textwidth]{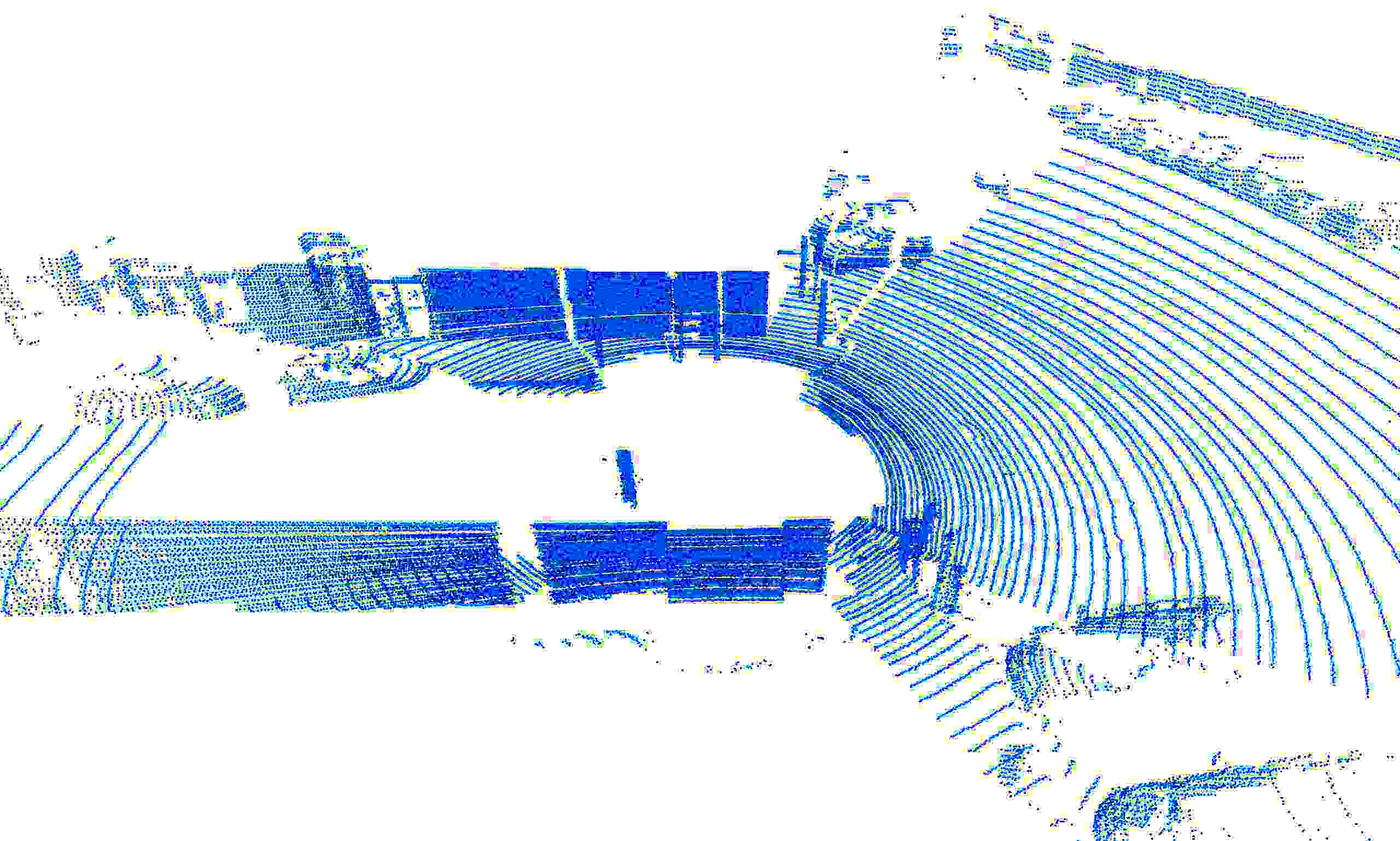}
    \put (0,55) {\colorbox{gray!30}{\scriptsize Ours: PSNR 73.08, Bitrate: 12.23}}
\end{overpic}
\begin{overpic}[width=0.32\textwidth]{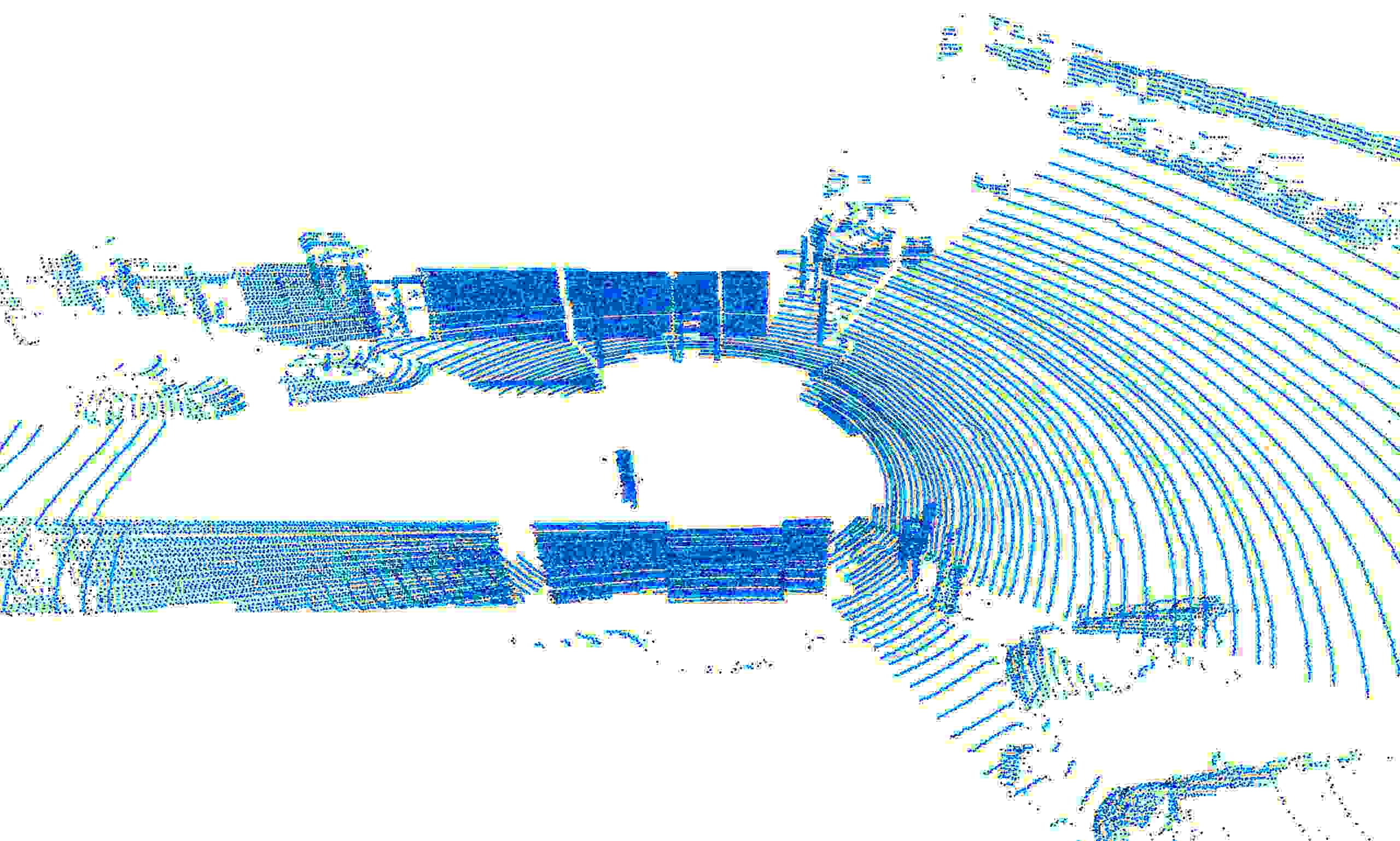}
    \put (0,55) {\colorbox{gray!30}{\scriptsize Draco: PSNR 69.81, Bitrate: 12.41}}
\end{overpic}
\end{center}

\begin{center}
\begin{overpic}[width=0.32\textwidth]{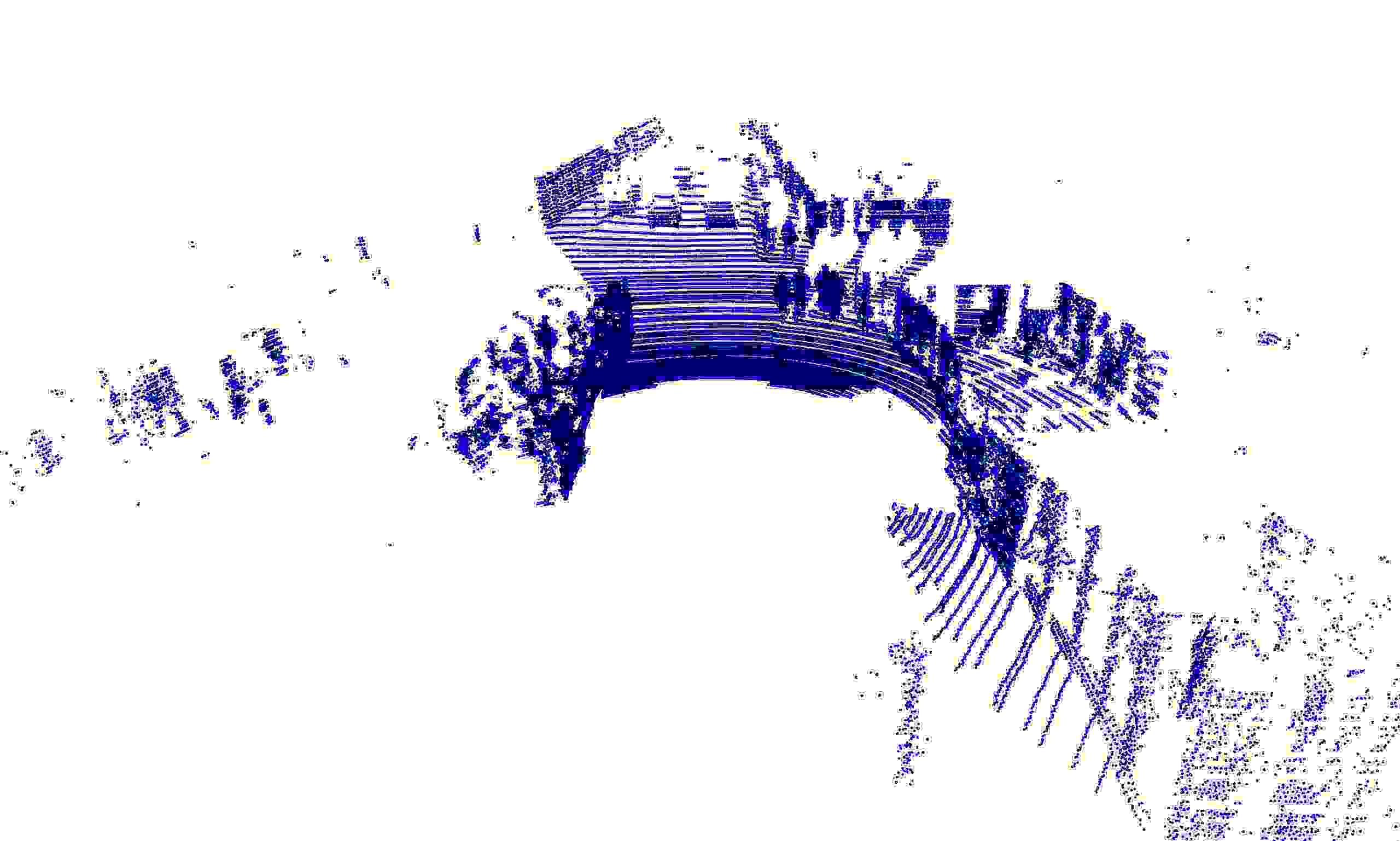}
    \put (0,55) {\colorbox{gray!30}{\scriptsize GT (NorthAmerica)}}
\end{overpic}
\begin{overpic}[width=0.32\textwidth]{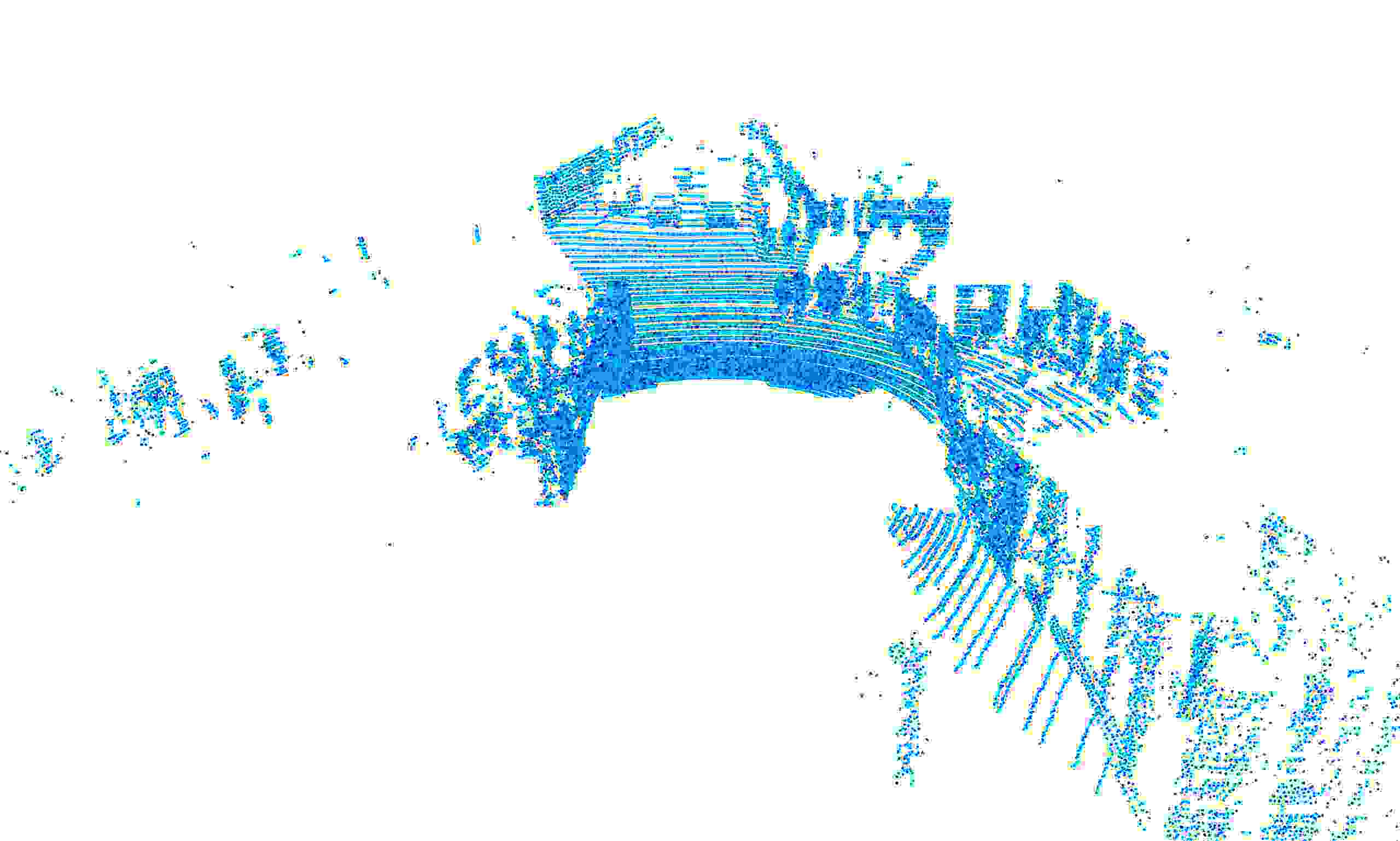}
    \put (0,55) {\colorbox{gray!30}{\scriptsize Ours: PSNR 68.42, Bitrate: 9.38}}
\end{overpic}
\begin{overpic}[width=0.32\textwidth]{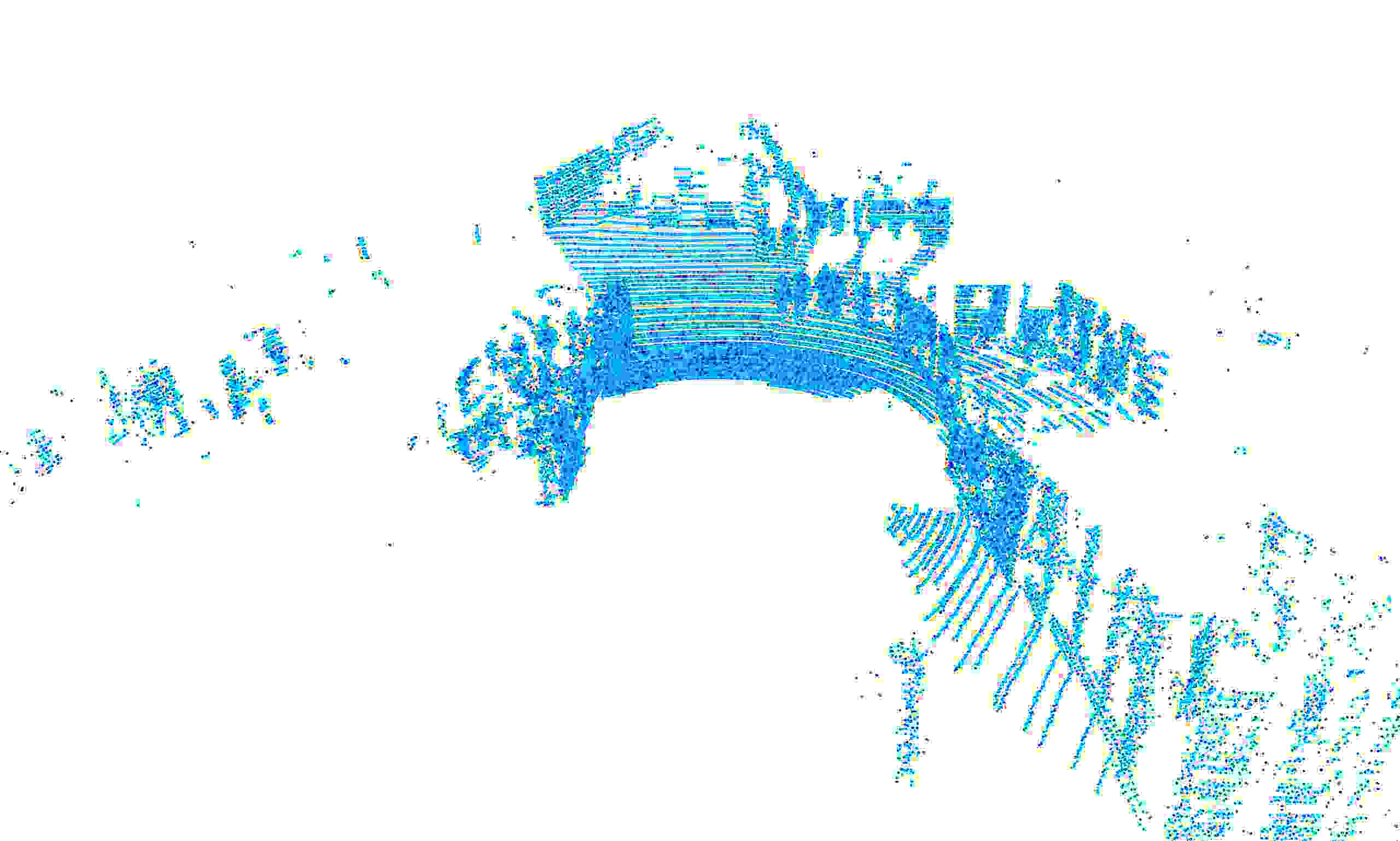}
    \put (0,55) {\colorbox{gray!30}{\scriptsize Draco: PSNR 67.59, Bitrate: 10.42}}
\end{overpic}
\end{center}

\begin{center}
\begin{overpic}[width=0.32\textwidth]{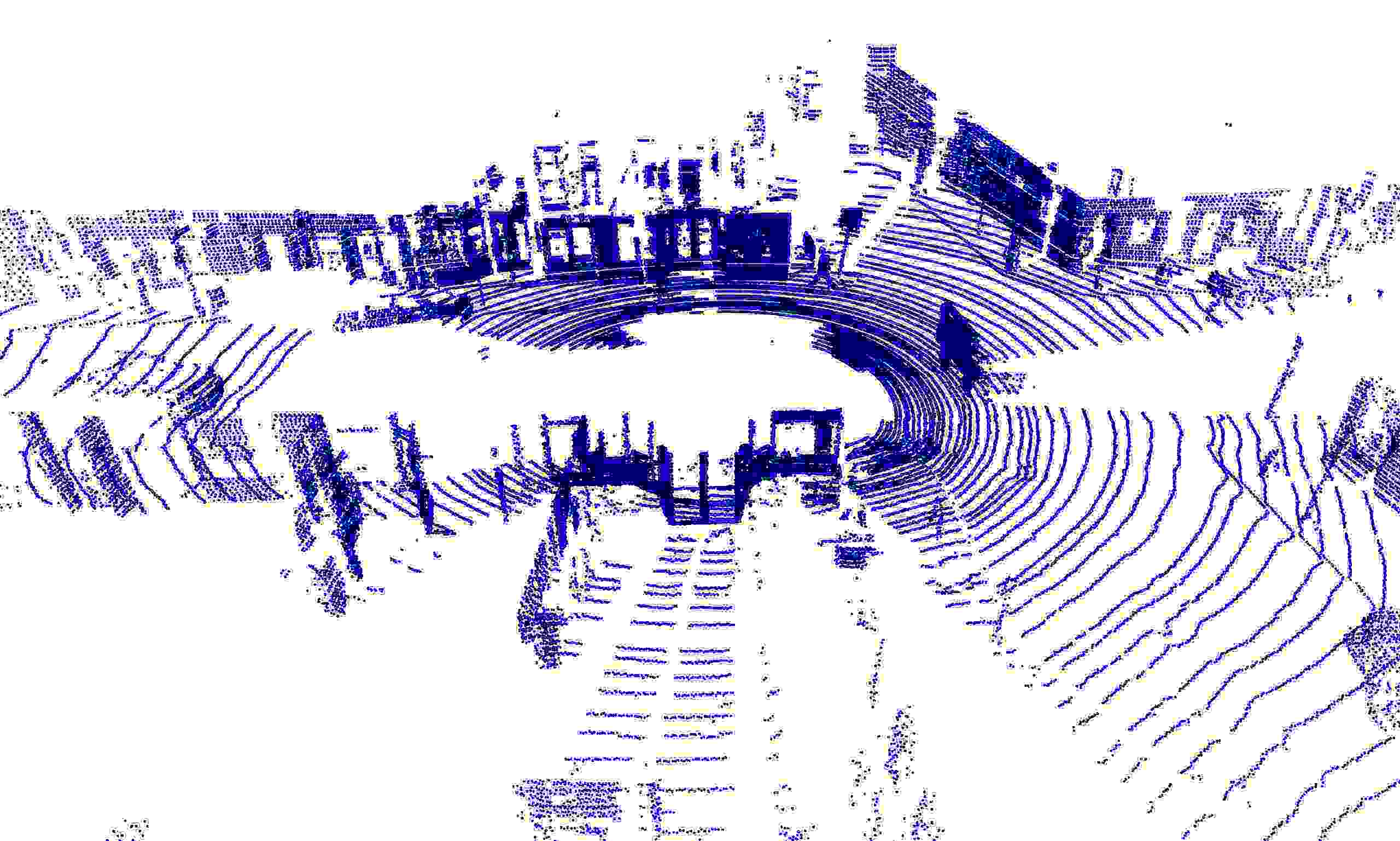}
    \put (0,55) {\colorbox{gray!30}{\scriptsize GT (NorthAmerica)}}
\end{overpic}
\begin{overpic}[width=0.32\textwidth]{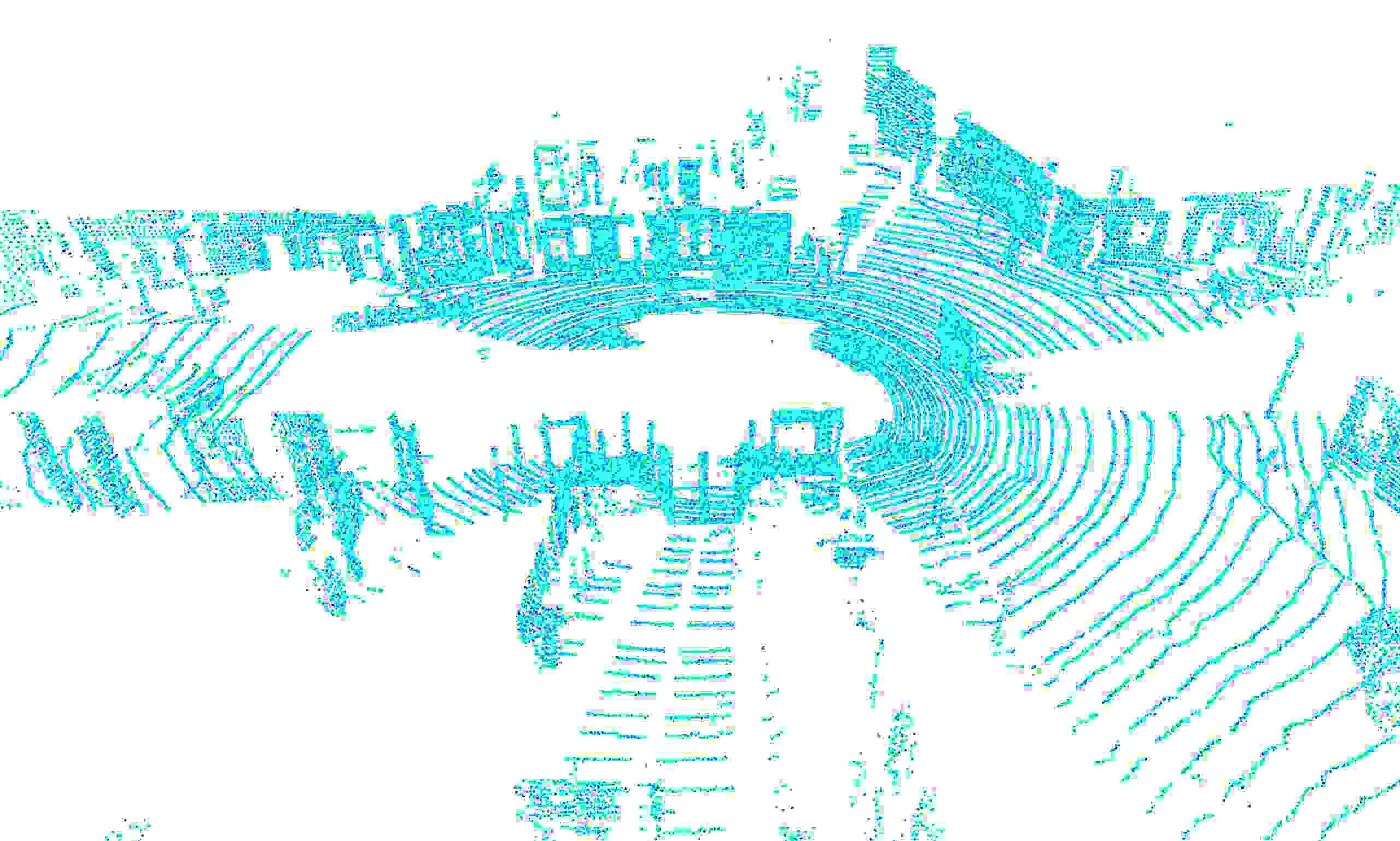}
    \put (0,55) {\colorbox{gray!30}{\scriptsize Ours: PSNR 62.48, Bitrate: 7.49}}
\end{overpic}
\begin{overpic}[width=0.32\textwidth]{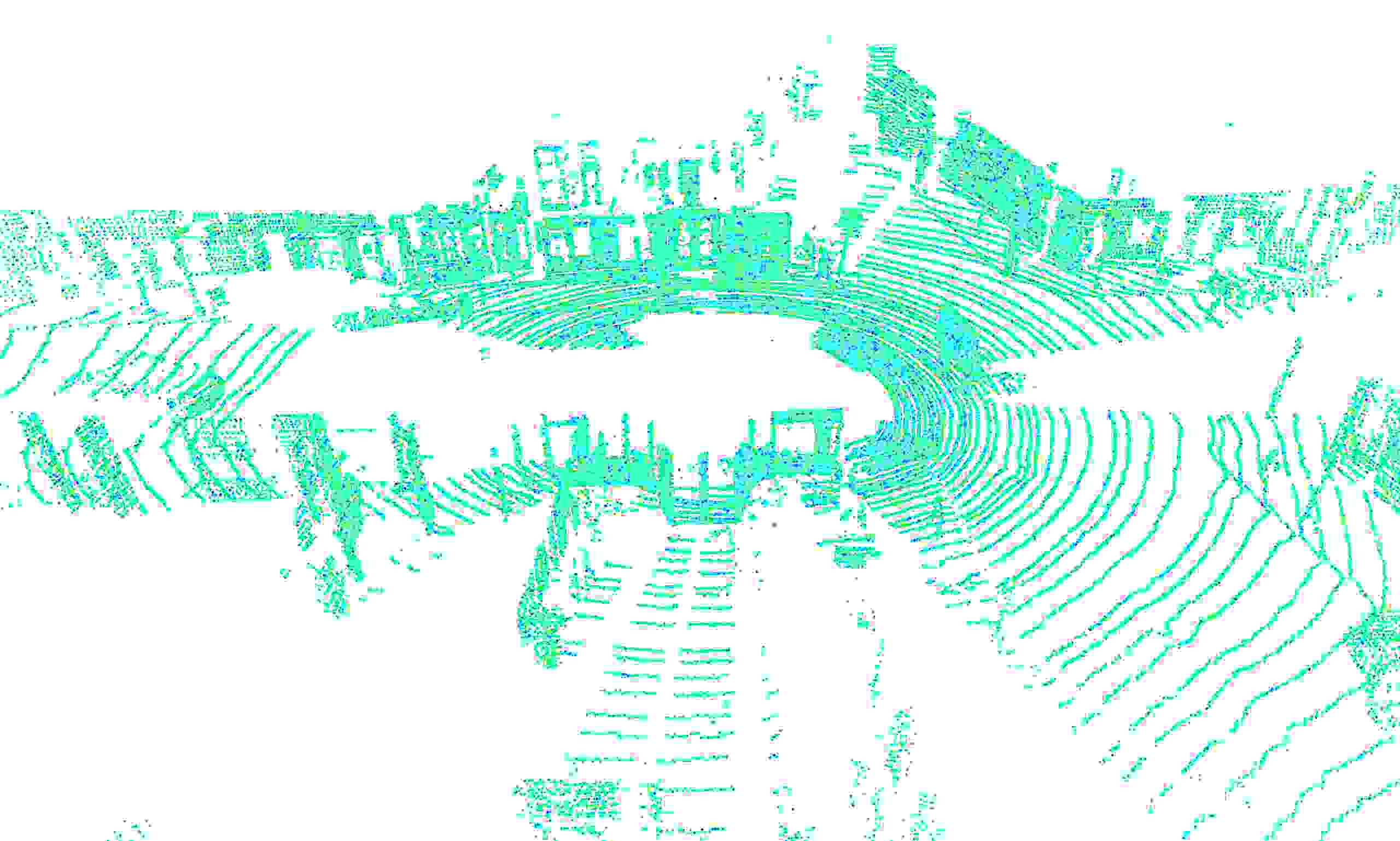}
    \put (0,55) {\colorbox{gray!30}{\scriptsize Draco: PSNR 58.85, Bitrate: 7.55}}
\end{overpic}
\end{center}

\begin{center}
\begin{overpic}[width=0.32\textwidth]{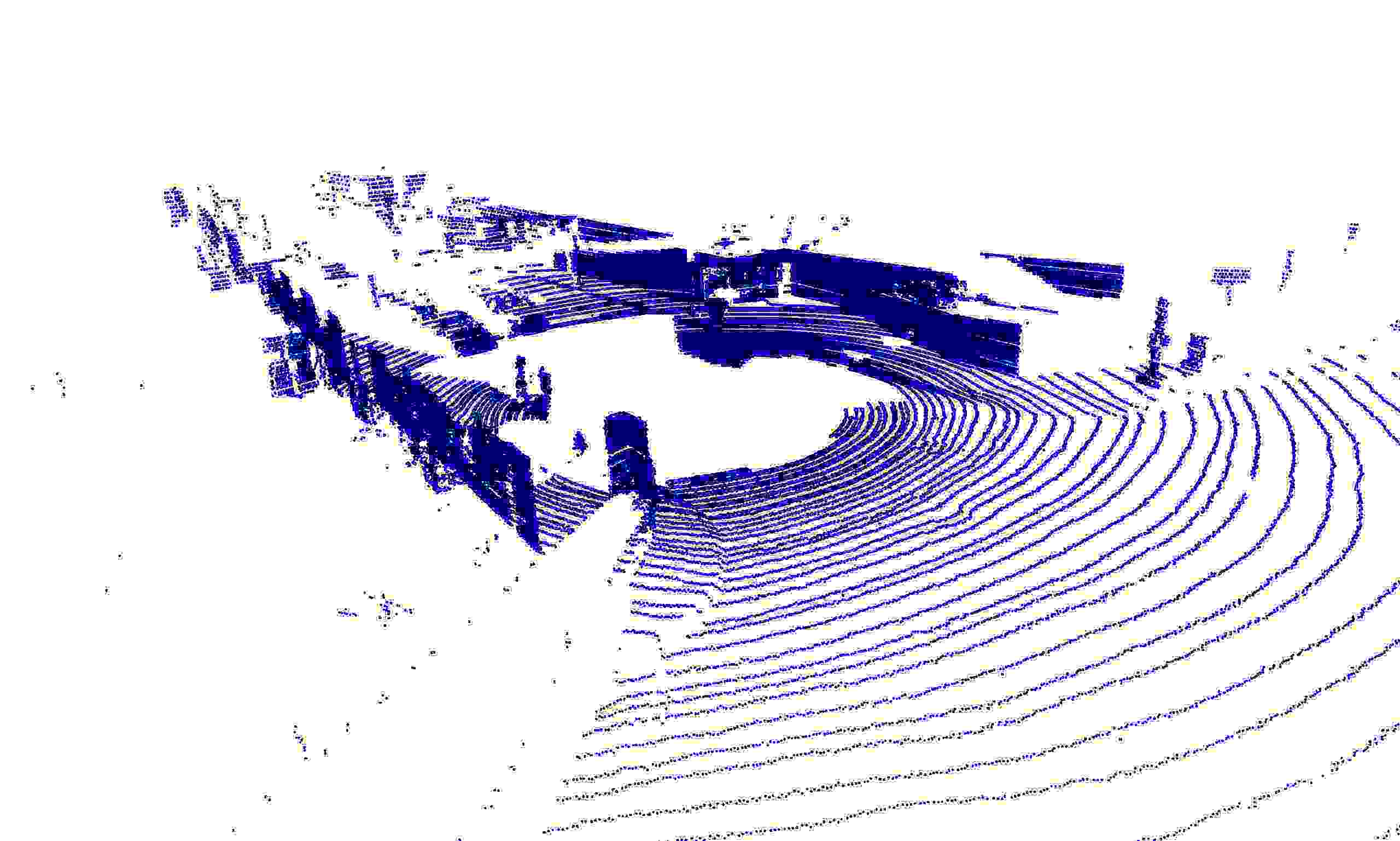}
    \put (0,55) {\colorbox{gray!30}{\scriptsize GT (NorthAmerica)}}
\end{overpic}
\begin{overpic}[width=0.32\textwidth]{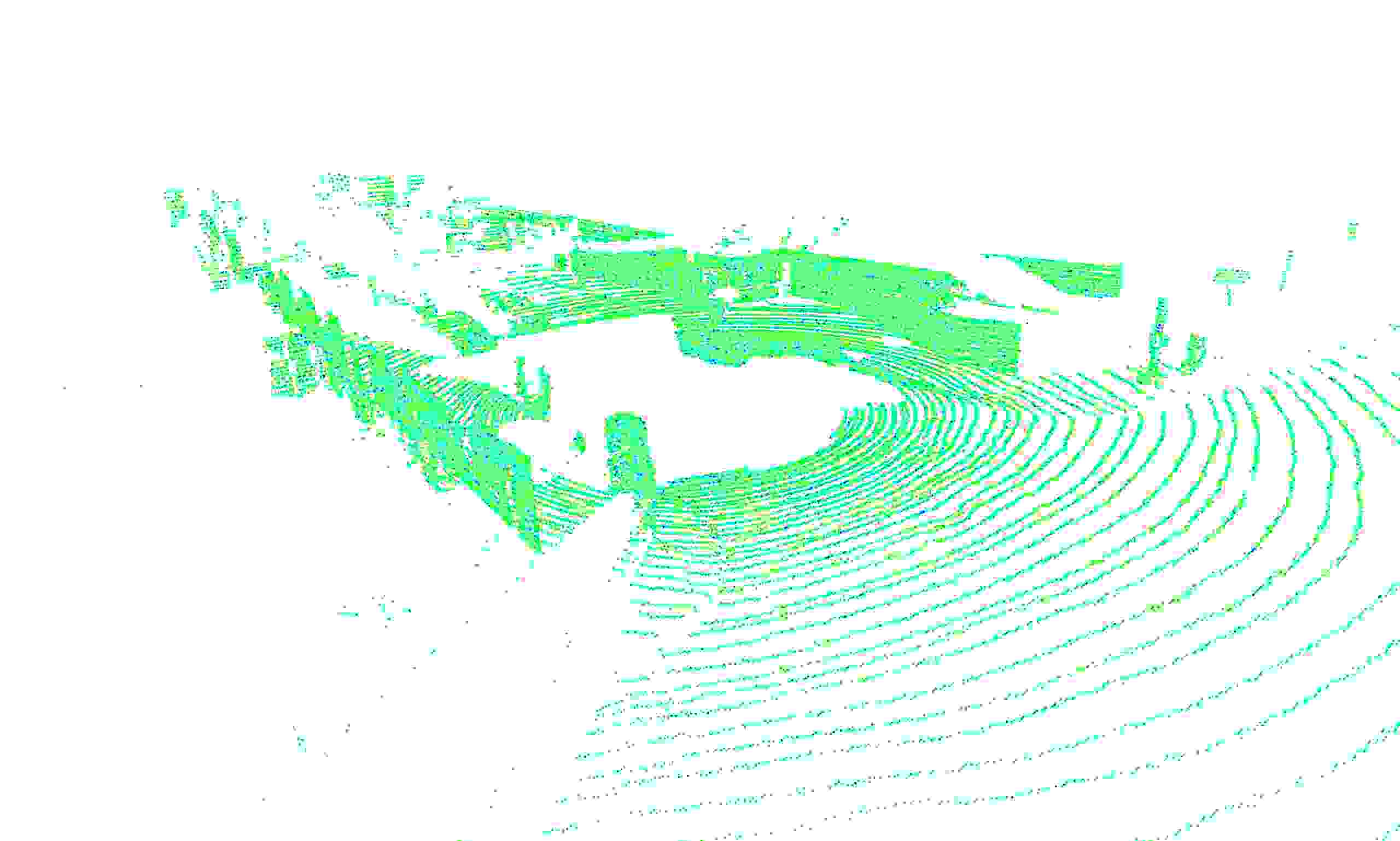}
    \put (0,55) {\colorbox{gray!30}{\scriptsize Ours: PSNR 71.94, Bitrate: 4.63}}
\end{overpic}
\begin{overpic}[width=0.32\textwidth]{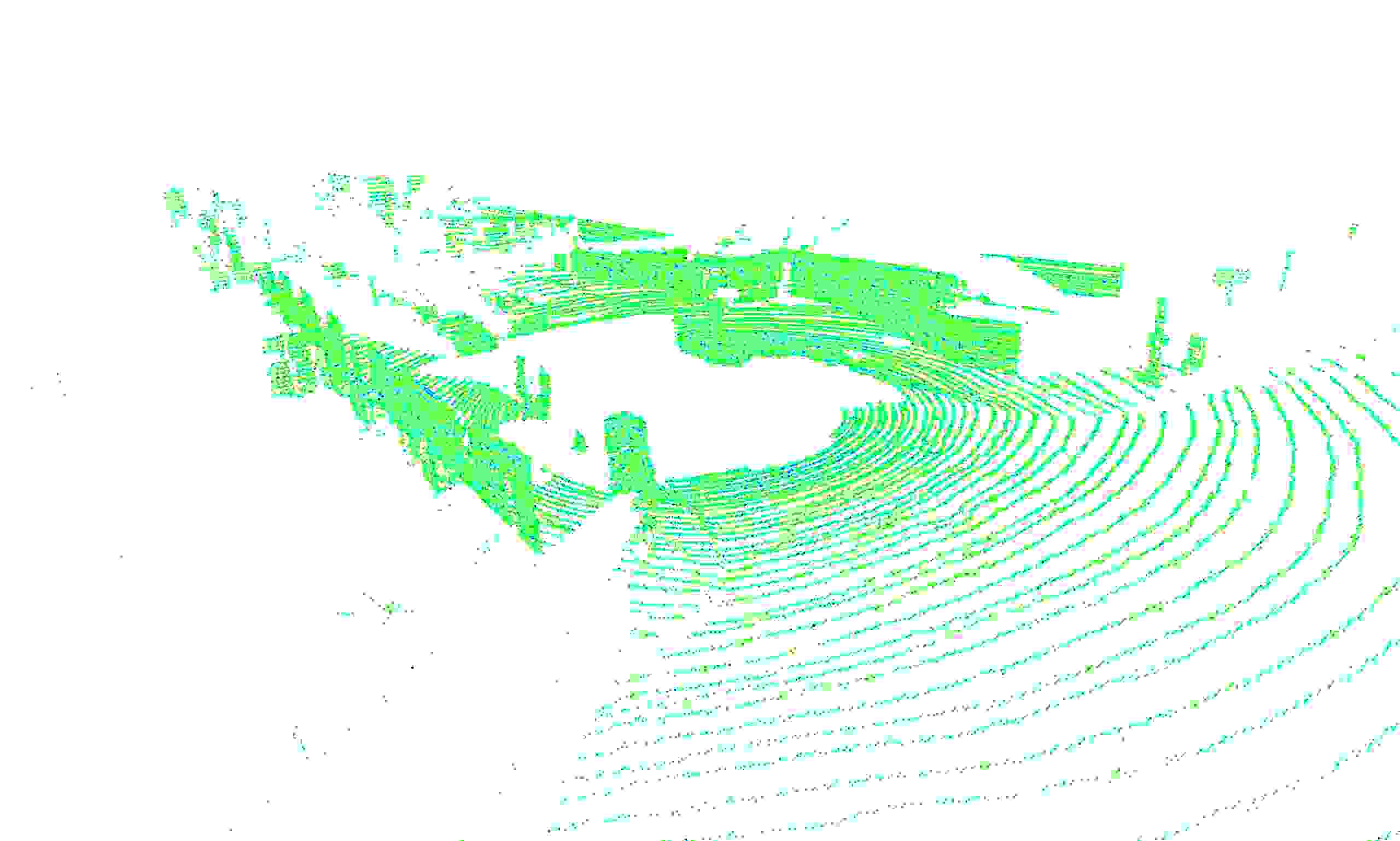}
    \put (0,55) {\colorbox{gray!30}{\scriptsize Draco: PSNR 71.03, Bitrate: 6.02}}
\end{overpic}
\end{center}

\begin{center}
\begin{overpic}[width=0.32\textwidth]{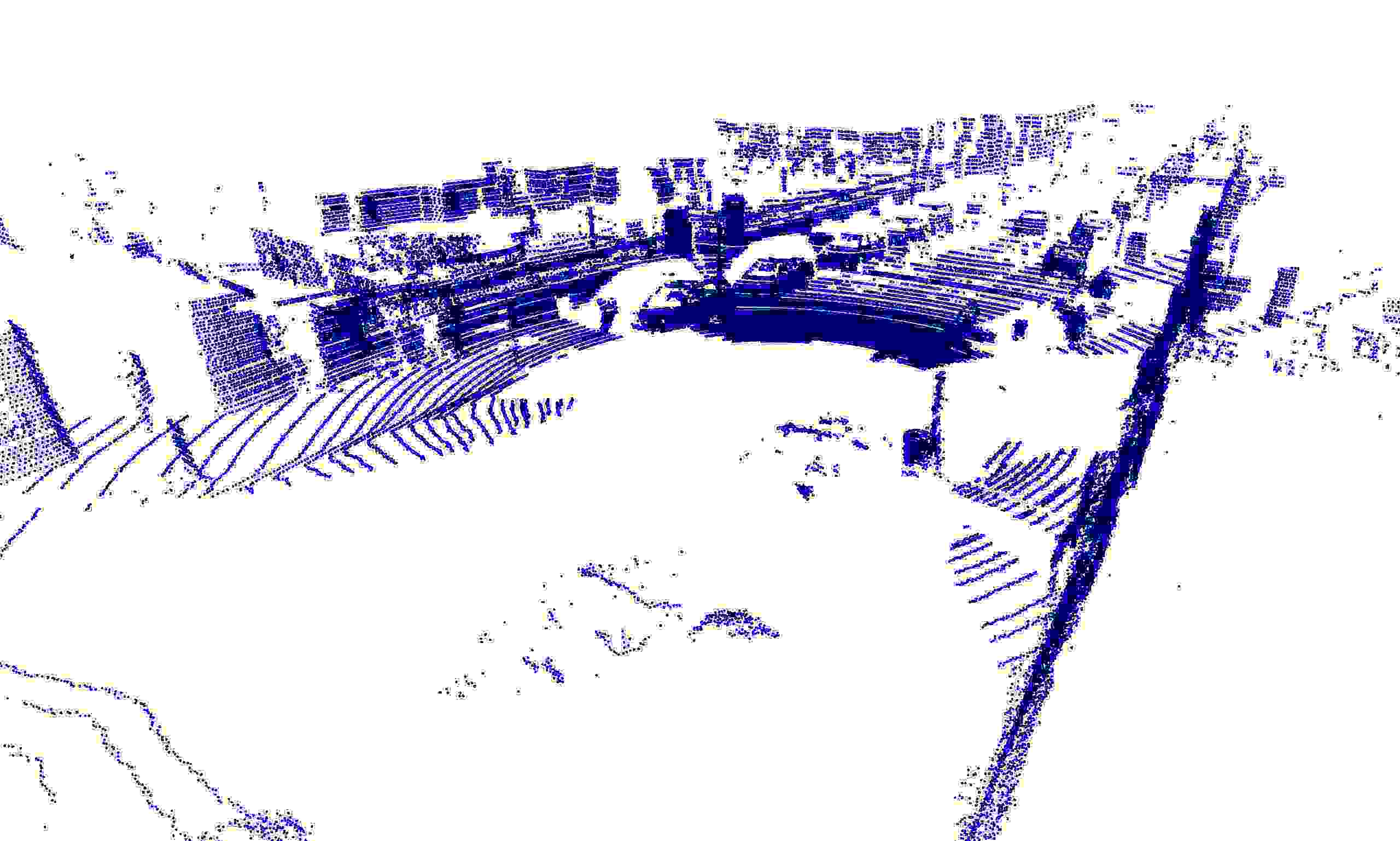}
    \put (0,55) {\colorbox{gray!30}{\scriptsize GT (NorthAmerica)}}
\end{overpic}
\begin{overpic}[width=0.32\textwidth]{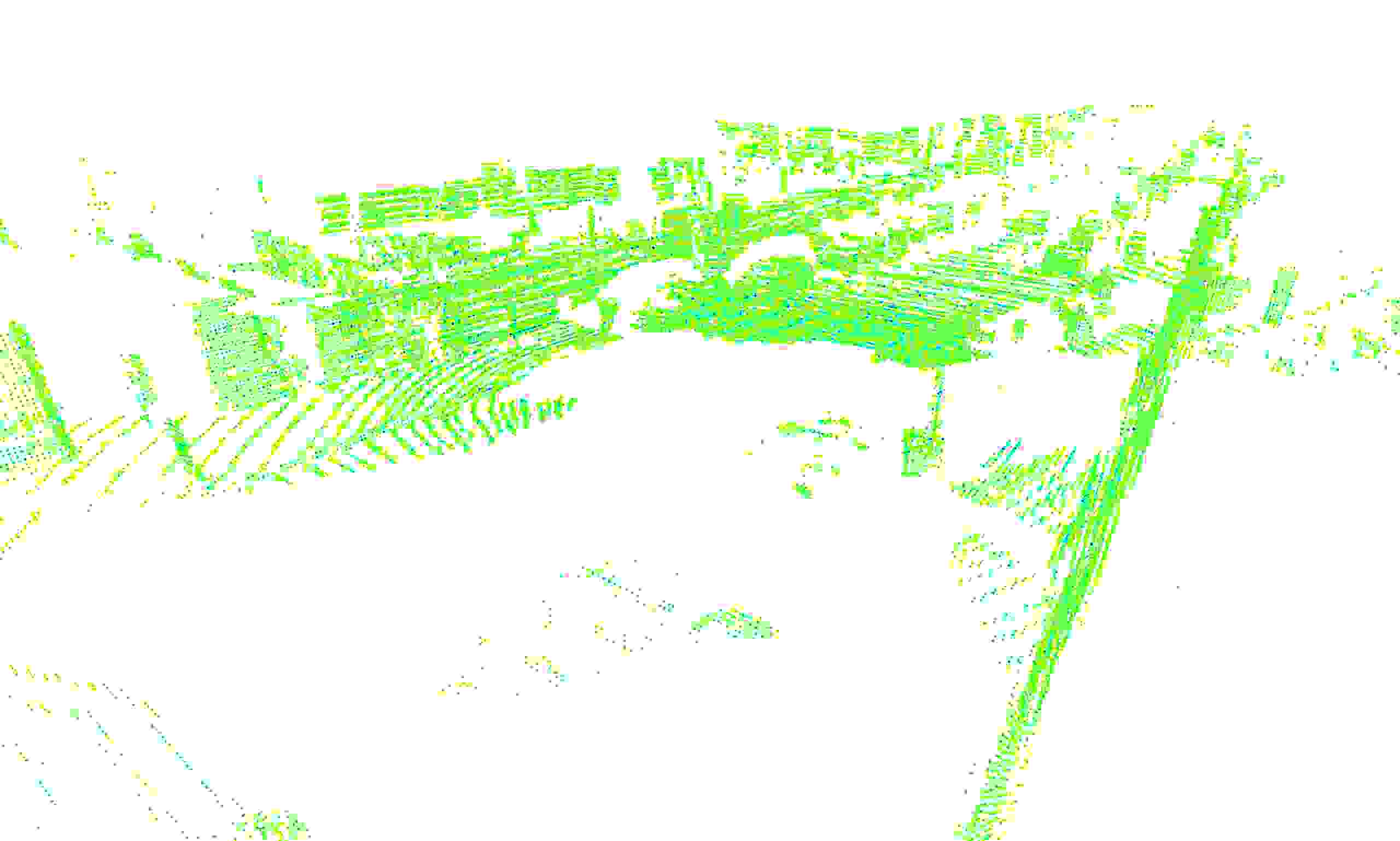}
    \put (0,55) {\colorbox{gray!30}{\scriptsize Ours: PSNR 46.49, Bitrate: 2.66}}
\end{overpic}
\begin{overpic}[width=0.32\textwidth]{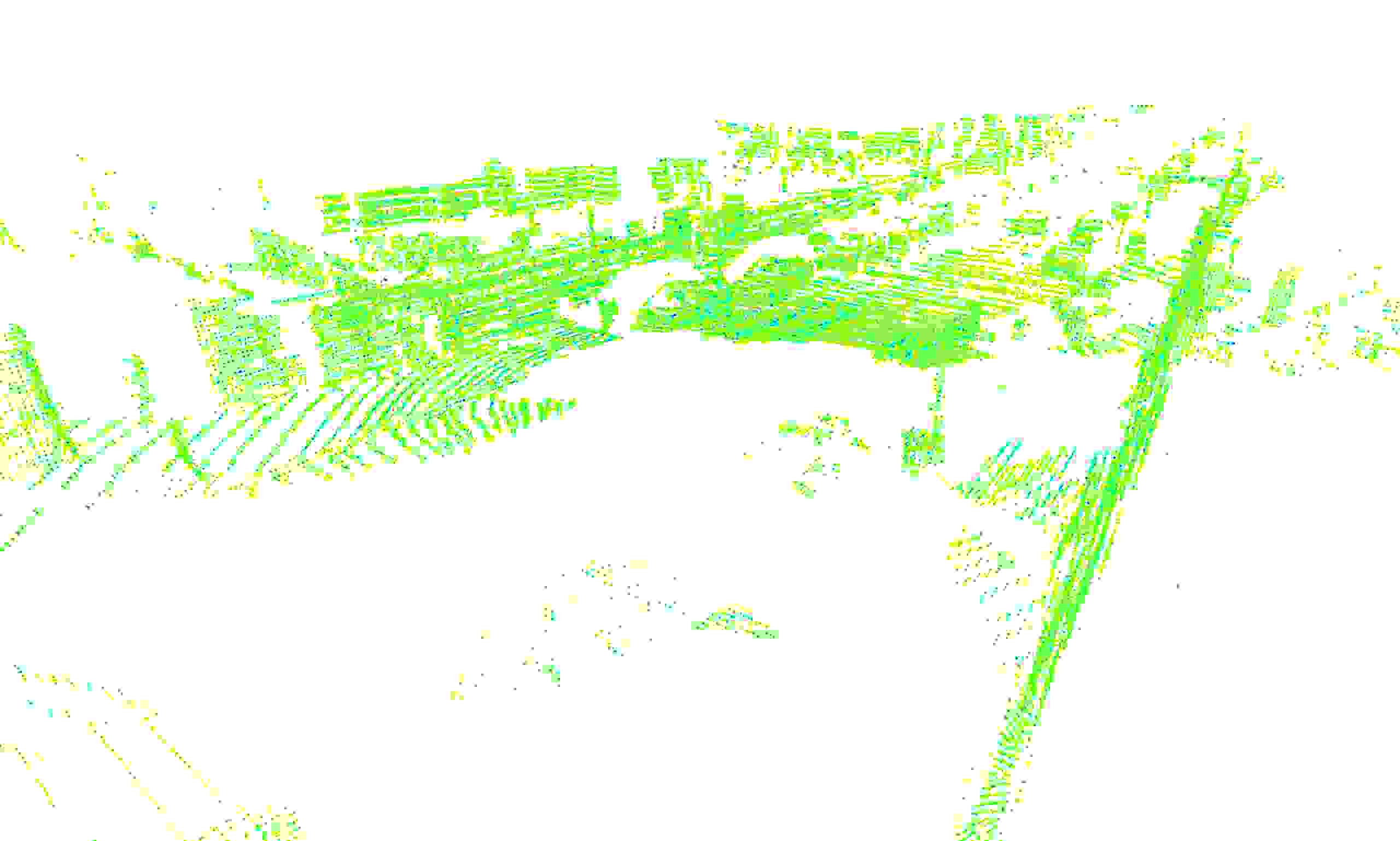}
    \put (0,55) {\colorbox{gray!30}{\scriptsize Draco: PSNR 45.49, Bitrate: 3.95}}
\end{overpic}
\end{center}

\begin{center}
\begin{overpic}[width=0.32\textwidth]{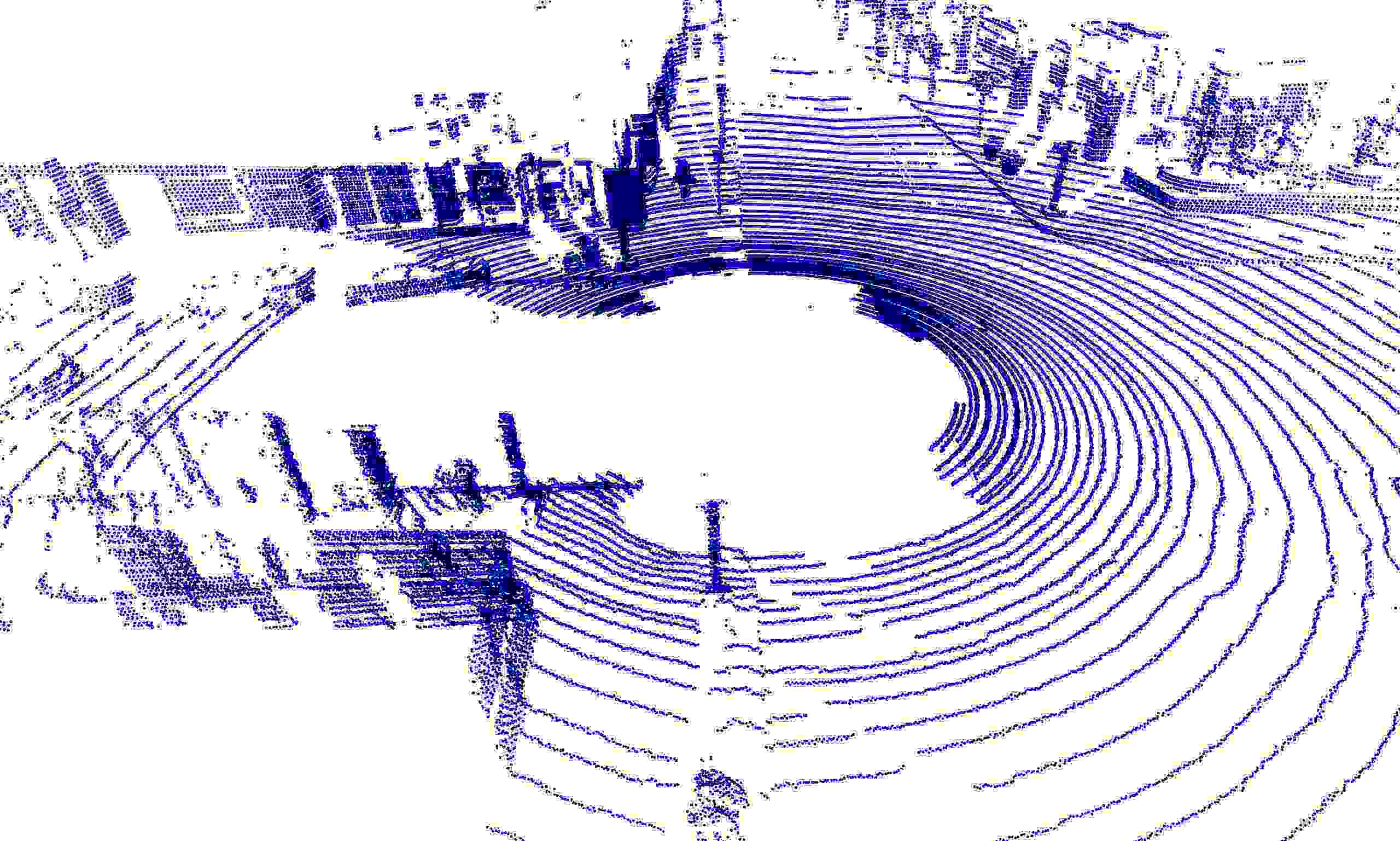}
    \put (0,55) {\colorbox{gray!30}{\scriptsize GT (NorthAmerica)}}
\end{overpic}
\begin{overpic}[width=0.32\textwidth]{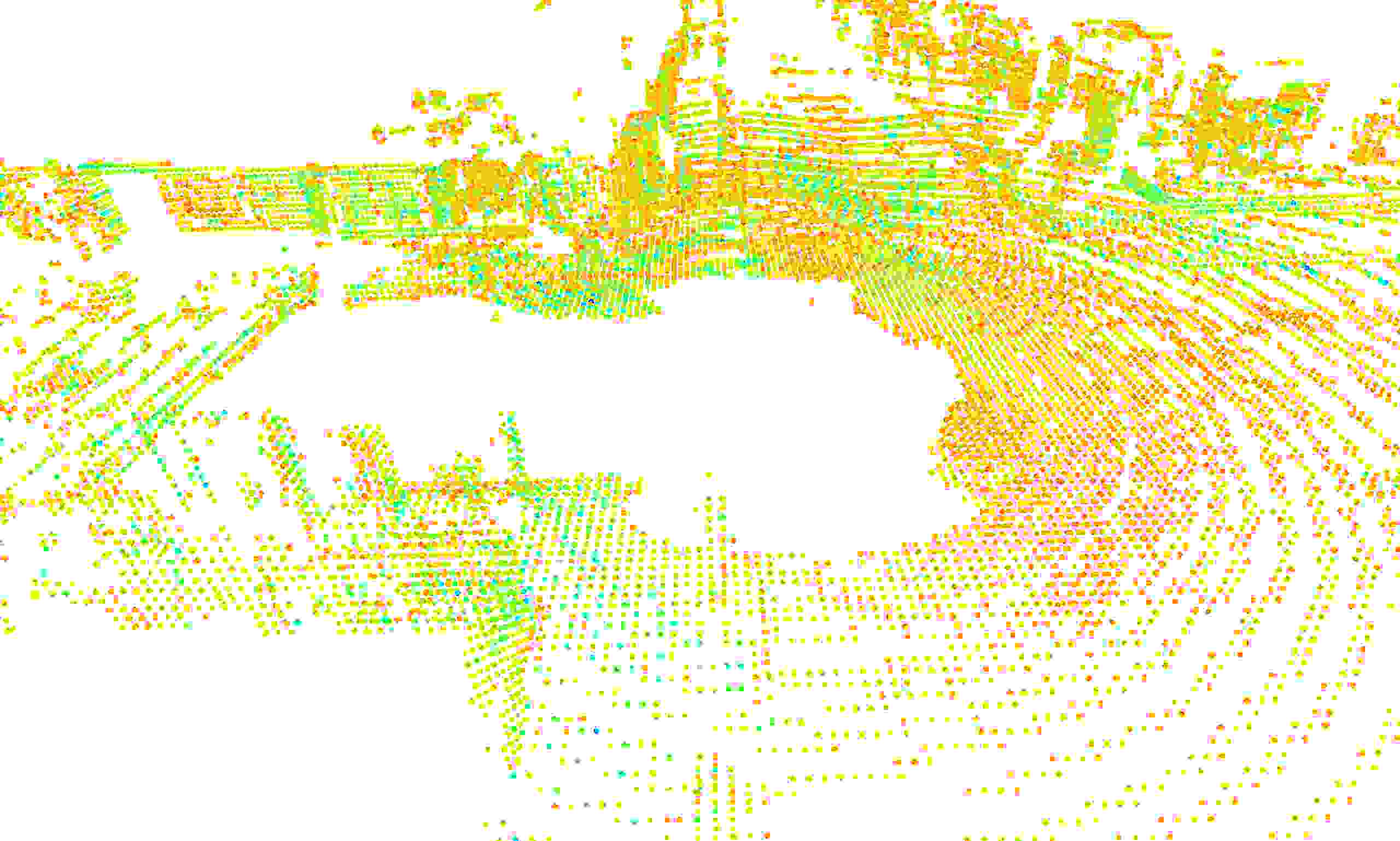}
    \put (0,55) {\colorbox{gray!30}{\scriptsize Ours: PSNR 48.61, Bitrate: 1.53}}
\end{overpic}
\begin{overpic}[width=0.32\textwidth]{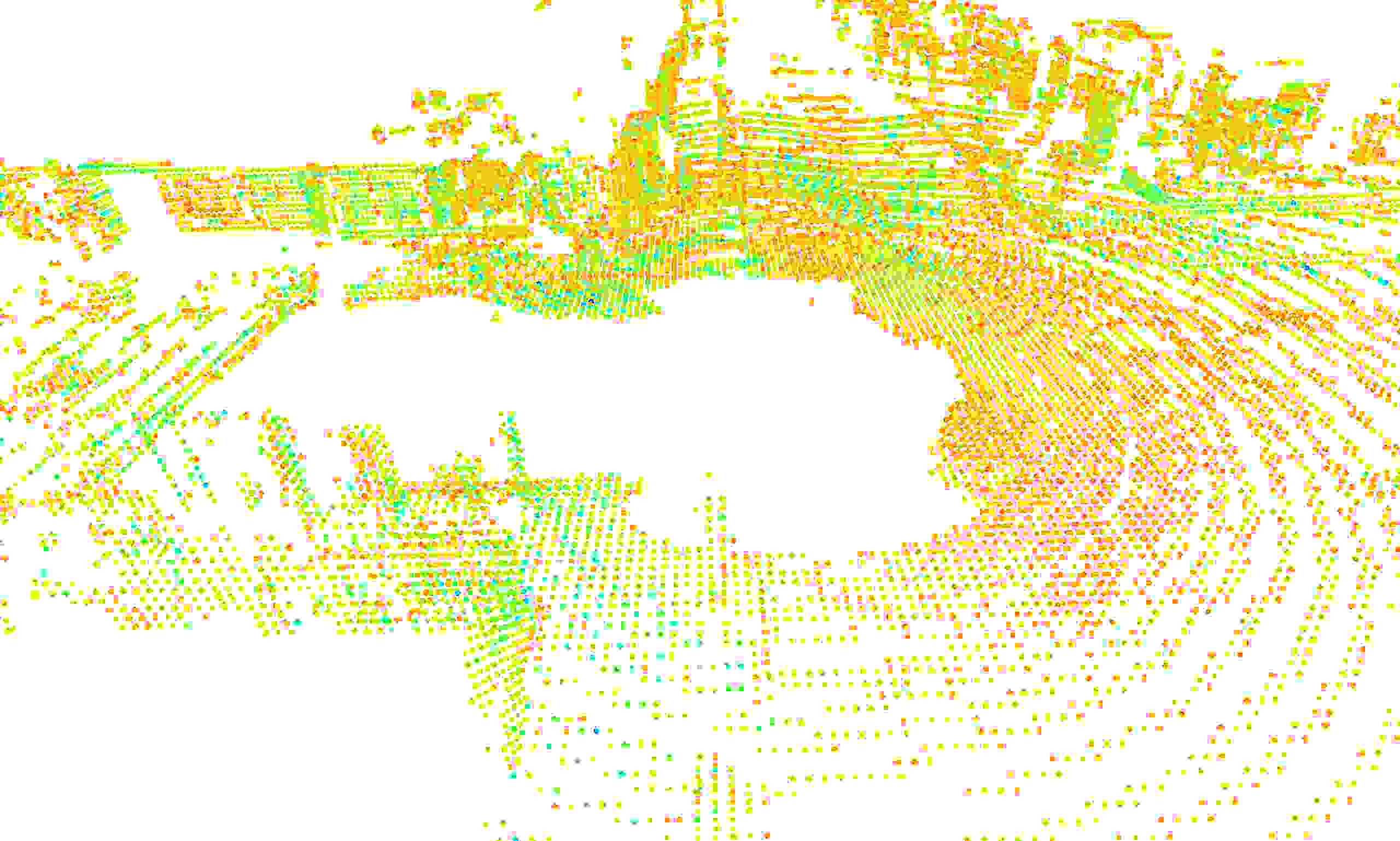}
    \put (0,55) {\colorbox{gray!30}{\scriptsize Draco: PSNR 47.28, Bitrate: 2.73}}
\end{overpic}
\end{center}

\vspace{-0.25in}
\begin{center}
\includegraphics[width=0.5\textwidth]{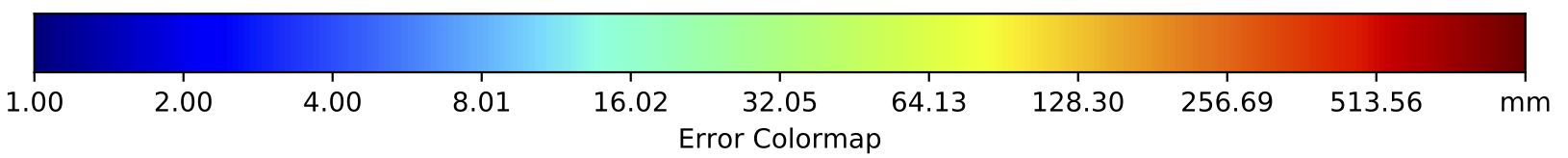}
\end{center}

\vspace{-0.25in}
\caption{Qualitative results of reconstruction quality for NorthAmerica.}
\label{fig:recon_na}
\end{figure*}

\begin{figure*}[h]
\vspace{-0.15in}
\begin{center}
\begin{overpic}[width=0.32\textwidth]{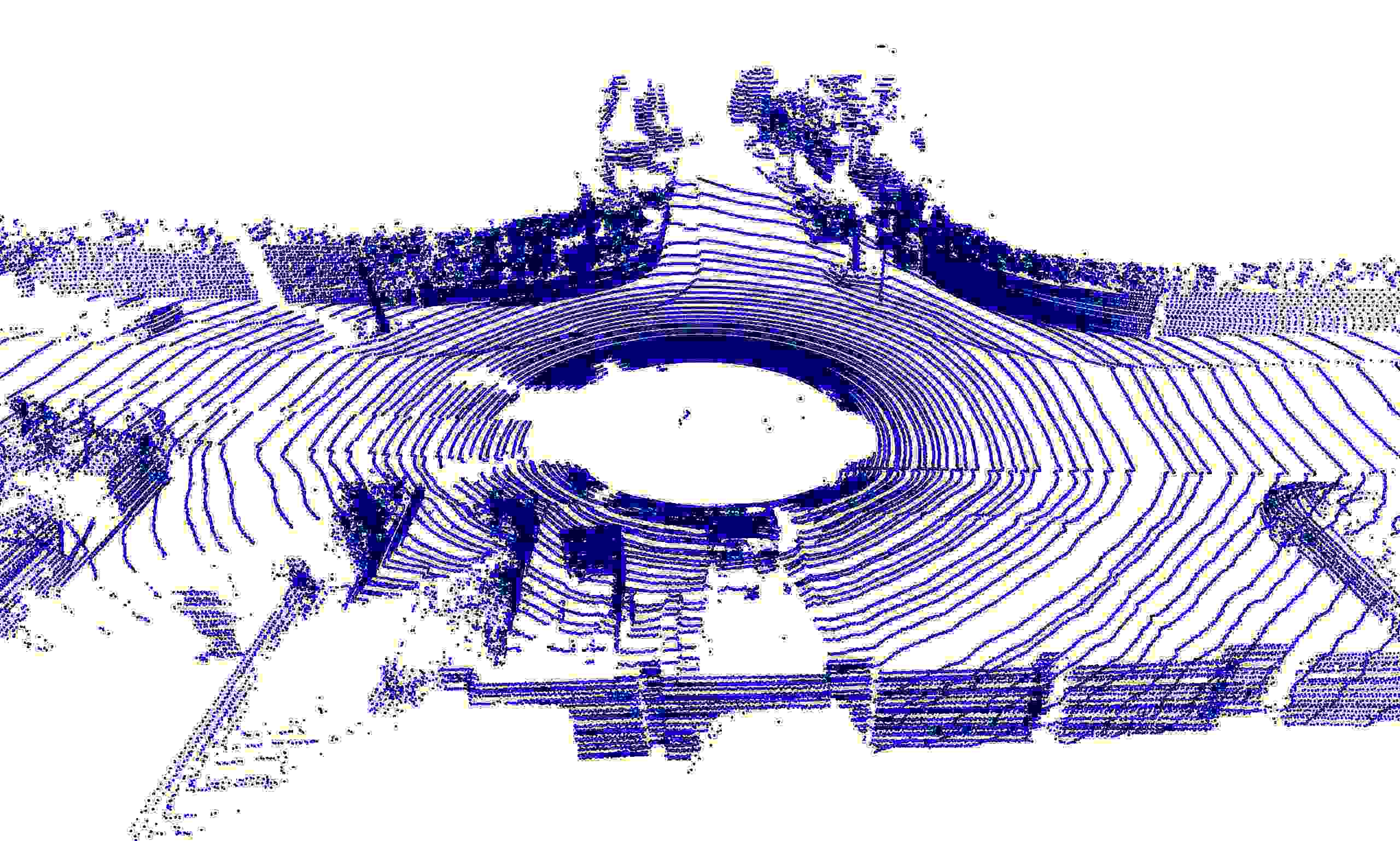}
    \put (0,55) {\colorbox{gray!30}{\scriptsize GT (KITTI)}}
\end{overpic}
\begin{overpic}[width=0.32\textwidth]{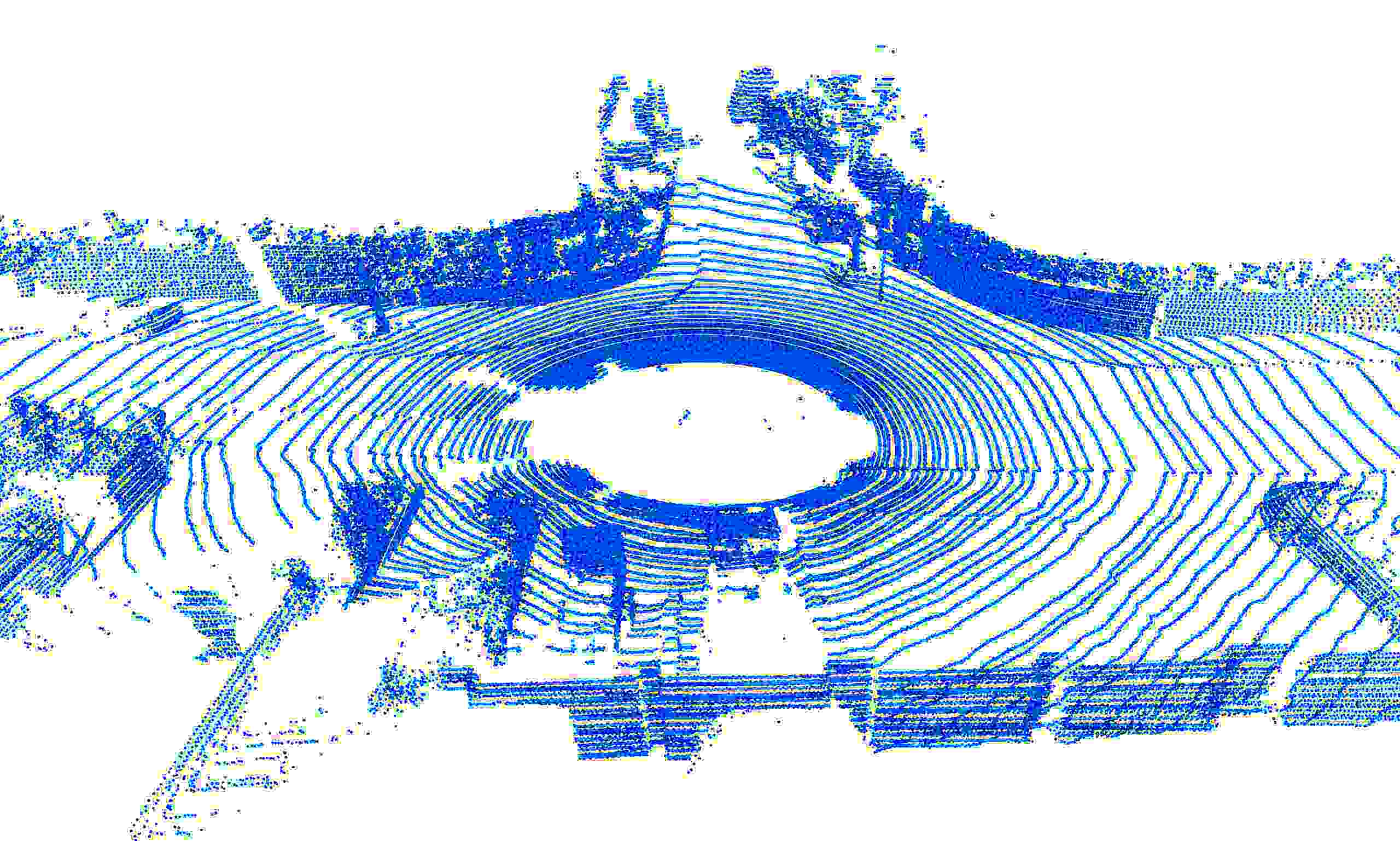}
    \put (0,55) {\colorbox{gray!30}{\scriptsize Ours: PSNR 65.29, Bitrate: 12.78}}
\end{overpic}
\begin{overpic}[width=0.32\textwidth]{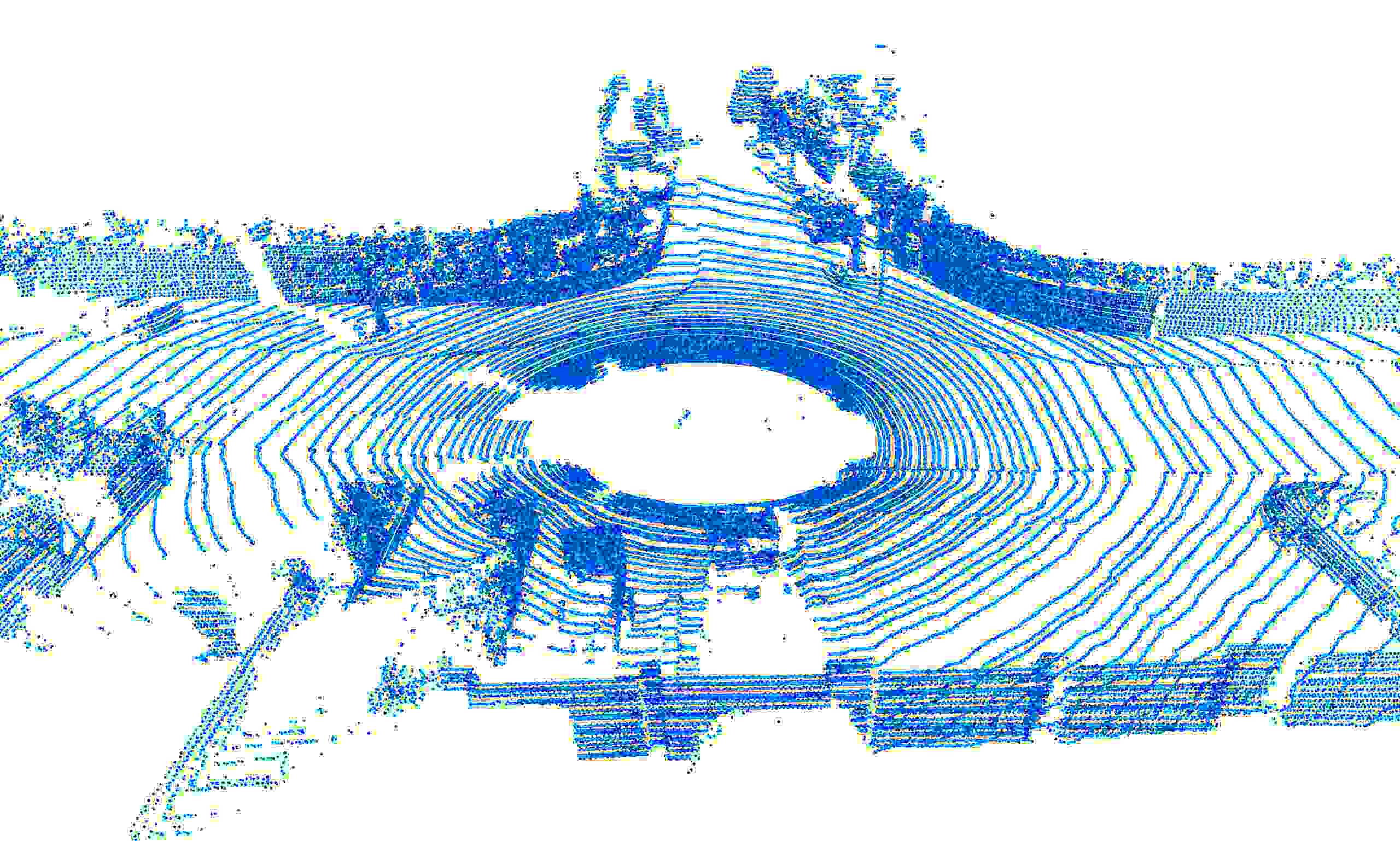}
    \put (0,55) {\colorbox{gray!30}{\scriptsize Draco: PSNR 62.46, Bitrate: 12.90}}
\end{overpic}
\end{center}

\begin{center}
\begin{overpic}[width=0.32\textwidth]{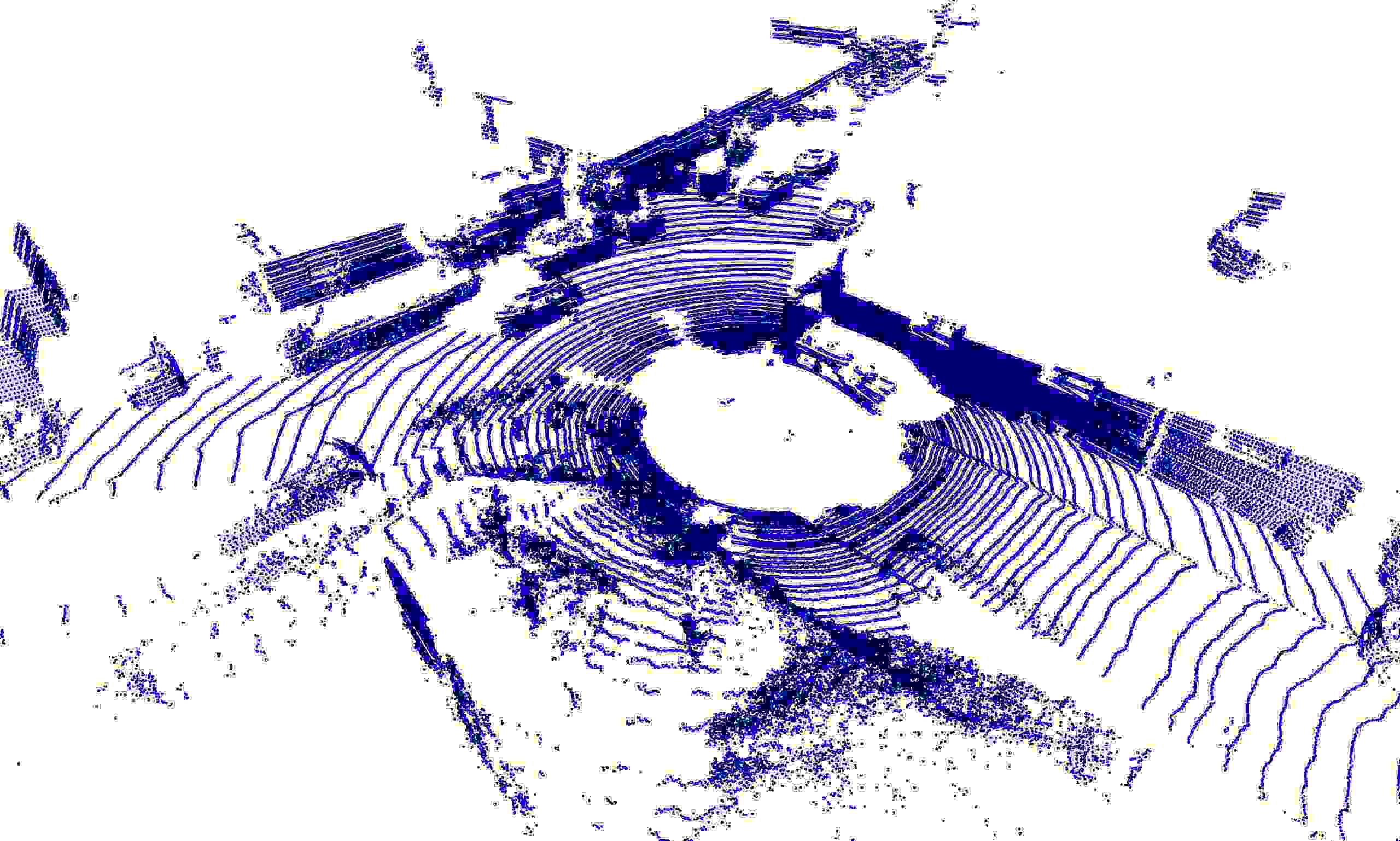}
    \put (0,55) {\colorbox{gray!30}{\scriptsize GT (KITTI)}}
\end{overpic}
\begin{overpic}[width=0.32\textwidth]{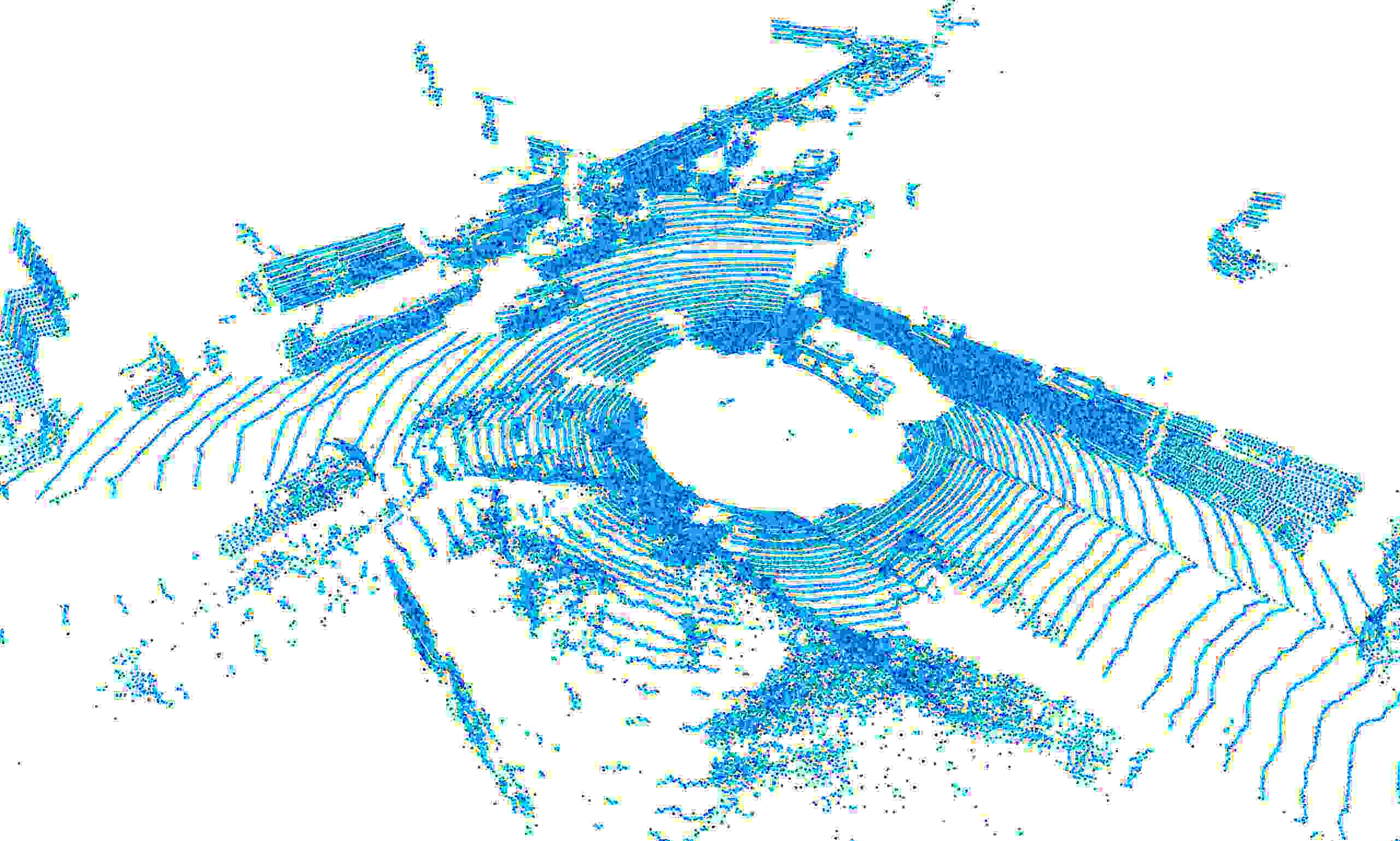}
    \put (0,55) {\colorbox{gray!30}{\scriptsize Ours: PSNR 69.99, Bitrate: 9.37}}
\end{overpic}
\begin{overpic}[width=0.32\textwidth]{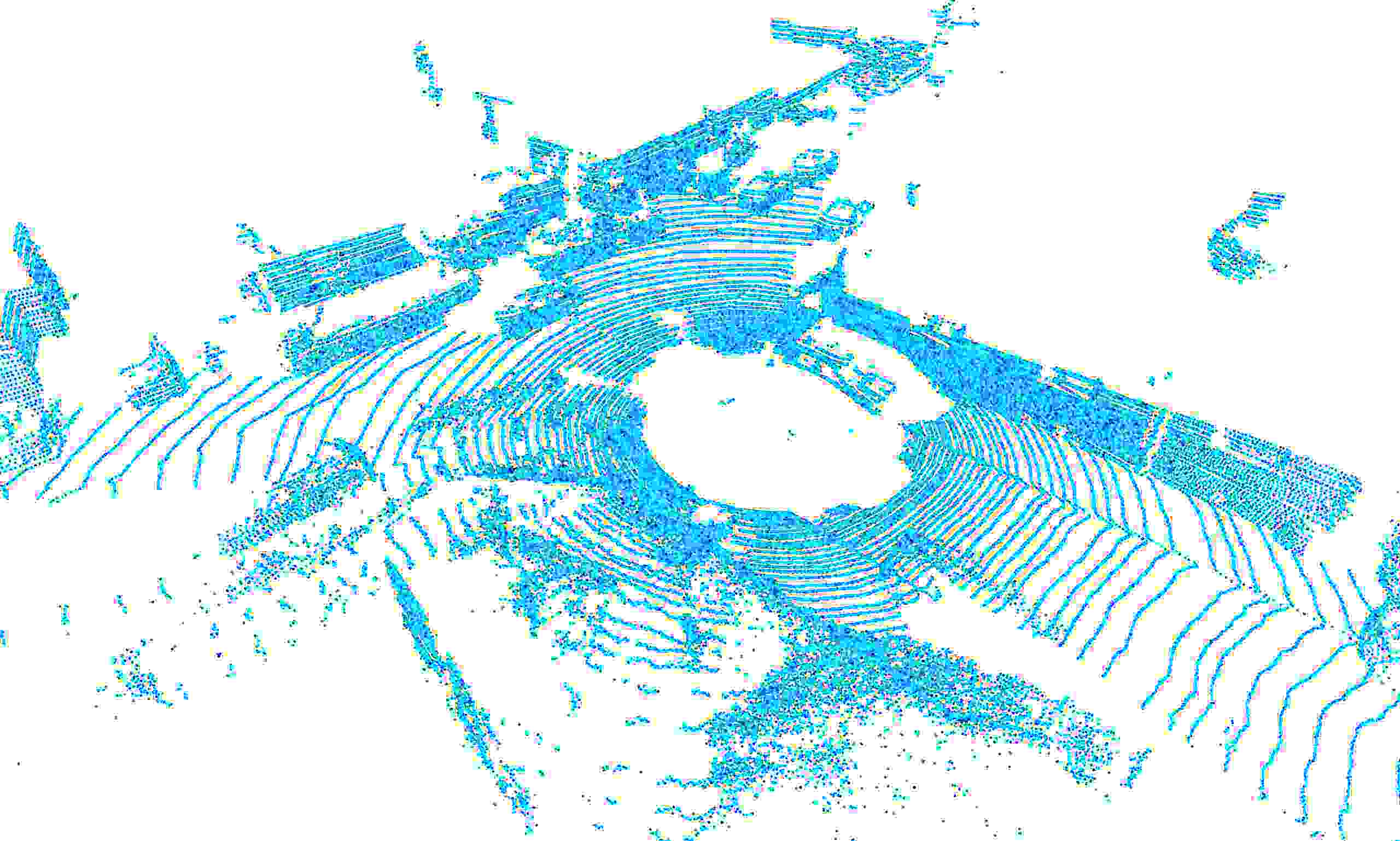}
    \put (0,55) {\colorbox{gray!30}{\scriptsize Draco: PSNR 66.64, Bitrate: 9.40}}
\end{overpic}
\end{center}

\begin{center}
\begin{overpic}[width=0.32\textwidth]{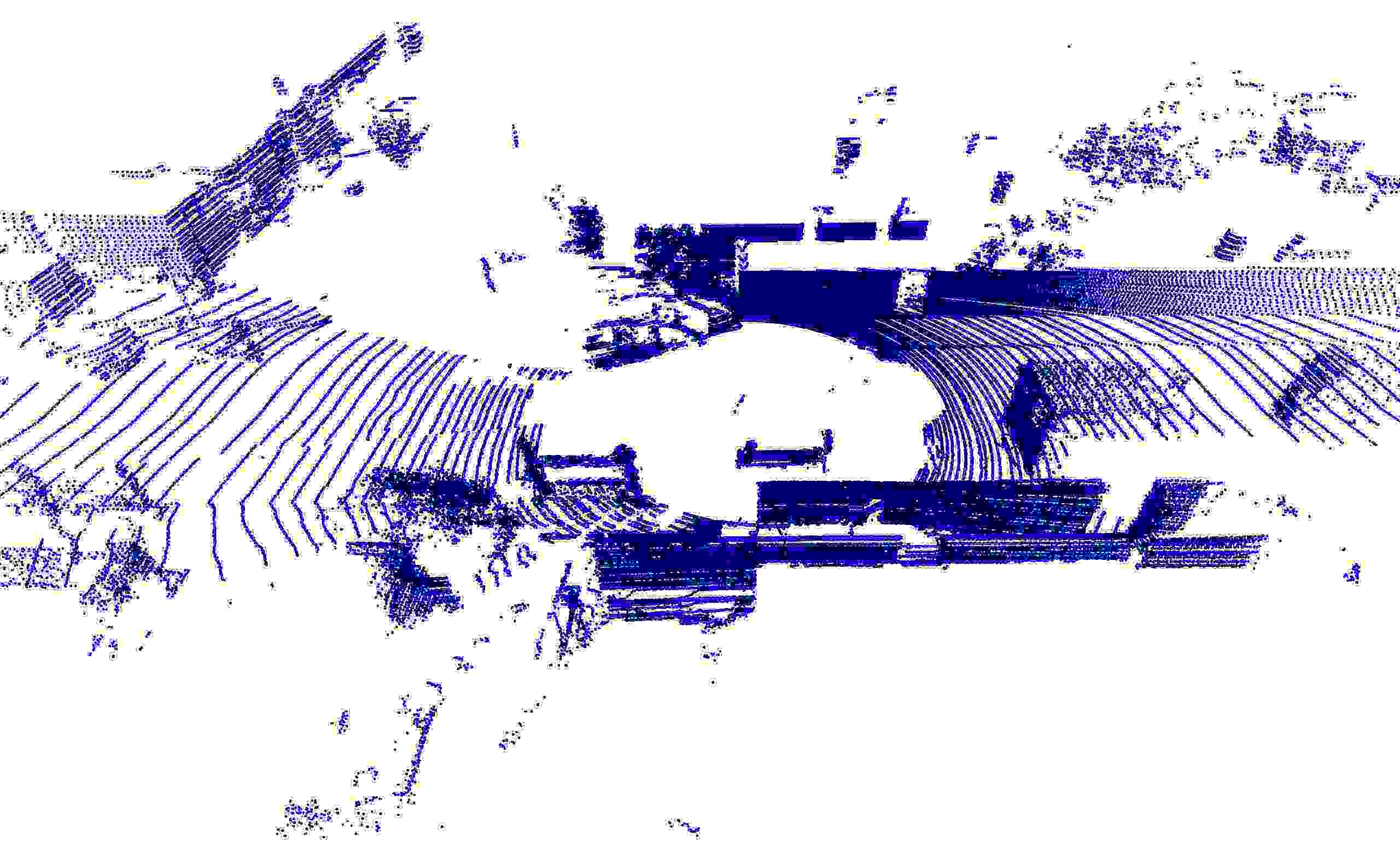}
    \put (0,55) {\colorbox{gray!30}{\scriptsize GT (KITTI)}}
\end{overpic}
\begin{overpic}[width=0.32\textwidth]{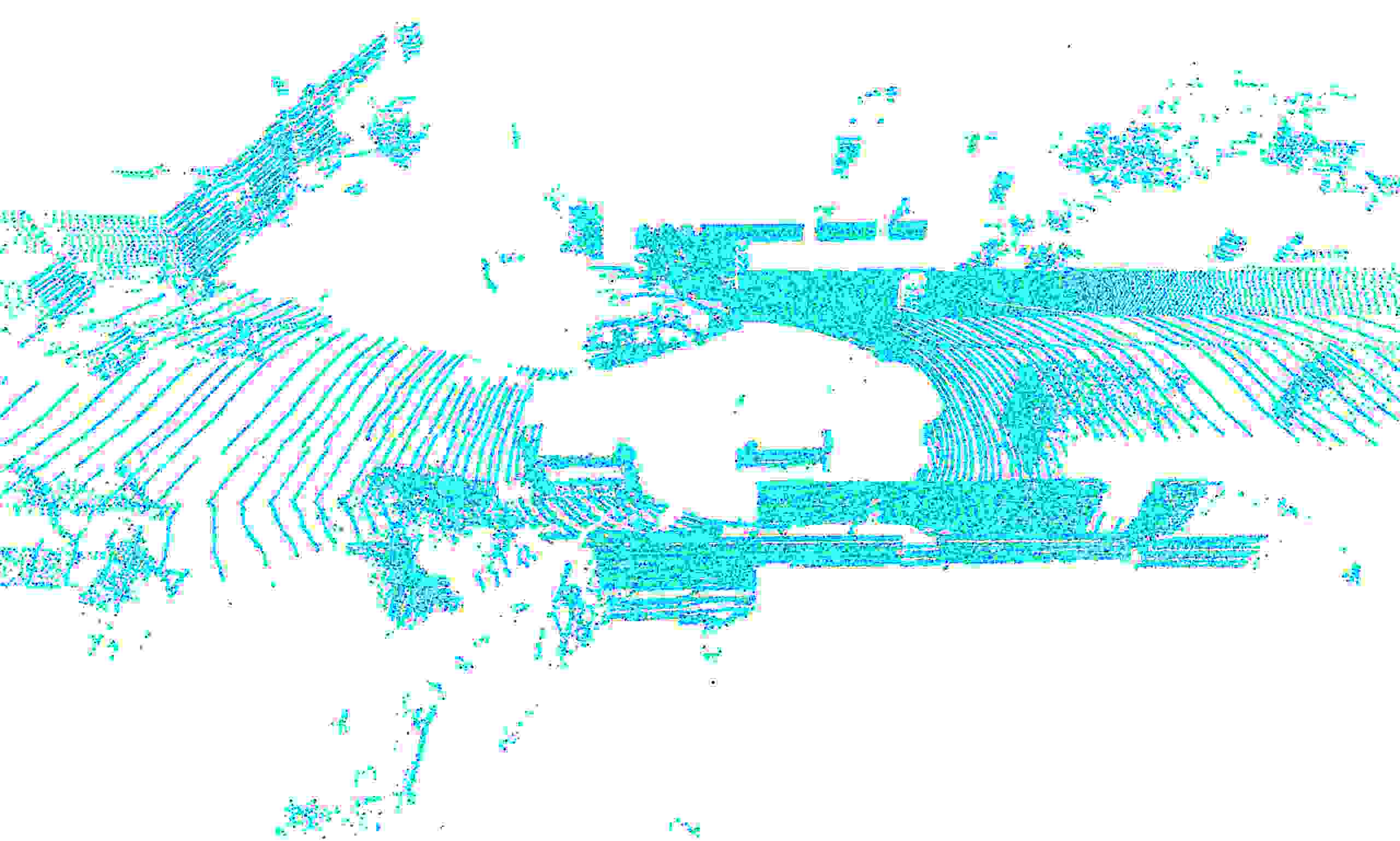}
    \put (0,55) {\colorbox{gray!30}{\scriptsize Ours: PSNR 57.59, Bitrate: 5.23}}
\end{overpic}
\begin{overpic}[width=0.32\textwidth]{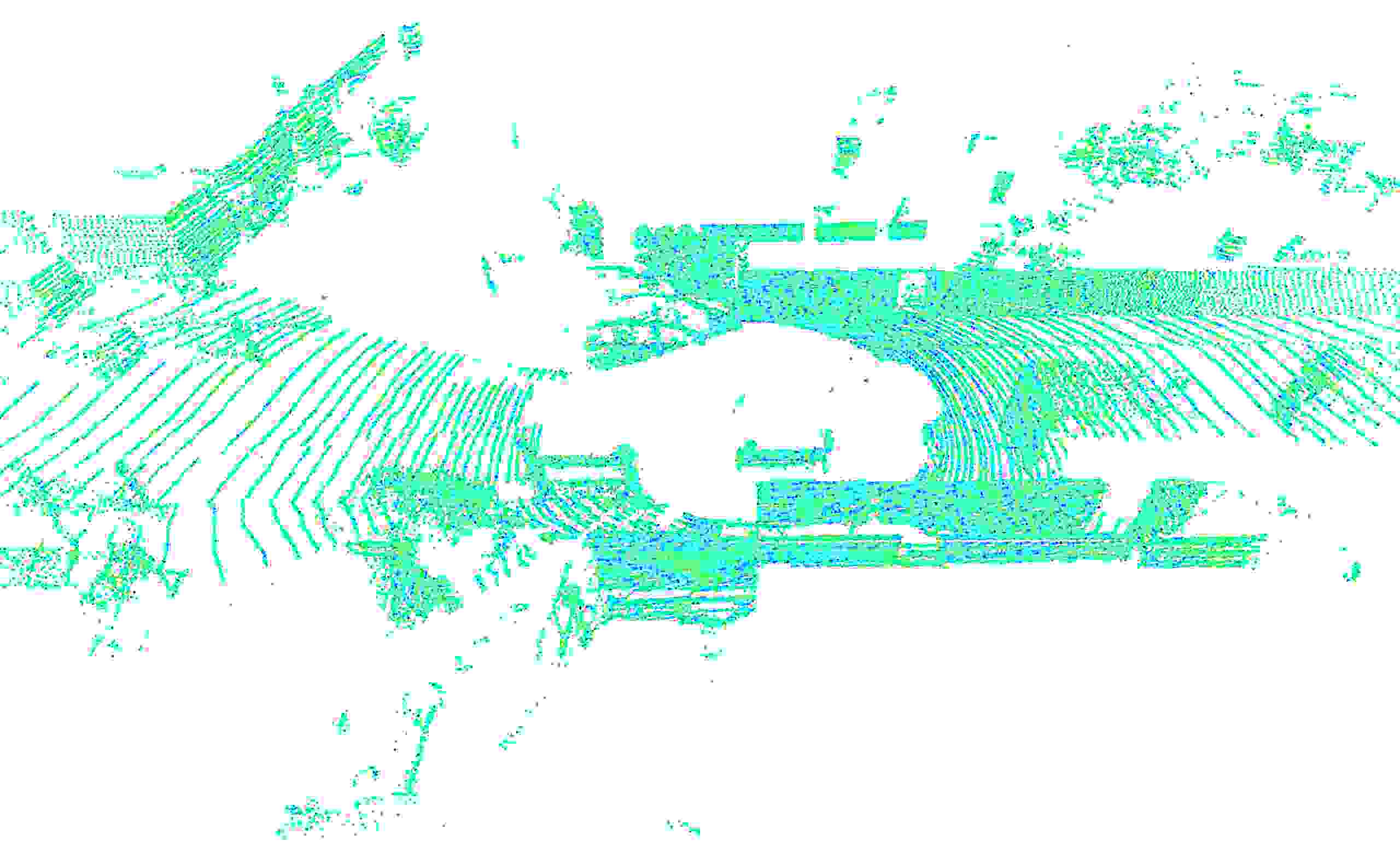}
    \put (0,55) {\colorbox{gray!30}{\scriptsize Draco: PSNR 53.42, Bitrate: 5.29}}
\end{overpic}
\end{center}

\begin{center}
\begin{overpic}[width=0.32\textwidth]{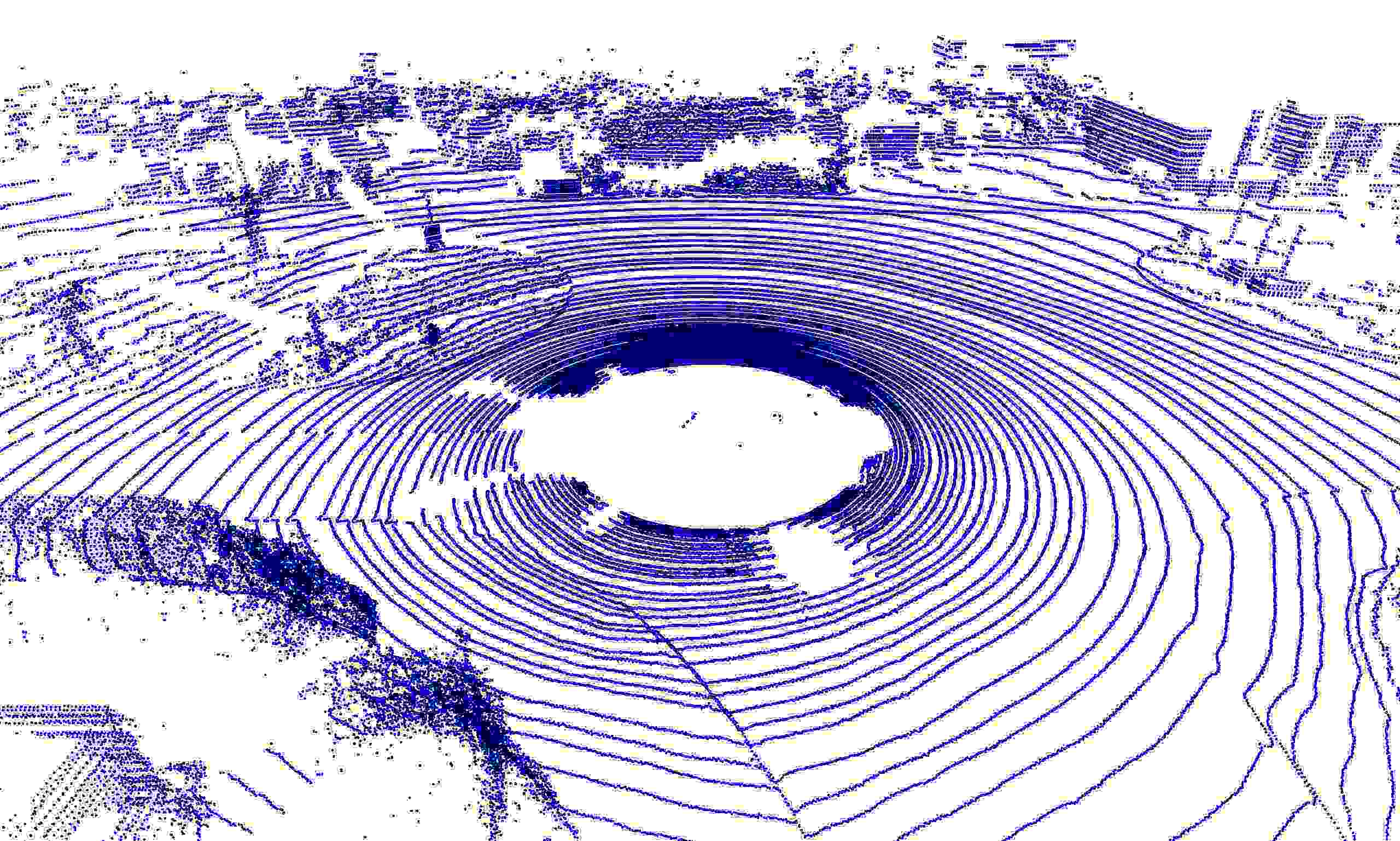}
    \put (0,55) {\colorbox{gray!30}{\scriptsize GT (KITTI)}}
\end{overpic}
\begin{overpic}[width=0.32\textwidth]{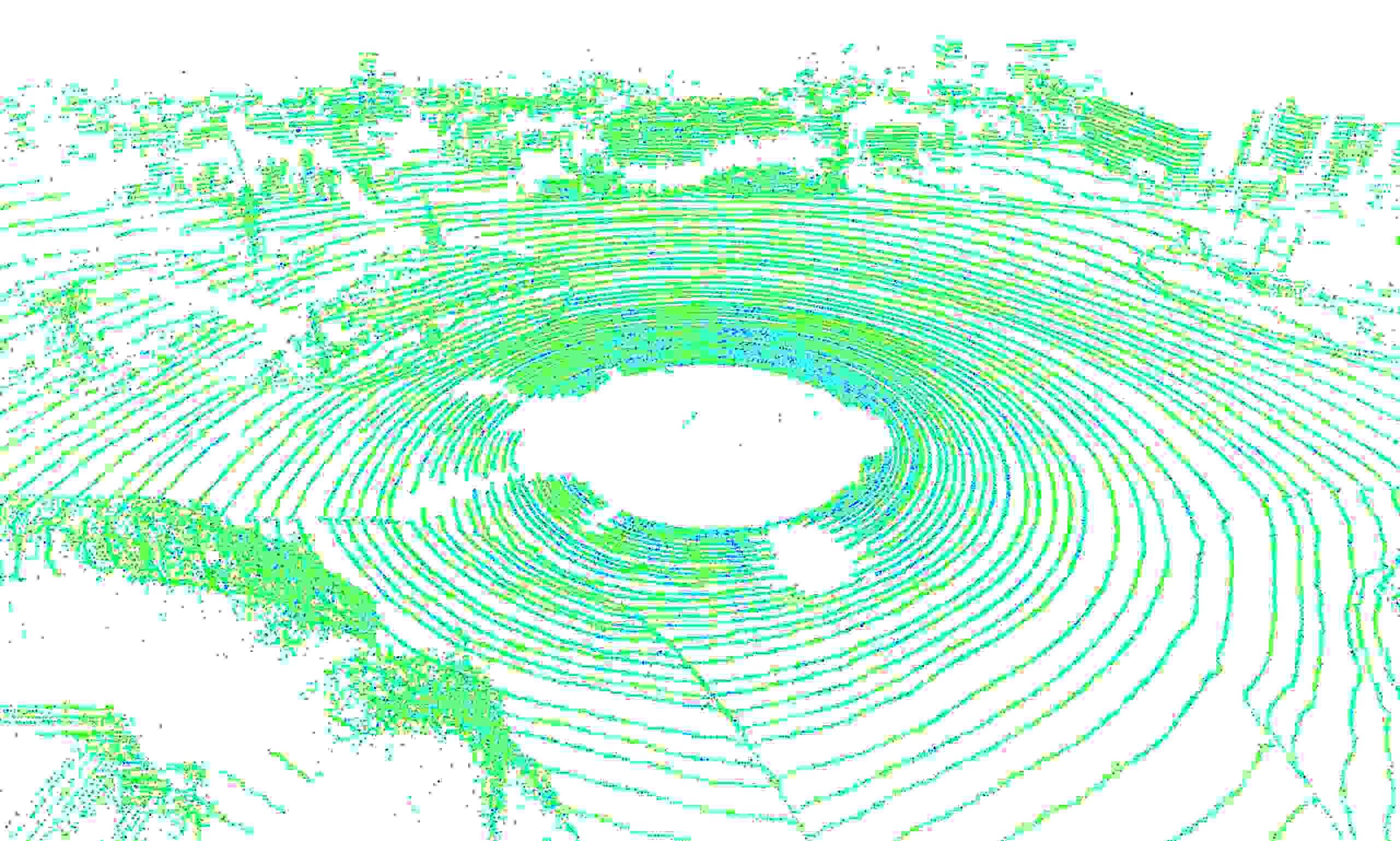}
    \put (0,55) {\colorbox{gray!30}{\scriptsize Ours: PSNR 52.41, Bitrate: 4.82}}
\end{overpic}
\begin{overpic}[width=0.32\textwidth]{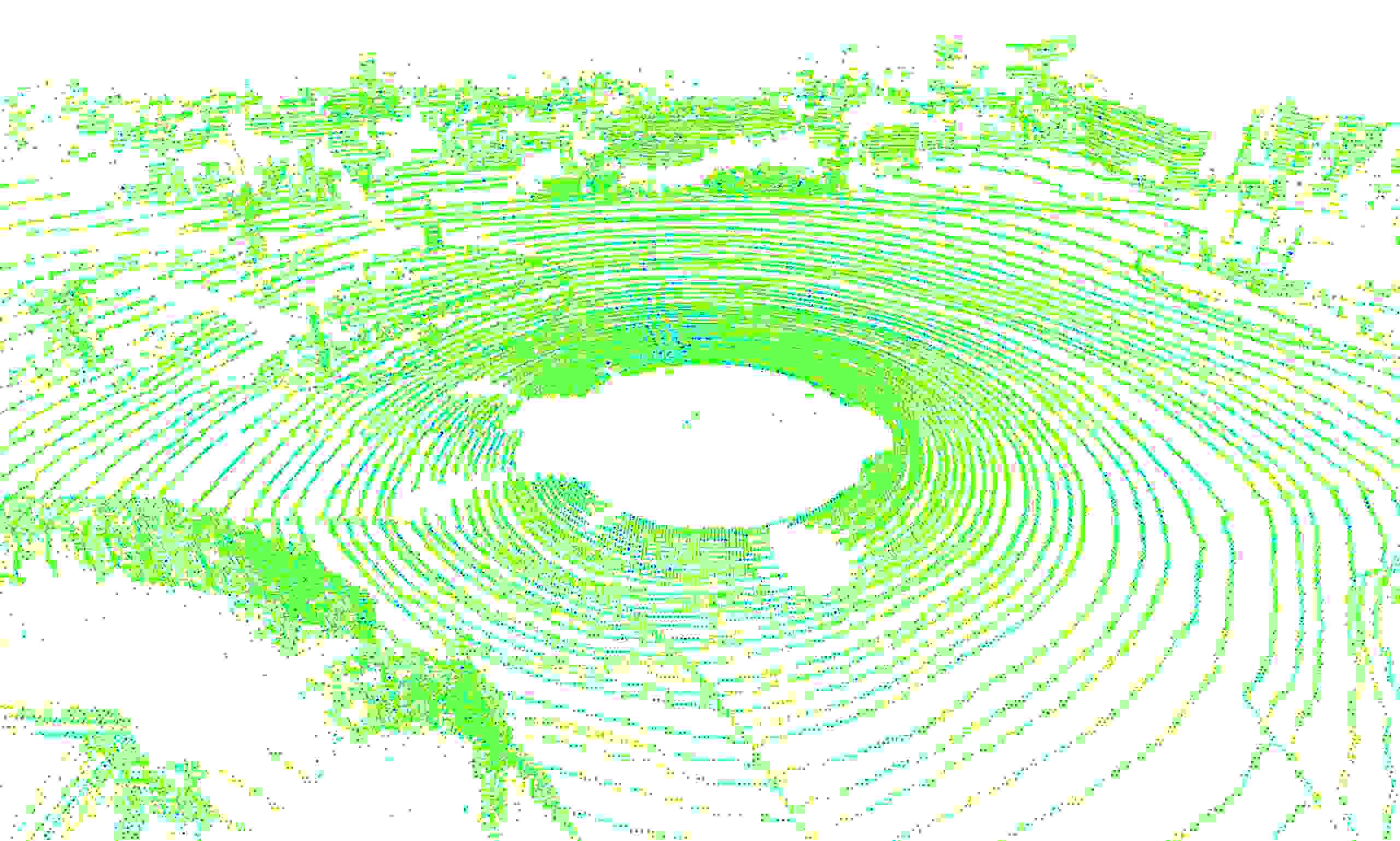}
    \put (0,55) {\colorbox{gray!30}{\scriptsize Draco: PSNR 48.00, Bitrate: 4.88}}
\end{overpic}
\end{center}

\begin{center}
\begin{overpic}[width=0.32\textwidth]{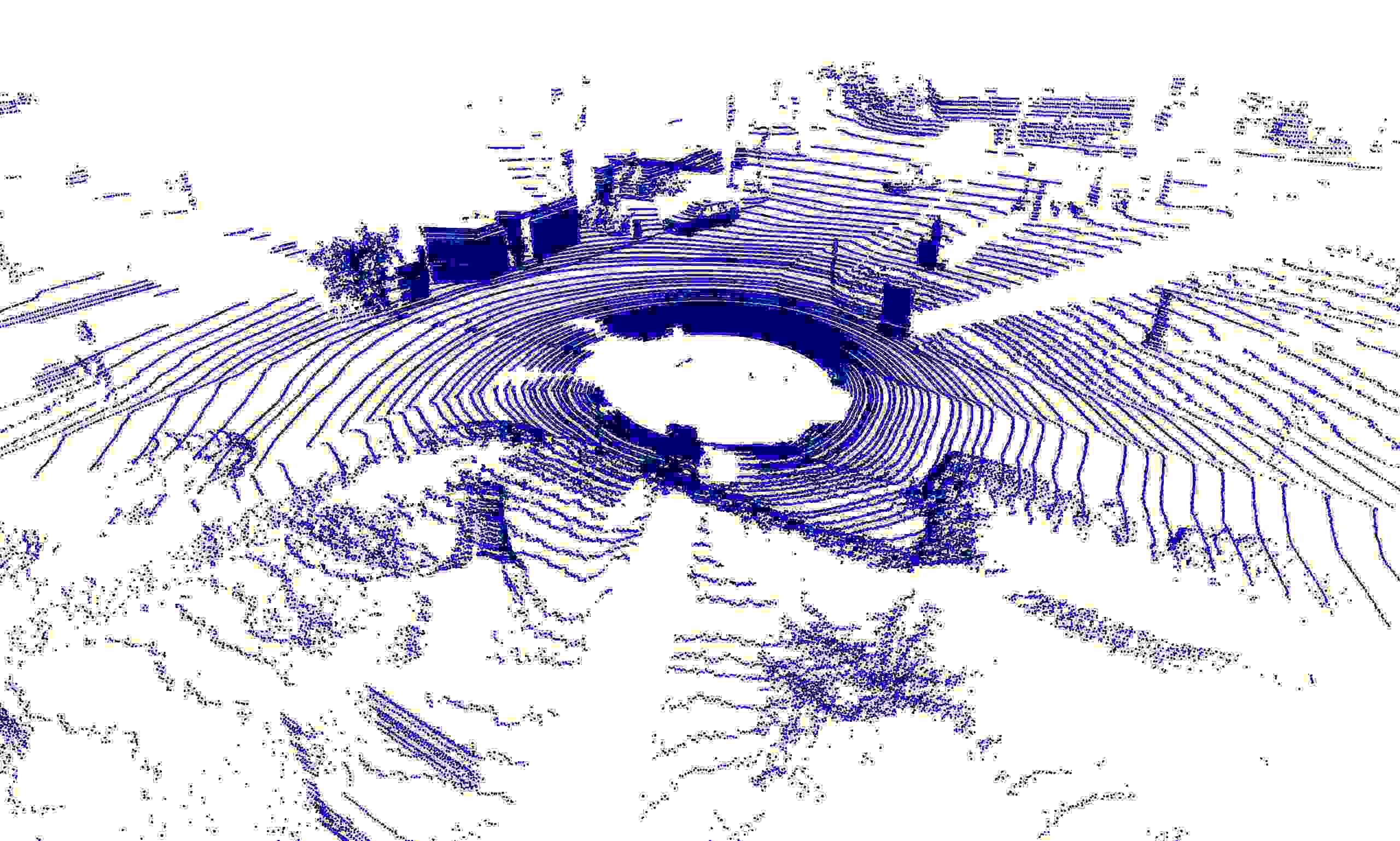}
    \put (0,55) {\colorbox{gray!30}{\scriptsize GT (KITTI)}}
\end{overpic}
\begin{overpic}[width=0.32\textwidth]{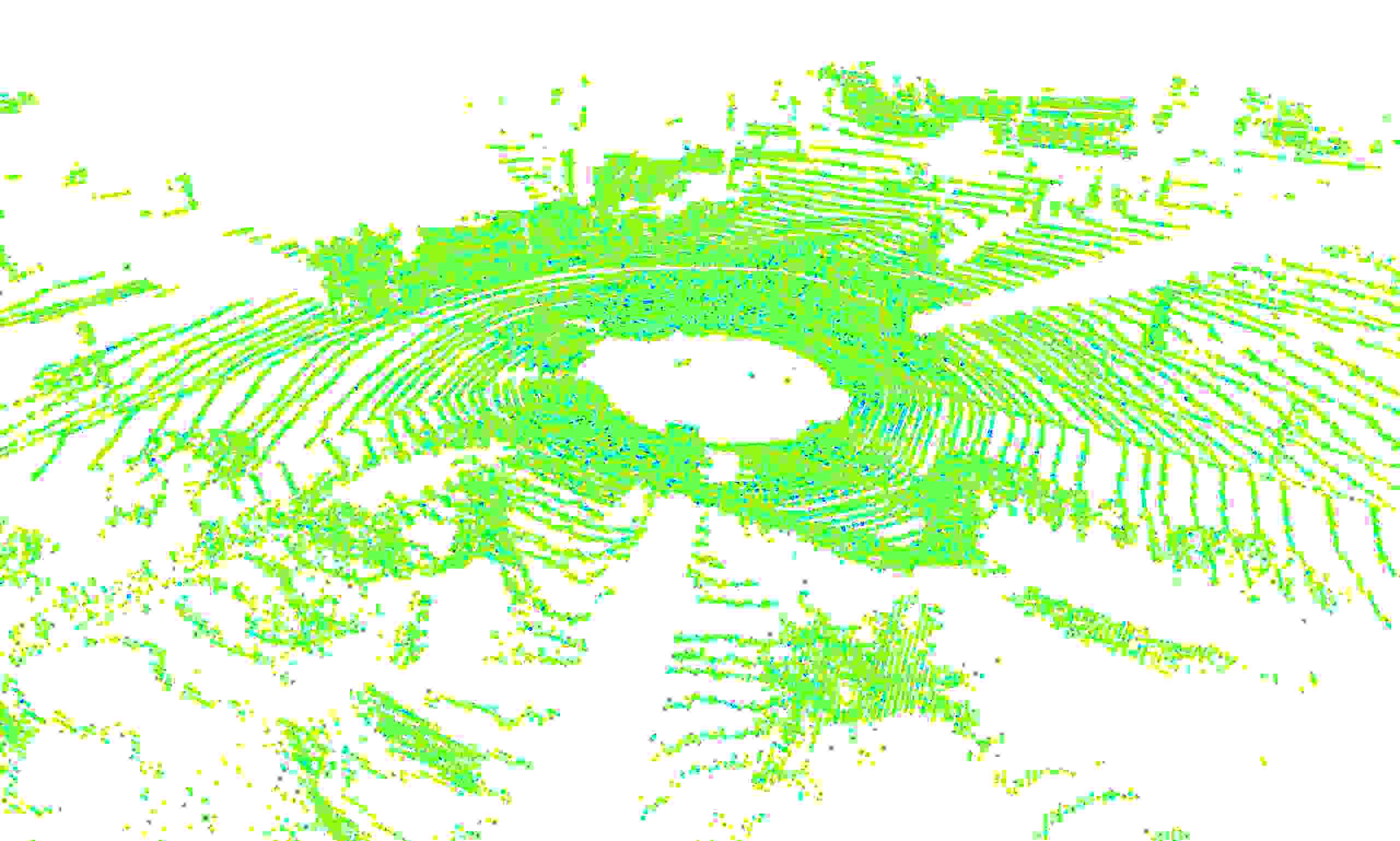}
    \put (0,55) {\colorbox{gray!30}{\scriptsize Ours: PSNR 55.06, Bitrate: 2.41}}
\end{overpic}
\begin{overpic}[width=0.32\textwidth]{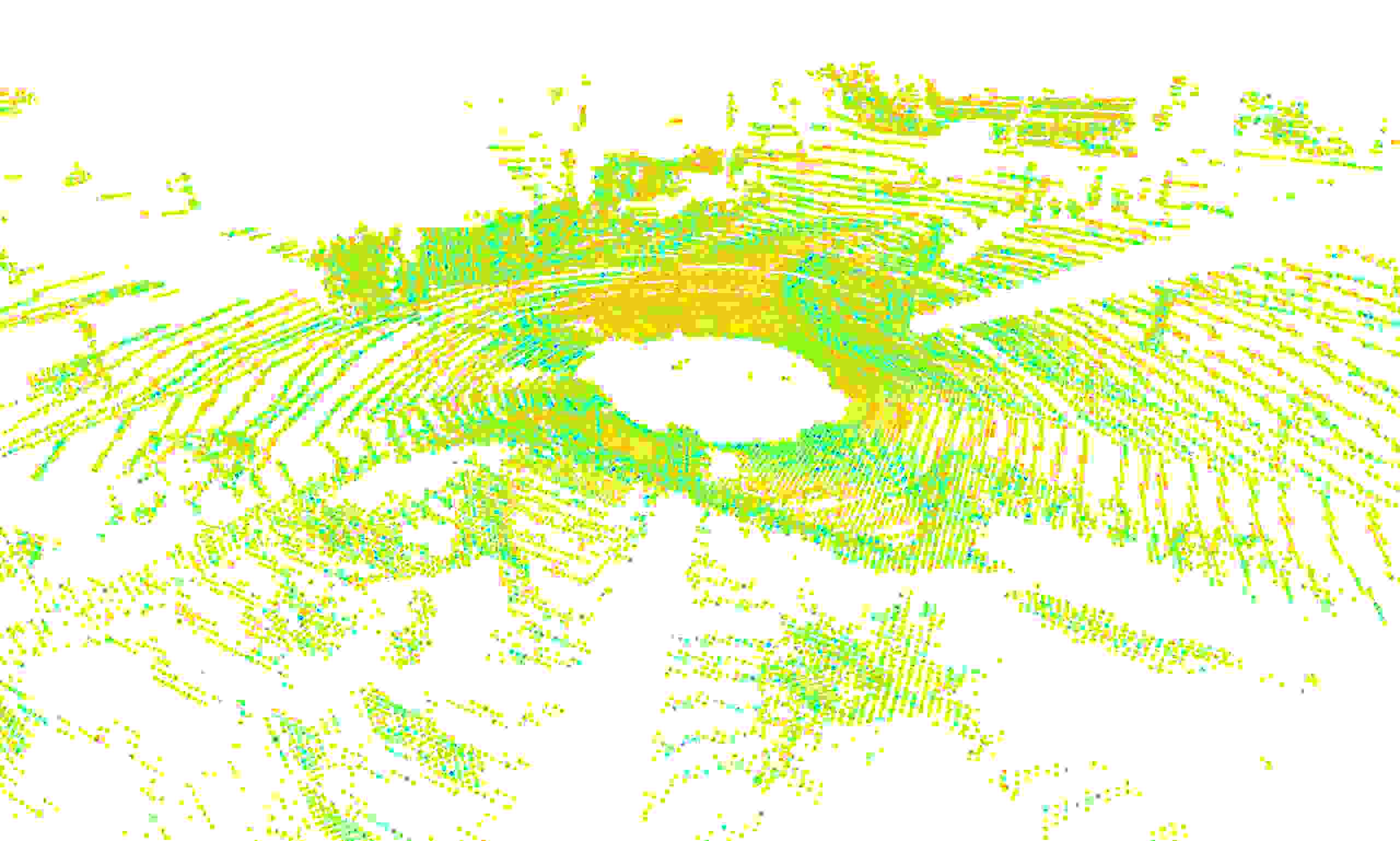}
    \put (0,55) {\colorbox{gray!30}{\scriptsize Draco: PSNR 50.24, Bitrate: 2.75}}
\end{overpic}
\end{center}

\begin{center}
\begin{overpic}[width=0.32\textwidth]{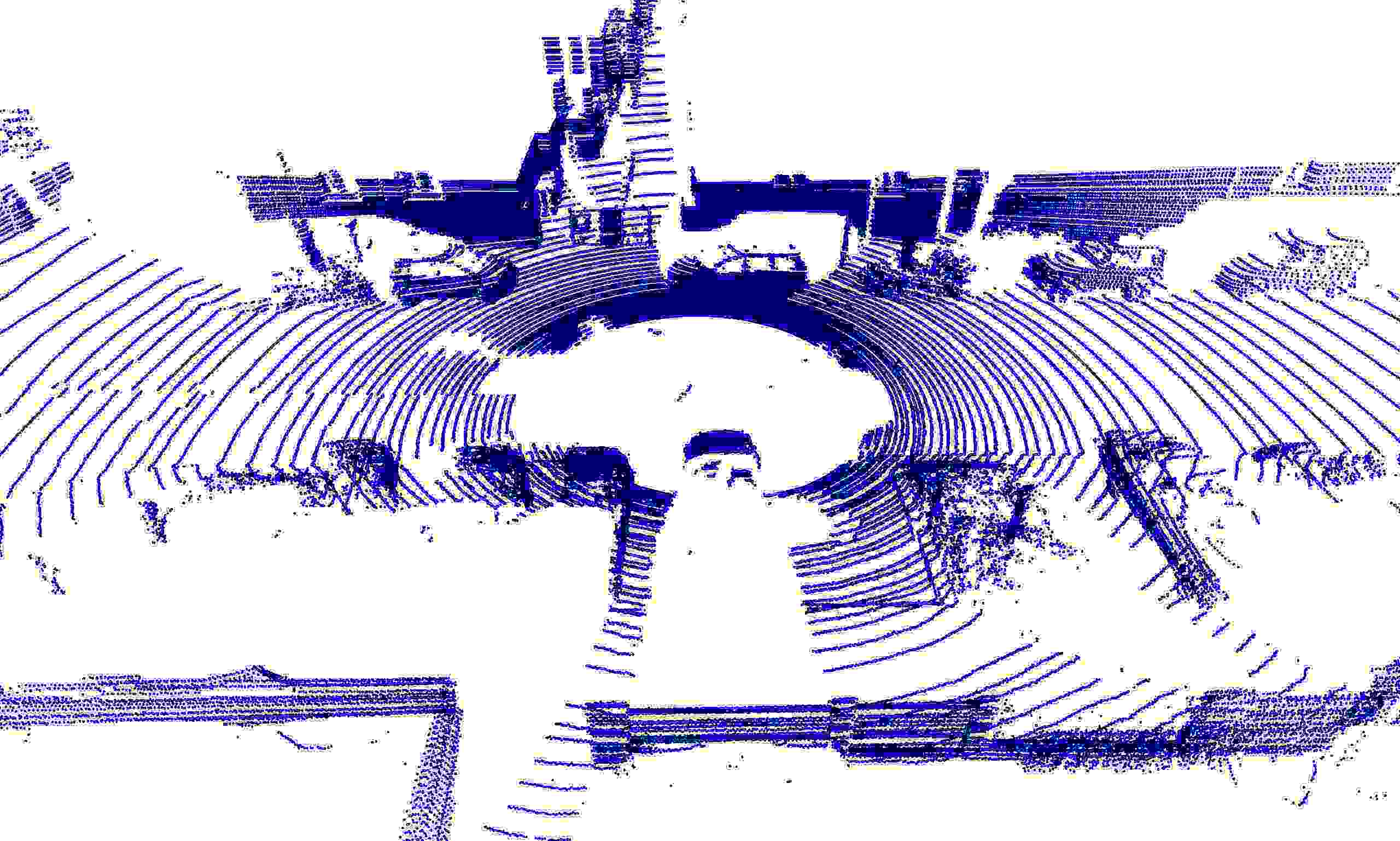}
    \put (0,55) {\colorbox{gray!30}{\scriptsize GT (KITTI)}}
\end{overpic}
\begin{overpic}[width=0.32\textwidth]{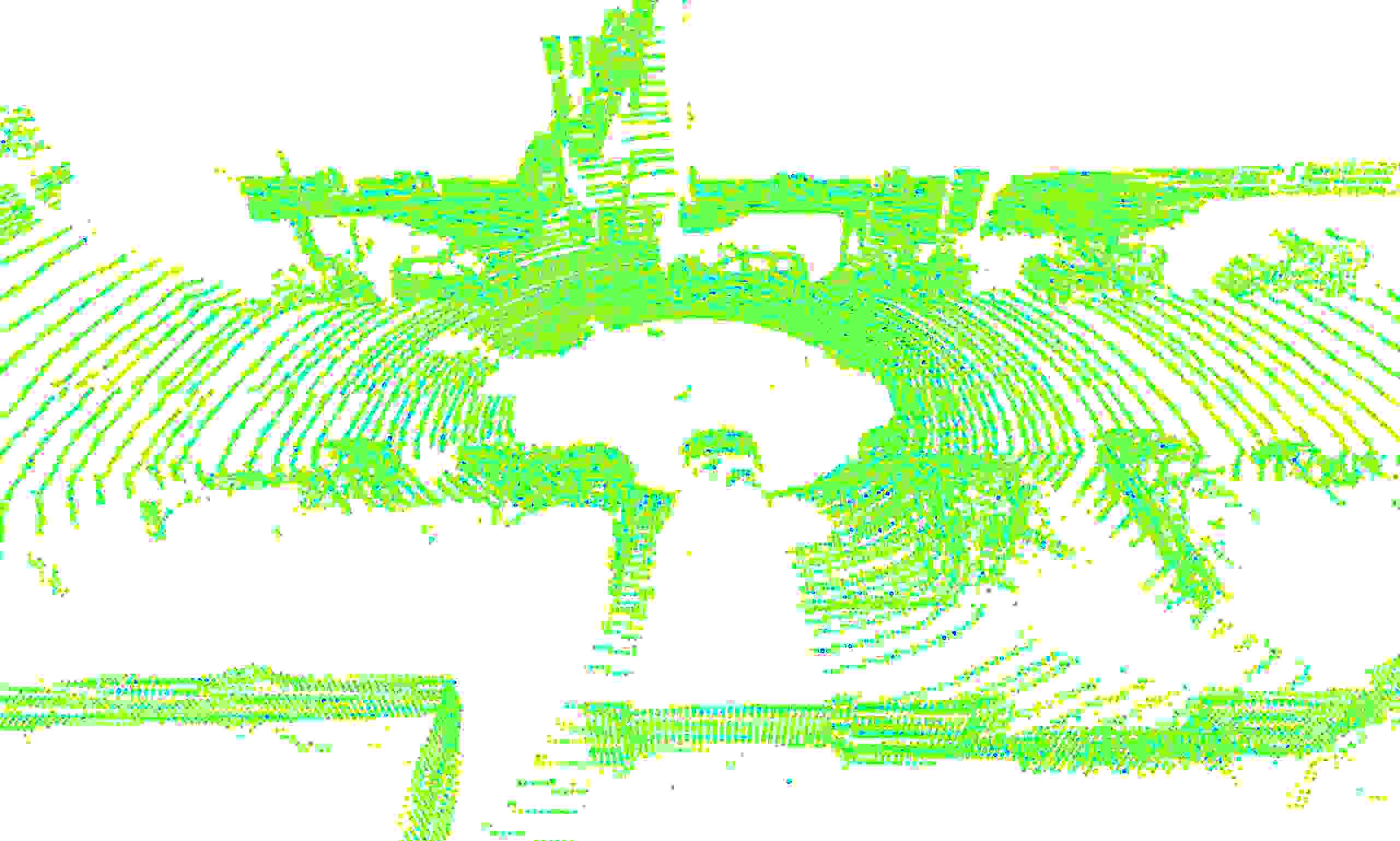}
    \put (0,55) {\colorbox{gray!30}{\scriptsize Ours: PSNR 43.67, Bitrate: 1.52}}
\end{overpic}
\begin{overpic}[width=0.32\textwidth]{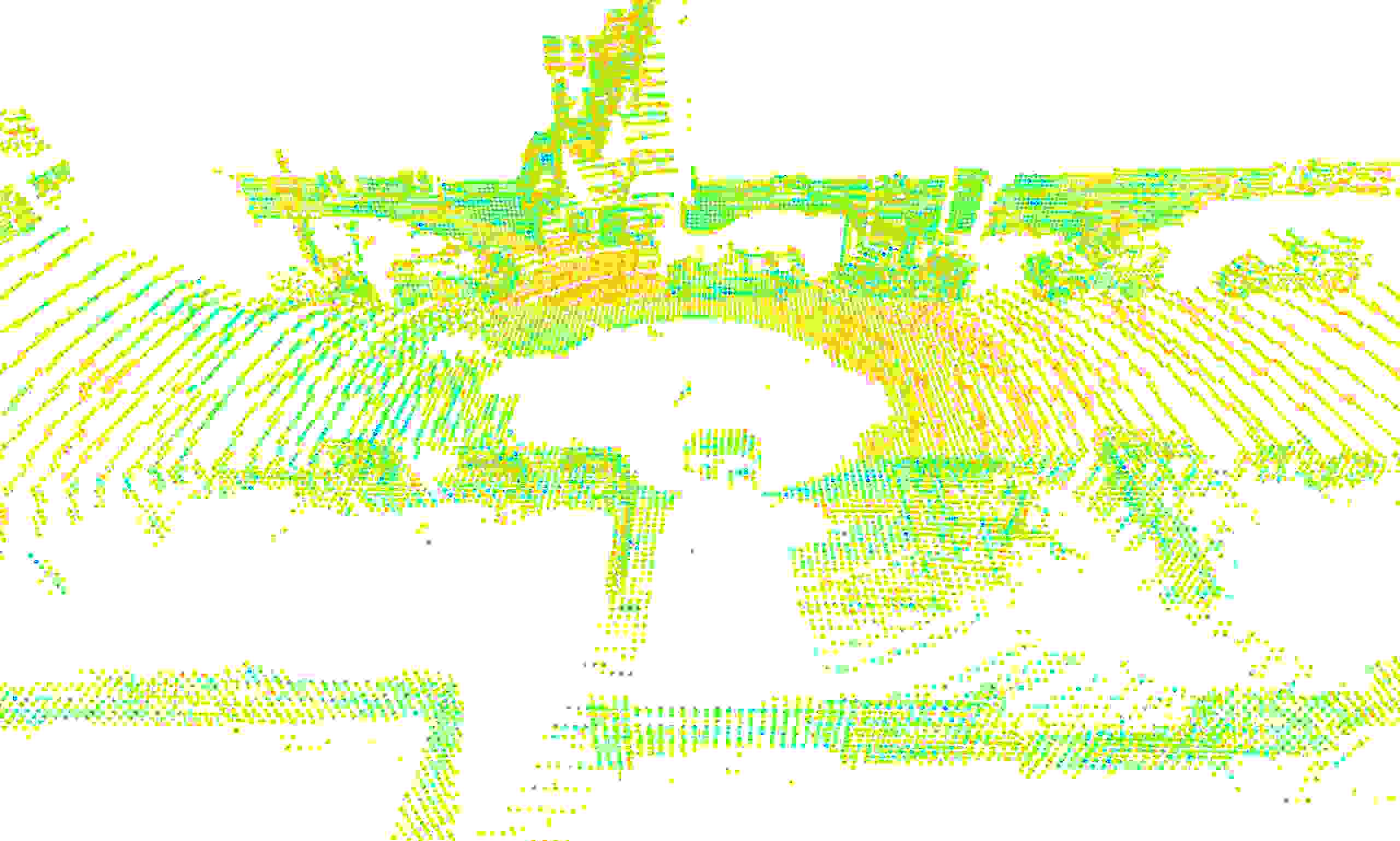}
    \put (0,55) {\colorbox{gray!30}{\scriptsize Draco: PSNR 39.09, Bitrate: 2.04}}
\end{overpic}
\end{center}

\vspace{-0.25in}
\begin{center}
\includegraphics[width=0.5\textwidth]{./figures/colormap.png}
\end{center}

\vspace{-0.25in}
\caption{Qualitative results of reconstruction quality for KITTI.}
\label{fig:recon_kitti}
\end{figure*}

\begin{figure*}[h]
\begin{center}
\begin{overpic}[width=0.32\textwidth]{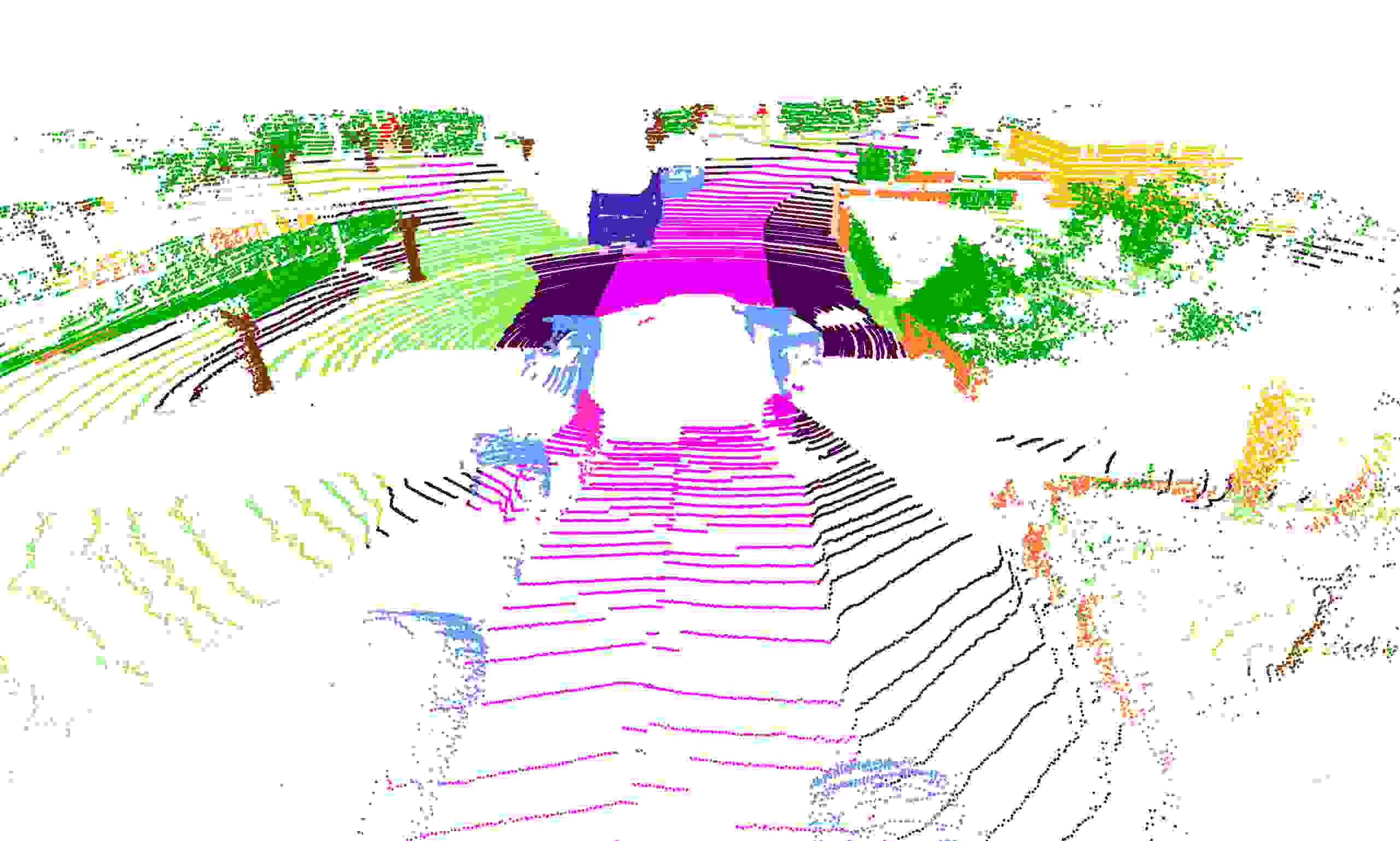}
    \put (0,55) {\colorbox{gray!30}{\scriptsize Oracle: IOU 40.86, Bitrate: 96.00}}
\end{overpic}
\begin{overpic}[width=0.32\textwidth]{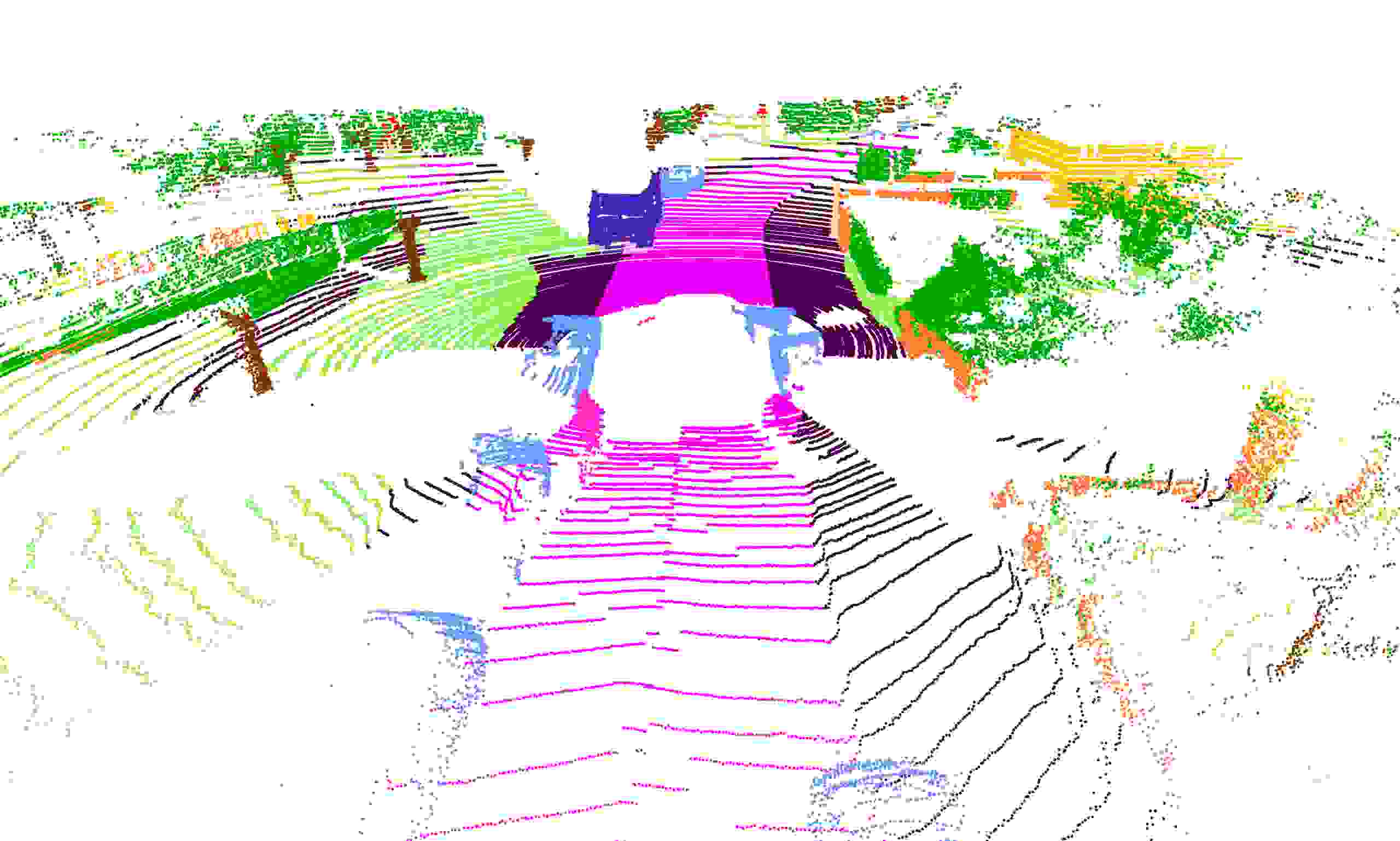}
    \put (0,55) {\colorbox{gray!30}{\scriptsize Ours: IOU 39.10, Bitrate: 13.37}}
\end{overpic}
\begin{overpic}[width=0.32\textwidth]{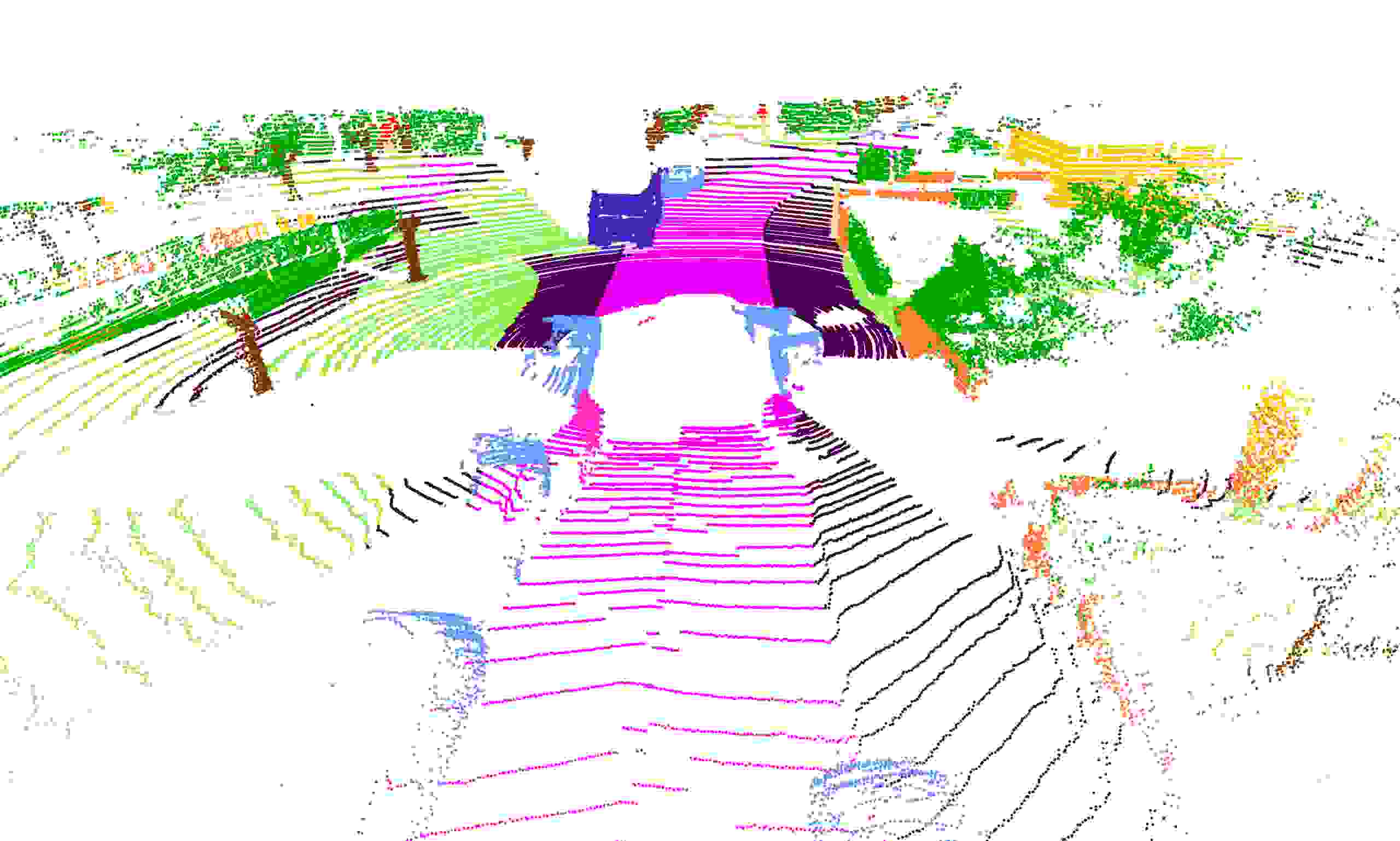}
    \put (0,55) {\colorbox{gray!30}{\scriptsize Draco: IOU 38.98, Bitrate: 15.63}}
\end{overpic}
\end{center}

\begin{center}
\begin{overpic}[width=0.32\textwidth]{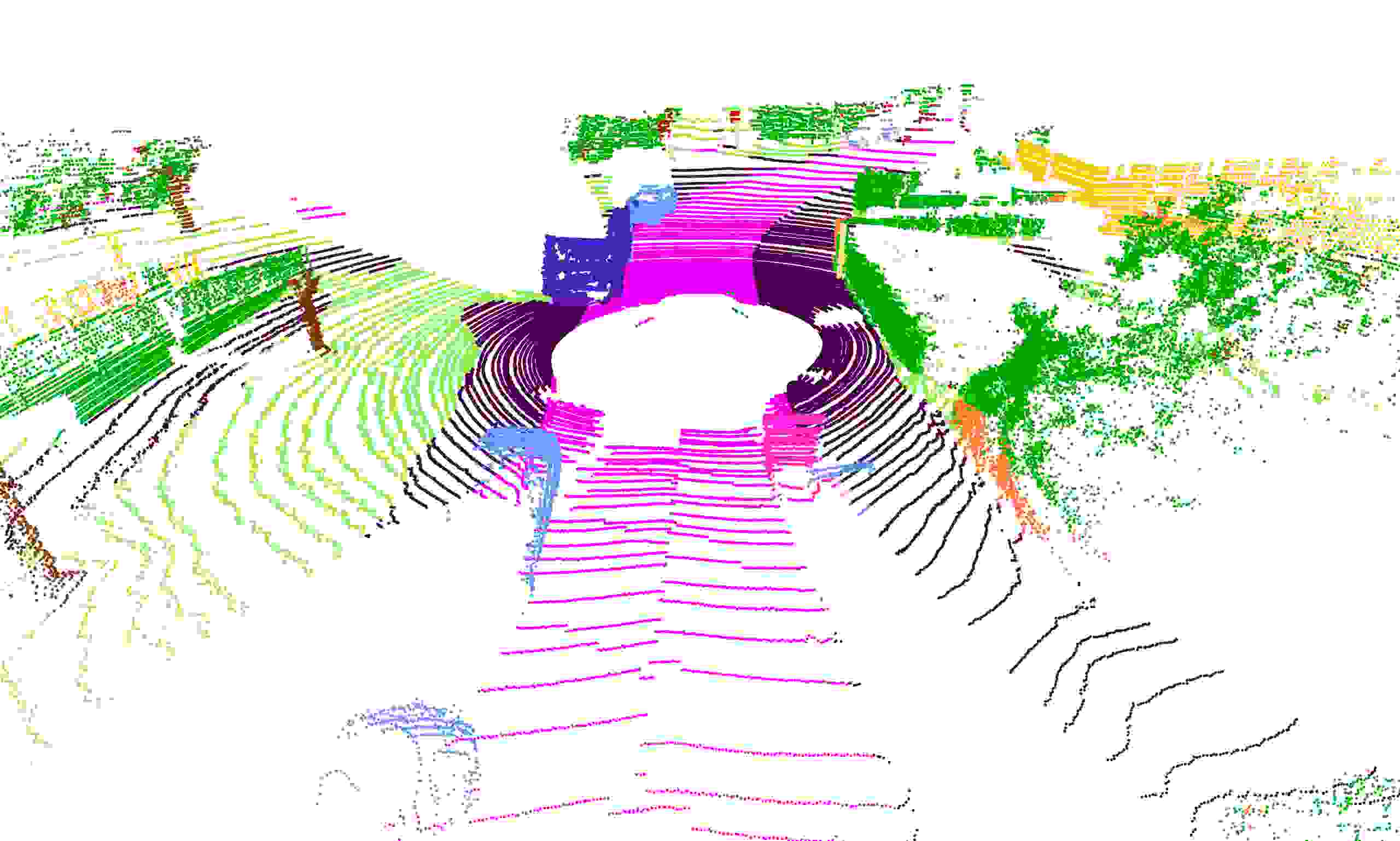}
    \put (0,55) {\colorbox{gray!30}{\scriptsize Oracle: IOU 38.39, Bitrate: 96.00}}
\end{overpic}
\begin{overpic}[width=0.32\textwidth]{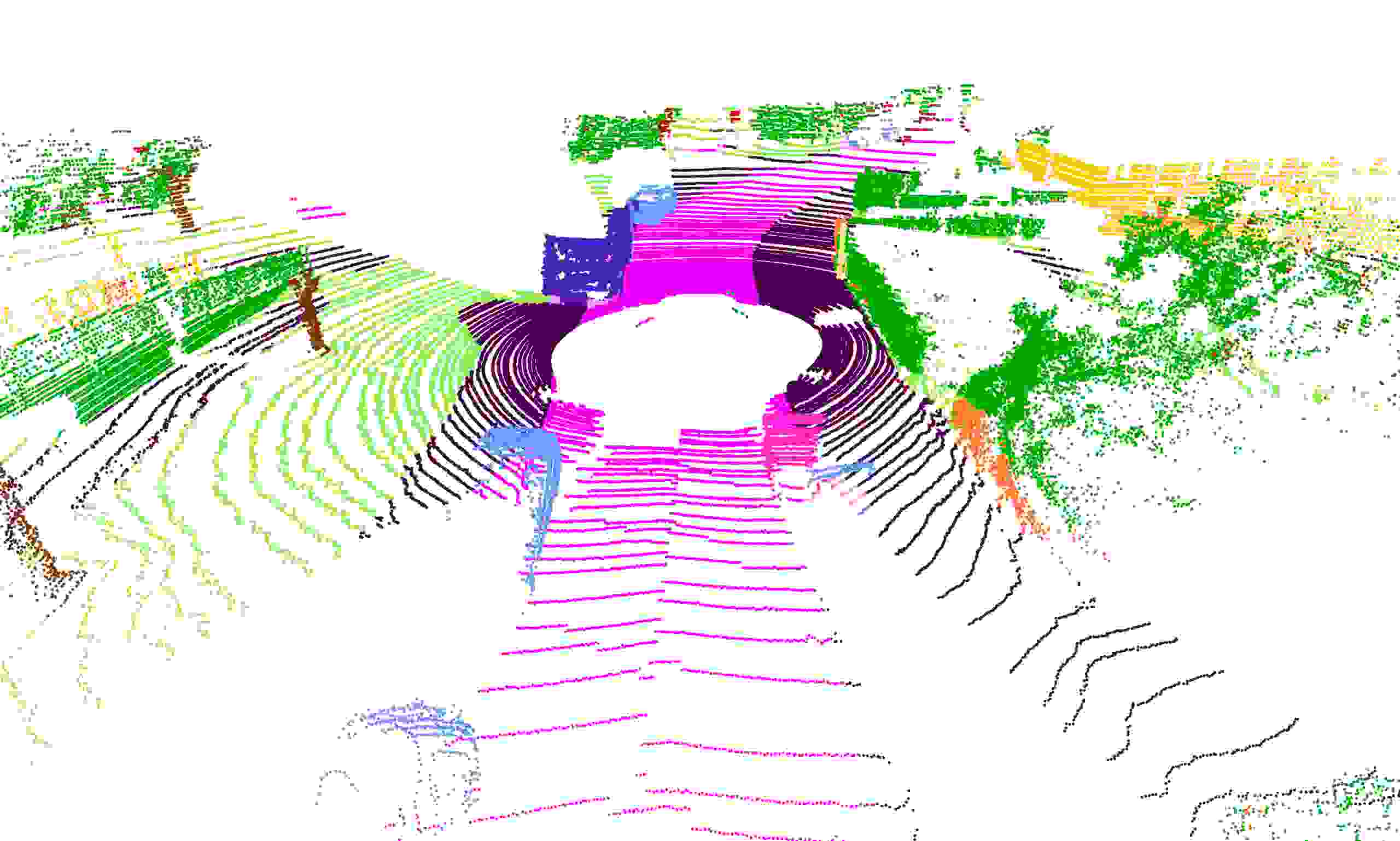}
    \put (0,55) {\colorbox{gray!30}{\scriptsize Ours: IOU 36.70, Bitrate: 10.37}}
\end{overpic}
\begin{overpic}[width=0.32\textwidth]{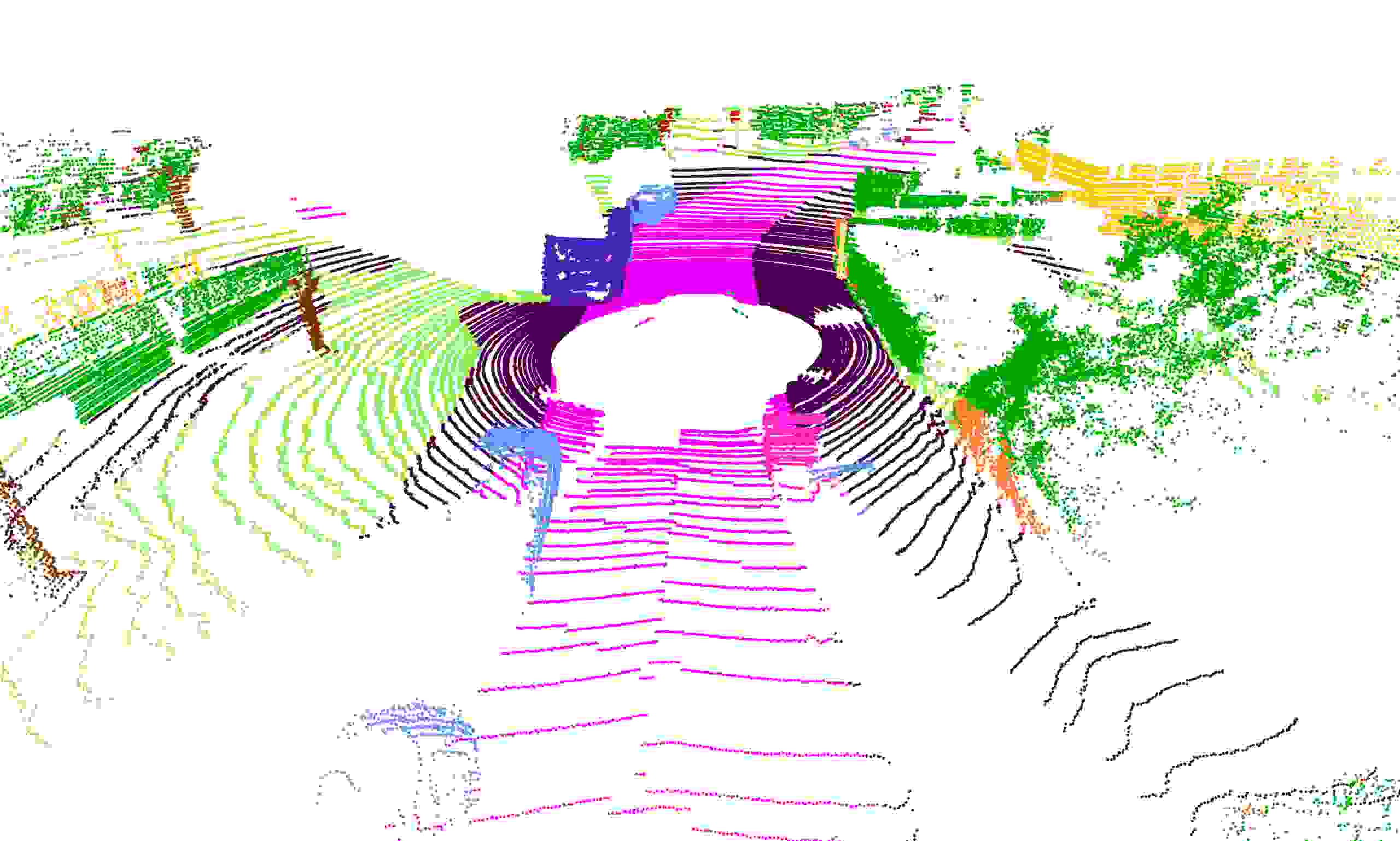}
    \put (0,55) {\colorbox{gray!30}{\scriptsize Draco: IOU 35.61, Bitrate: 12.70}}
\end{overpic}
\end{center}

\begin{center}
\begin{overpic}[width=0.32\textwidth]{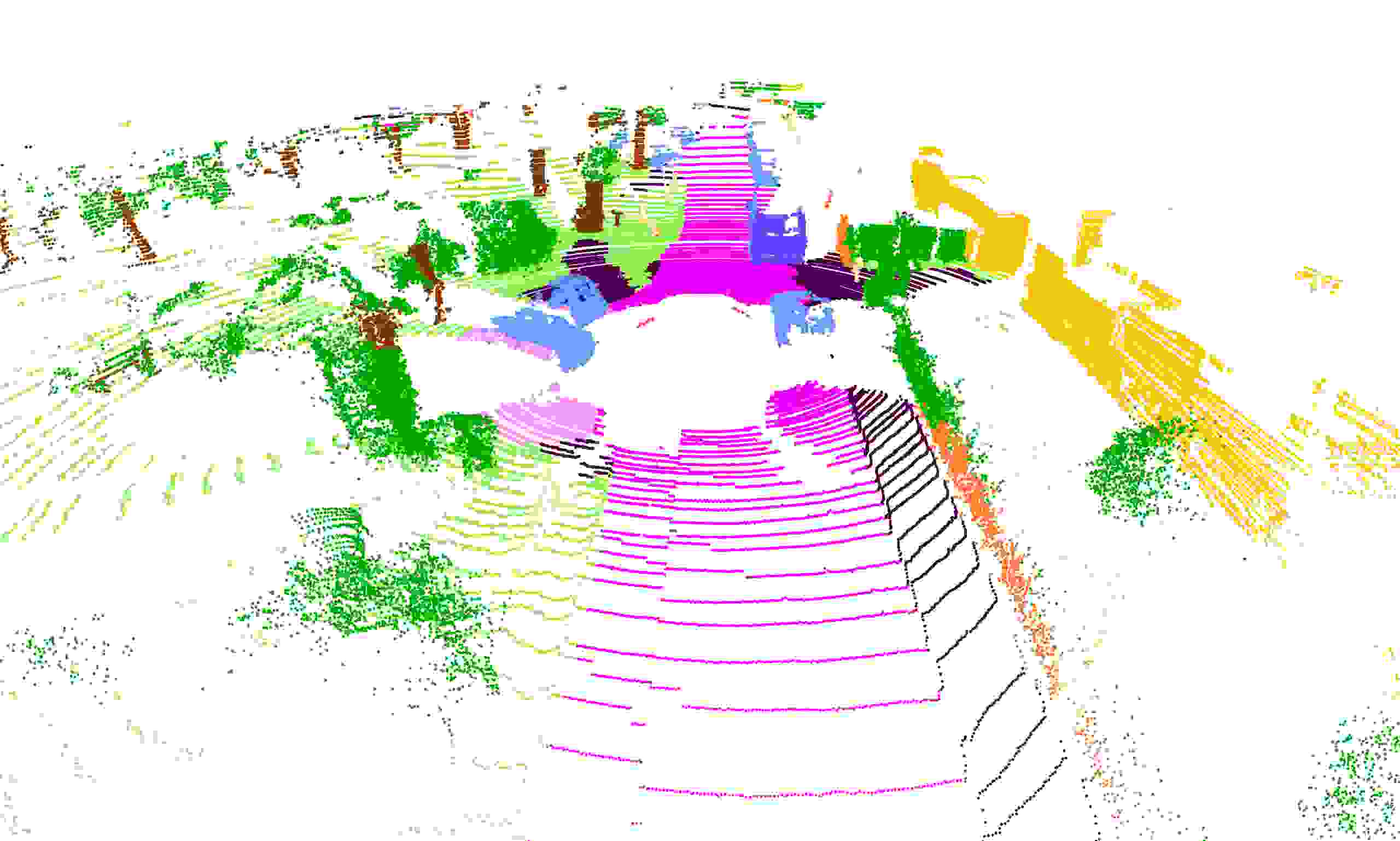}
    \put (0,55) {\colorbox{gray!30}{\scriptsize Oracle: IOU 35.52, Bitrate: 96.00}}
\end{overpic}
\begin{overpic}[width=0.32\textwidth]{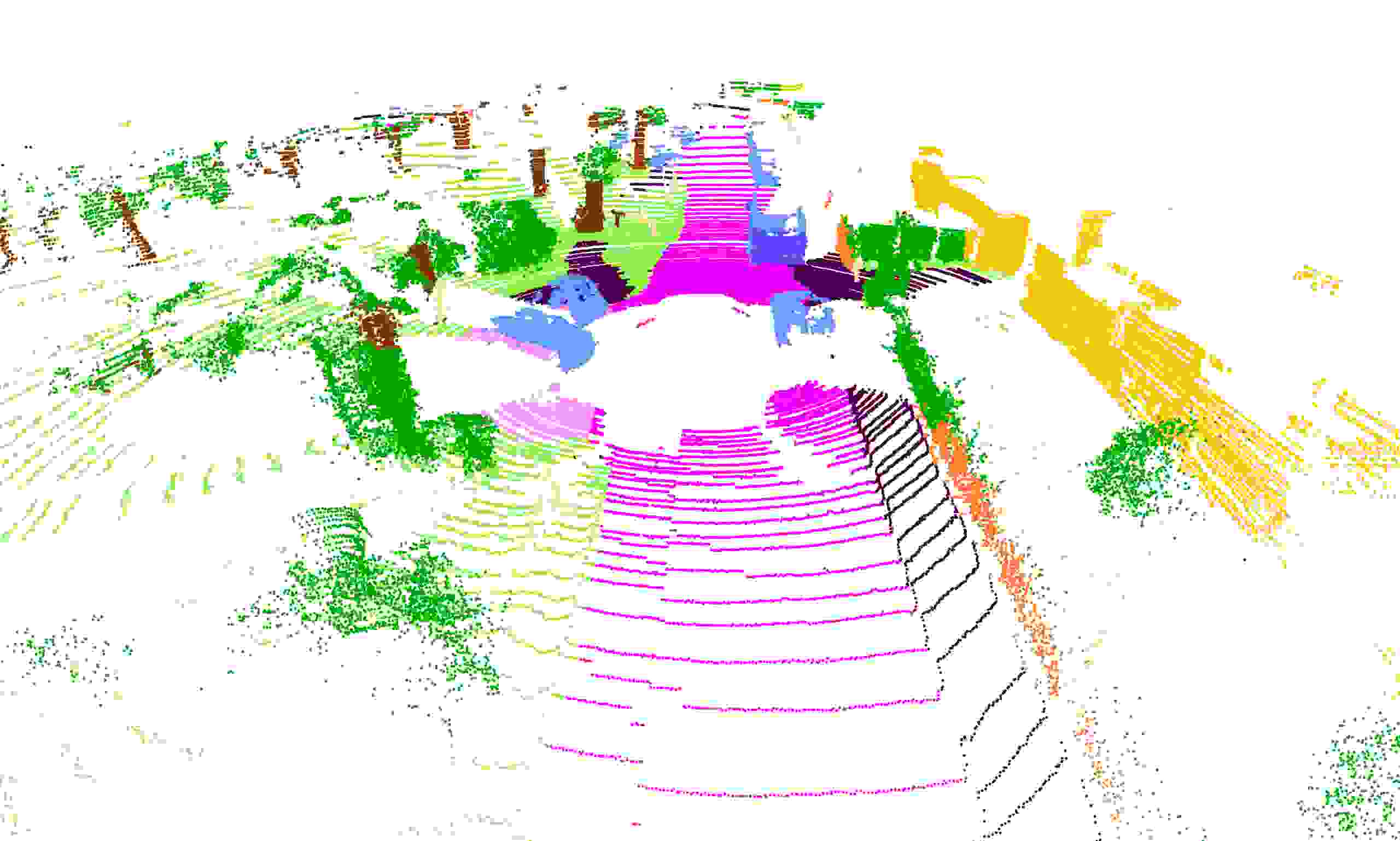}
    \put (0,55) {\colorbox{gray!30}{\scriptsize Ours: IOU 35.75, Bitrate: 6.87}}
\end{overpic}
\begin{overpic}[width=0.32\textwidth]{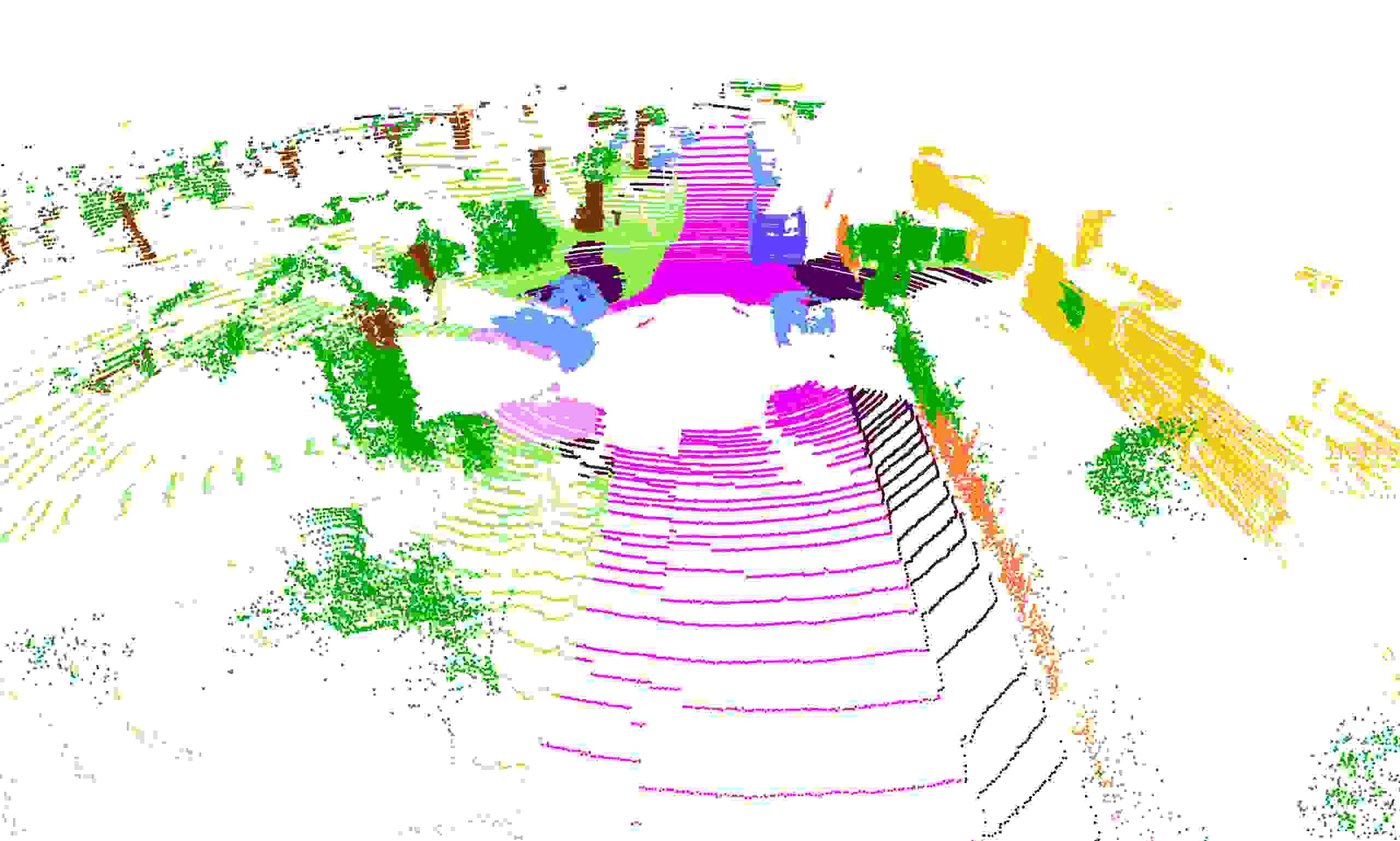}
    \put (0,55) {\colorbox{gray!30}{\scriptsize Draco: IOU 35.08, Bitrate: 9.30}}
\end{overpic}
\end{center}

\begin{center}
\begin{overpic}[width=0.32\textwidth]{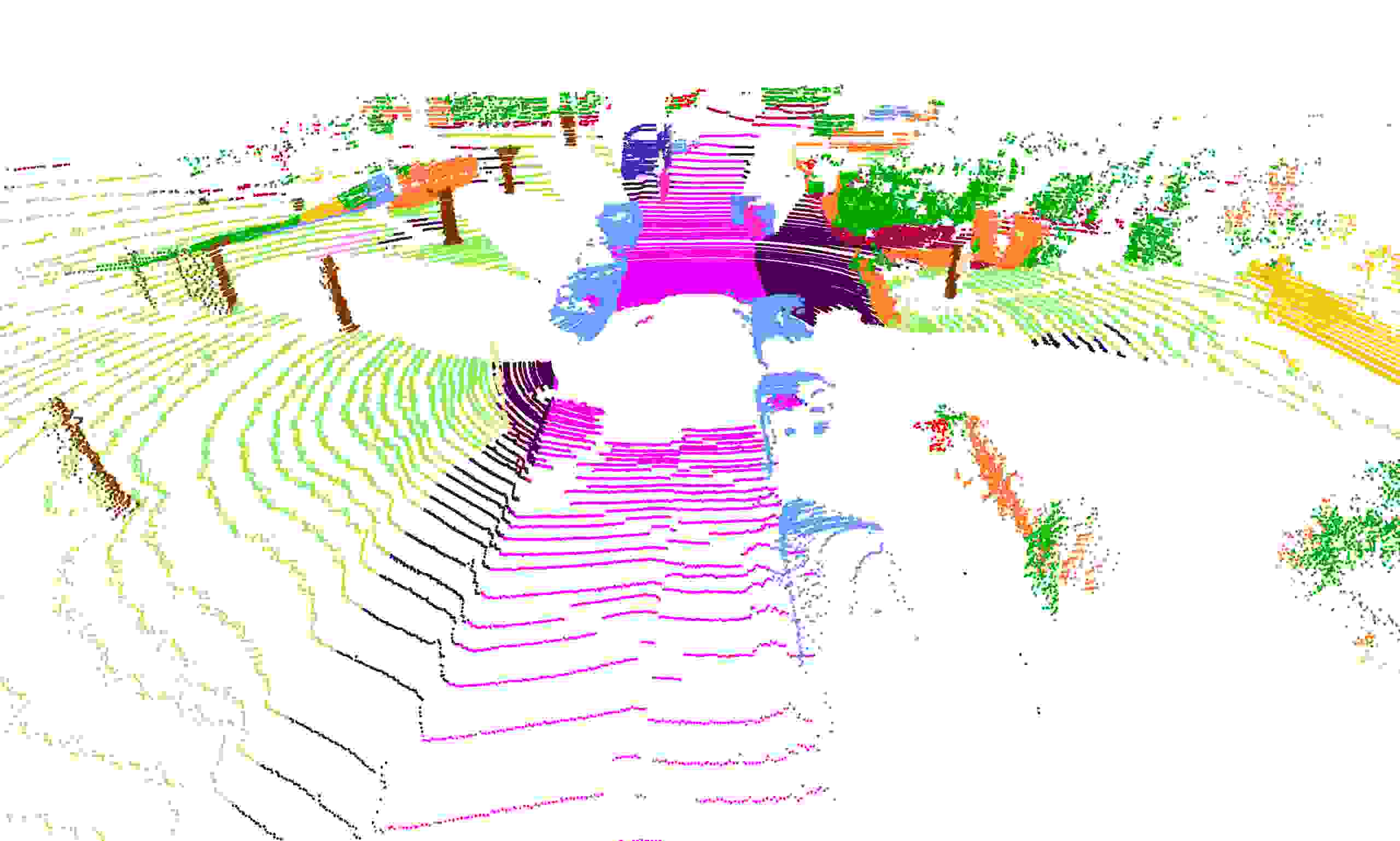}
    \put (0,55) {\colorbox{gray!30}{\scriptsize Oracle: IOU 39.41, Bitrate: 96.00}}
\end{overpic}
\begin{overpic}[width=0.32\textwidth]{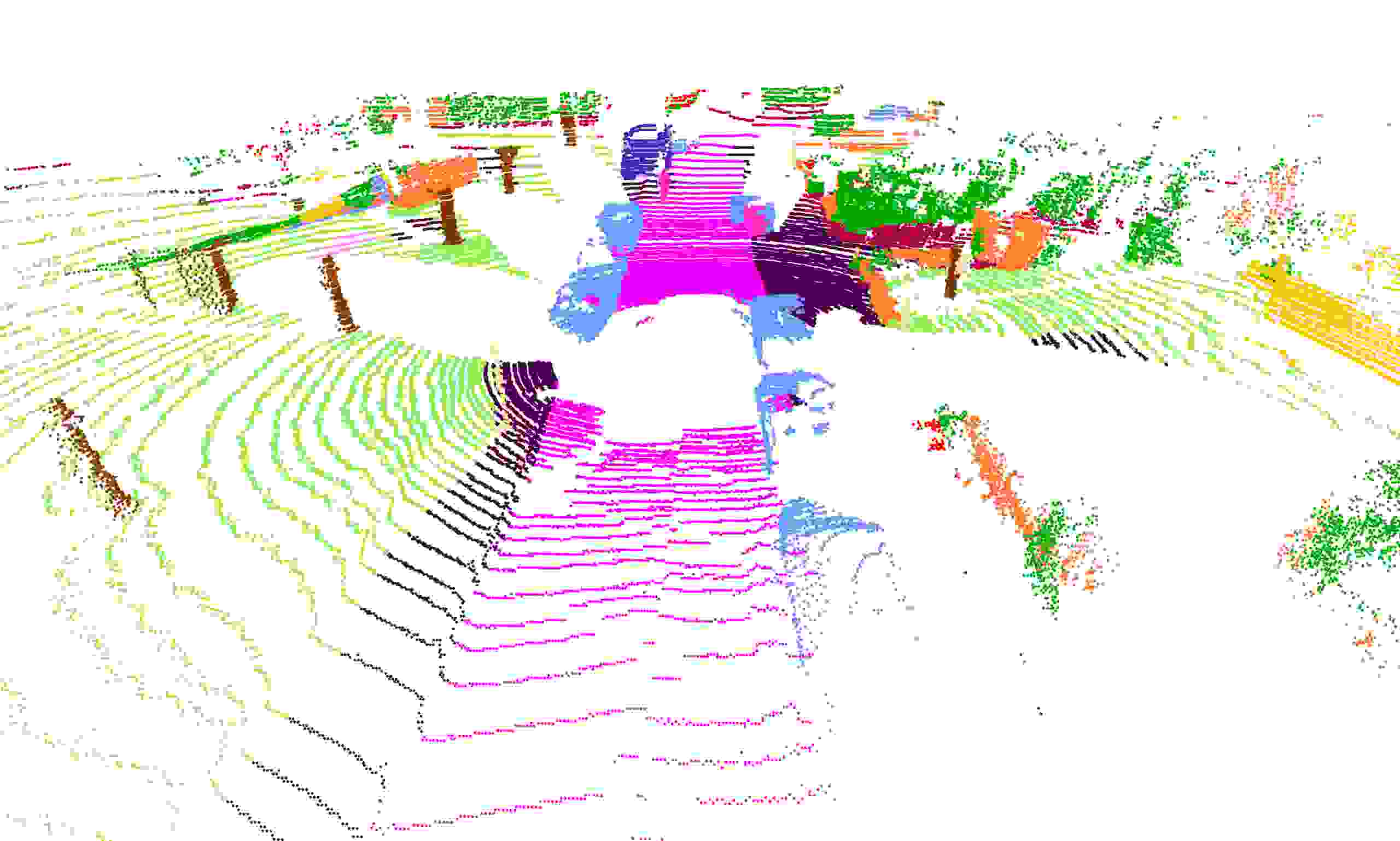}
    \put (0,55) {\colorbox{gray!30}{\scriptsize Ours: IOU 29.92, Bitrate: 4.70}}
\end{overpic}
\begin{overpic}[width=0.32\textwidth]{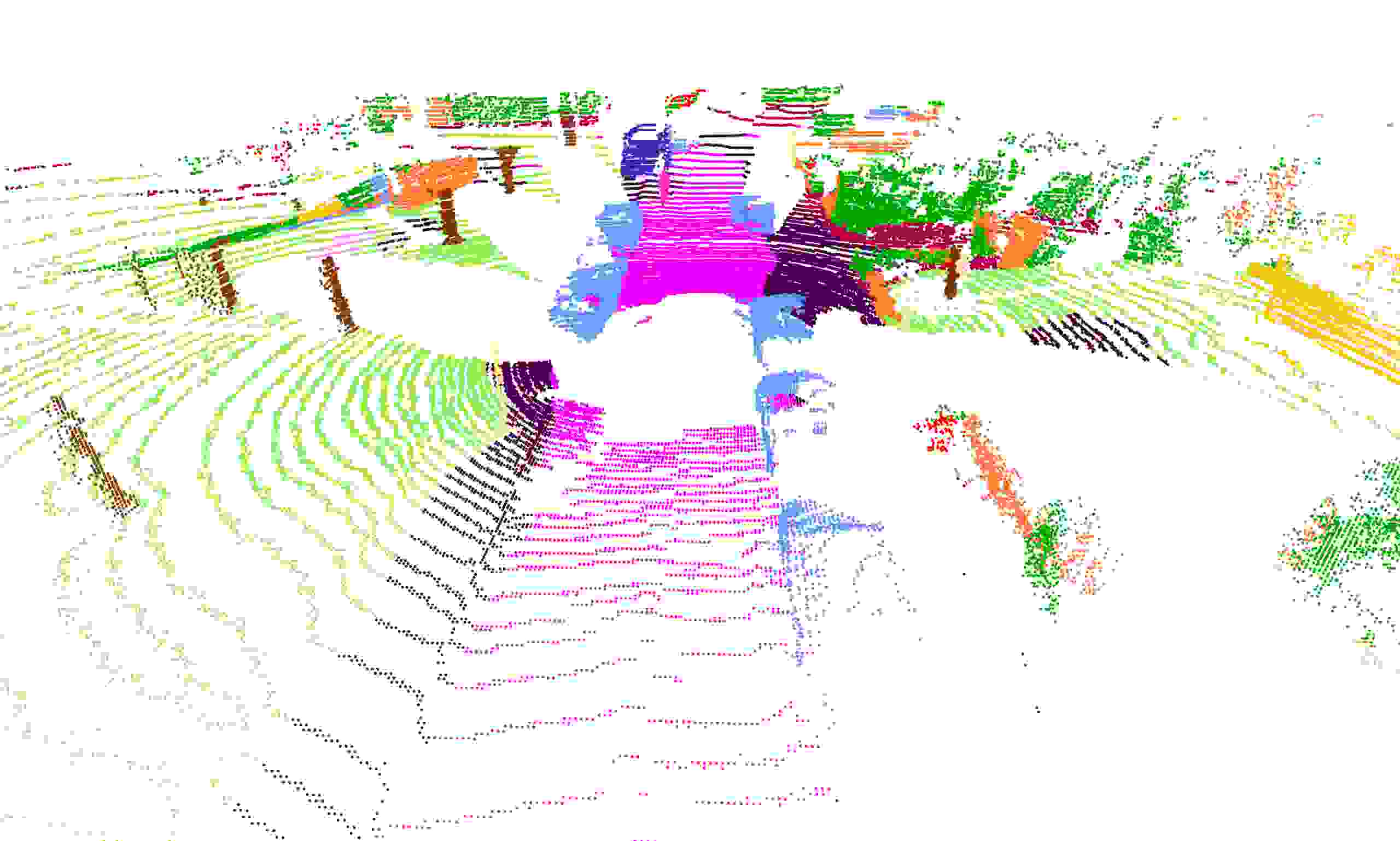}
    \put (0,55) {\colorbox{gray!30}{\scriptsize Draco: IOU 23.23, Bitrate: 4.70}}
\end{overpic}
\end{center}

\begin{center}
\begin{overpic}[width=0.32\textwidth]{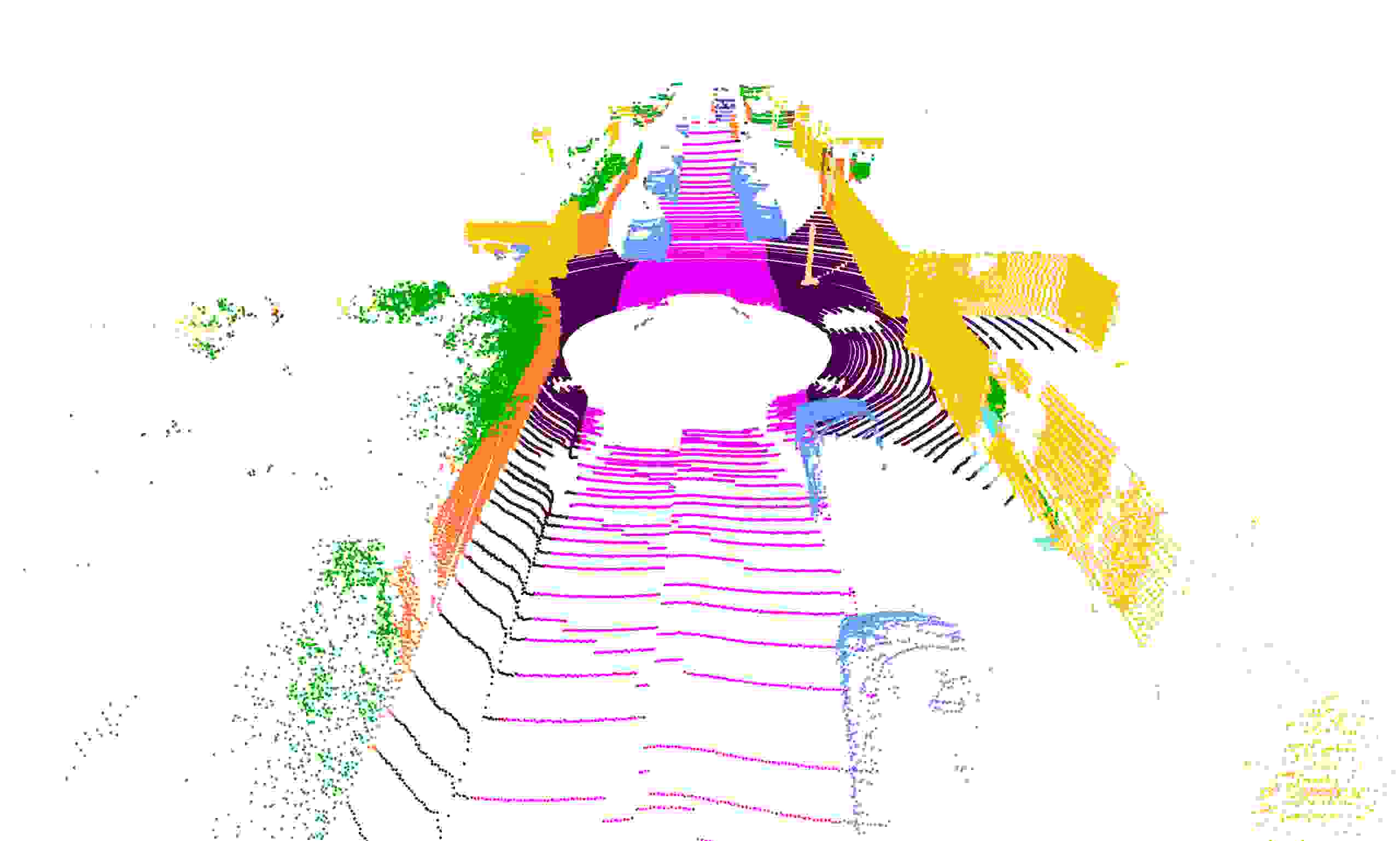}
    \put (0,55) {\colorbox{gray!30}{\scriptsize Oracle: IOU 35.20, Bitrate: 96.00}}
\end{overpic}
\begin{overpic}[width=0.32\textwidth]{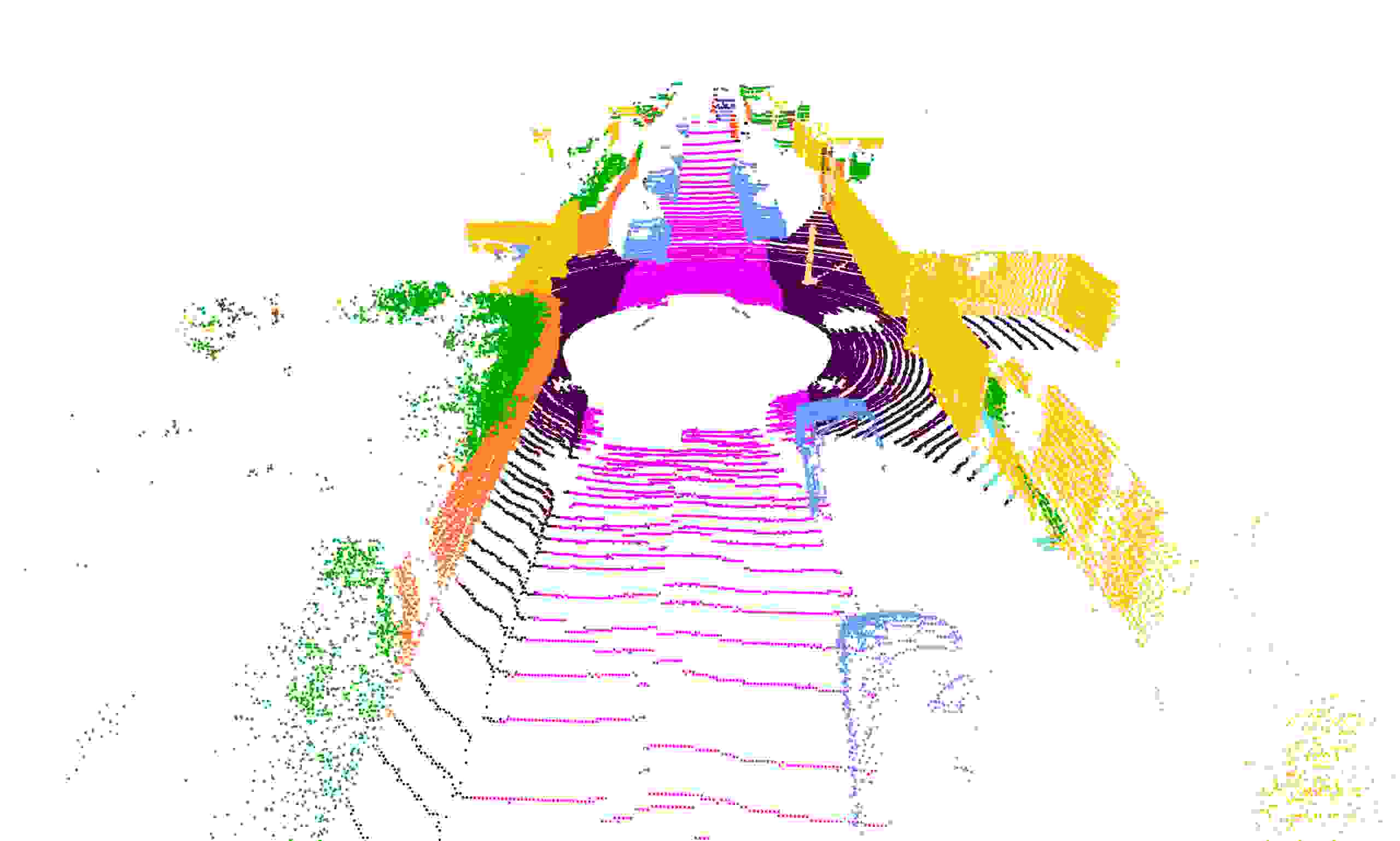}
    \put (0,55) {\colorbox{gray!30}{\scriptsize Ours: IOU 26.38, Bitrate: 2.56}}
\end{overpic}
\begin{overpic}[width=0.32\textwidth]{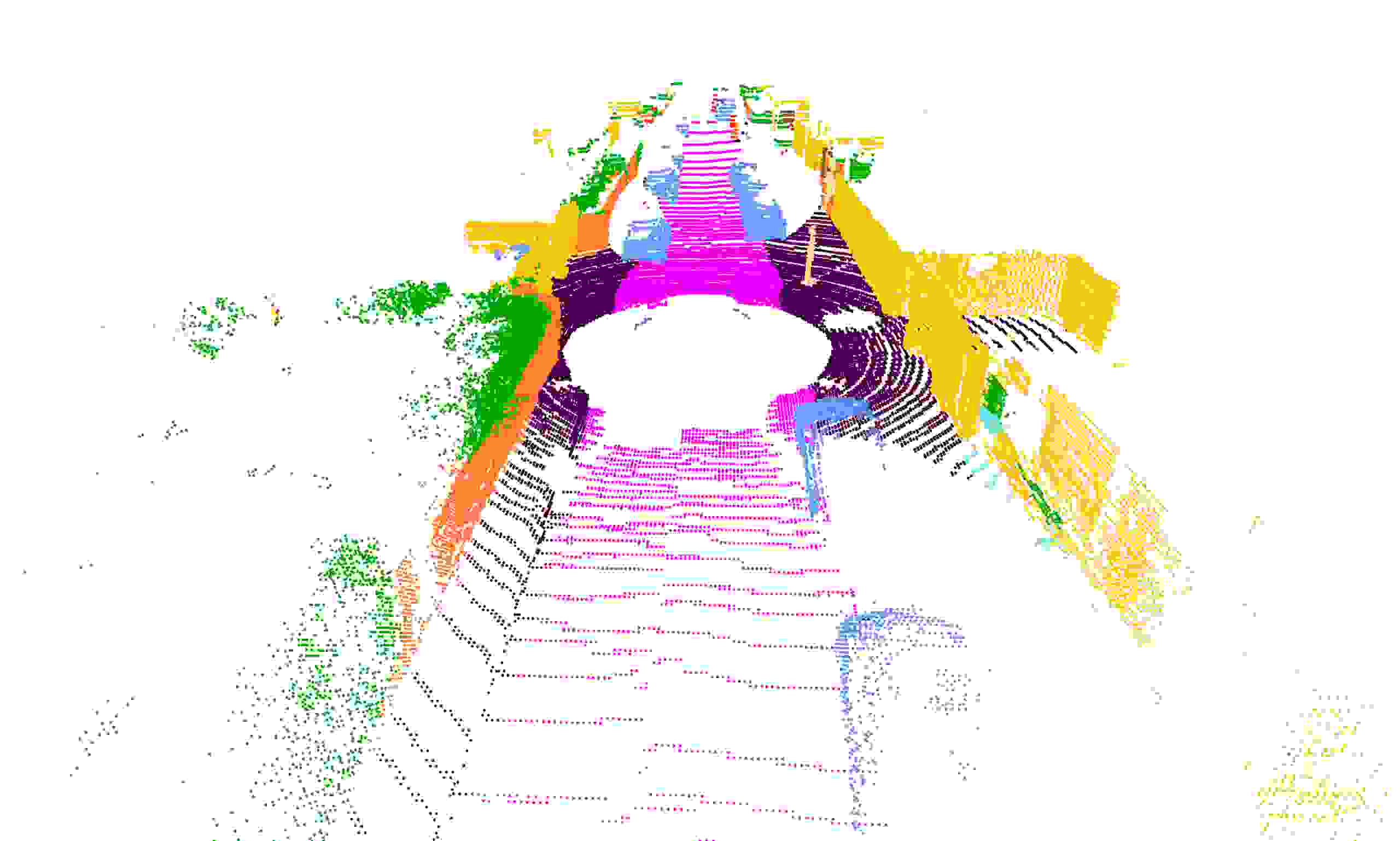}
    \put (0,55) {\colorbox{gray!30}{\scriptsize Draco: IOU 21.02, Bitrate: 2.67}}
\end{overpic}
\end{center}

\begin{center}
\begin{overpic}[width=0.32\textwidth]{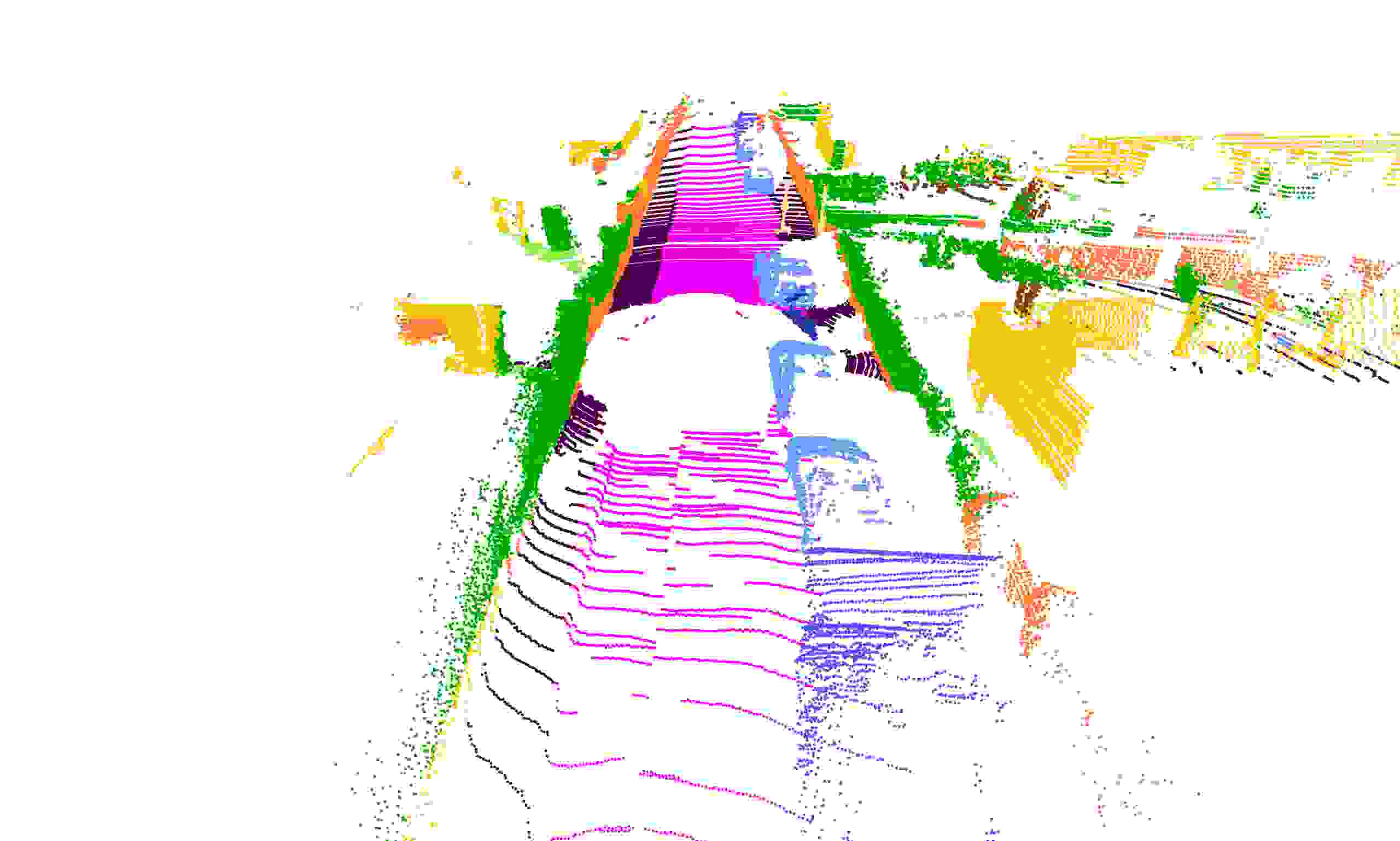}
    \put (0,55) {\colorbox{gray!30}{\scriptsize Oracle: IOU 34.81, Bitrate: 96.00}}
\end{overpic}
\begin{overpic}[width=0.32\textwidth]{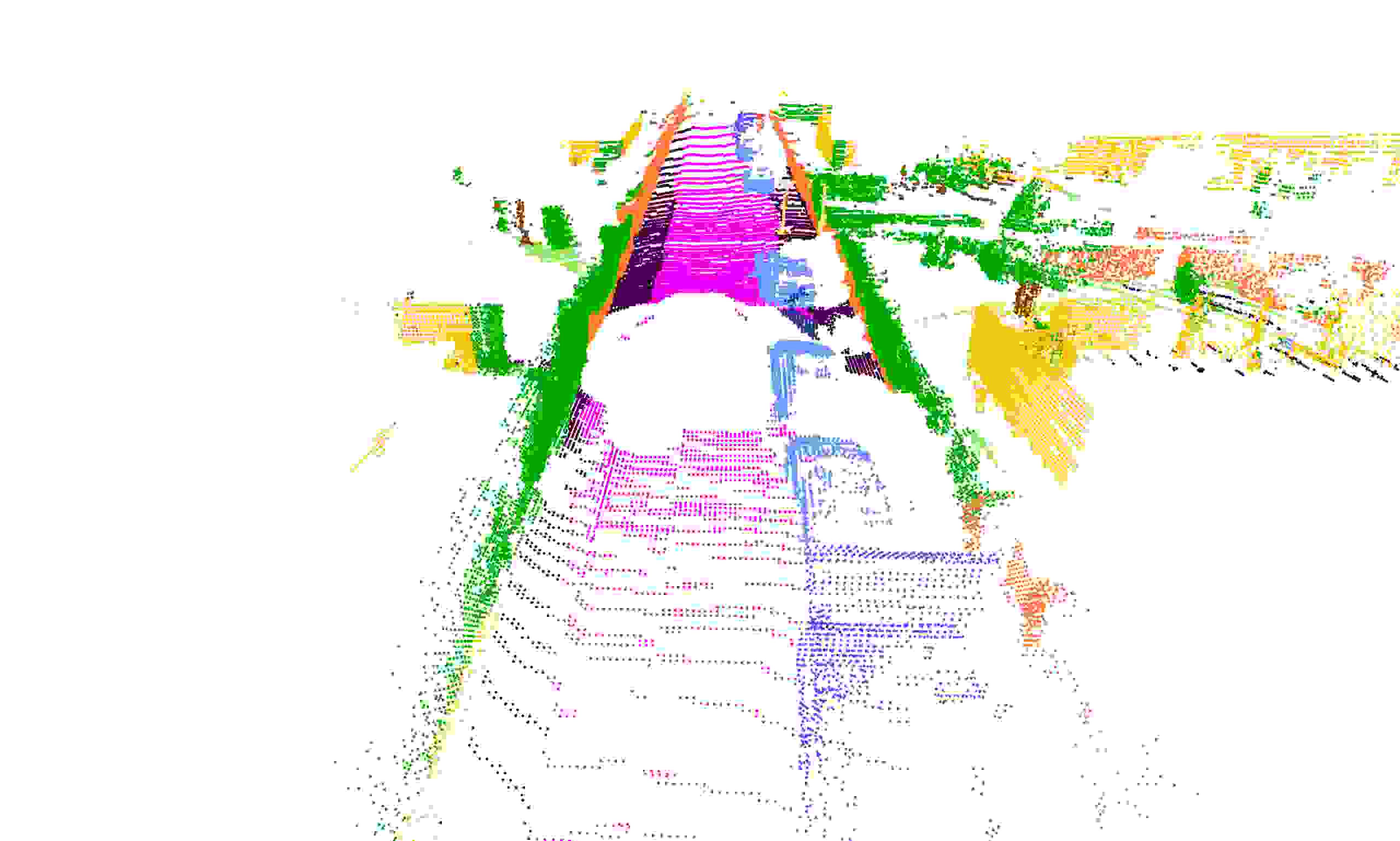}
    \put (0,55) {\colorbox{gray!30}{\scriptsize Ours: IOU 23.04, Bitrate: 1.62}}
\end{overpic}
\begin{overpic}[width=0.32\textwidth]{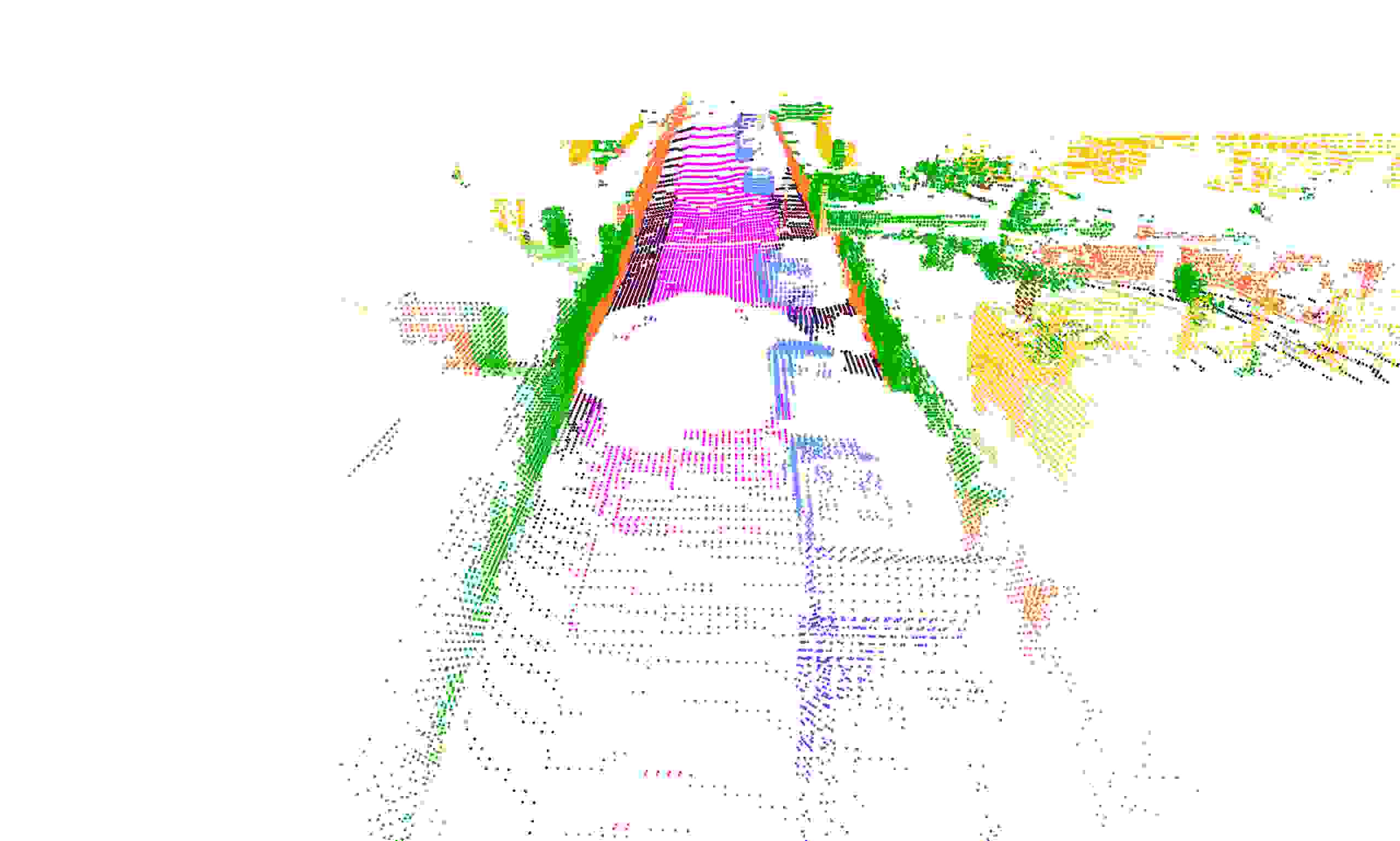}
    \put (0,55) {\colorbox{gray!30}{\scriptsize Draco: IOU 18.55, Bitrate: 1.96}}
\end{overpic}
\end{center}

\caption{Qualitative results of semantic segmentation for KITTI. IOU is averaged over all classes.}
\label{fig:semantic_kitti}
\end{figure*}

\begin{figure*}[h]
\begin{center}
\begin{overpic}[width=0.32\textwidth]{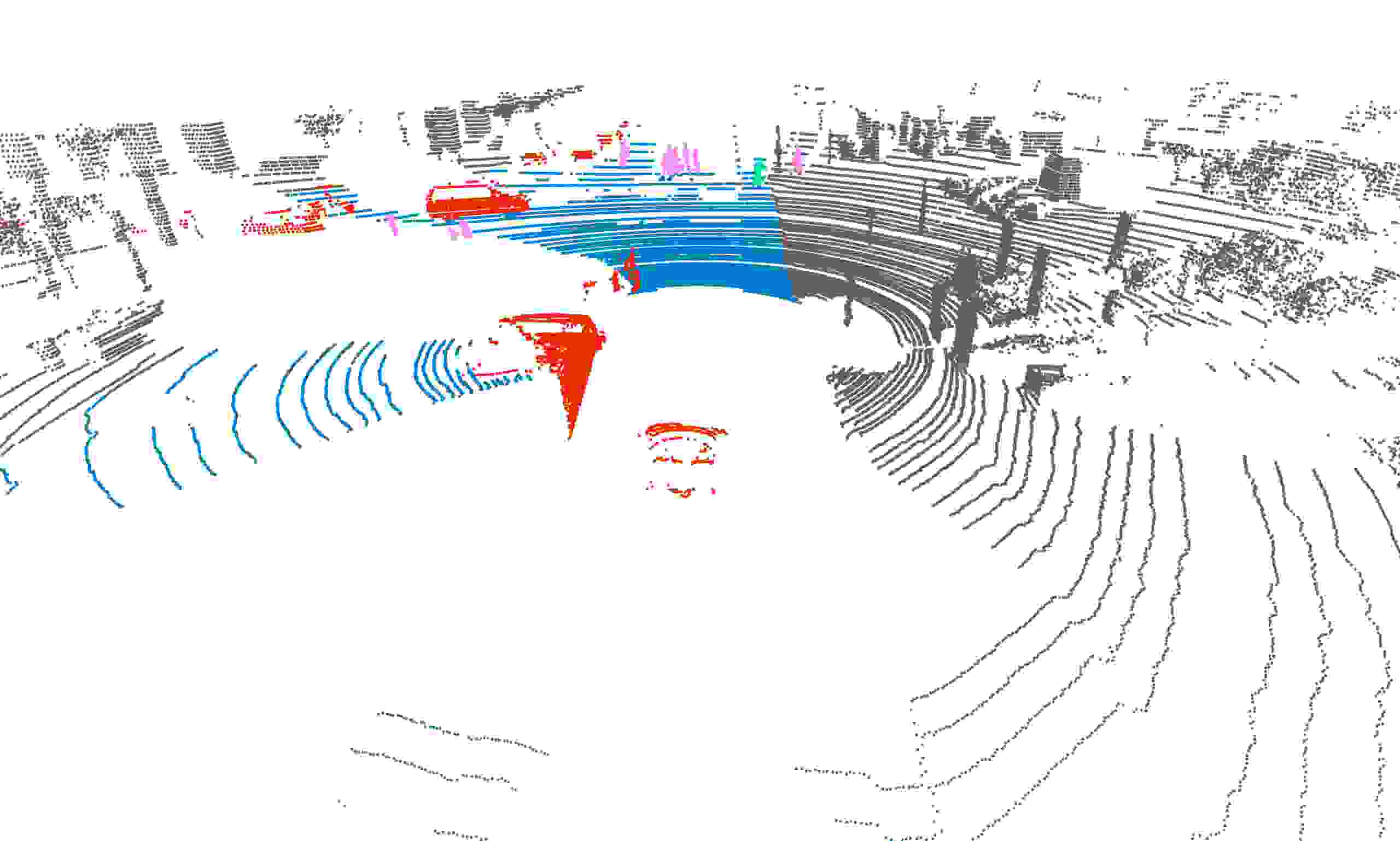}
    \put (0,55) {\colorbox{gray!30}{\scriptsize Oracle: IOU 99.55, Bitrate: 96.00}}
\end{overpic}
\begin{overpic}[width=0.32\textwidth]{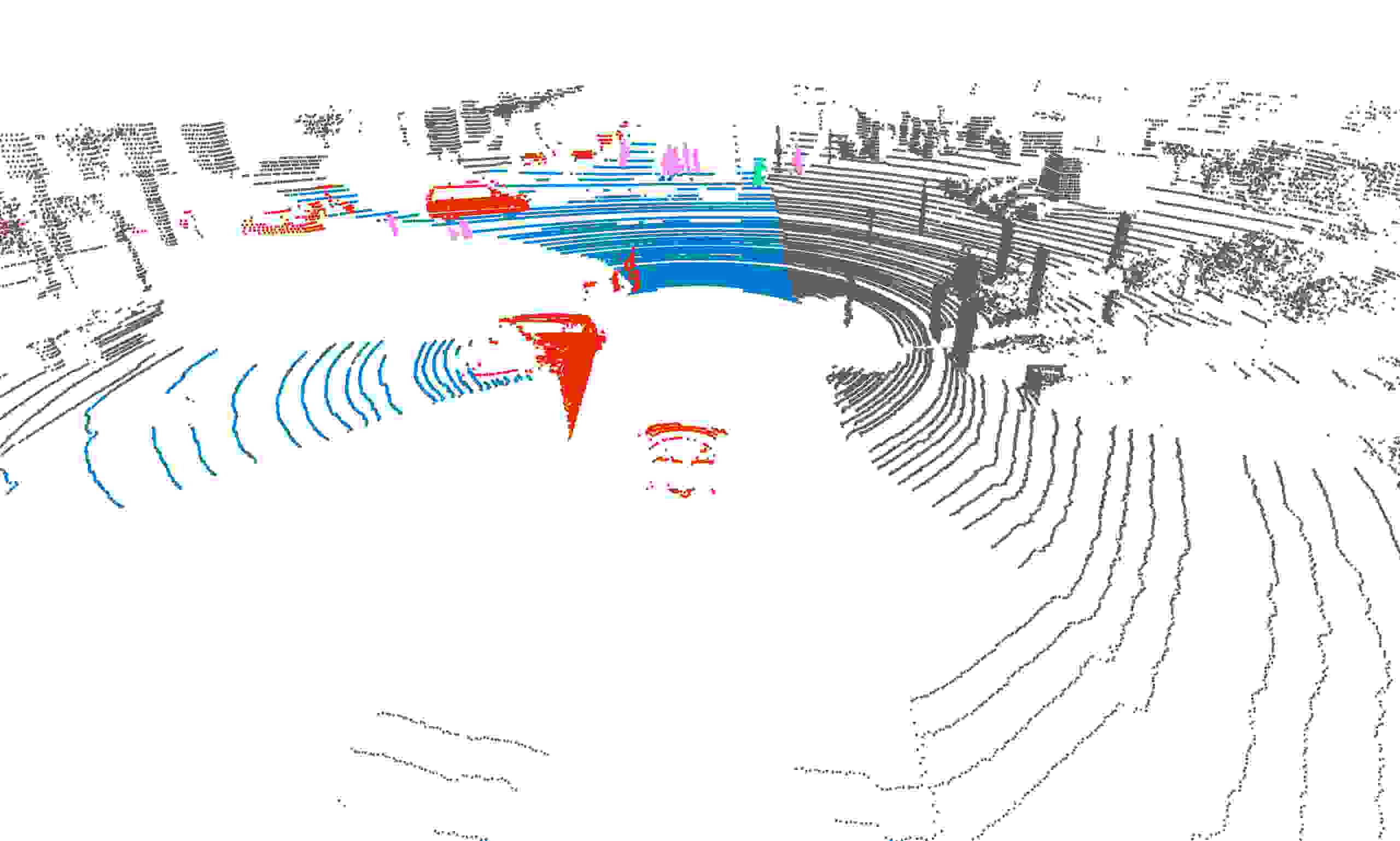}
    \put (0,55) {\colorbox{gray!30}{\scriptsize Ours: IOU 96.77, Bitrate: 14.87}}
\end{overpic}
\begin{overpic}[width=0.32\textwidth]{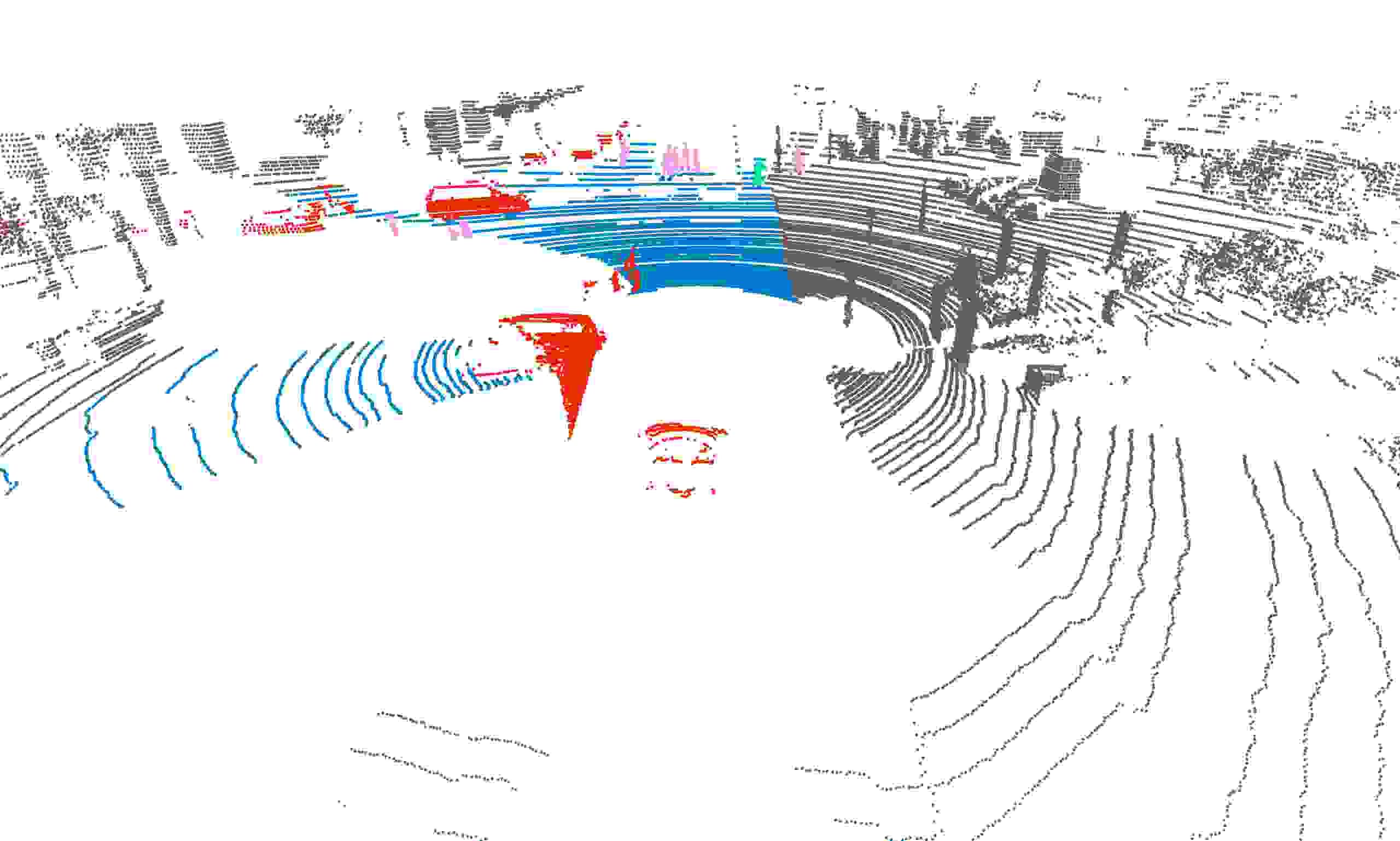}
    \put (0,55) {\colorbox{gray!30}{\scriptsize Draco: IOU 95.64, Bitrate: 16.46}}
\end{overpic}
\end{center}

\begin{center}
\begin{overpic}[width=0.32\textwidth]{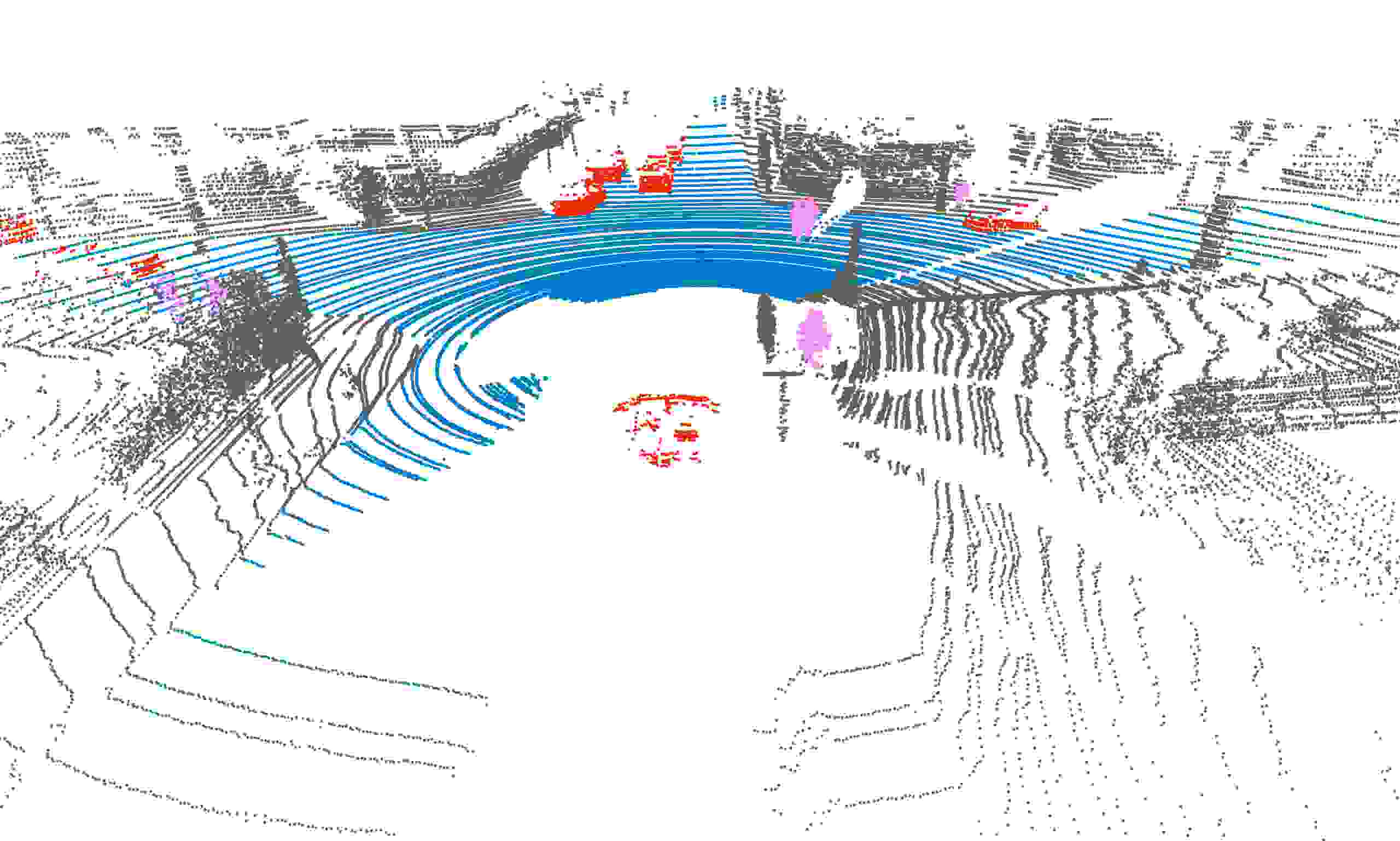}
    \put (0,55) {\colorbox{gray!30}{\scriptsize Oracle: IOU 97.21, Bitrate: 96.00}}
\end{overpic}
\begin{overpic}[width=0.32\textwidth]{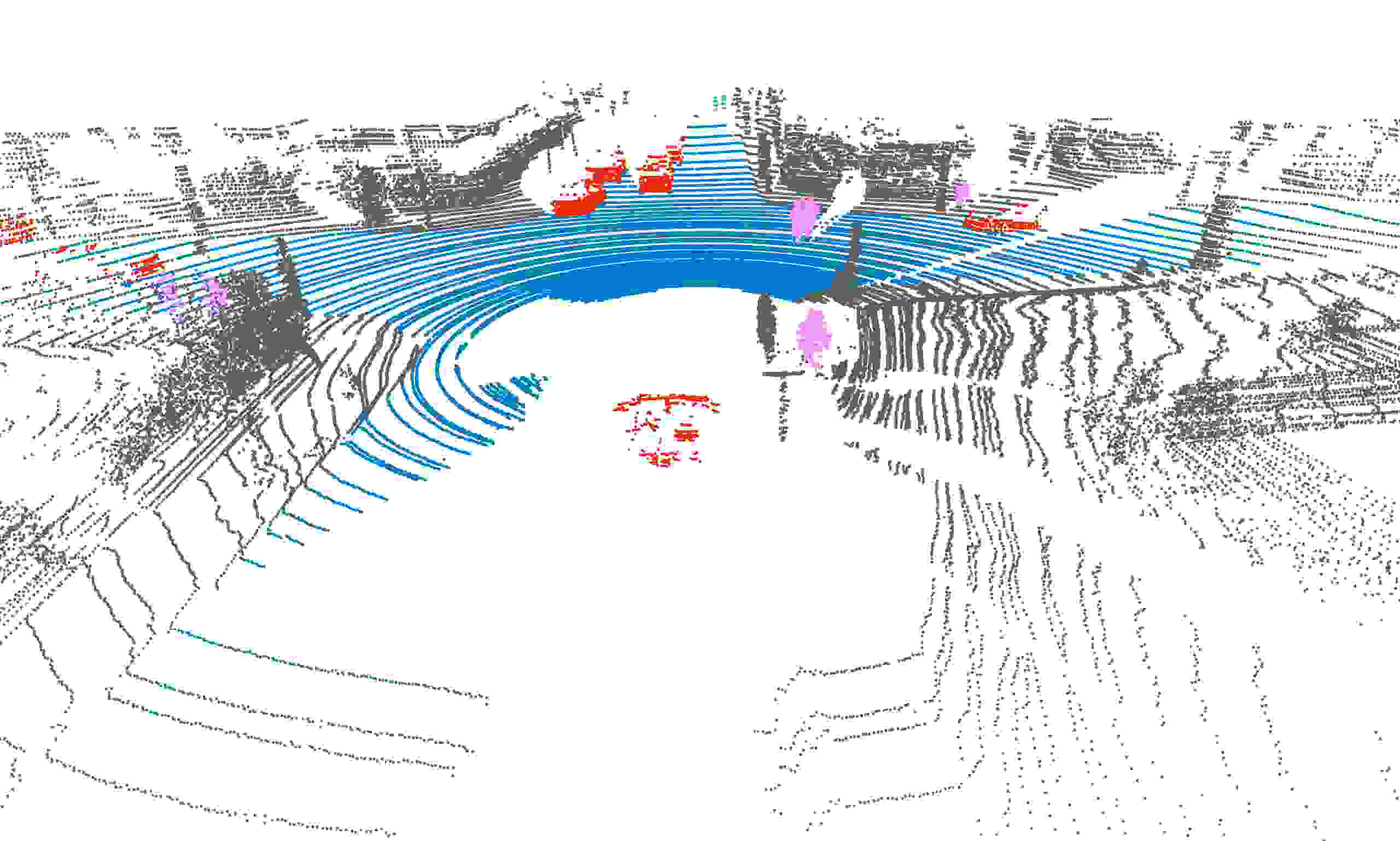}
    \put (0,55) {\colorbox{gray!30}{\scriptsize Ours: IOU 92.50, Bitrate: 11.68}}
\end{overpic}
\begin{overpic}[width=0.32\textwidth]{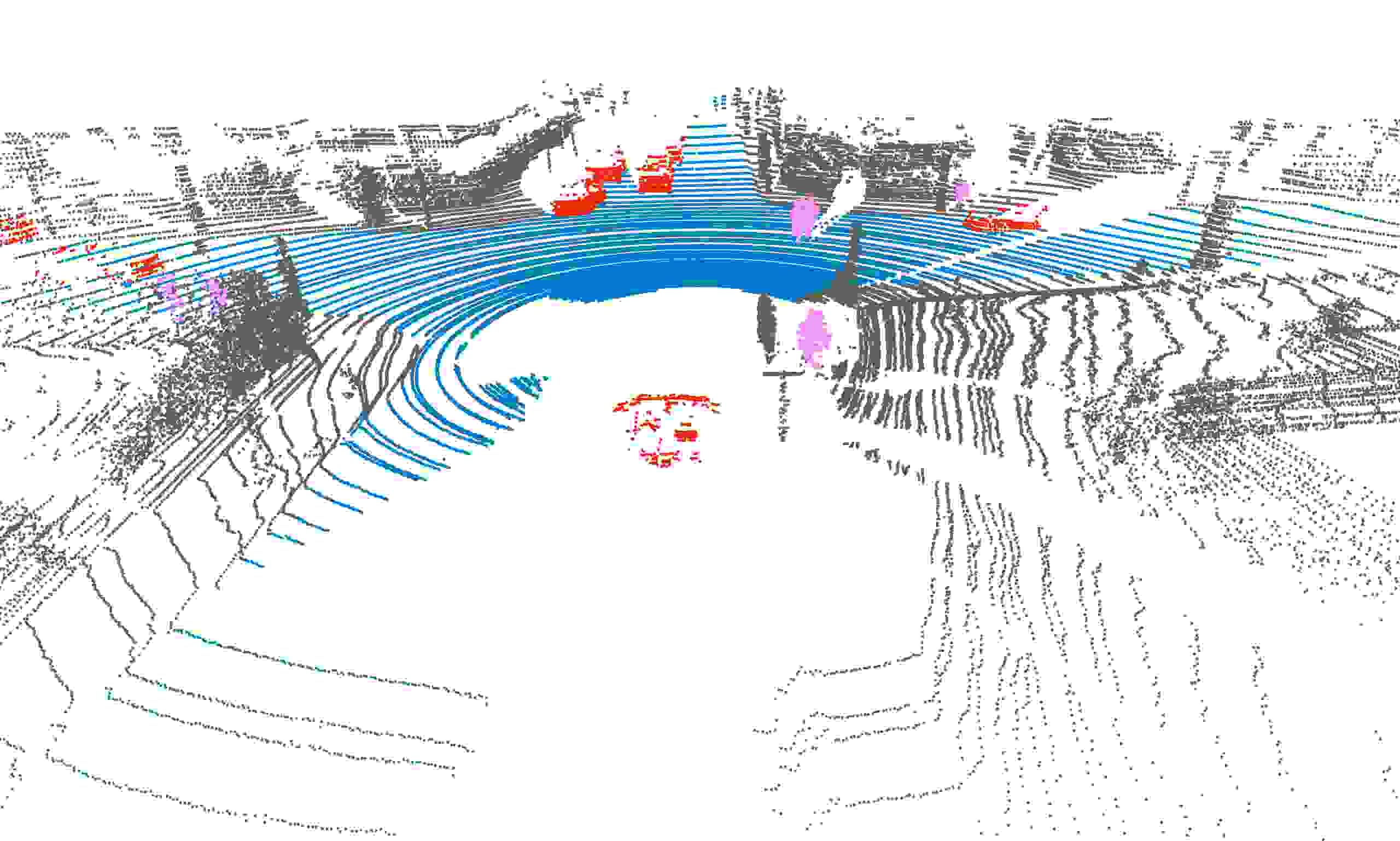}
    \put (0,55) {\colorbox{gray!30}{\scriptsize Draco: IOU 91.28, Bitrate: 12.95}}
\end{overpic}
\end{center}

\begin{center}
\begin{overpic}[width=0.32\textwidth]{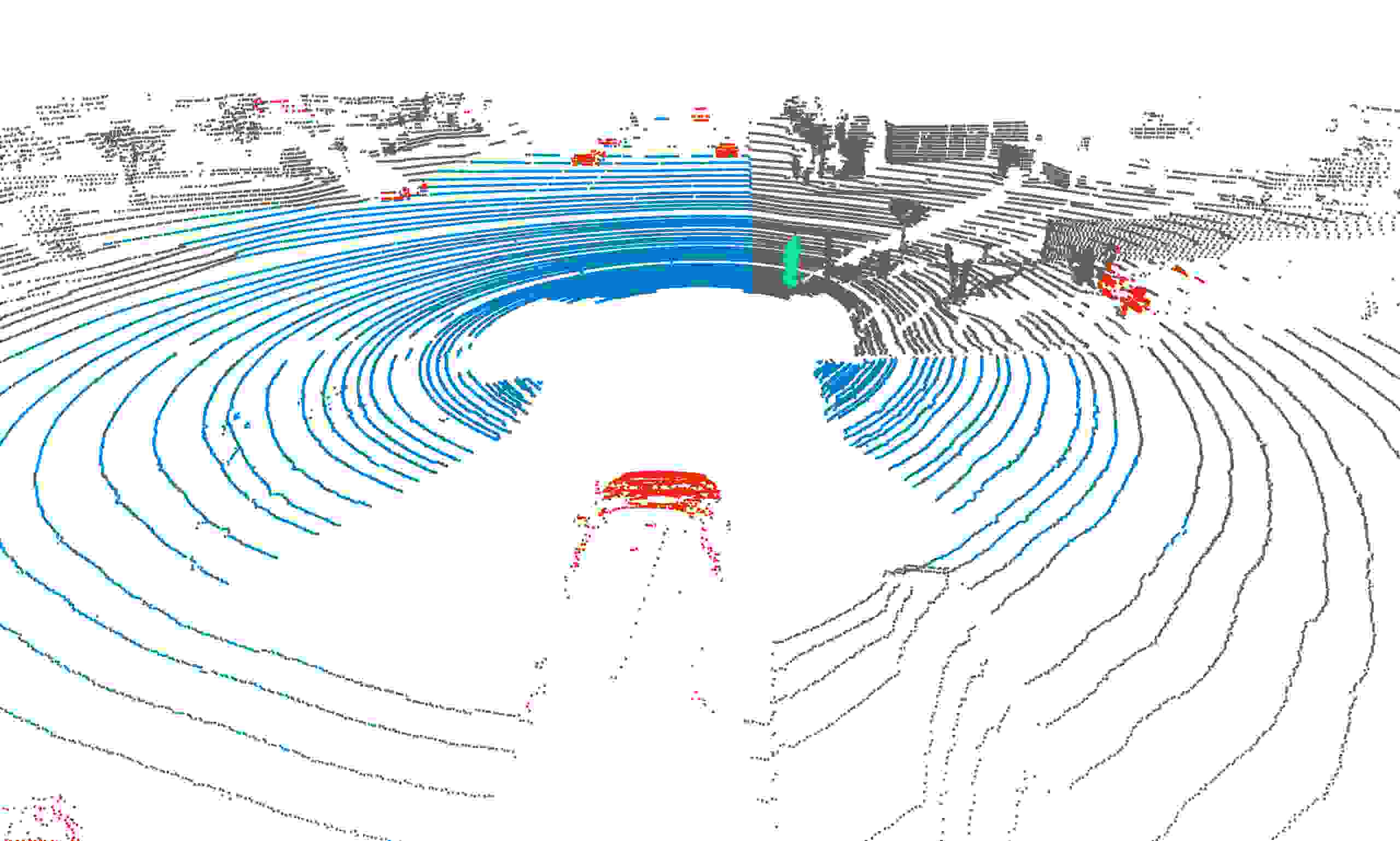}
    \put (0,55) {\colorbox{gray!30}{\scriptsize Oracle: IOU 98.09, Bitrate: 96.00}}
\end{overpic}
\begin{overpic}[width=0.32\textwidth]{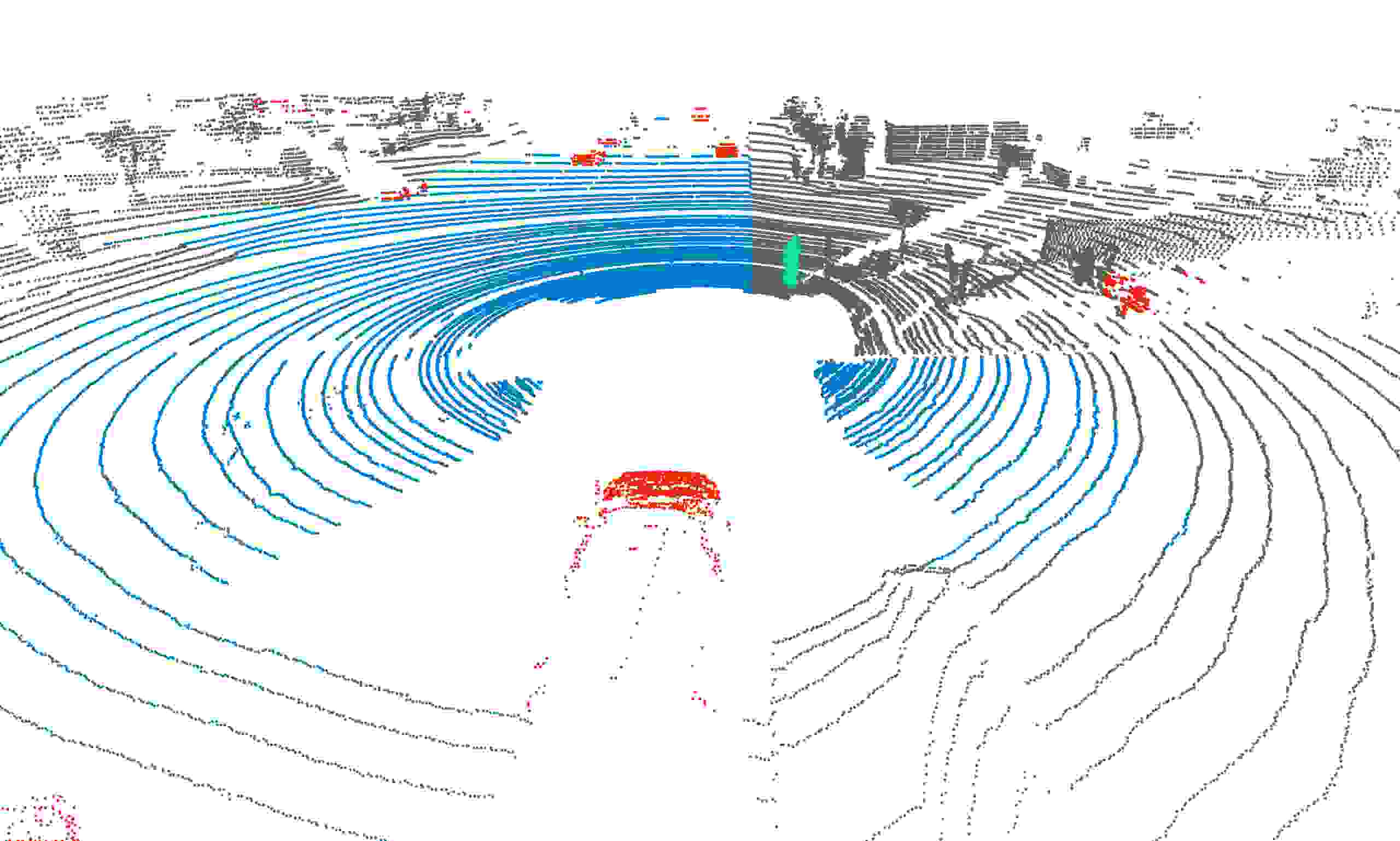}
    \put (0,55) {\colorbox{gray!30}{\scriptsize Ours: IOU 86.44, Bitrate: 8.62}}
\end{overpic}
\begin{overpic}[width=0.32\textwidth]{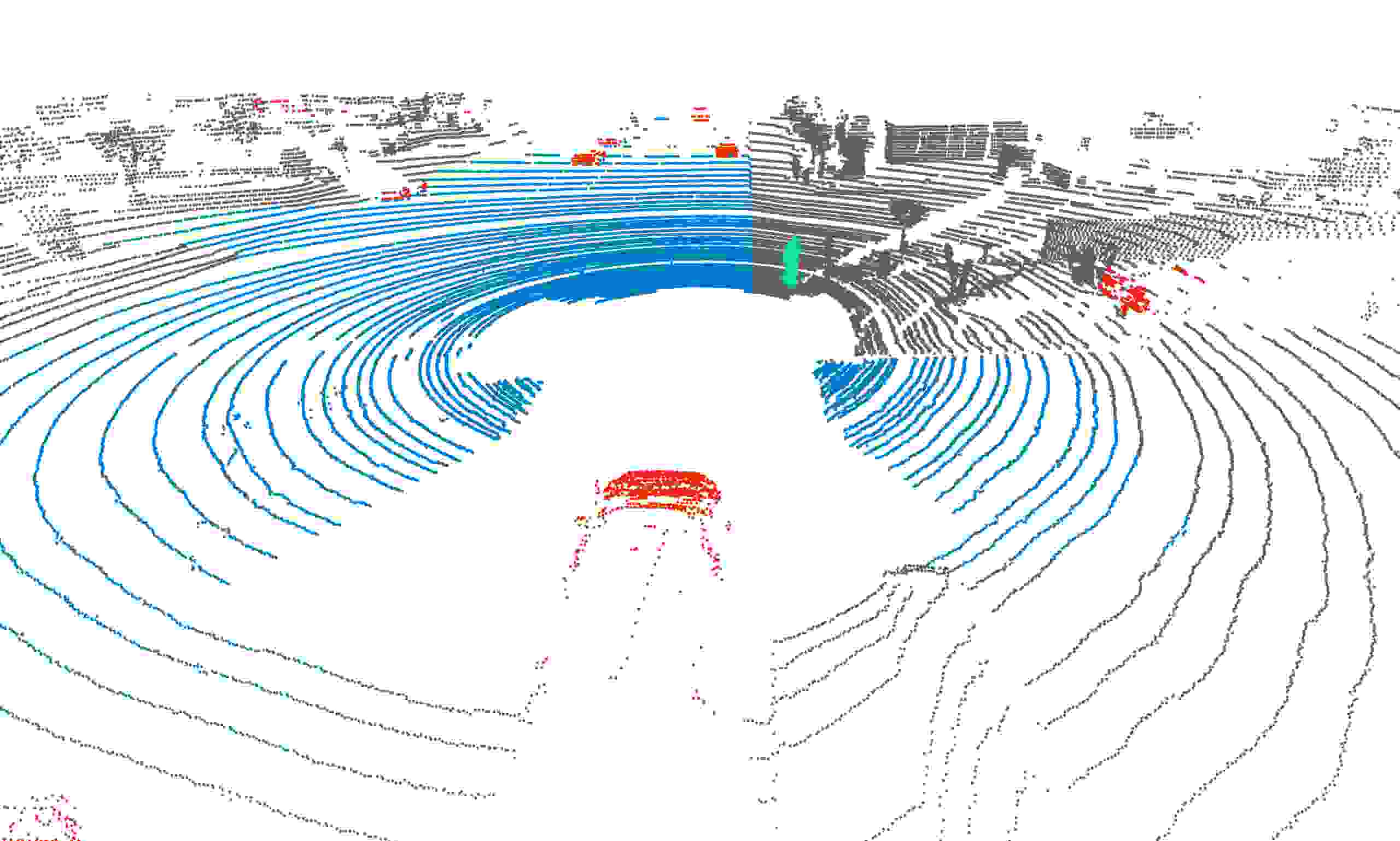}
    \put (0,55) {\colorbox{gray!30}{\scriptsize Draco: IOU 82.93, Bitrate: 10.21}}
\end{overpic}
\end{center}

\begin{center}
\begin{overpic}[width=0.32\textwidth]{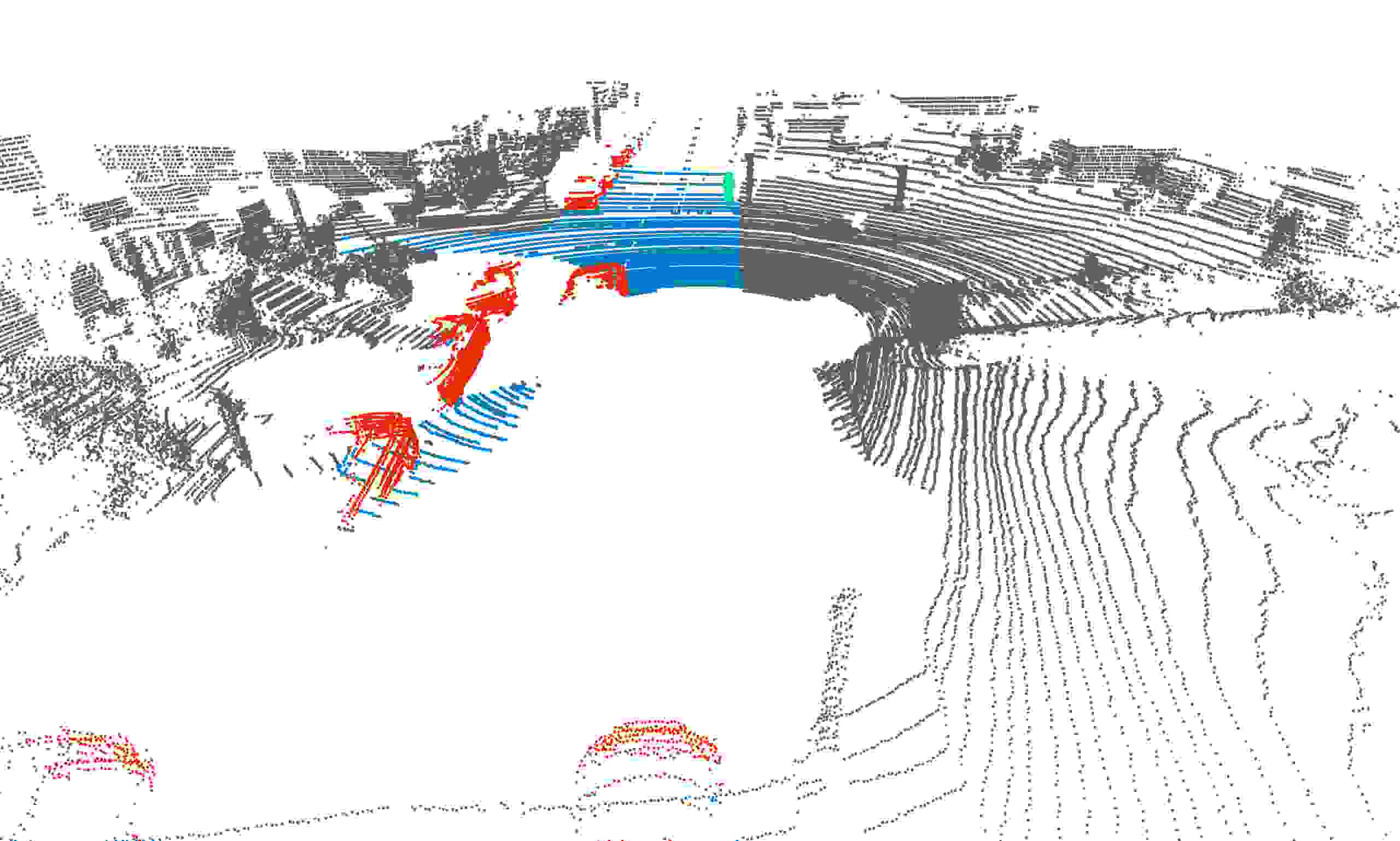}
    \put (0,55) {\colorbox{gray!30}{\scriptsize Oracle: IOU 98.01, Bitrate: 96.00}}
\end{overpic}
\begin{overpic}[width=0.32\textwidth]{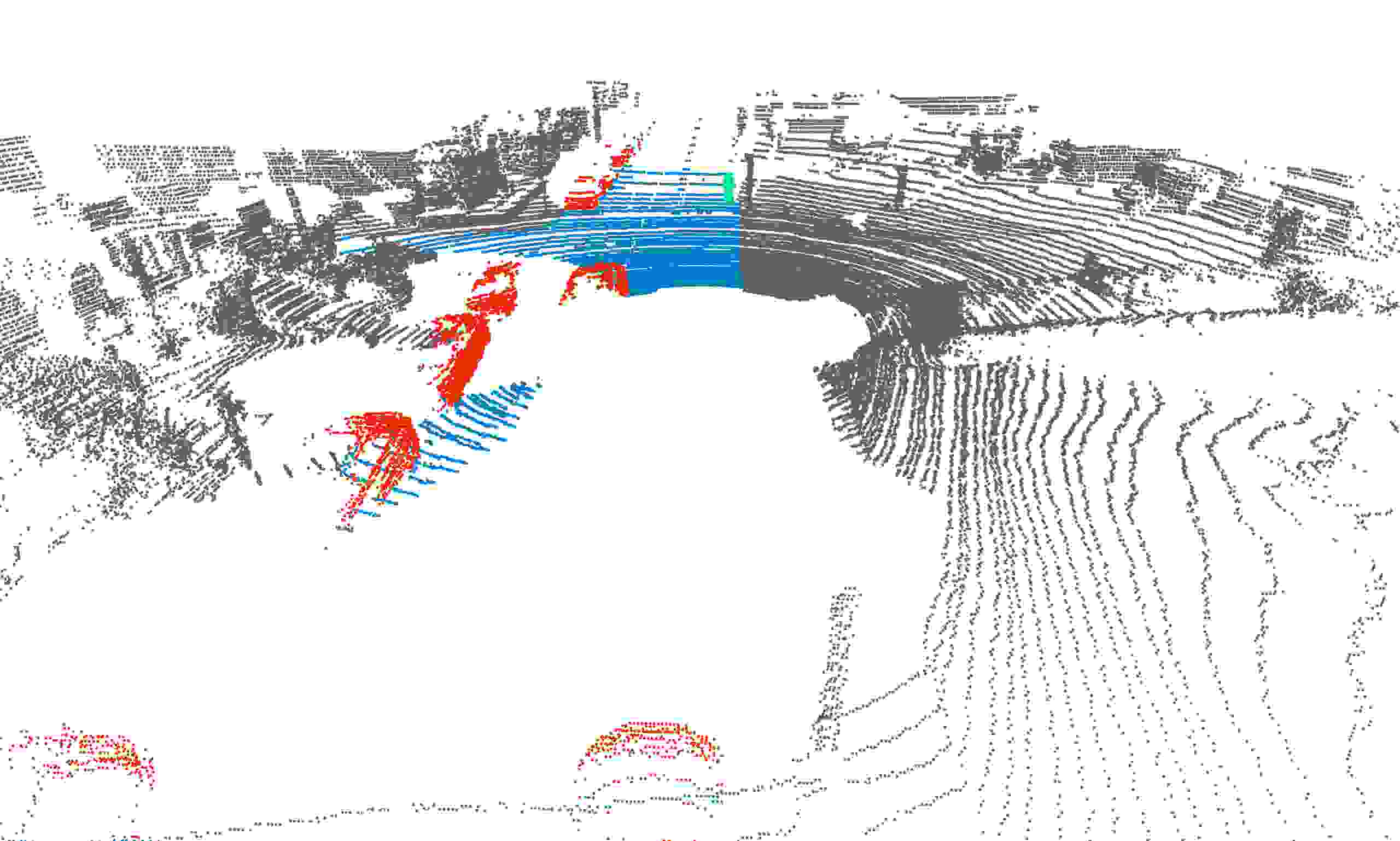}
    \put (0,55) {\colorbox{gray!30}{\scriptsize Ours: IOU 81.95, Bitrate: 5.52}}
\end{overpic}
\begin{overpic}[width=0.32\textwidth]{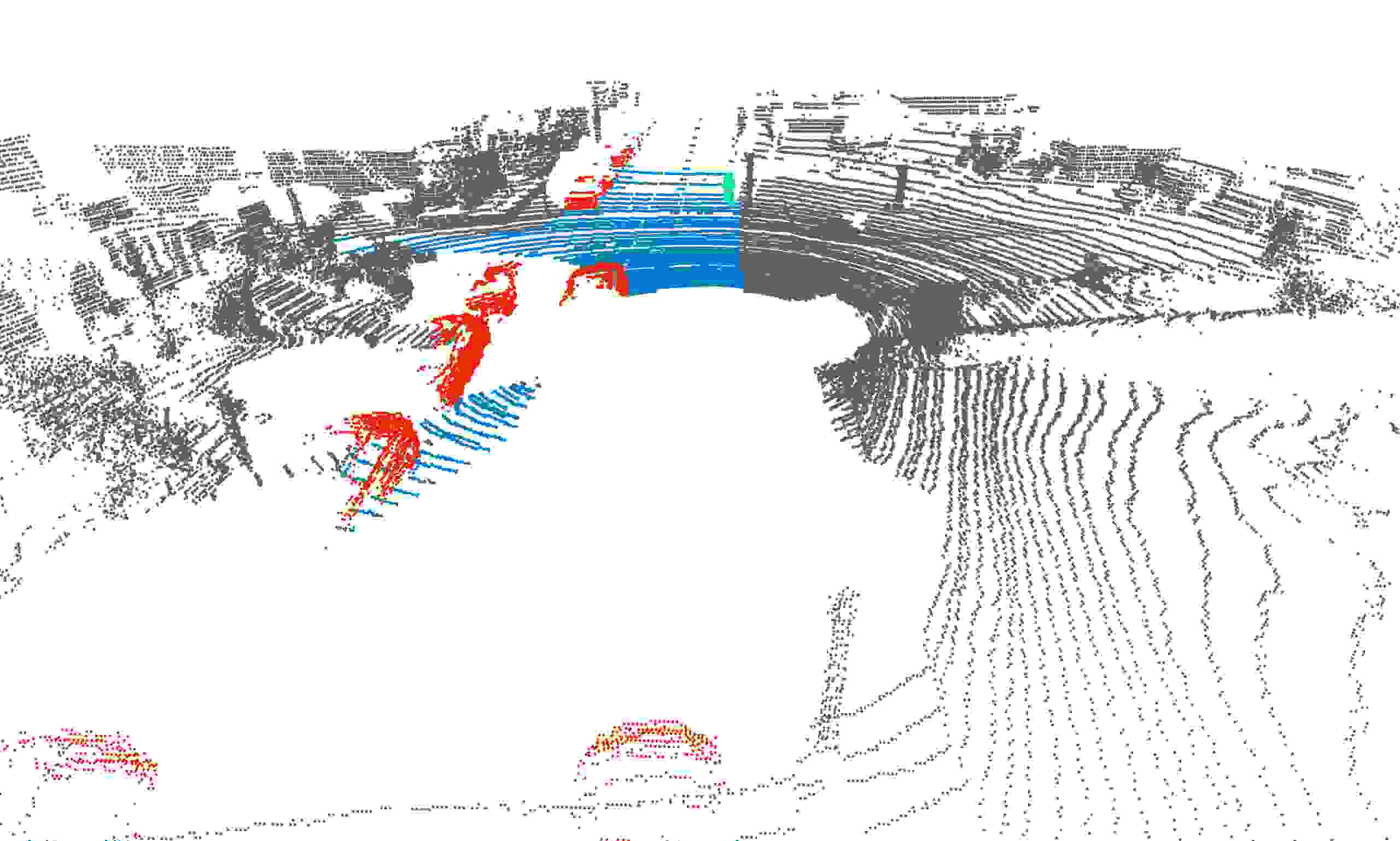}
    \put (0,55) {\colorbox{gray!30}{\scriptsize Draco: IOU 80.61, Bitrate: 6.99}}
\end{overpic}
\end{center}

\begin{center}
\begin{overpic}[width=0.32\textwidth]{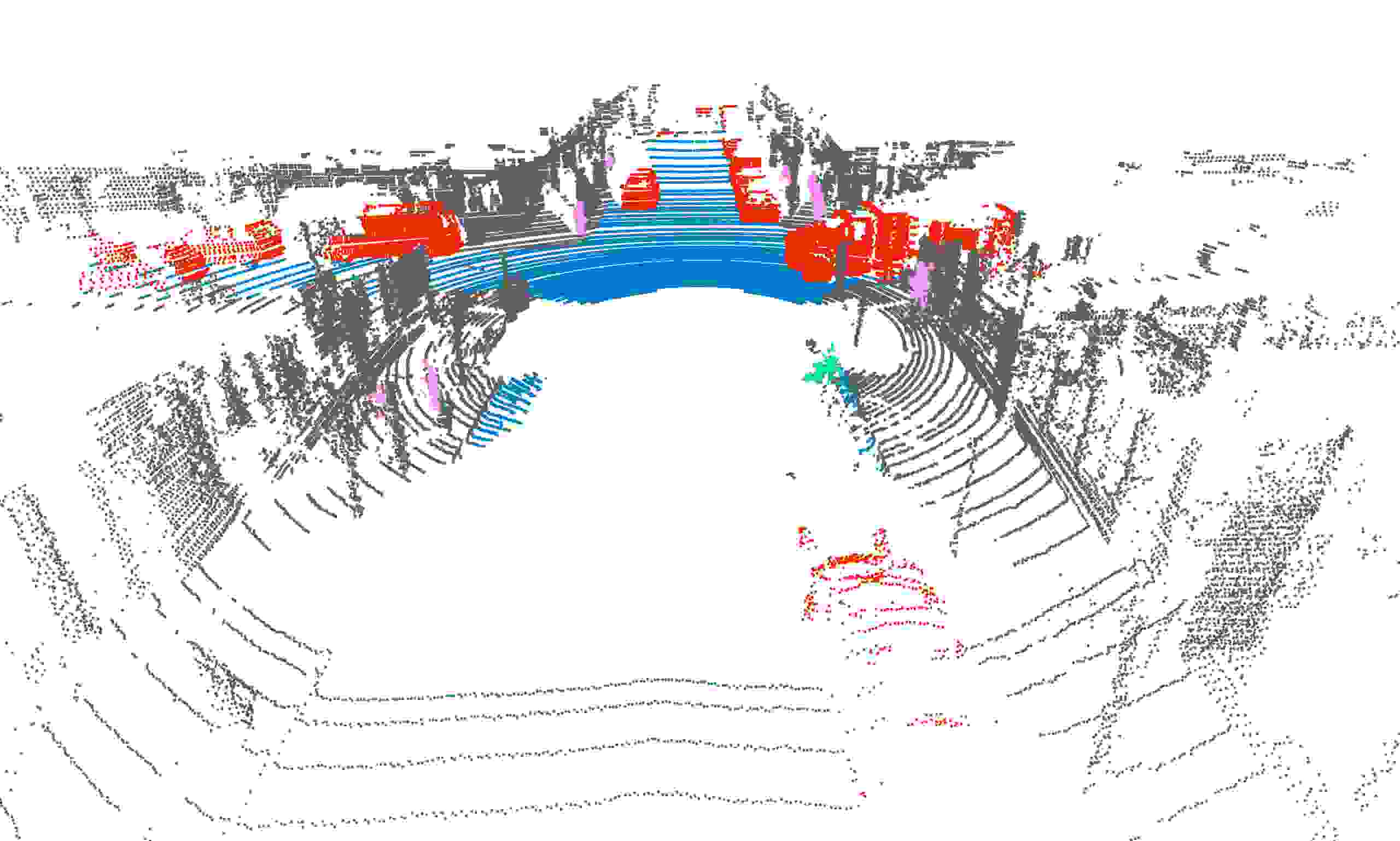}
    \put (0,55) {\colorbox{gray!30}{\scriptsize Oracle: IOU 92.31, Bitrate: 96.00}}
\end{overpic}
\begin{overpic}[width=0.32\textwidth]{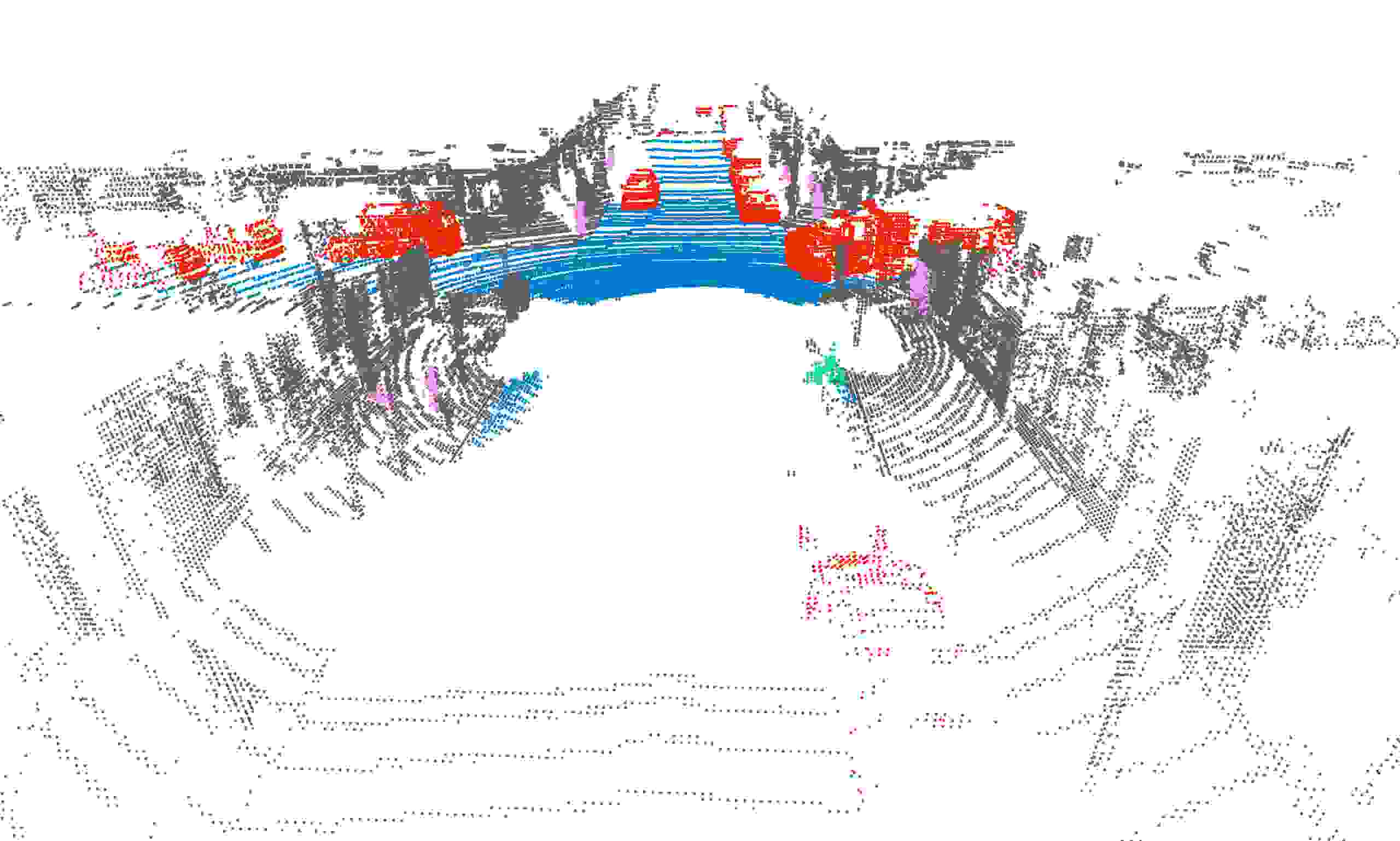}
    \put (0,55) {\colorbox{gray!30}{\scriptsize Ours: IOU 58.62, Bitrate: 2.67}}
\end{overpic}
\begin{overpic}[width=0.32\textwidth]{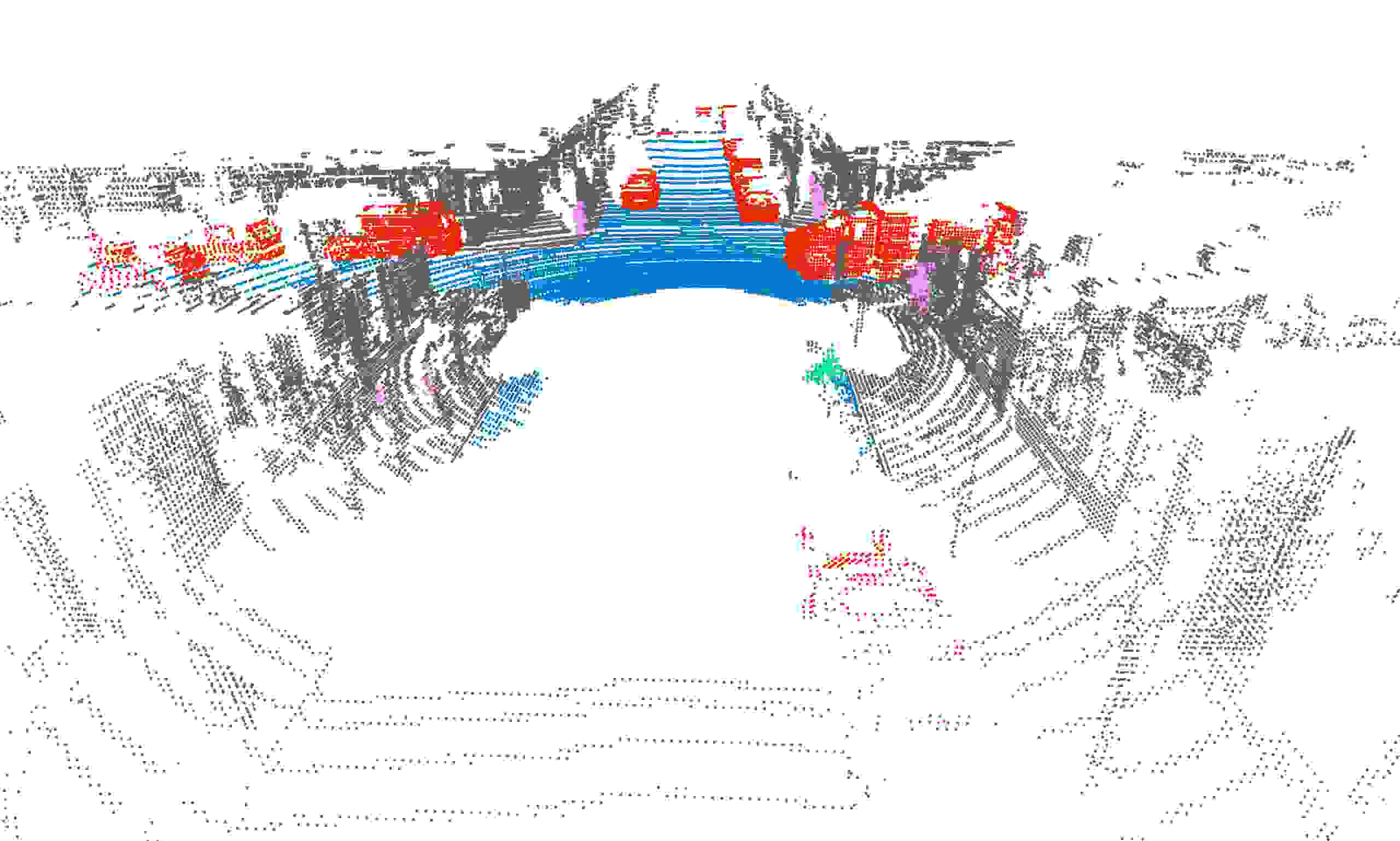}
    \put (0,55) {\colorbox{gray!30}{\scriptsize Draco: IOU 57.18, Bitrate: 3.99}}
\end{overpic}
\end{center}

\begin{center}
\begin{overpic}[width=0.32\textwidth]{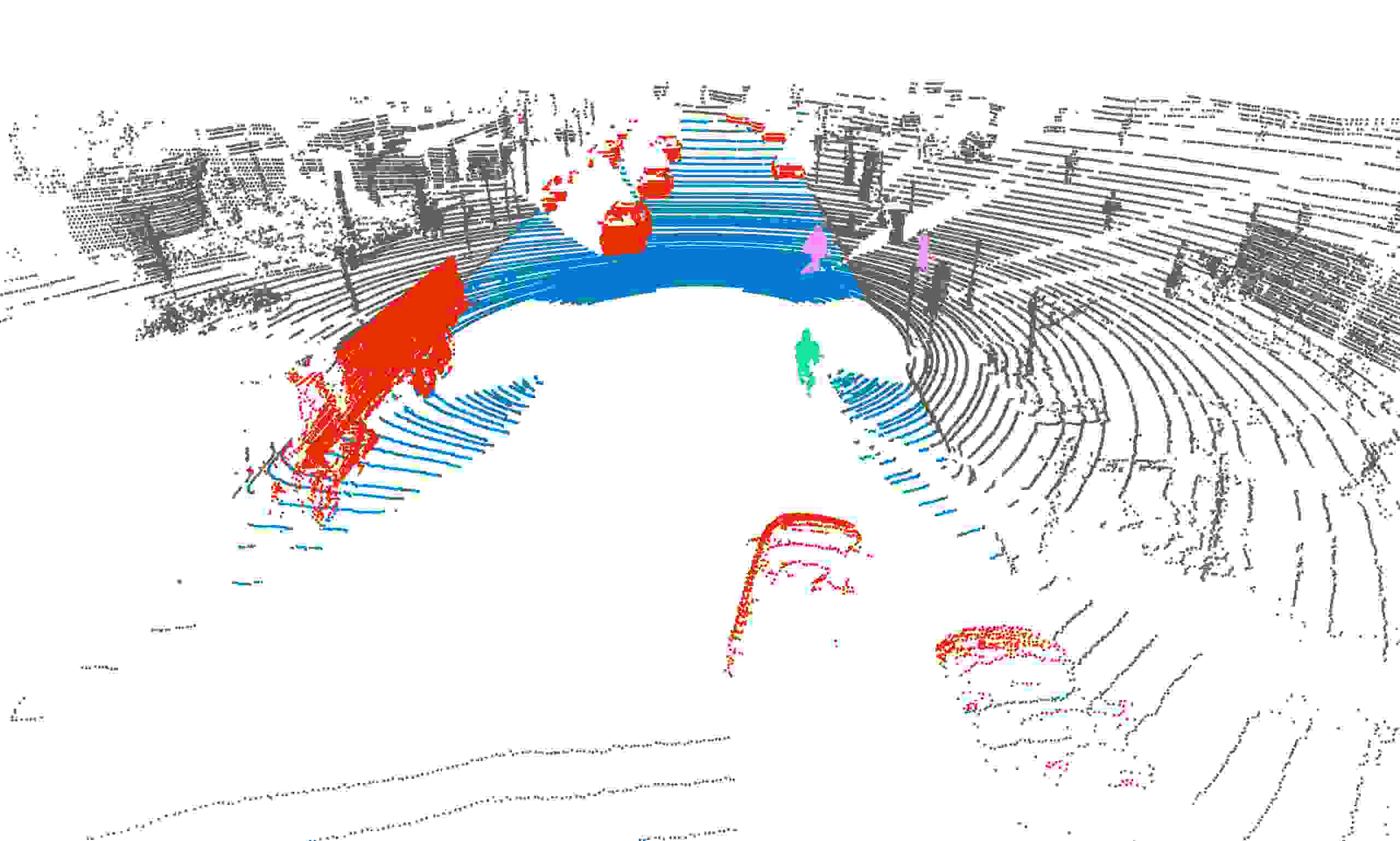}
    \put (0,55) {\colorbox{gray!30}{\scriptsize Oracle: IOU 95.53, Bitrate: 96.00}}
\end{overpic}
\begin{overpic}[width=0.32\textwidth]{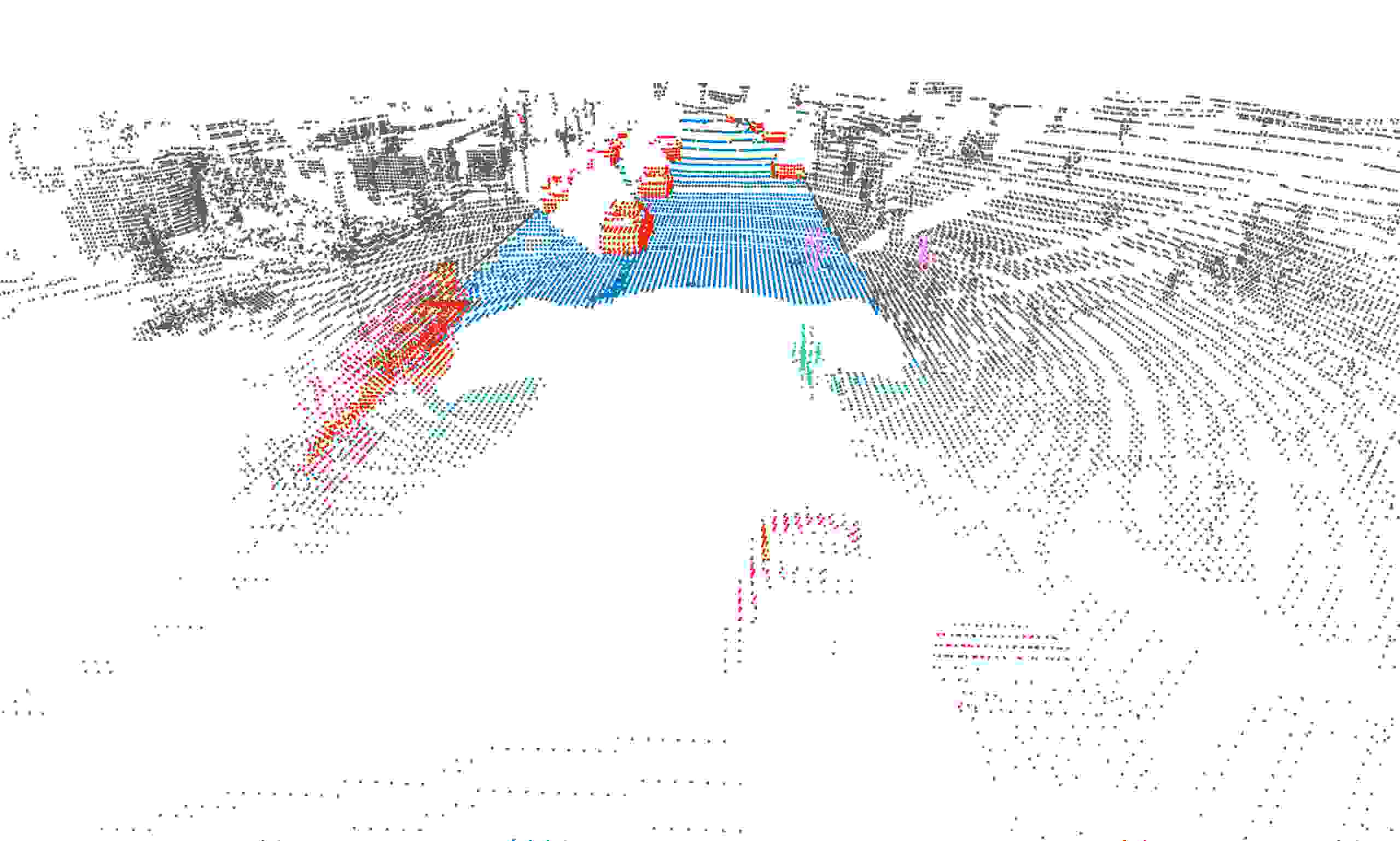}
    \put (0,55) {\colorbox{gray!30}{\scriptsize Ours: IOU 30.87, Bitrate: 1.78}}
\end{overpic}
\begin{overpic}[width=0.32\textwidth]{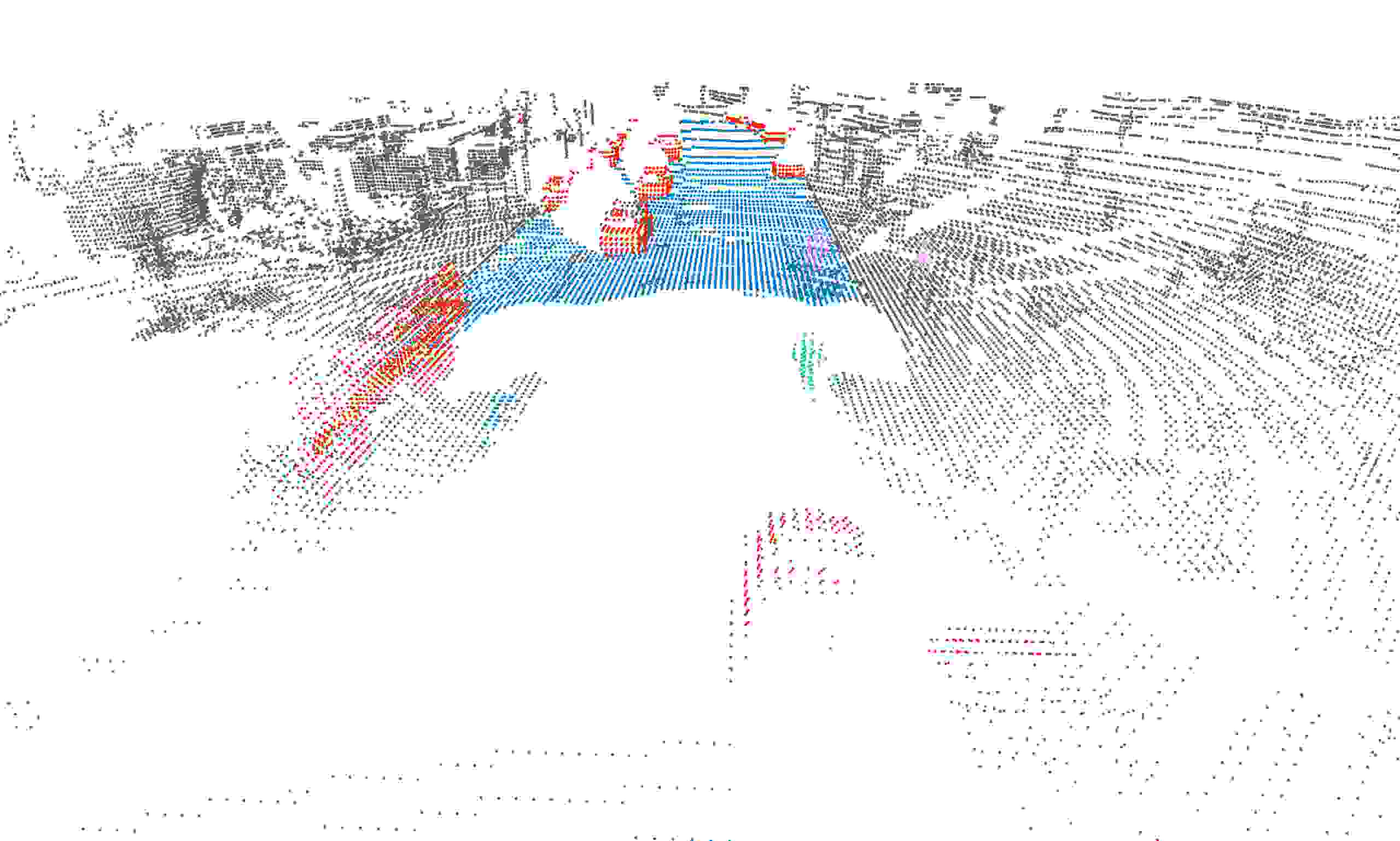}
    \put (0,55) {\colorbox{gray!30}{\scriptsize Draco: IOU 25.28, Bitrate: 2.90}}
\end{overpic}
\end{center}

\caption{Qualitative results of semantic segmentation for NorthAmerica. IOU is averaged over all classes.}
\label{fig:semantic_na}
\end{figure*}

\begin{figure*}[h]
\begin{center}
\begin{overpic}[width=0.32\textwidth]{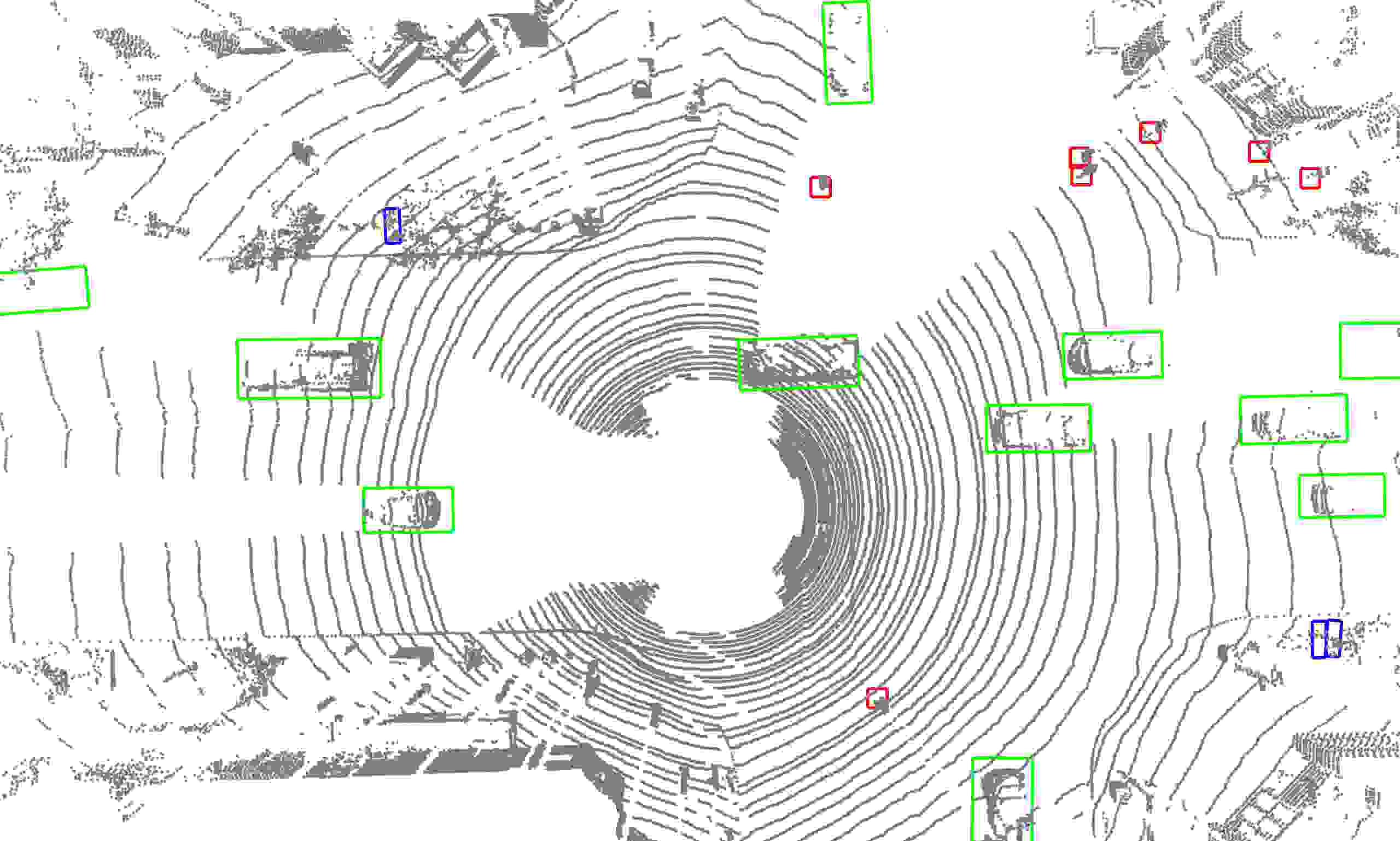}
    \put (0,55) {\colorbox{gray!30}{\scriptsize Oracle: AP: 93.02, Bitrate: 96.00}}
\end{overpic}
\begin{overpic}[width=0.32\textwidth]{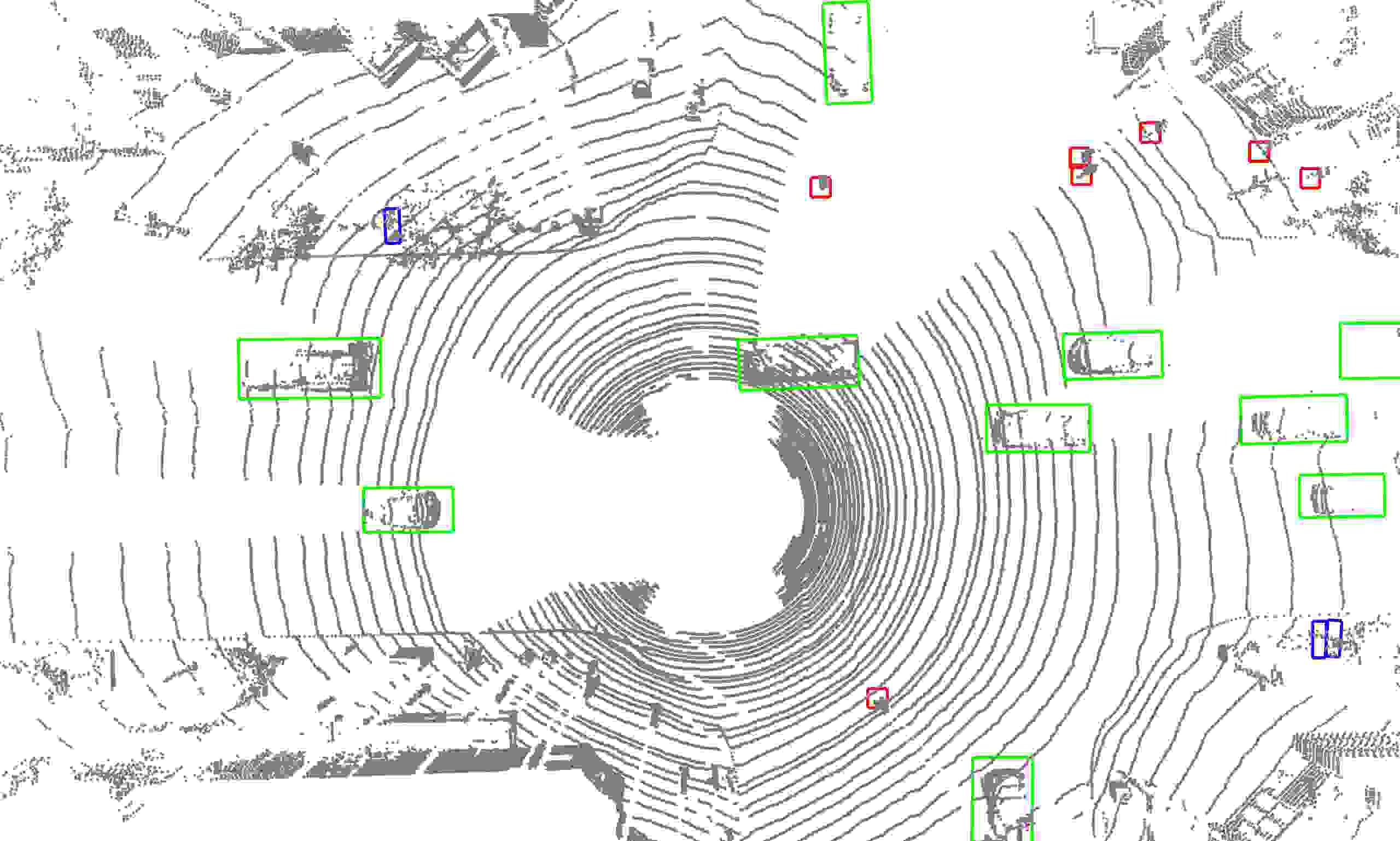}
    \put (0,55) {\colorbox{gray!30}{\scriptsize Ours: AP: 93.00, Bitrate: 14.64}}
\end{overpic}
\begin{overpic}[width=0.32\textwidth]{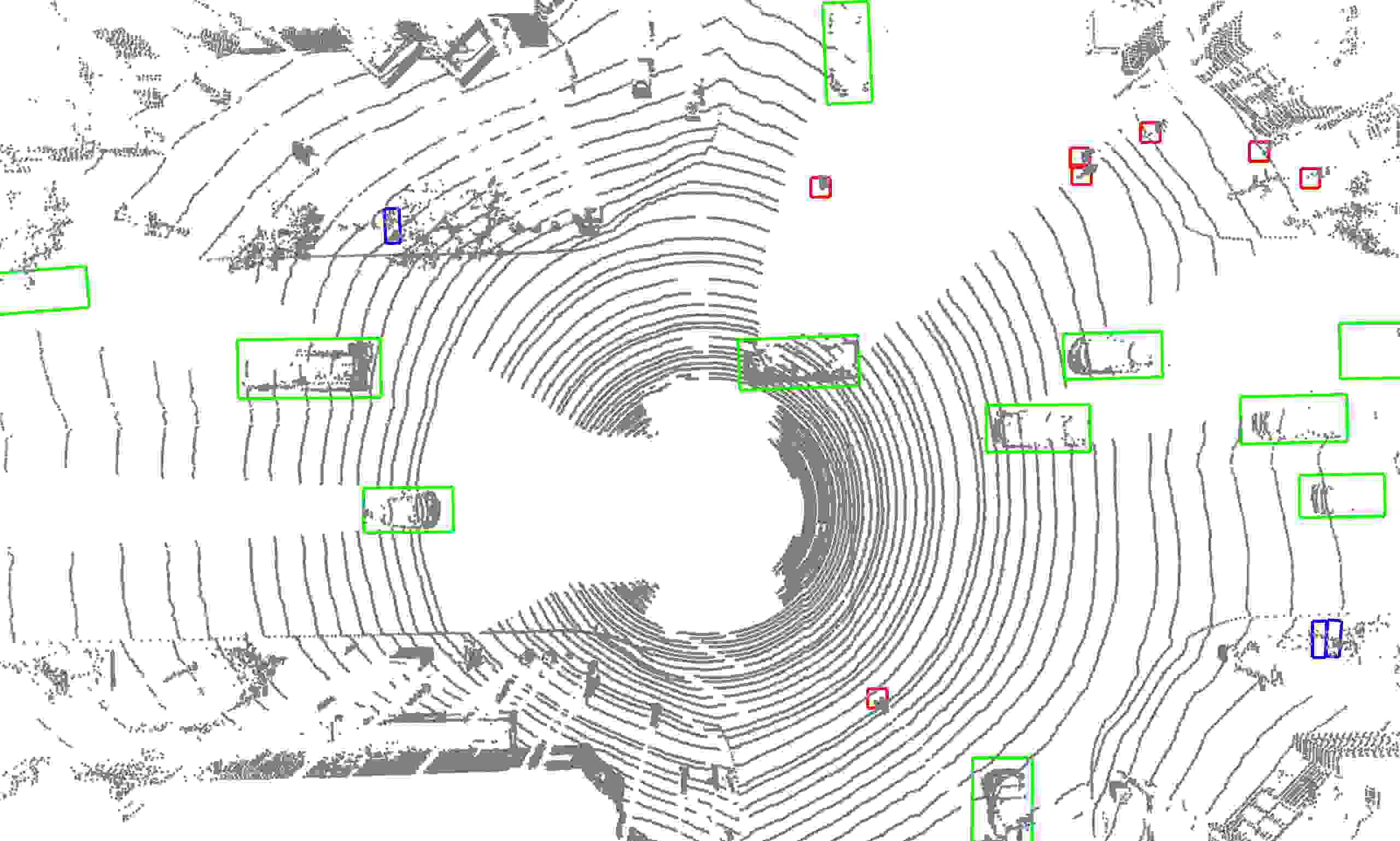}
    \put (0,55) {\colorbox{gray!30}{\scriptsize Draco: AP: 92.78, Bitrate: 18.87}}
\end{overpic}
\end{center}

\begin{center}
\begin{overpic}[width=0.32\textwidth]{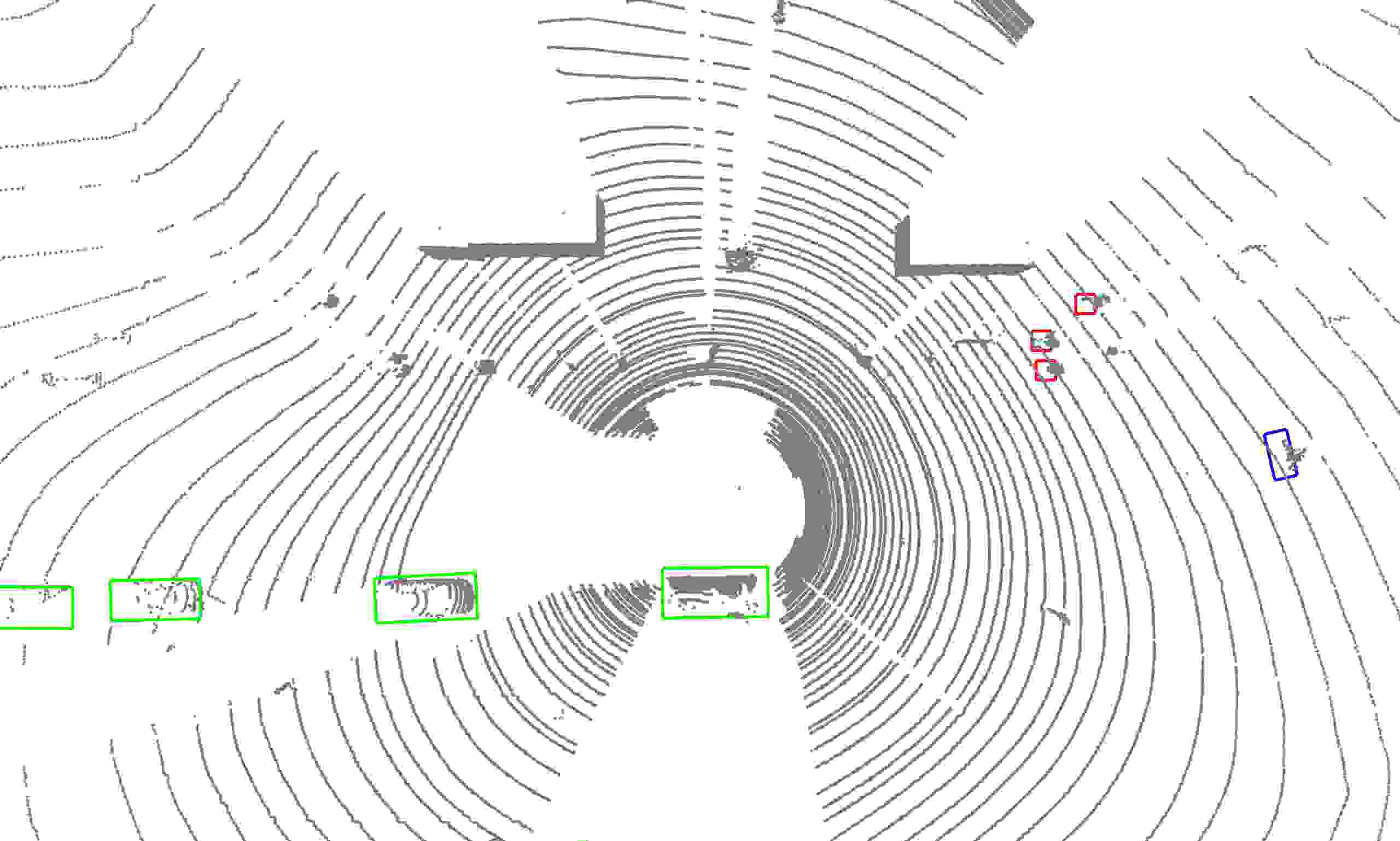}
    \put (0,55) {\colorbox{gray!30}{\scriptsize Oracle: AP: 88.49, Bitrate: 96.00}}
\end{overpic}
\begin{overpic}[width=0.32\textwidth]{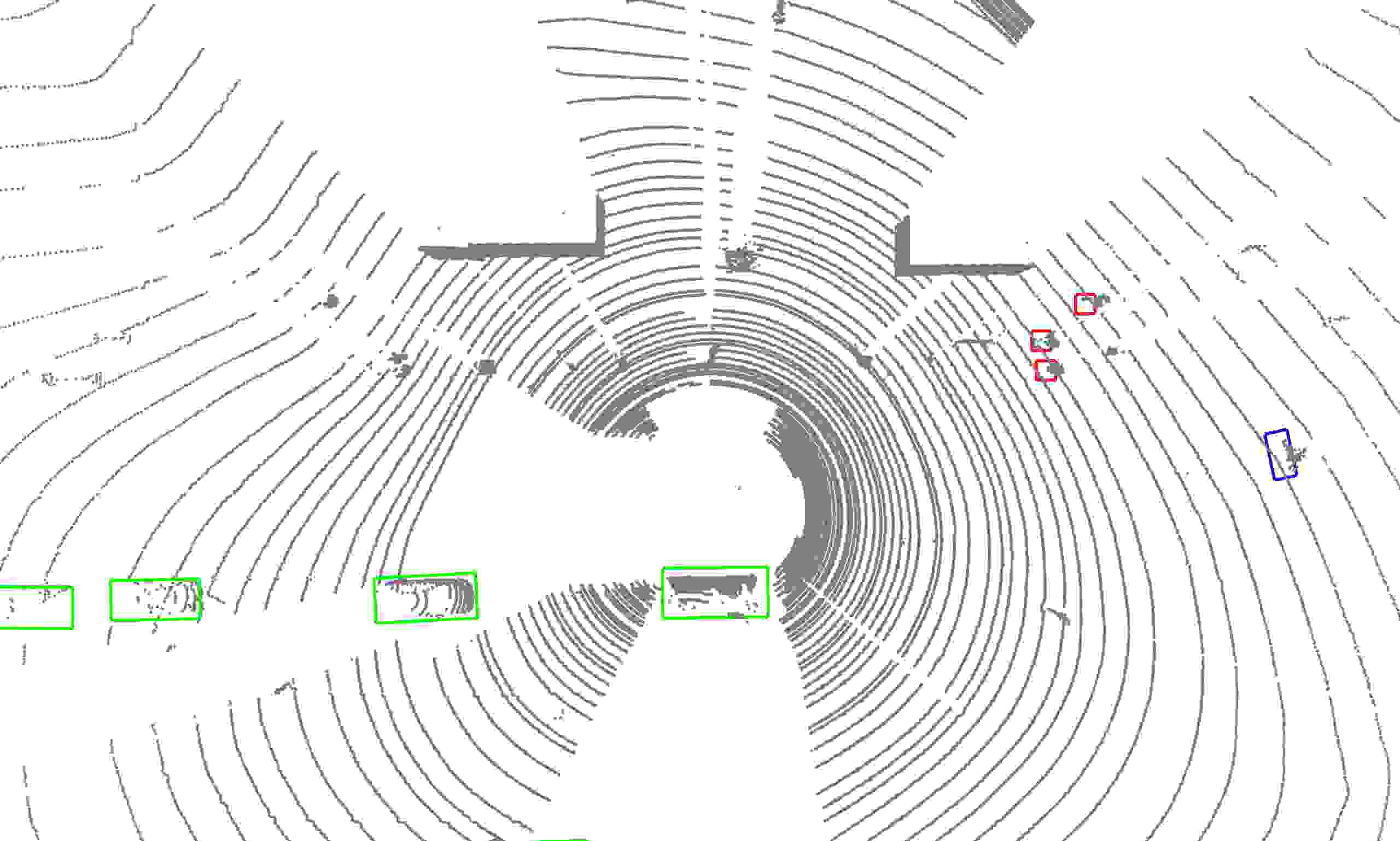}
    \put (0,55) {\colorbox{gray!30}{\scriptsize Ours: AP: 88.06, Bitrate: 10.72}}
\end{overpic}
\begin{overpic}[width=0.32\textwidth]{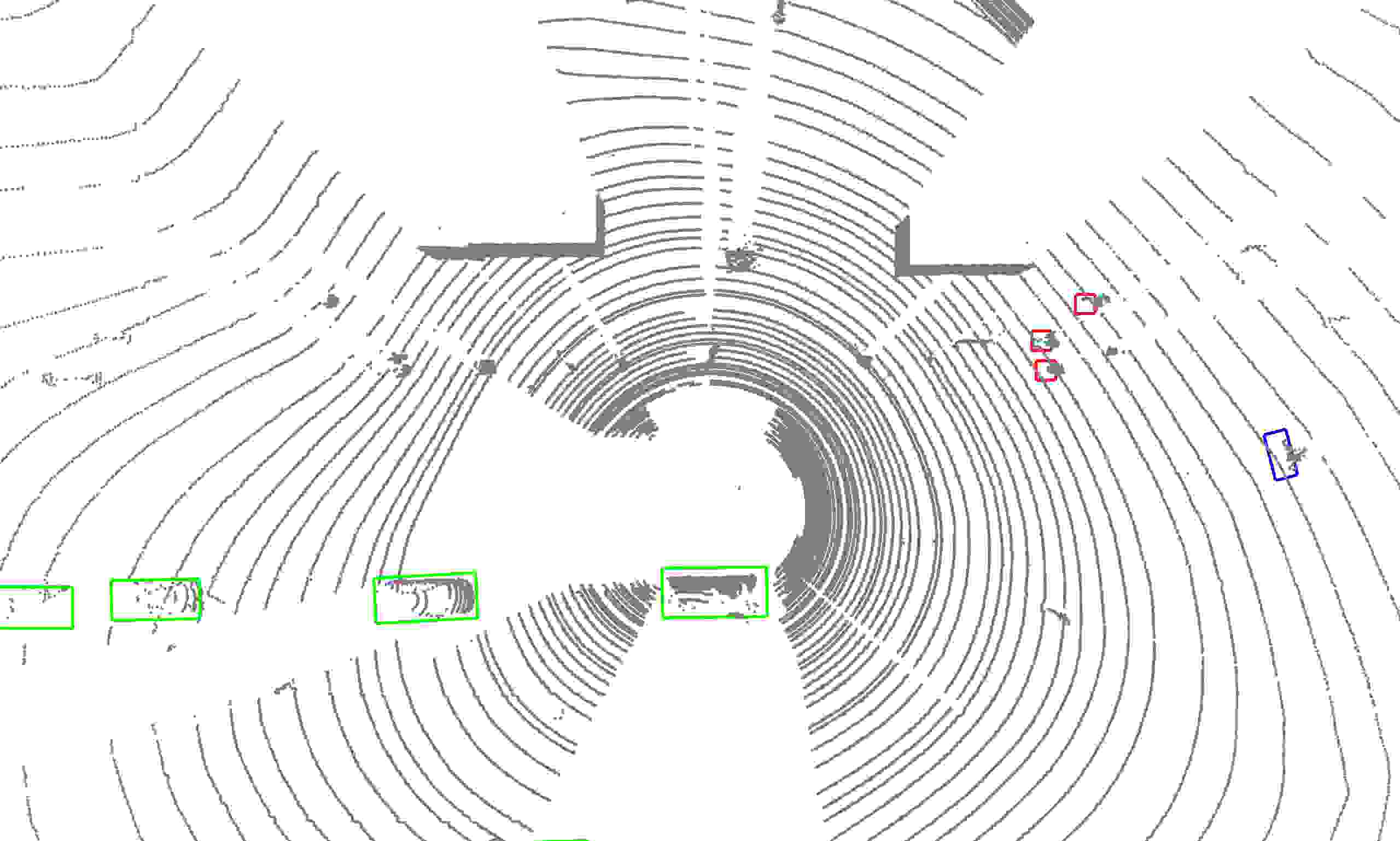}
    \put (0,55) {\colorbox{gray!30}{\scriptsize Draco: AP: 88.27, Bitrate: 11.97}}
\end{overpic}
\end{center}

\begin{center}
\begin{overpic}[width=0.32\textwidth]{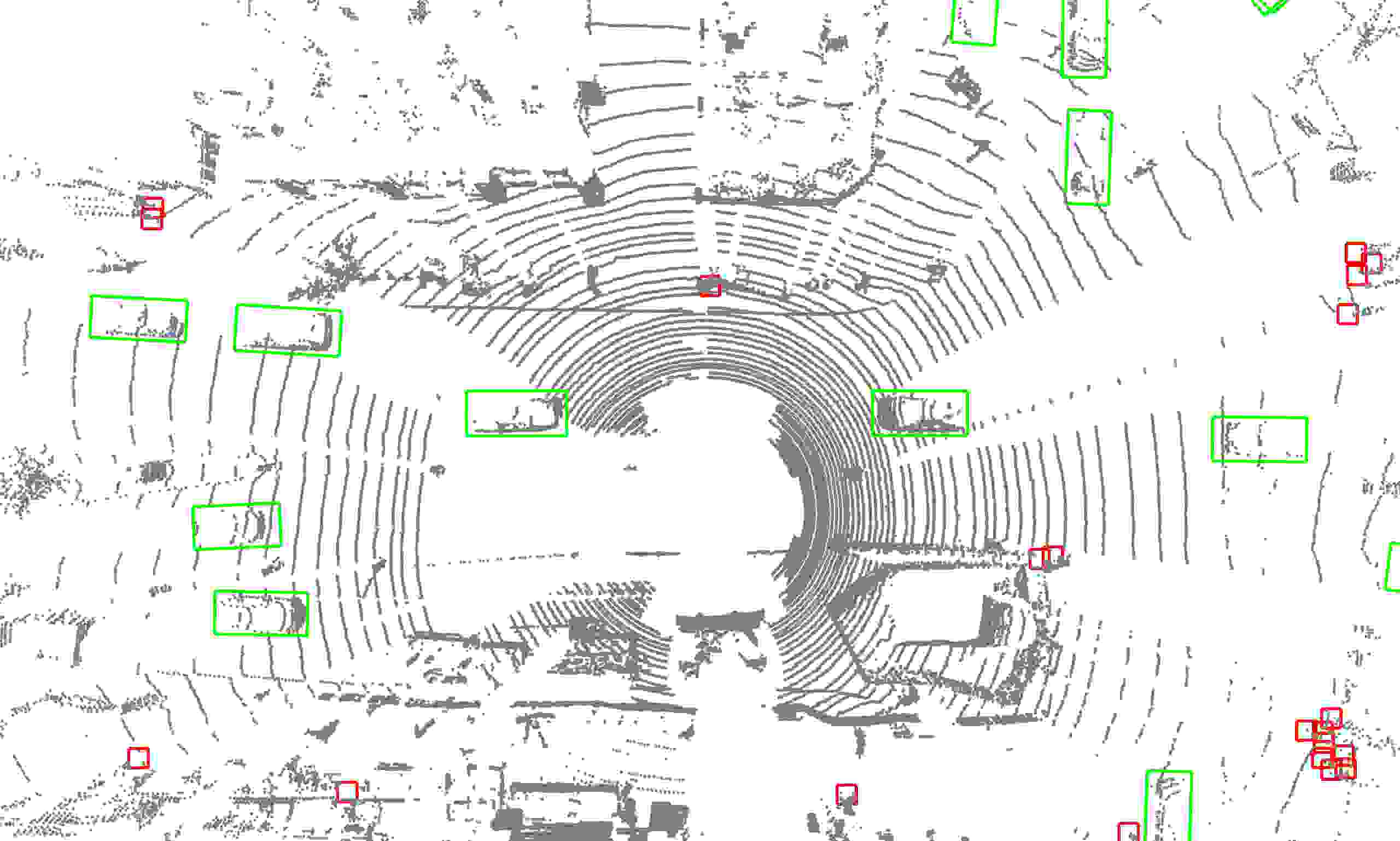}
    \put (0,55) {\colorbox{gray!30}{\scriptsize Oracle: AP: 87.85, Bitrate: 96.00}}
\end{overpic}
\begin{overpic}[width=0.32\textwidth]{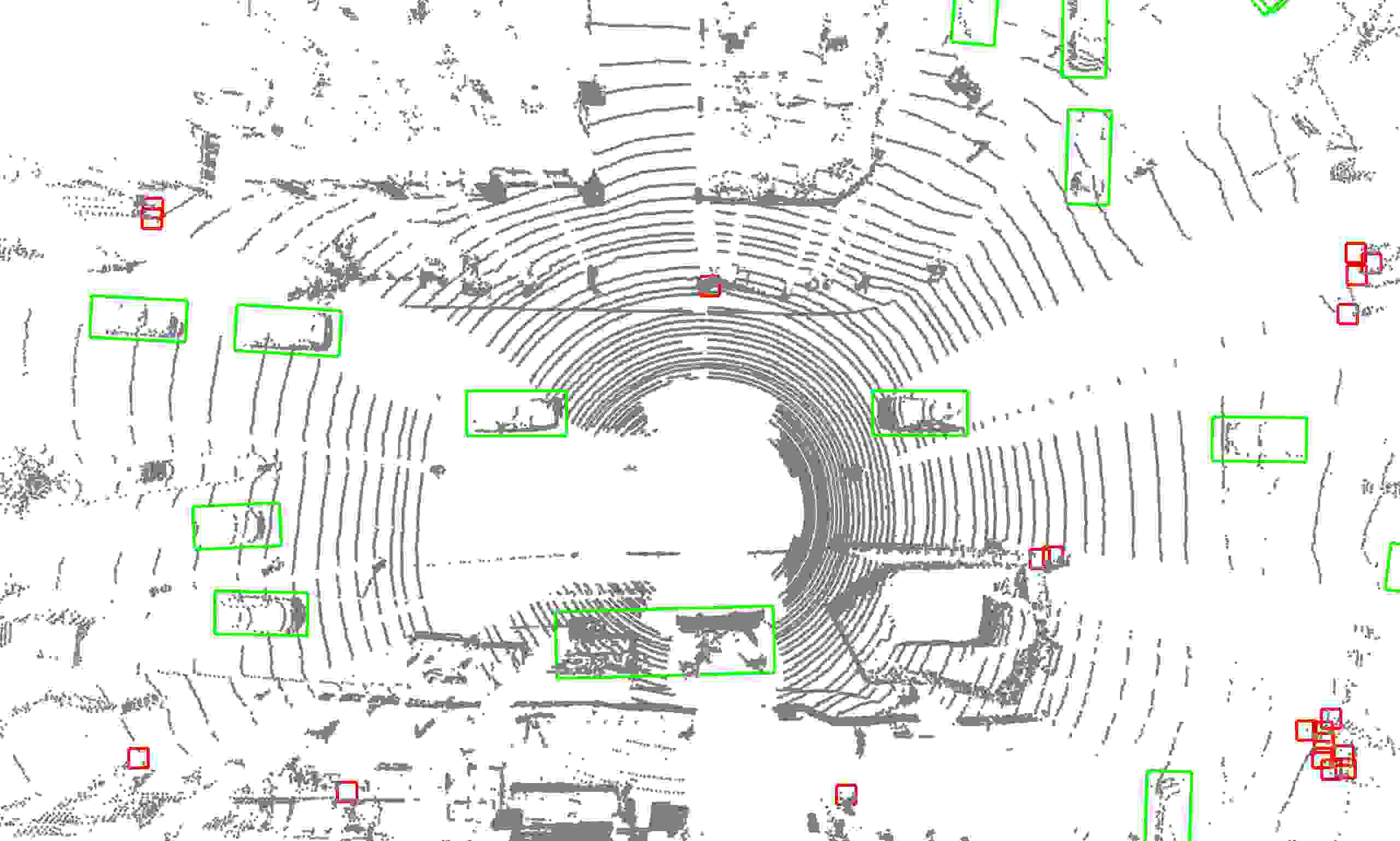}
    \put (0,55) {\colorbox{gray!30}{\scriptsize Ours: AP: 87.50, Bitrate: 8.52}}
\end{overpic}
\begin{overpic}[width=0.32\textwidth]{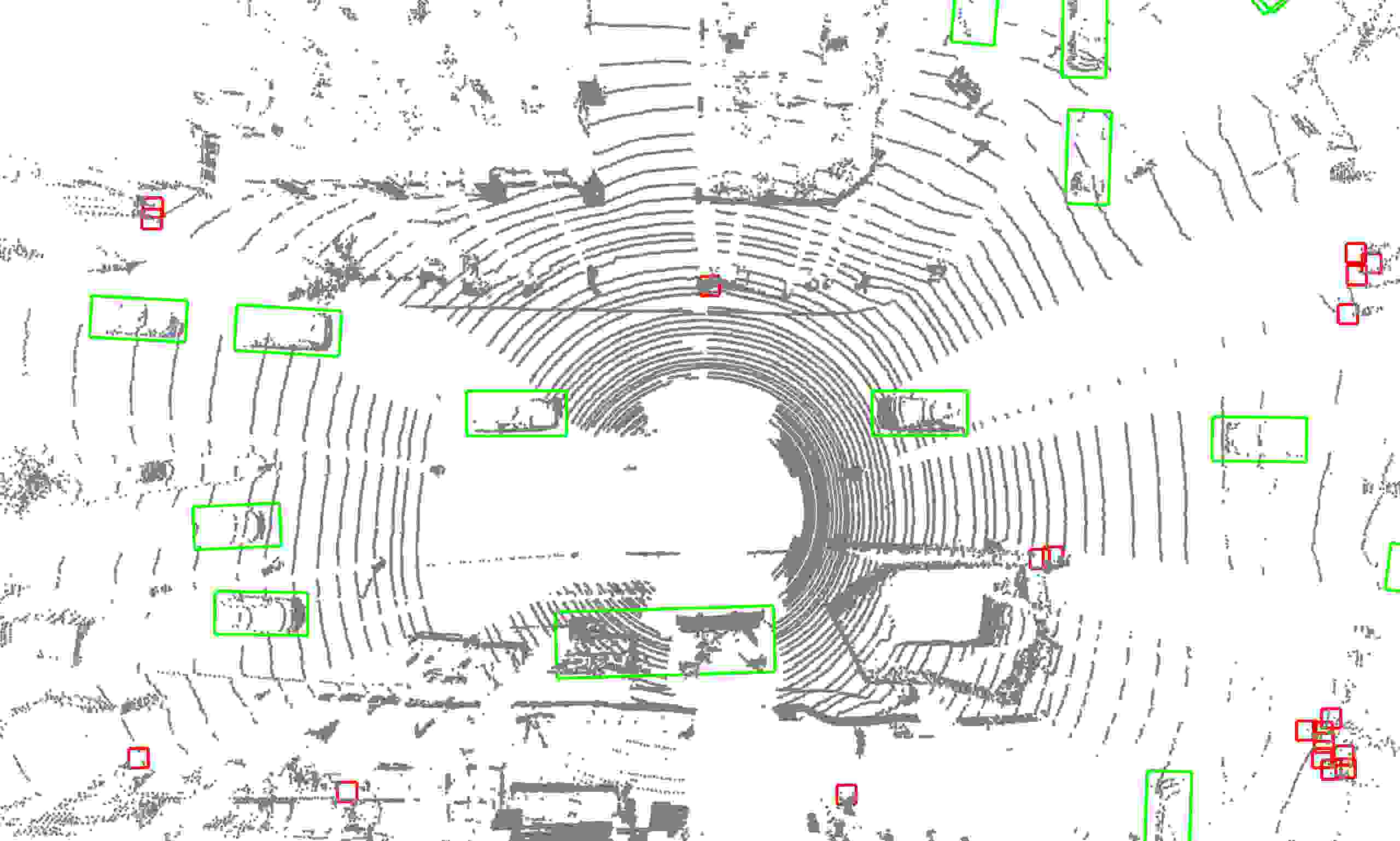}
    \put (0,55) {\colorbox{gray!30}{\scriptsize Draco: AP: 86.95, Bitrate: 10.16}}
\end{overpic}
\end{center}

\begin{center}
\begin{overpic}[width=0.32\textwidth]{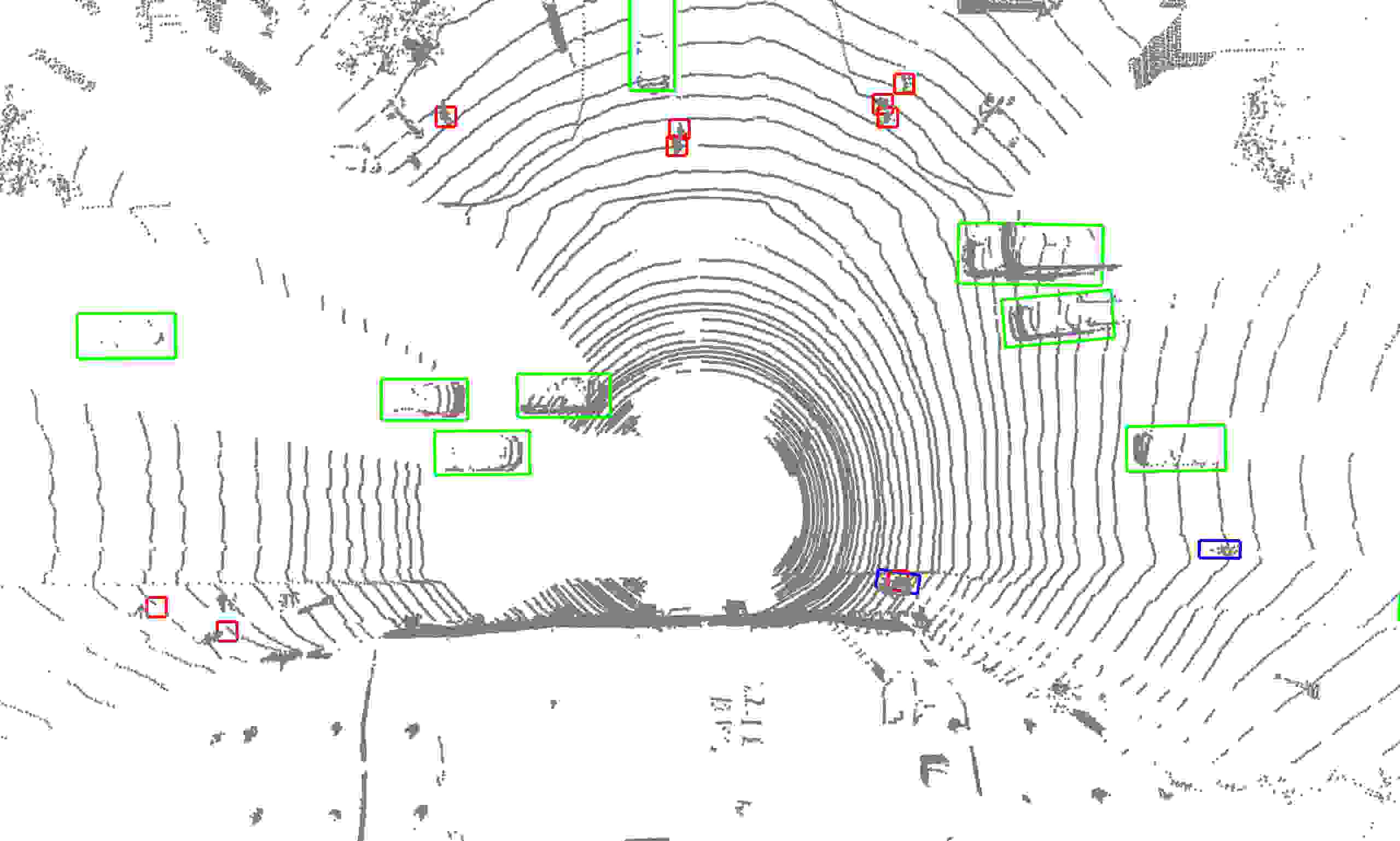}
    \put (0,55) {\colorbox{gray!30}{\scriptsize Oracle: AP: 89.79, Bitrate: 96.00}}
\end{overpic}
\begin{overpic}[width=0.32\textwidth]{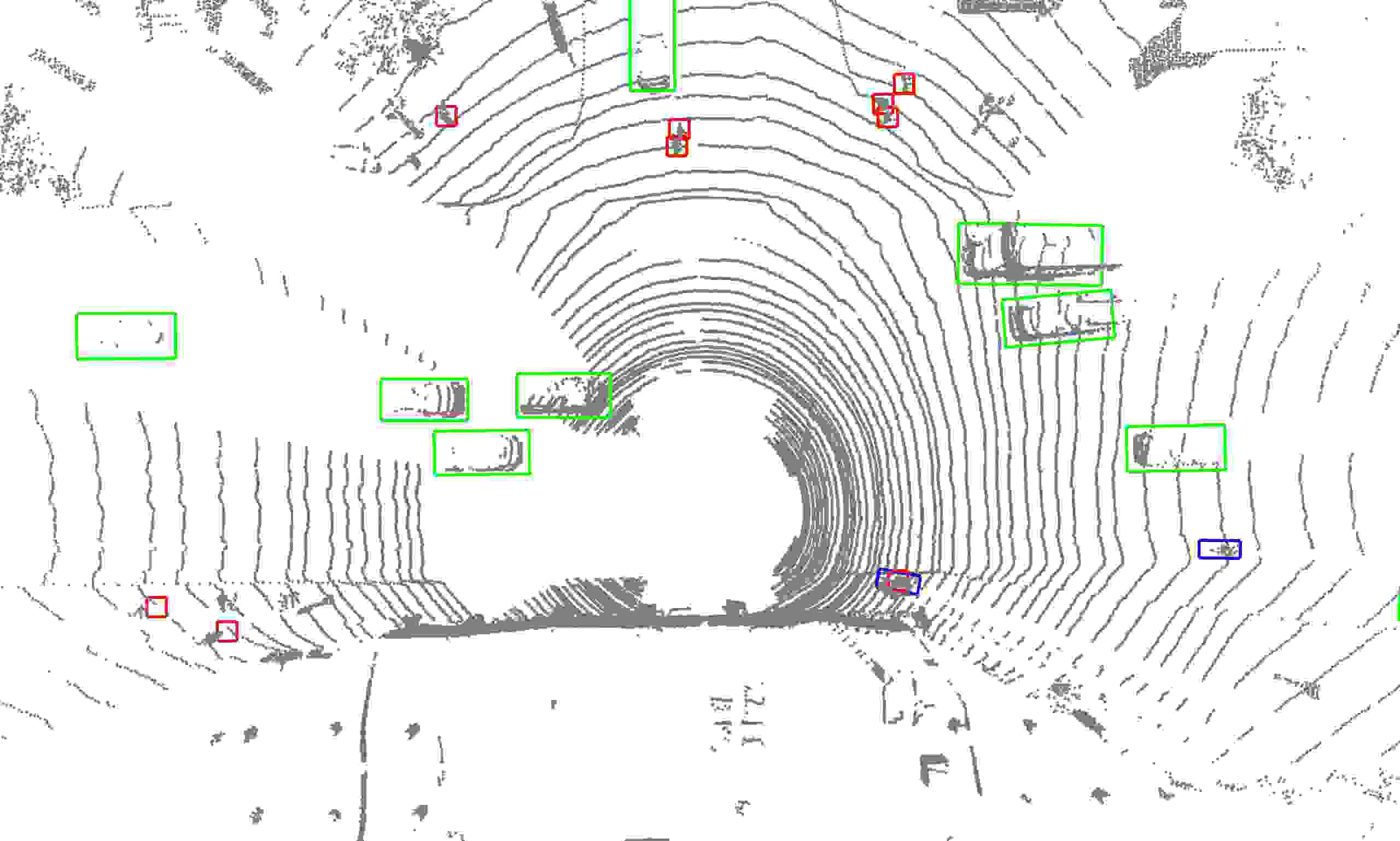}
    \put (0,55) {\colorbox{gray!30}{\scriptsize Ours: AP: 89.39, Bitrate: 5.51}}
\end{overpic}
\begin{overpic}[width=0.32\textwidth]{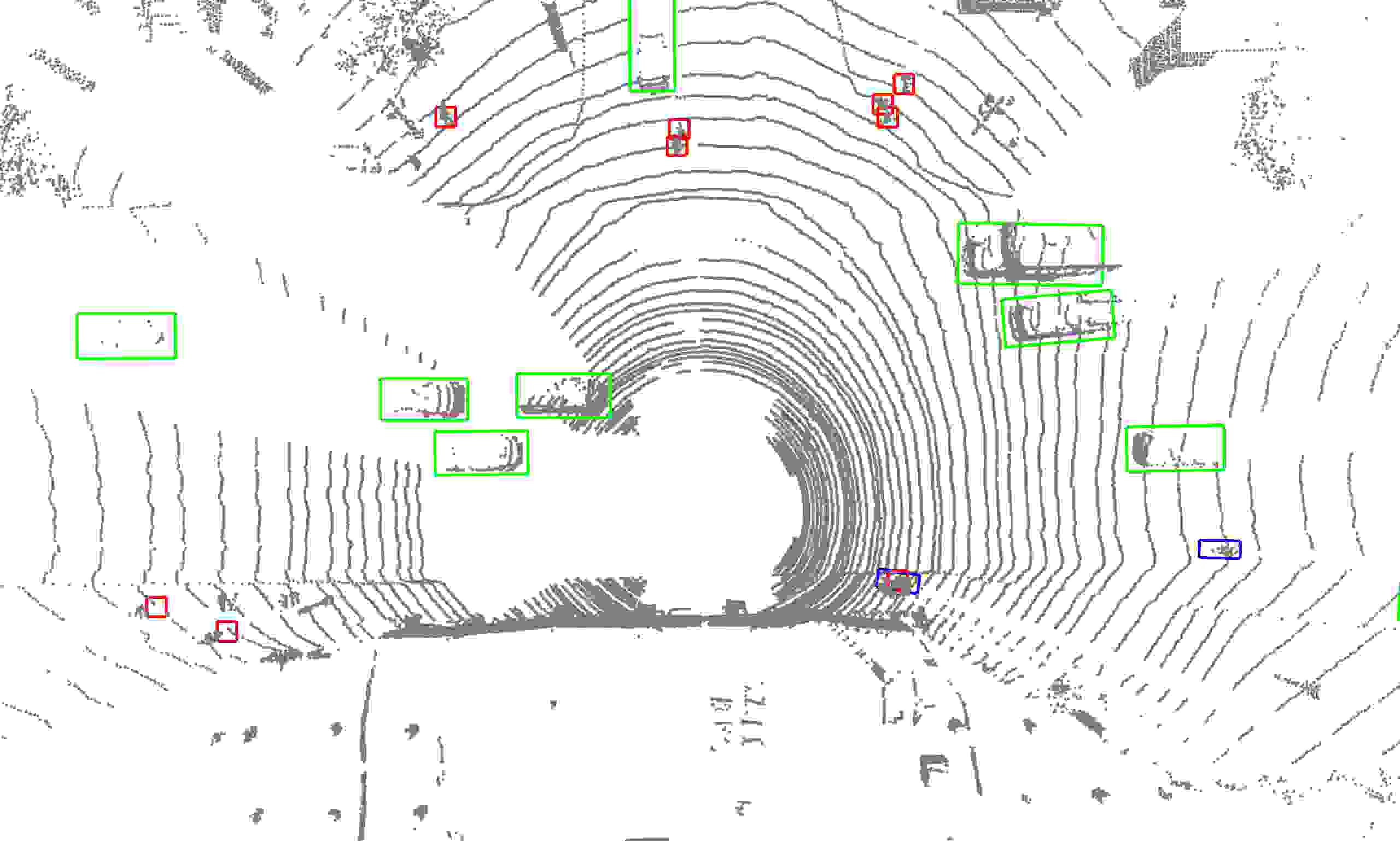}
    \put (0,55) {\colorbox{gray!30}{\scriptsize Draco: AP: 89.26, Bitrate: 6.94}}
\end{overpic}
\end{center}

\begin{center}
\begin{overpic}[width=0.32\textwidth]{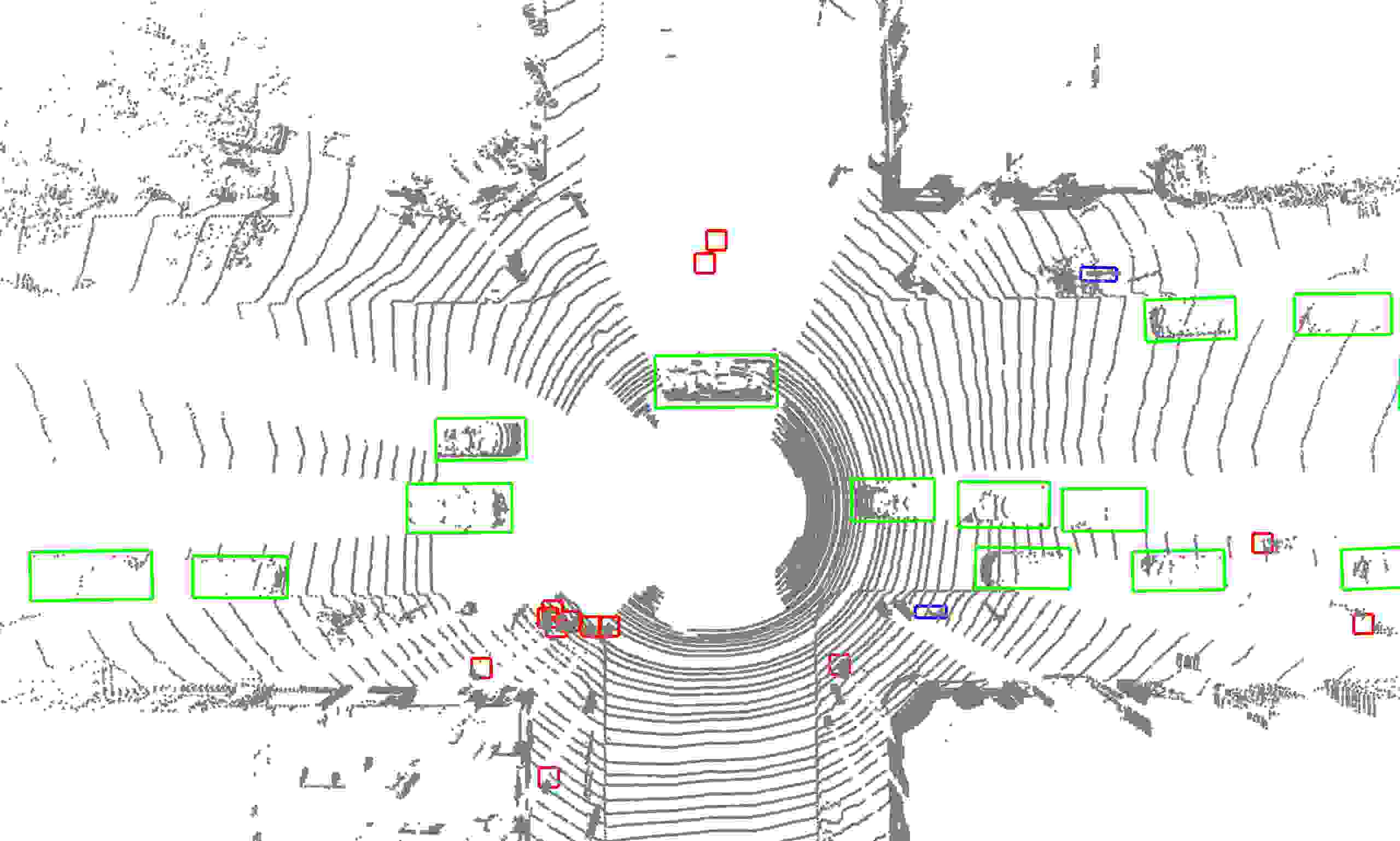}
    \put (0,55) {\colorbox{gray!30}{\scriptsize Oracle: AP: 86.71, Bitrate: 96.00}}
\end{overpic}
\begin{overpic}[width=0.32\textwidth]{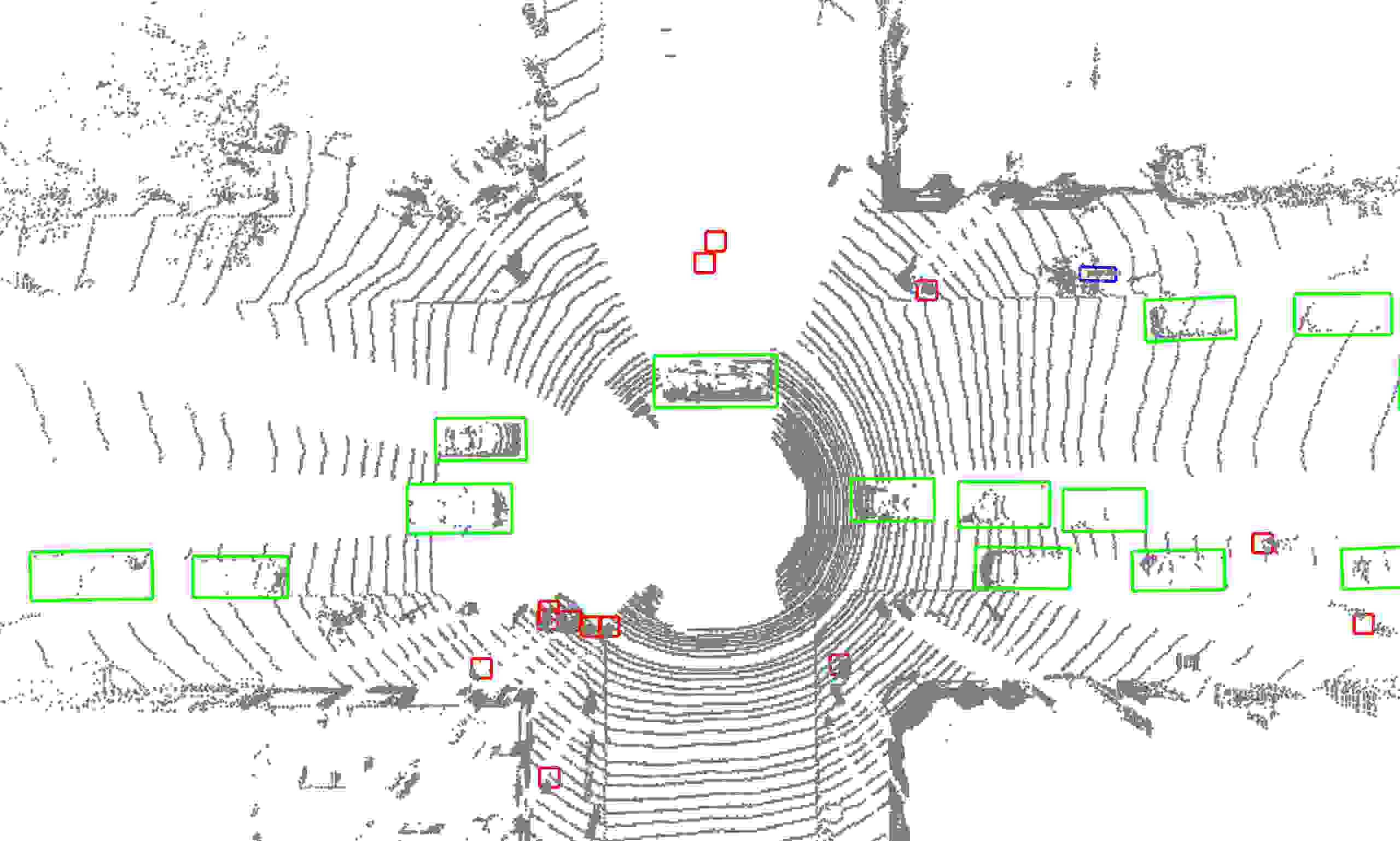}
    \put (0,55) {\colorbox{gray!30}{\scriptsize Ours: AP: 86.17, Bitrate: 3.39}}
\end{overpic}
\begin{overpic}[width=0.32\textwidth]{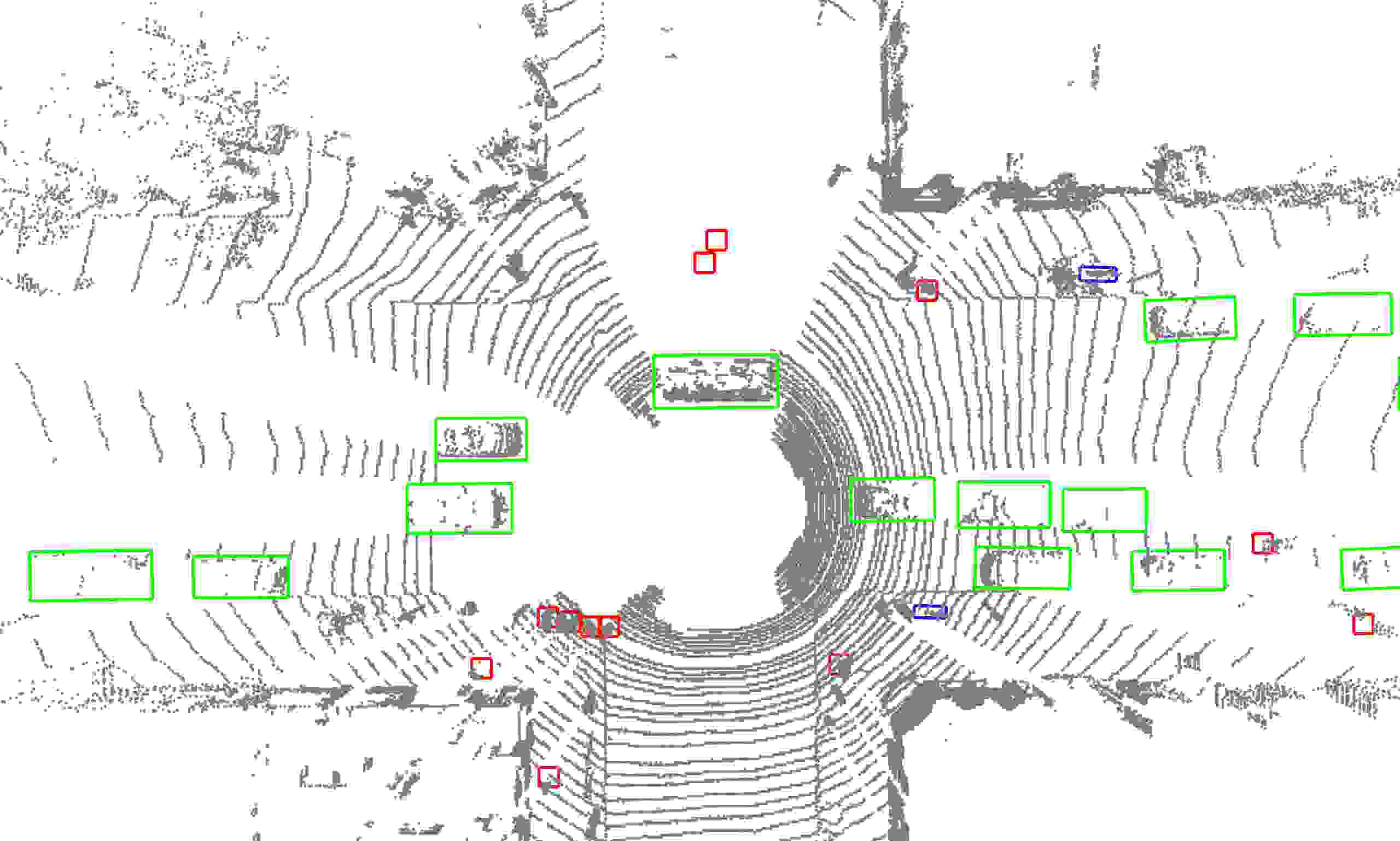}
    \put (0,55) {\colorbox{gray!30}{\scriptsize Draco: AP: 85.27, Bitrate: 5.00}}
\end{overpic}
\end{center}

\begin{center}
\begin{overpic}[width=0.32\textwidth]{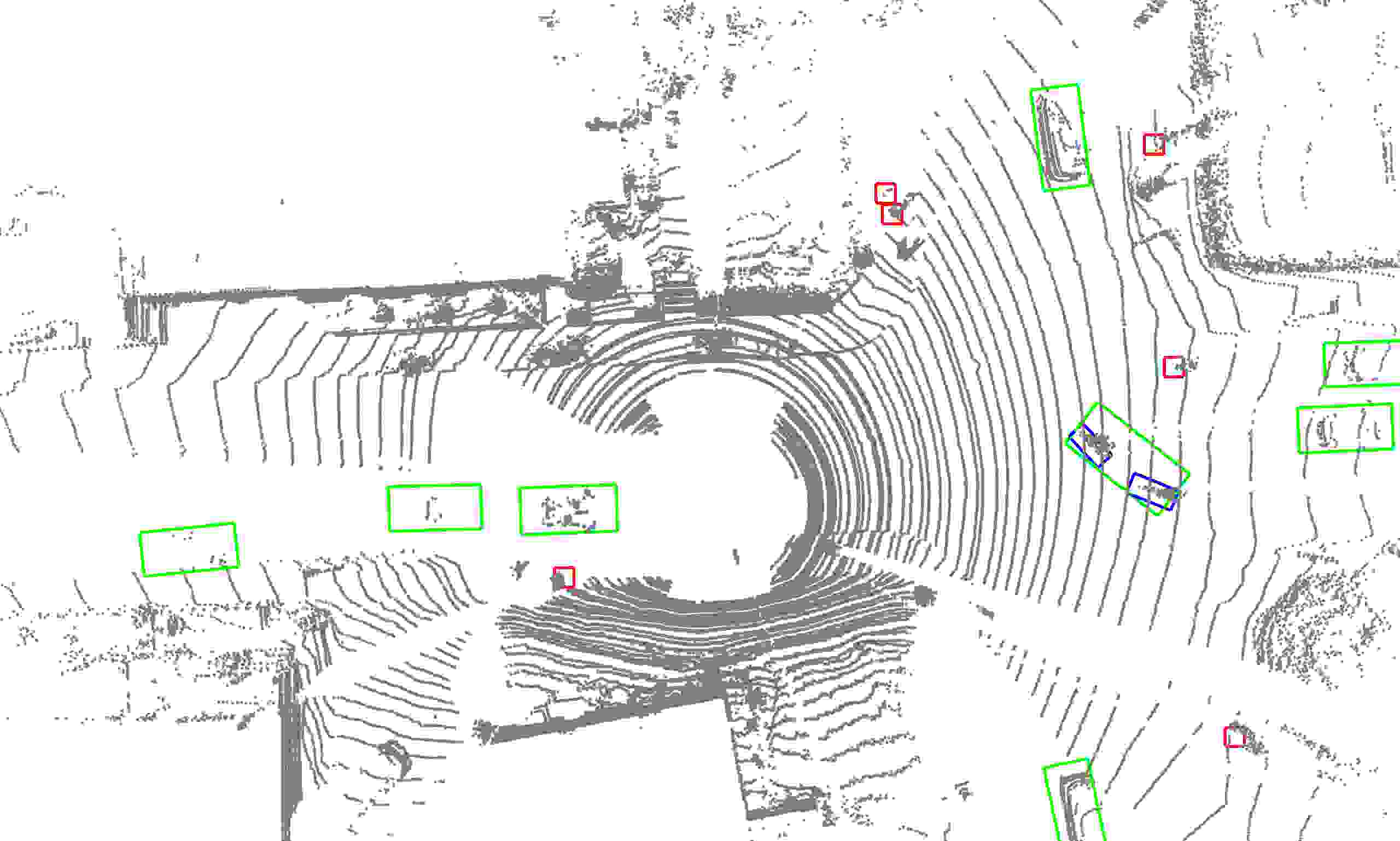}
    \put (0,55) {\colorbox{gray!30}{\scriptsize Oracle: AP: 88.08, Bitrate: 96.00}}
\end{overpic}
\begin{overpic}[width=0.32\textwidth]{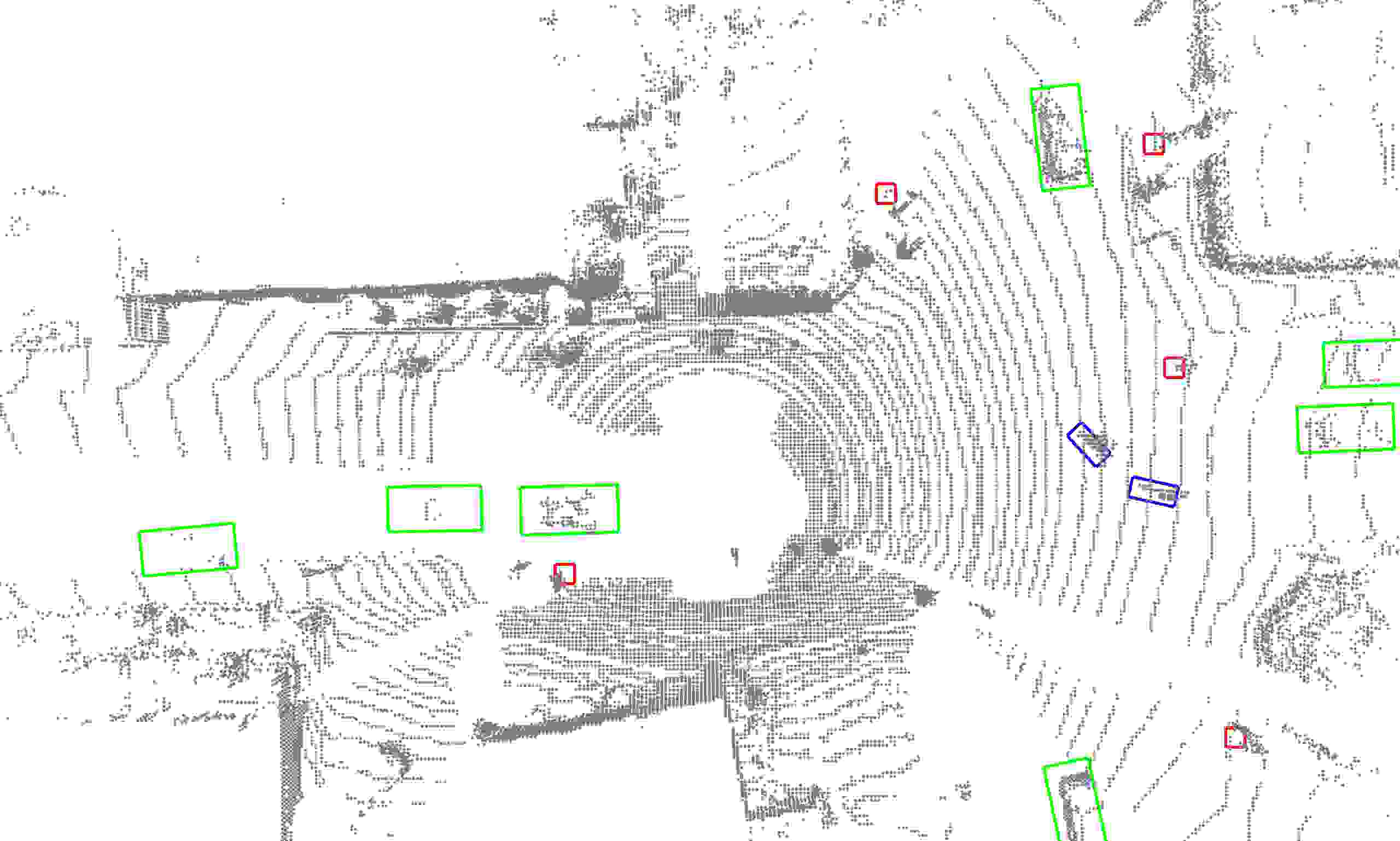}
    \put (0,55) {\colorbox{gray!30}{\scriptsize Ours: AP: 87.30, Bitrate: 1.39}}
\end{overpic}
\begin{overpic}[width=0.32\textwidth]{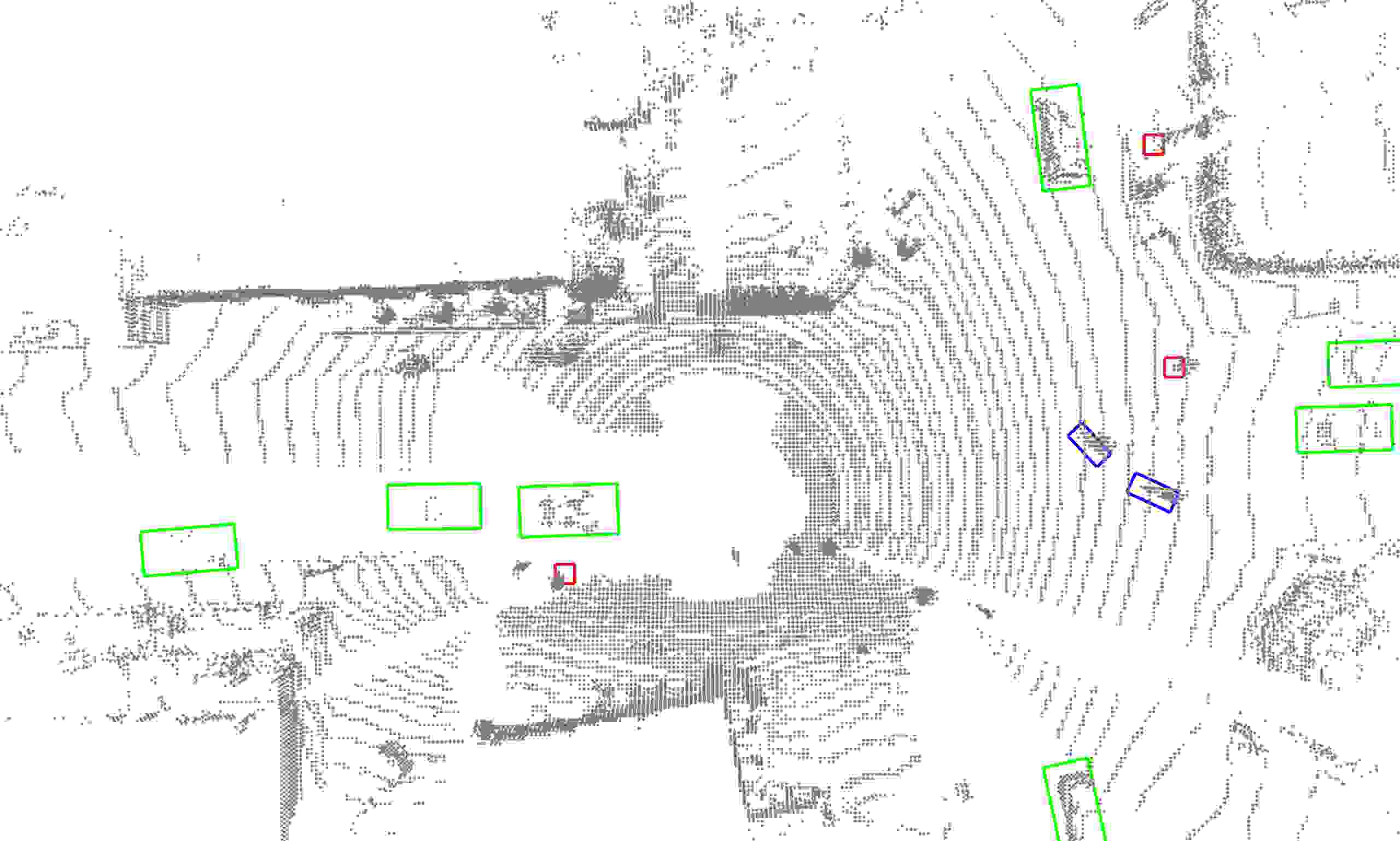}
    \put (0,55) {\colorbox{gray!30}{\scriptsize Draco: AP: 83.80, Bitrate: 2.42}}
\end{overpic}
\end{center}

\caption{Qualitative results of object detection for NorthAmerica. AP is averaged over vehicle, motorbike, and pedestrian classes.}
\label{fig:obj_na}
\end{figure*}

We exhibit an extensive array of qualitative results that compare our method
against Draco across a spectrum of bitrates.
In Fig.~\ref{fig:recon_na} and \ref{fig:recon_kitti}, we show the
reconstruction quality of our method versus Draco.
Then, in Fig.~\ref{fig:semantic_kitti} and \ref{fig:semantic_na}, we show
their respective downstream semantic segmentation performance.
Finally, in Fig.~\ref{fig:obj_na}, we show their respective downstream object
detection performance.
As indicated in these figures, our model can attain better results than Draco
at comparable---or even lower---bitrates.

\section{Change Log}

\paragraph{ArXiv v2:}
We corrected our definitions of the reconstruction metrics:
the symmetric point-to-point Chamfer distance and the symmetric
point-to-plane PSNR.
We also corrected our estimates of OctSqueeze's decoding runtime.

%
%

%
%

%
%

{\small
	\bibliographystyle{ieee_fullname}
	\bibliography{egbib}
}

%% file: sections/abstract.tex
\begin{abstract}
We present a novel deep compression algorithm to reduce the memory footprint of LiDAR point clouds.
Our method exploits the sparsity and structural redundancy between points to reduce the bitrate.
Towards this goal, we first encode the LiDAR points into an octree, a data-efficient structure suitable for sparse point clouds.
We then design a tree-structured conditional entropy model that models the probabilities of the octree symbols to encode the octree into a compact bitstream.
We validate the effectiveness of our method over two large-scale datasets.
The results demonstrate that our approach reduces the bitrate by 10-20\% at the same reconstruction quality, compared to the previous state-of-the-art.
Importantly, we also show that for the same bitrate, our approach outperforms other compression algorithms when performing downstream 3D segmentation and detection tasks using compressed representations.
Our algorithm can be used to reduce the onboard and offboard storage of LiDAR points for applications such as self-driving cars, where a single vehicle captures 84 billion points per day.
\end{abstract}

%% file: sections/intro.tex
\vspace{-5mm}
\section{Introduction}

In the past few decades, we have witnessed artificial intelligence revolutionizing robotic perception. Robots powered by these AI algorithms often utilize a plethora of different sensors to perceive and interact with the world. In particular, 3D sensors such as LiDAR and structured light cameras have proven to be crucial for many types of robots, such as self-driving cars, indoor rovers, robot arms, and drones, thanks to their ability to accurately capture the 3D geometry of a scene. These sensors produce a significant amount of data: a single Velodyne HDL-64 LiDAR sensor generates over 100,000 points per sweep, resulting in over 84 billion points per day. This enormous quantity of raw sensor data brings challenges to onboard and offboard storage as well as real-time communication.
Hence, it is necessary to develop an efficient compression method for 3D point clouds.

\begin{figure*}
\vspace{-0.15in}
\begin{center}
\includegraphics[width=\textwidth]{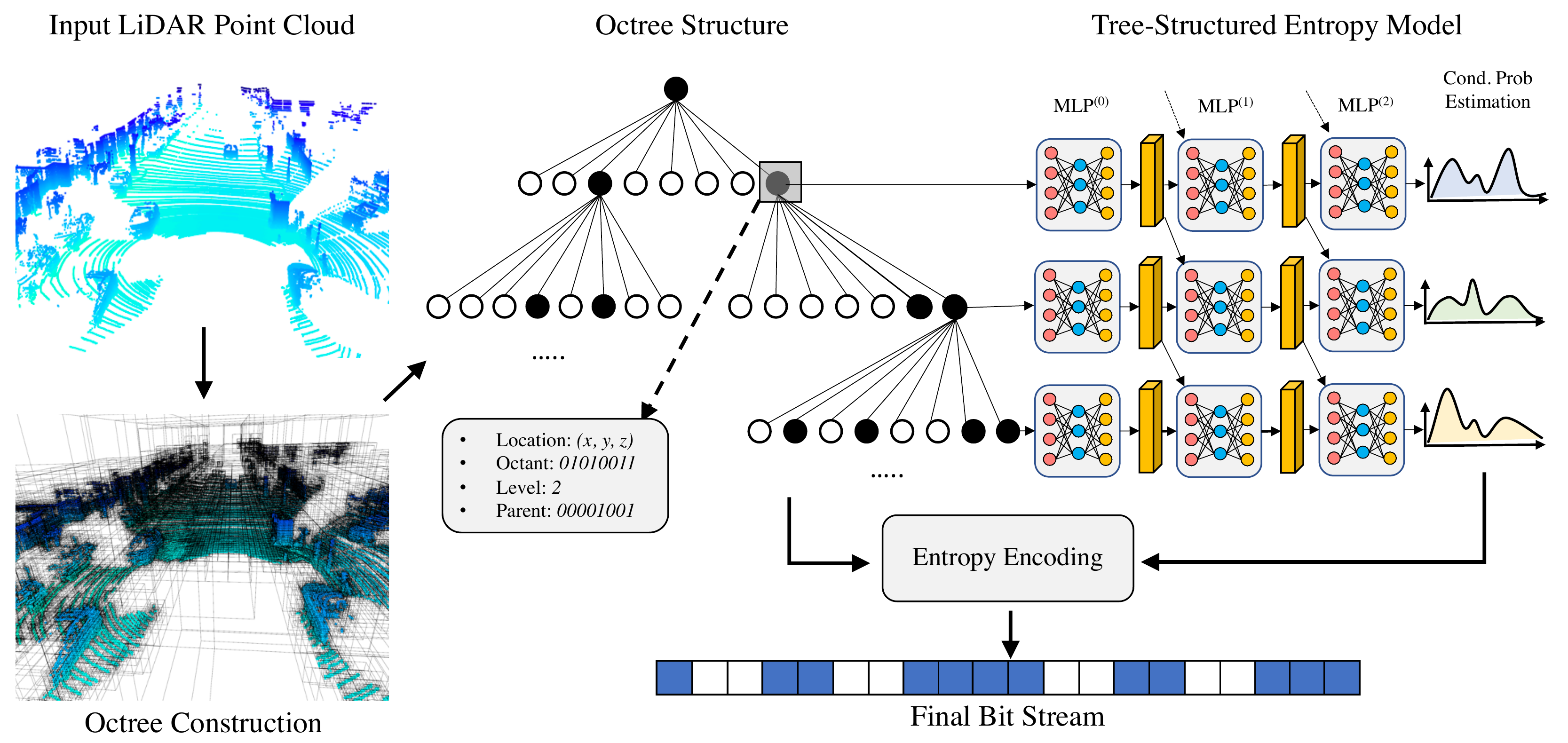}
\end{center}
\vspace{-0.2in}
\caption{
   The overview of the proposed octree-structured entropy model for LiDAR point
   cloud compression.
   The input point cloud, received as a $n \times 3$ float array, is quantized
   to $k$ bits by scaling to $[0, 2^k)$ and rounding down.
   An octree is constructed from the quantized point cloud.
   Each node is represented by an 8-bit occupancy symbol.
   We apply a tree-structured conditional entropy model on top of the octree
   to estimate the probability of each symbol conditioned on prior context.
   Finally, we use the estimated probability to encode the serialized symbols into the
   final compressed bitstream.
}
\label{fig:overall}
\vspace{-0.15in}
\end{figure*}

Raw 3D point clouds are represented as unstructured $n\times 3$ matrices at float precision. This uncompressed data representation does not exploit the fact that the geometry of the scene is usually well structured. Prior works have approached point cloud compression by using data structures such as KD-trees \cite{kdtree} and octrees \cite{meagher1982} to encode a point cloud's structure. Quantization is exploited to further reduce storage. However, there remains a massive quantity of redundant information hidden in these representations, such as repetitive local structures, planar surfaces, or object categories with a strong shape prior, such as cars and humans. In theory, this redundant information can be exploited during compression to reduce the bitrate even further. However, this has not yet been exploited to its full potential in point cloud compression.

The recent success of deep neural networks in image and video compression brings a new paradigm towards structured data compression for point clouds. These approaches typically contain three steps: 1) encode the data into a hidden representation through a convolutional neural network; 2) quantize the hidden features; and 3) learn an entropy model to reduce the bitstream further through entropy coding. The key to the learned entropy model is encoding context information to improve the predictability of a symbol's occurrence, which directly increases its compressibility. However, it is non-trivial to apply these deep compression algorithms directly on a LiDAR point cloud, as it is sparse and non-grid structured. Hence, there are two major challenges that we need to address: 1) What is a memory-efficient data structure to represent LiDAR while exploiting its sparsity? 2) How can we train a deep entropy model to encode the representation to bitstreams efficiently? 

In this work, we propose a novel deep learning model for LiDAR point cloud compression. Our approach first exploits the efficient and self-adaptive octree structure to get an initial encoding of the raw point cloud. We then learn a tree-structured deep conditional entropy model over each intermediate node of the tree, incorporating both the prior and the context of the scene simultaneously to help predict the node symbols. The predicted probabilities from our learned entropy model are then passed to an entropy coder to encode the serialized symbols into the final bitstream.

We evaluate the performance of our approach over two challenging LiDAR point cloud datasets comprising of complicated urban traffic scenes, namely the KITTI \cite{behley_semantickitti} and NorthAmerica datasets. Our results show that the proposed model outperforms all state-of-the-art methods in terms of both reconstruction quality and downstream task performance. At the same reconstruction quality, our bitrate is 10-20\% lower than the previous state-of-the-art. 

%% file: sections/related.tex
\section{Related Work}

\subsection{Point Cloud Compression}
Tree structures are the primary methods used in prior point cloud compression algorithms.
Numerous approaches store data in an octree and perform entropy coding with hand-crafted entropy models such as adaptive histograms, parent context \cite{garcia2018}, and estimations based on planar approximations \cite{schnabel2006} or neighbour proximity \cite{huang2018}. To exploit temporal redundancies in point cloud streams, Kammerl \etal \cite{kammerl2012} encode the xor differences between successive octree representations and Mekuria \etal \cite{mekuria2017} use ICP to encode blocks with rigid transformations. Both methods use range coding with empirical histograms for entropy coding. The advantage of the octree structure is that it can model arbitrary point clouds in a hierarchical structure, which provide a natural progressive coding---if the octree is traversed in breadth-first order, then decoding can stop at any time; the longer the decoding, the finer the precision of the point cloud reconstruction. A related structure is utilized in Google's open-source compression software \textit{Draco} \cite{draco} which uses a KD-tree compression method \cite{devillers2000}. All of the above approaches do not leverage deep learning.

Besides tree structures, point clouds can be represented as regular voxel grids~\cite{quach2019,huang2019}.
These methods use voxel-based convolutional autoencoders which can learn surface representations of point clouds but struggles with large-scale sparse data.
Moreover, since the geometry of a LiDAR scan can be represented by a panorama image with one channel for distance, point clouds can also be represented as range images, and compressed via image compression techniques. For example, Houshiar \etal \cite{houshiar2015} use conventional image compressors such as JPEG, PNG, and TIFF to compress LiDAR range images.

\begin{figure*}
\vspace{-0.15in}
\begin{center}
\includegraphics[width=0.245\textwidth]{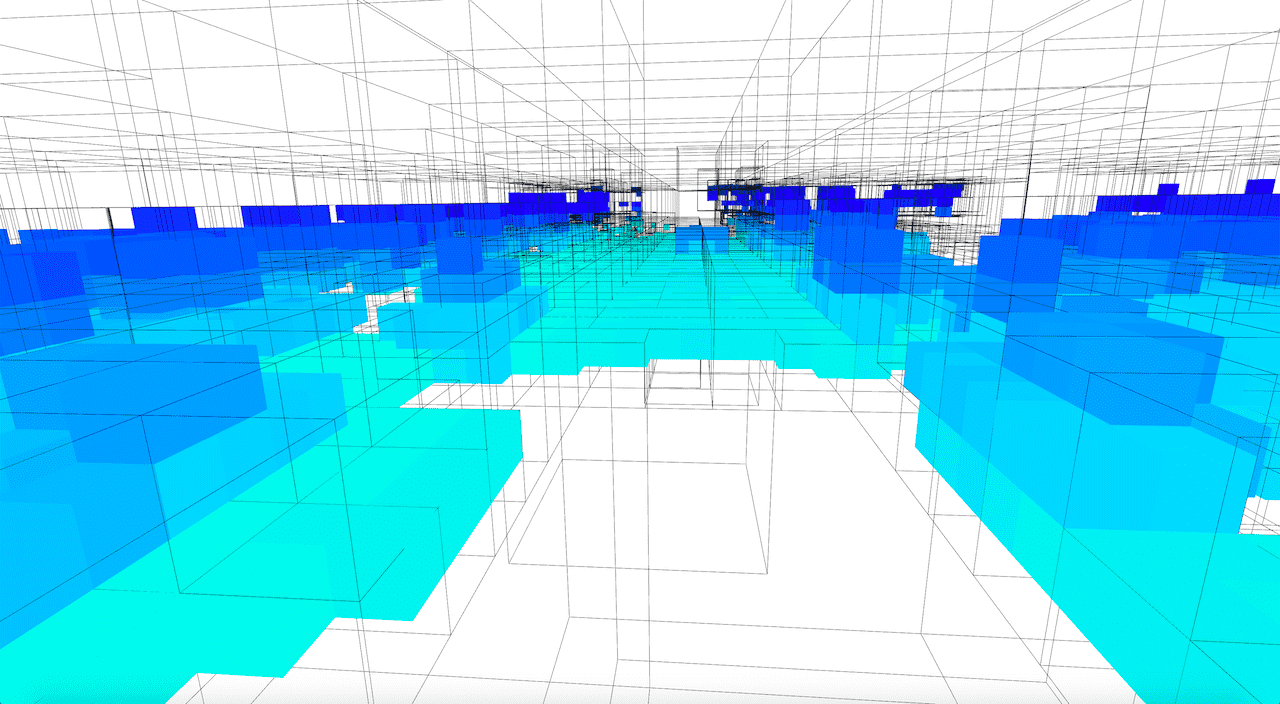}
\includegraphics[width=0.245\textwidth]{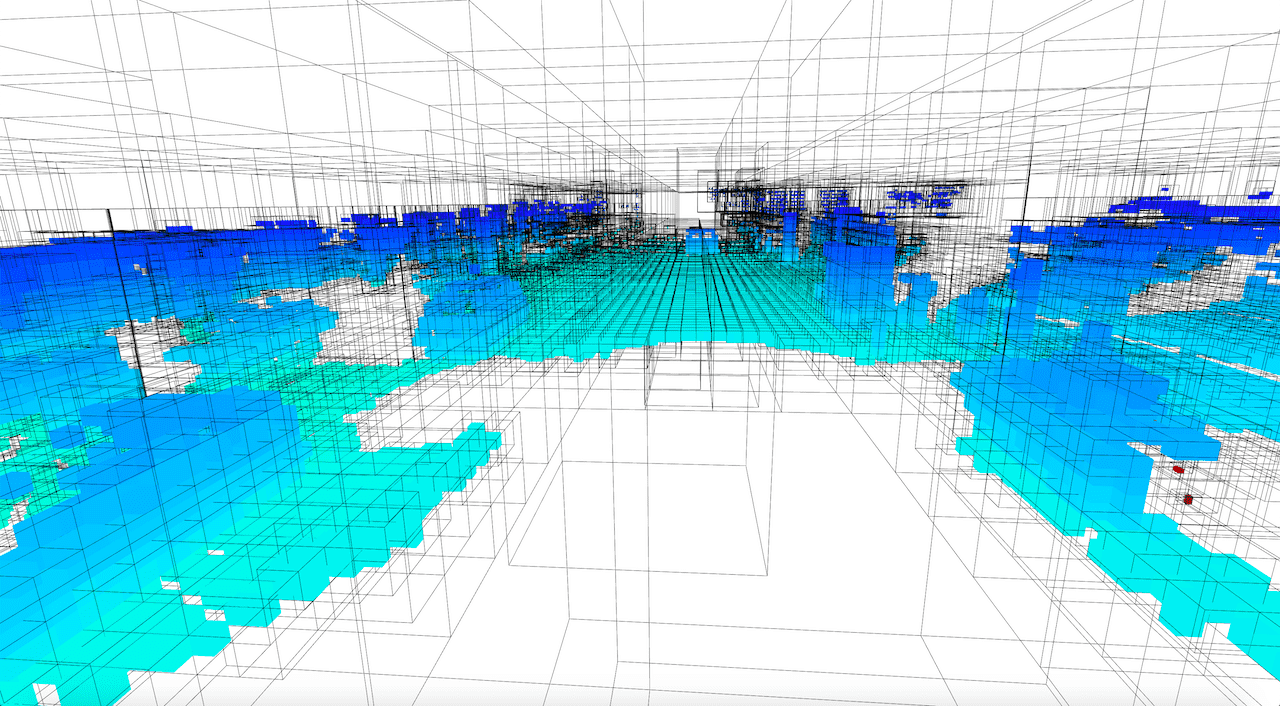}
\includegraphics[width=0.245\textwidth]{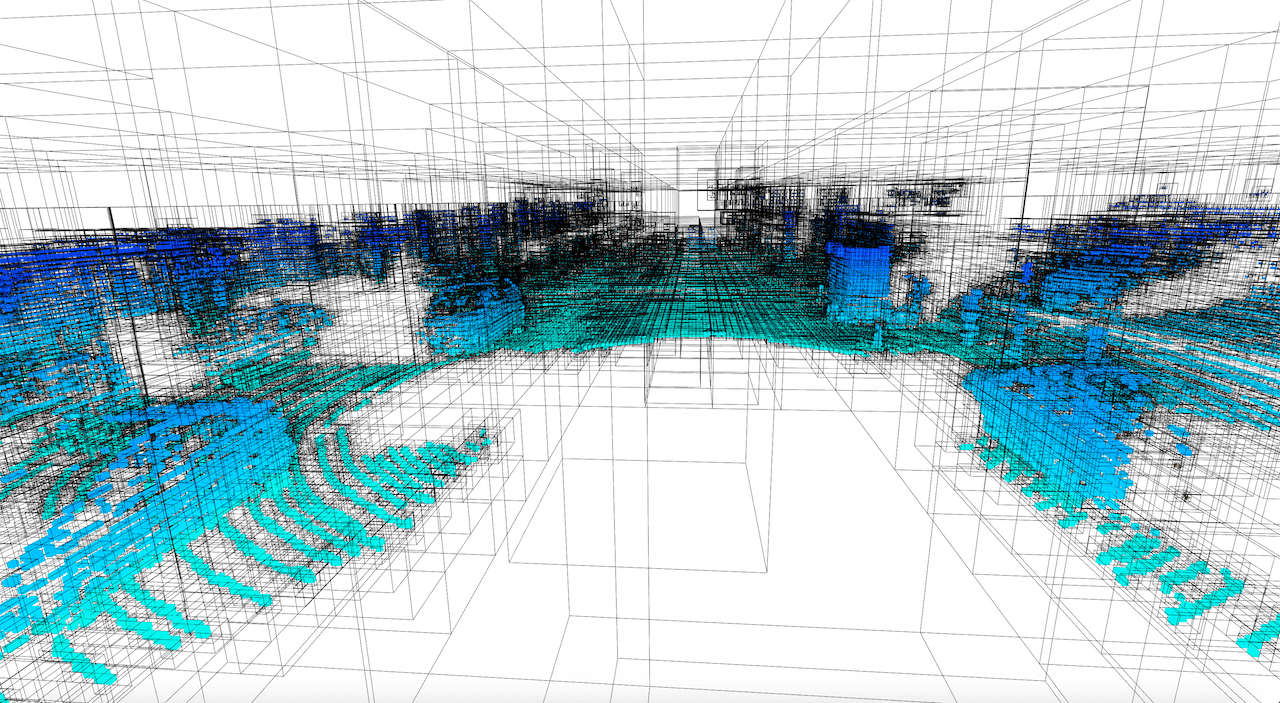}
\includegraphics[width=0.245\textwidth]{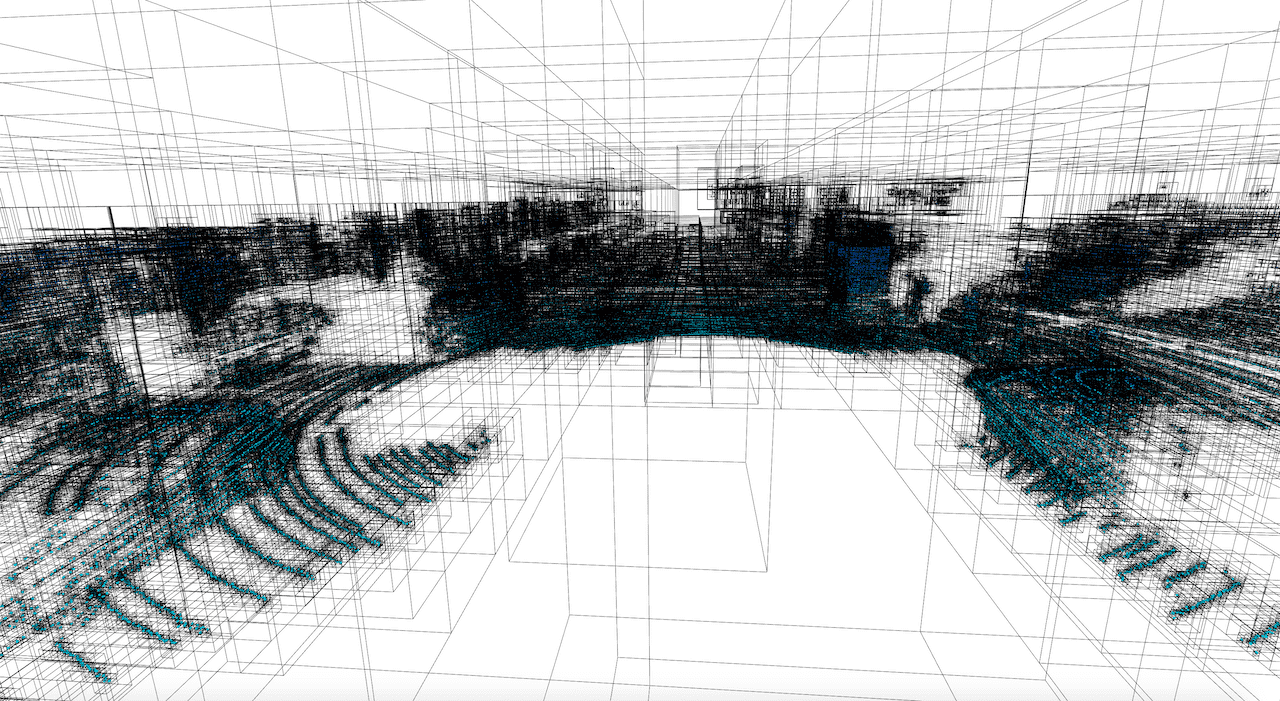}
\end{center}
\vspace{-0.2in}
\caption{Construction of octree structures to represent a point cloud. Max depth of the octree (from left to right): 8, 10, 12, 14.}
\label{fig:octree}
\vspace{-0.15in}
\end{figure*}

\subsection{Deep Learning on Point Clouds}

Inspired by recent successes in the image domain, researchers have developed a
flurry of new deep learning methods for point cloud data.
One class of methods uses deep convolutional neural networks to process
voxel representations of the 3D point cloud
\cite{wu2015,maturana2015,qi2016,zhou2018,zhang2018,yang2018}.
These approaches, however, require large memory footprints and thus
induce a trade-off between input resolution and model capacity.
To address this shortcoming, \cite{ren2018,graham2018} propose to use sparse
operators on the point cloud's voxel representation and \cite{su2015,meyer2019}
propose to process 2D projections of the point cloud instead.

Another line of work tackles this problem by directly operating on the point cloud,
thus leveraging its sparsity to sidestep this trade-off.
PointNet \cite{qi2017a} uses multi-layer perceptrons to extract
features from individual points and then pools them into a global feature.
As PointNet cannot capture local structures in the point cloud,
a number of follow-up works have proposed to hierarchically
aggregate local information \cite{qi2017b,wang2018,wu2019,xiong2019,thomas2019}.
These methods can be viewed as graph neural networks that operate on graphs
defined by each point's local neighbourhood; \eg, $ k $-nearest neighbors graph.
Other possible graphs include KD-trees~\cite{klokov2017} and
octrees~\cite{riegler2017,wang2017}.
Inspired by the success of these graph-structured networks, we designed an entropy
model that operates on an octree's serialized byte streams but exploits
its structure to encode contextual information.

\subsection{Deep Image and Video Compression}

The field of image and video compression is extensive and has been well-explored over the past few decades, ranging from lossless image formats (PNG, TIFF), to lossy image codecs (JPEG, BPG), to video codecs (AVC/H.264, HEVC/H.265). In recent years, there has been a rapid increase in learned image and video compression methods \cite{toderici_fullimgcomp, balle_imgcomp_end2end, balle_varhyperprior, mentzer_condprobimg, theis_imgcomp_ae, wu_vidinterpolation, rippel_learnedvidcomp, lu_dvc, habibian_rdauto}, which exploit concepts from traditional codecs and the power of deep neural networks. These approaches typically use deep convolutional autoencoders to apply nonlinear transforms to traditional components of the compression pipeline, from transform coding used in JPEG to motion compensation used in video codecs. Moreover, many approaches use separate neural nets to model the entropy of the image/video latent codes as a tight lower bound of the bitrate; this model is then used during entropy coding to losslessly compress the symbols into bits. Such approaches have included fully factorized models \cite{balle_imgcomp_end2end, theis_imgcomp_ae}, encoding ``side information" as latent variables for entropy prediction \cite{balle_varhyperprior, mentzer_lossless, lu_dvc} as well as using autoregressive models (e.g. PixelCNN \cite{oord_pixelrnn}) to model pixel-level conditional distributions \cite{mentzer_condprobimg, toderici_fullimgcomp, habibian_rdauto, wu_vidinterpolation}. Inspired by these approaches towards entropy modeling, we aim to apply these insights towards the compression of point clouds.

%% file: sections/method.tex
\newcommand{\bc}{\mathbf{c}}
\newcommand{\bx}{\mathbf{x}}
\newcommand{\bw}{\mathbf{w}}
\newcommand{\an}{\mathrm{an}}
\newcommand{\pa}{\mathrm{pa}}

\section{Octree-Structured Entropy Model}

{In this work we tackle the problem of \textit{lossy compression} on 3D LiDAR point clouds}. Our goal is to reduce the storage footprint of our encodings as much as possible while preserving reconstruction quality.
Towards this goal, we propose a novel, octree-structured compression method using a \textit{deep entropy model}.

Specifically, we firstly quantize and encode a LiDAR point cloud into an octree. Each node of the tree uses an 8-bit symbol to encode the occupancy of its children.
We then serialize the octree into an intermediate, uncompressed bytestream of symbols.
For each node, we select a set of context features that are available during decoding time. We then feed these context features into our tree-structured deep entropy model,
which is trained to predict the probability of each symbol's presence given the context input.
These probabilities are then directly fed into arithmetic encoding with the symbol bytestream to produce the final bitstream,
where the bitrate is approximately measured by the cross-entropy of these probabilities with the actual symbol distribution.
Our overall approach is shown in Fig. \ref{fig:overall}.

\subsection{Octree Structure}
Two difficulties in LiDAR point cloud compression are the sparsity of the data and the lack of structure in a raw point cloud.
Space-partitioning data structures, such as octrees and KD-trees, effectively provide a representation for 3D spatial data while keeping sparsity in mind,
as their memory usage scales with the number of points in the cloud compared to voxel representations which scale with the cloud's bounding volume.
In addition, tree structures give implicit levels of detail which can be used for progressive decoding.
We choose to use an octree as the base data structure for quantization due to its memory efficiency and ease of construction and serialization.

\paragraph{Bit Representation:}
An octree \cite{meagher1982} stores point clouds by recursively partitioning the input space into equal octants and storing occupancy in a tree structure.
Each intermediate node of the octree contains a 8-bit symbol to store the occupancy of its eight child nodes, with each bit corresponding to a specific child.
Each leaf contains a single point and stores additional information to represent the position of the point relative to the cell corner.
The size of leaf information is adaptive and depends on the level.
An octree with $k$ levels can store $k$ bits of precision by keeping the last $k - i$ bits of each of the $(x, y, z)$  coordinates for a child on the $i$-th level of the octree.
The resolution increases as the number of levels in the octree increases.
The advantage of such a representation is twofold: firstly, only non-empty cells are further subdivided and encoded, which makes the data structure adapt to different levels of sparsity; secondly, the occupancy symbol per node is a tight bit representation.
Fig.~\ref{fig:octree} shows the partial construction of the octree structure at different levels from a KITTI point cloud \cite{geiger_kitti}.

\paragraph{Serialization:}
Using a breadth-first or depth-first traversal, an octree can be serialized into two intermediate uncompressed bytestreams of occupancy codes and leaf-node offsets. The original tree can be completely reconstructed from these streams. We note that serialization is a lossless scheme in the sense that offsets and occupancy information are all exactly preserved. Thus the only lossy procedure is due to quantization during construction of the octree. Consequently, octree-based compression schemes are lossy up to this quantization error, which gives an upper bound on the distortion ratio.

We use the occupancy serialization format during our entropy coding stage, detailed in  Sec.~\ref{sec:entropy_coding} and Sec.~\ref{sec:entropy_coder}. During range decoding of a given occupancy code, we note that information such as node depth, parent occupancy, and spatial locations of the current octant are already known given prior knowledge of the traversal format. Hence we incorporate this information as a context $\bc_i$ for each node that we can use during entropy coding.

\subsection{A Deep Entropy Model for Entropy Coding} \label{sec:entropy_coding}

The serialized occupancy bytestream of the octree can be further losslessly encoded into a shorter bit-stream through entropy coding.
Entropy encoding is theoretically grounded in information theory.
Specifically, an entropy model estimates the probability of occurrence of a given symbol; the probabilities can be adaptive given available context information.
A key intuition behind entropy coding is that symbols that are predicted with higher probability can be encoded with fewer bits, achieving higher compression rates.
\begin{figure*}[t]
\vspace{-0.15in}
\begin{center}
\includegraphics[height=0.25\textwidth]{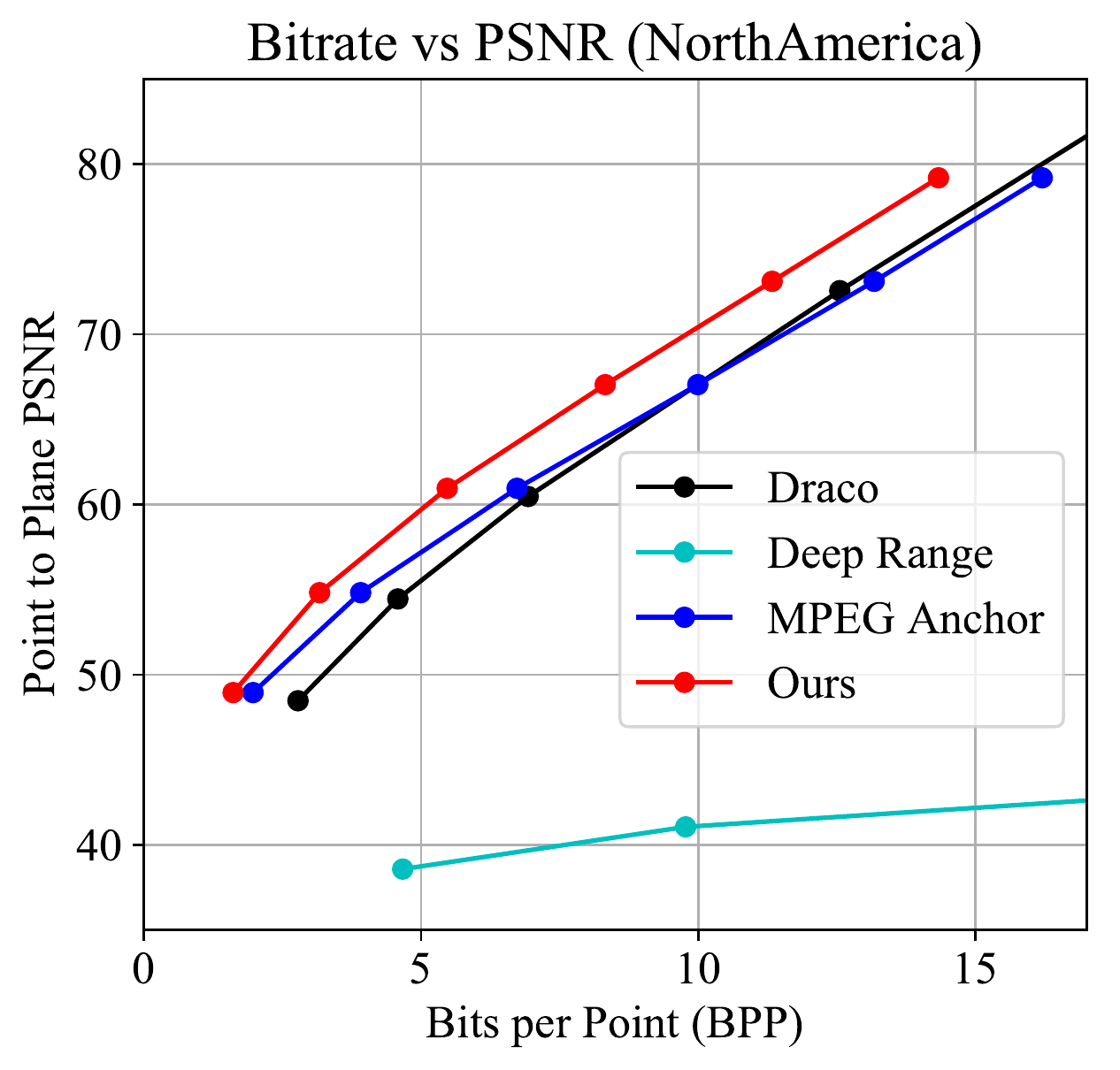}
\includegraphics[height=0.25\textwidth]{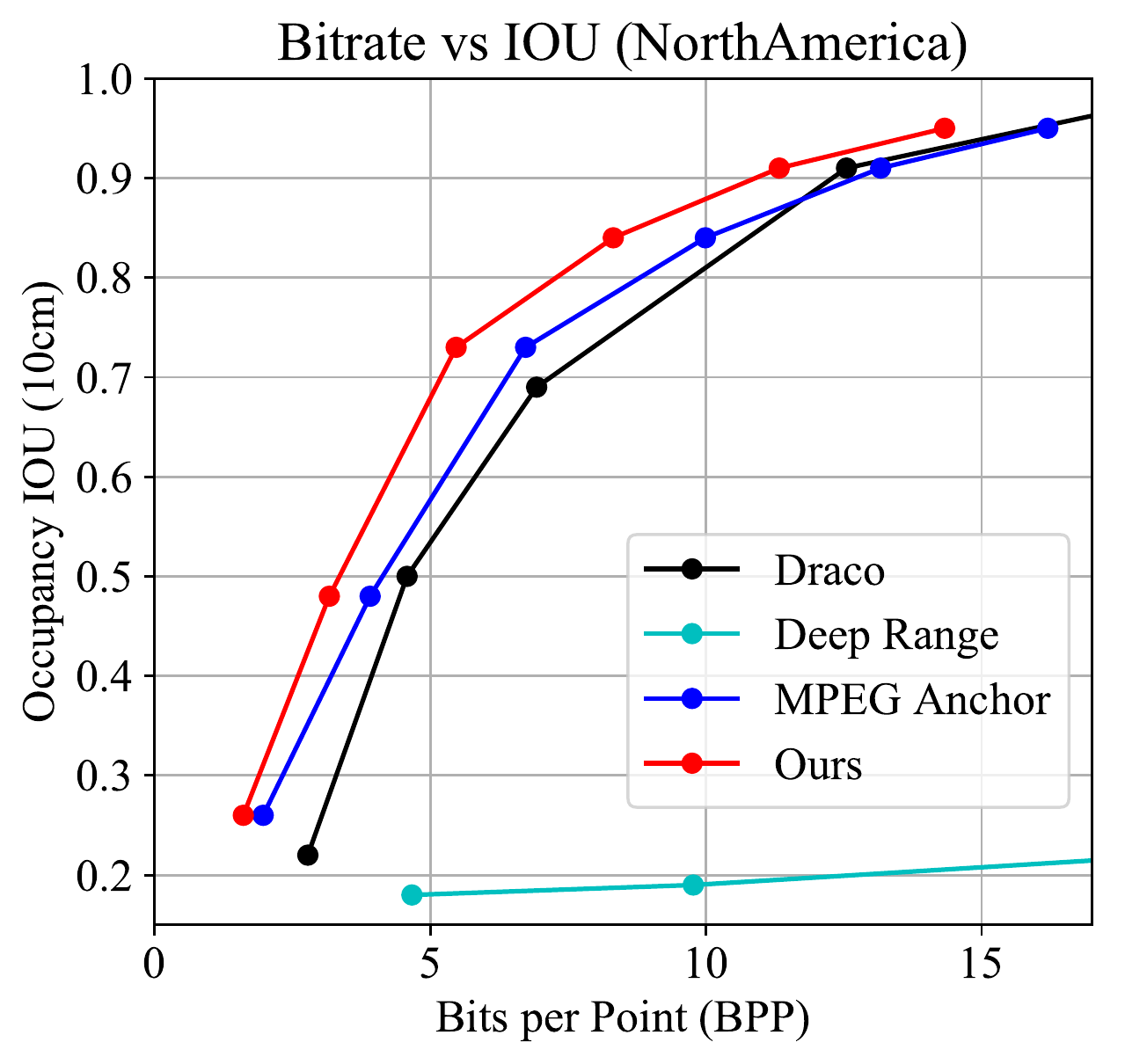}
\includegraphics[height=0.25\textwidth]{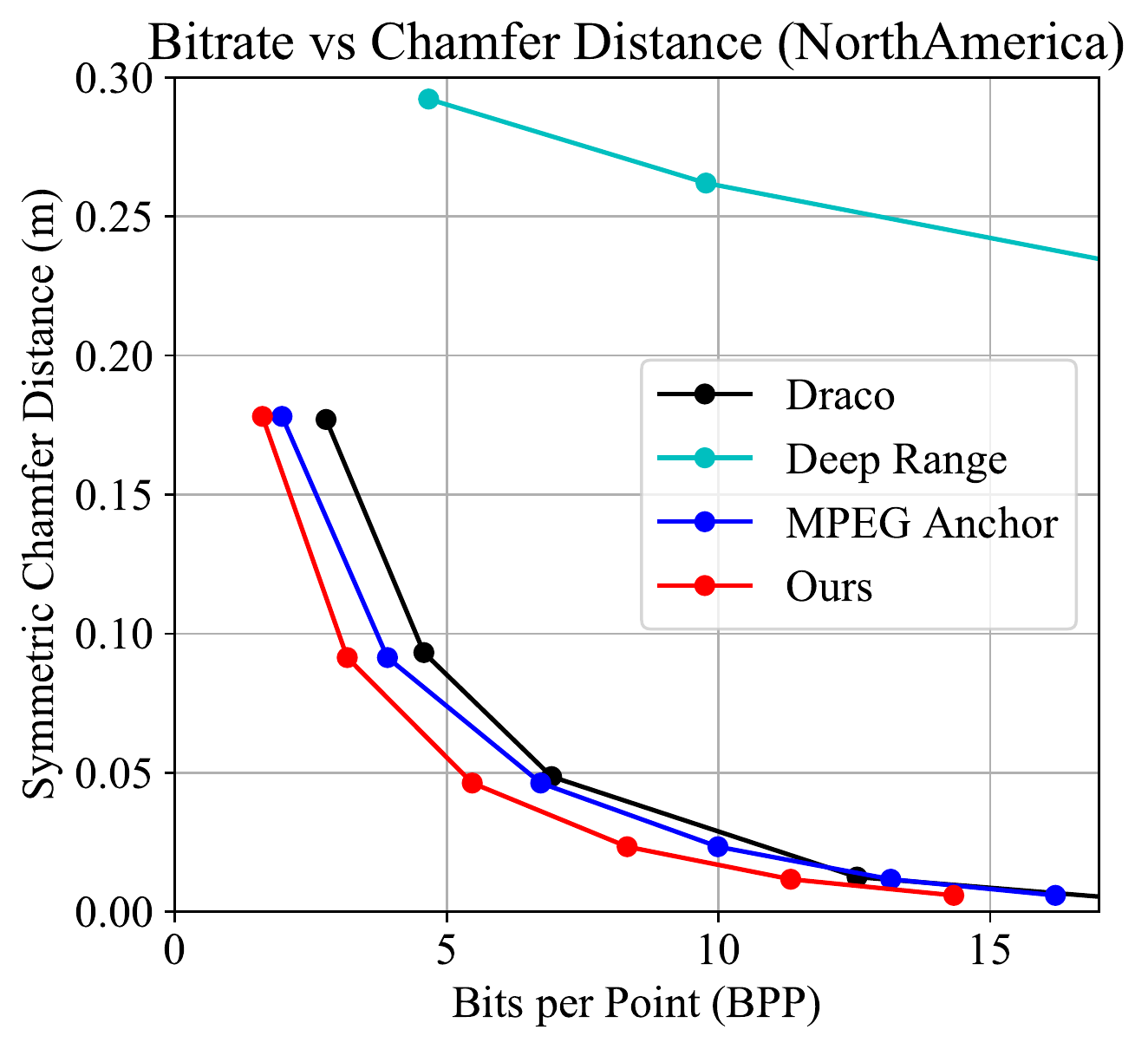}
\end{center}
\vspace{-0.25in}
\begin{center}
\includegraphics[height=0.25\textwidth]{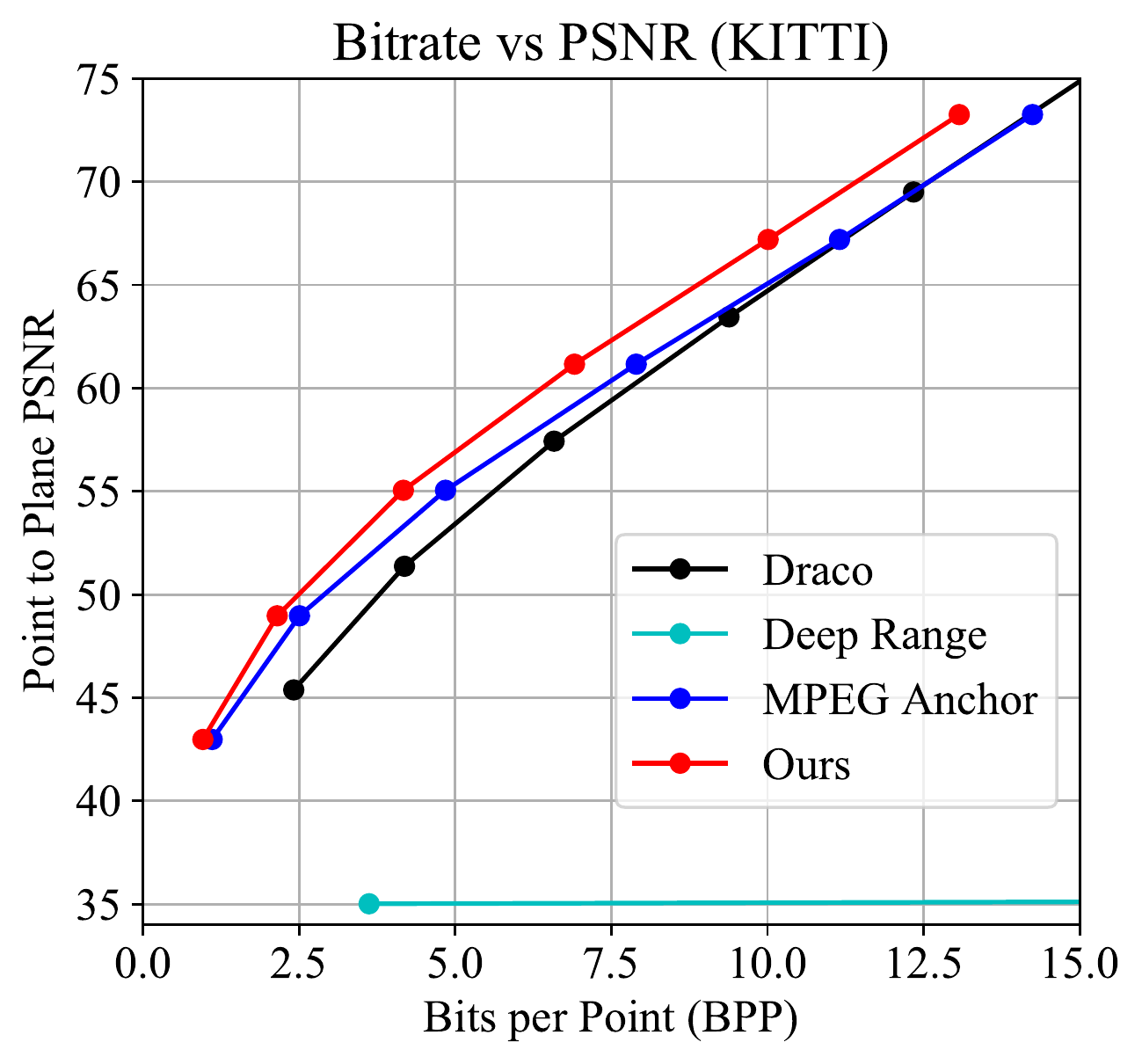}
\includegraphics[height=0.25\textwidth]{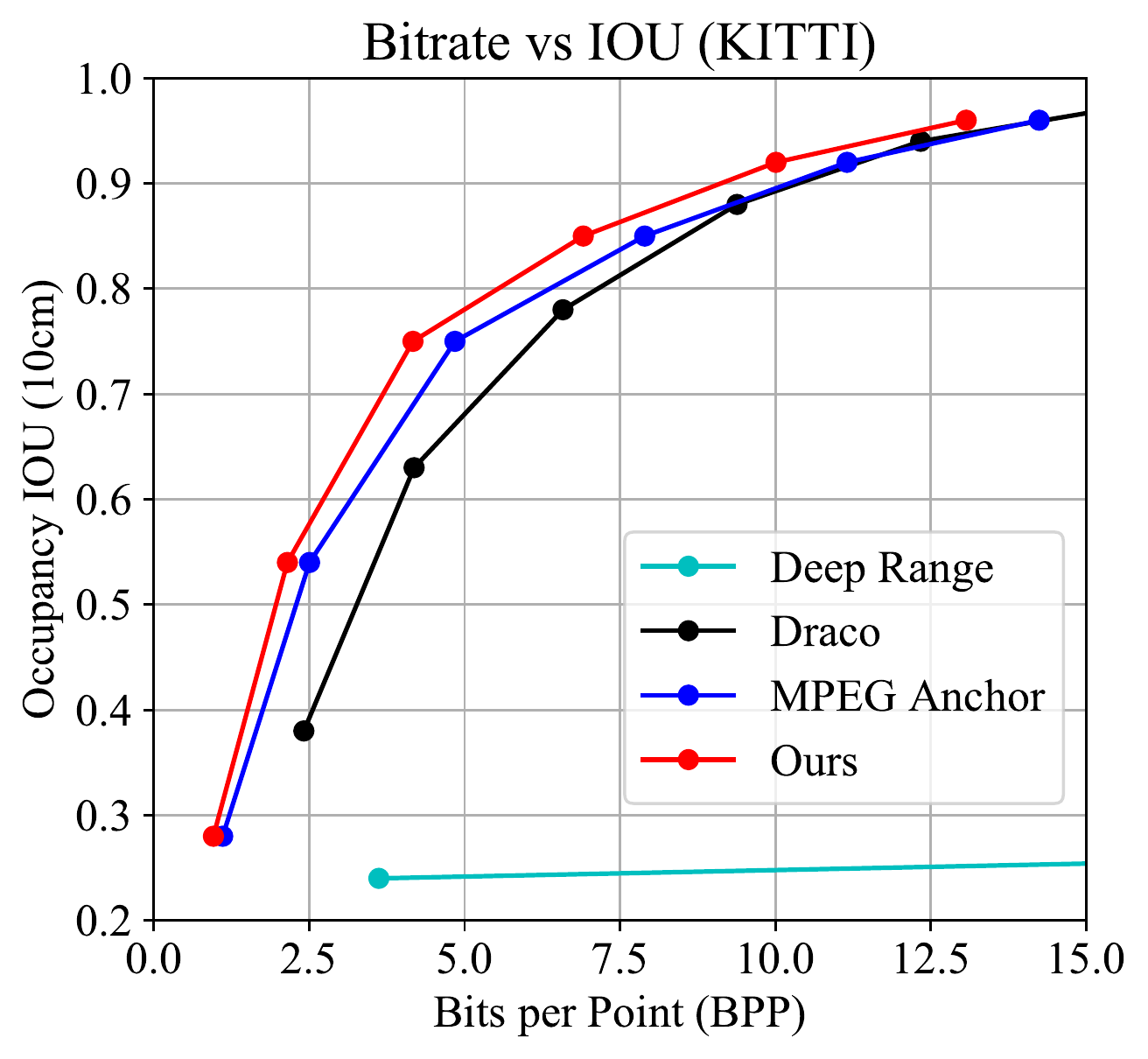}
\includegraphics[height=0.25\textwidth]{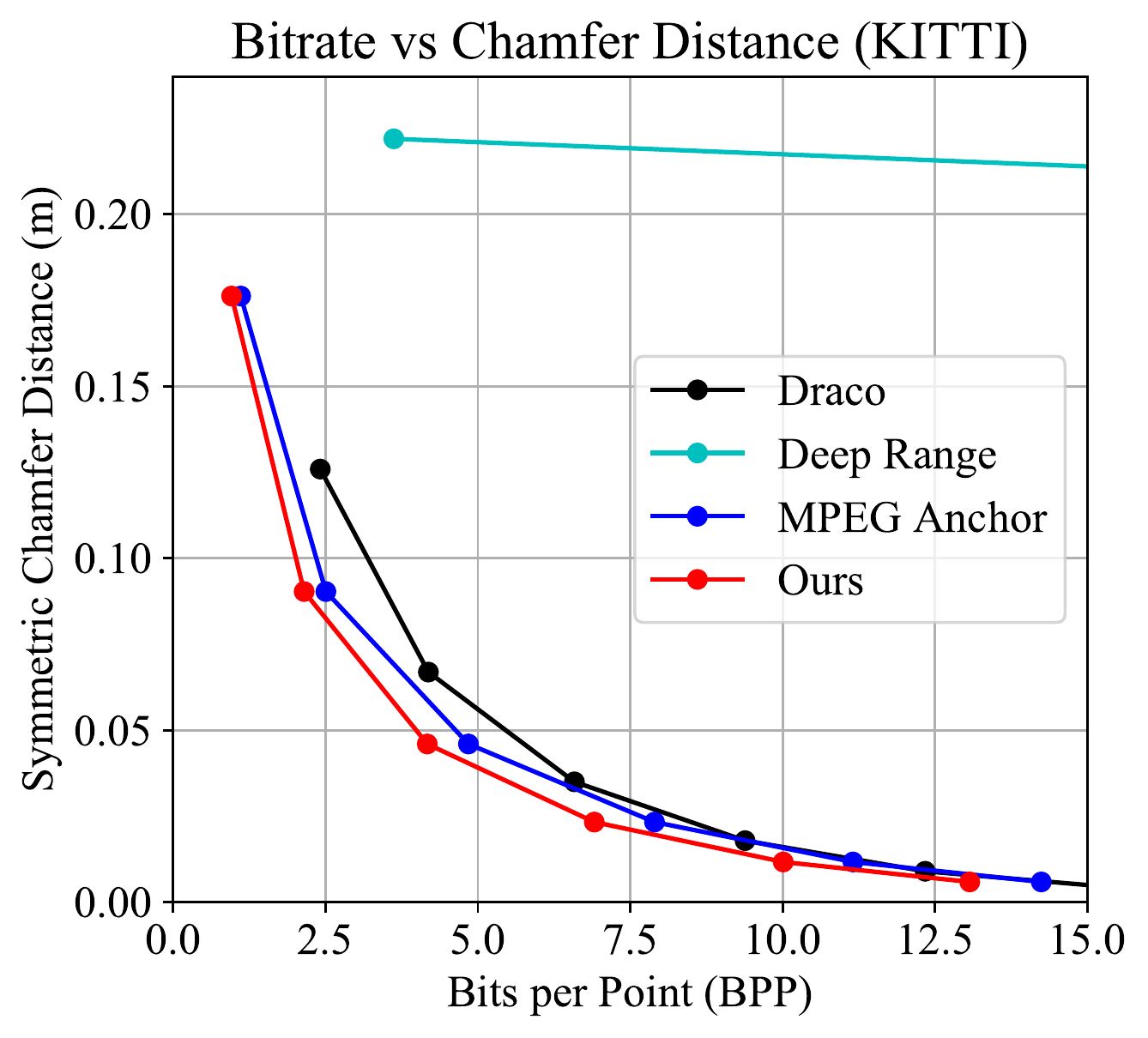}
\end{center}
\vspace{-0.2in}
\caption{
    Quantitative results on NorthAmerica and KITTI.
    From left to right: point-to-plane PSNR, IOU, and Chamfer distance.
}
\label{fig:quant_na}
\vspace{-0.15in}
\end{figure*}

Existing entropy models on octree structures tend to either lack the ability to accurately represent the data in the case of adaptive histograms \cite{schnabel2006, kammerl2012}, or require very long decoding times in the case of geometric predictions \cite{huang2018}.
Moreover, these entropy models do not fully utilize the hierarchical octree structure to encode geometric priors of the scene to facilitate entropy prediction.
Inspired by the success of using deep entropy models in image and video compression, we propose a deep network which models the entropy of the serialized octree data during entropy coding.
Our approach extends  prior methods in the sense that we better utilize the contextual information over the octree structure for prediction through an end-to-end learnable density estimation network.
\paragraph{Formulation:}
Given the sequence of occupancy 8-bit symbols $\mathbf{x} = [x_1, x_2 \dotsc x_n]$, the goal of an entropy model is to learn an estimated distribution $q(\mathbf{x})$ such that it minimizes the cross-entropy with the actual distribution of the symbols
$p(\mathbf{x})$:
\begin{equation}\label{equ:cross_entropy}
H(p, q) = \mathbb{E}_{\mathbf{x} \sim p}[-\log_2 q(\mathbf{x})]
\end{equation}
According to Shannon's source coding theorem \cite{shannon_lowerbound}, the cross-entropy between $q(\mathbf{x})$ and $p(\mathbf{x})$ provides a tight lower bound on the bitrate achievable by arithmetic or range coding algorithms \cite{witten_arithmeticcoding}; the better $q(\mathbf{x})$ approximates $p(\mathbf{x})$, the lower the true bitrate. We thus train to minimize the cross-entropy loss between the model’s predicted distribution $q$ and the distribution of training data.

\paragraph{Entropy Model:} \label{models}
We now describe the formulation of our entropy model over the octree structure $\mathbf{x}$. We factorize $q(\mathbf{x})$ into a product of conditional probabilities of each individual occupancy symbol $x_i$ as follows:
\begin{equation}\label{eq:mlp}
q(\mathbf{x}) = \prod_i q_i(x_i \mid \bx_{\an(i)}, \bc_i; \bw).
\end{equation}
where $\bx_{\an(i)} = \{x_{\pa(i)}, x_{\pa(\pa(i))}, ..., x_{\pa(...(\pa(i)))}\}$ with $| \bx_{\an(i)} | \leq K$
is the set of ancestor nodes of a given node $i$, up to a given order $K$,
and $\bw$ is the weights parametrizing our entropy model.
Here, $\bc_i$ is the context information that is available as prior knowledge during encoding/decoding of $x_i$,
such as octant index, spatial location of the octant, level in the octree, parent occupancy, \etc.
These models take advantage of the tree structure to gather both the information from nodes at coarser levels and the context information available at the current node. Intuitively, conditioning on ancestor nodes can help to reduce the entropy for the current node prediction, since it is easier to predict the finer geometry structure at the current node when the coarse structure represented by ancestor nodes is already known. Context information such as location information help to reduce entropy even further by capturing the prior structure of the scene.
For instance, in the setting of using LiDAR in the self-driving scenario, an occupancy node 0.5 meters above the LiDAR sensor is unlikely to be occupied.

\paragraph{Architecture:}
Our proposed entropy architecture models $q_i(x_i \mid \bx_{\an(i)}, \bc_i; \bw)$ by first extracting an independent contextual embedding for each $x_i$, and then performing progressive   aggregation of contextual embeddings to incorporate ancestral information $\bx_{\an(i)}$ for a given node.

For a given intermediate octree node $x_i$, the input context feature $\bc_i$ includes the node's location, octant, level, and parent (see Fig.~\ref{fig:overall}).
Specifically, `location' is the node's 3D location encoded as a vector in $ \mathbb{R}^3 $,
`octant' is its octant index encoded as an integer in $ \{0, \ldots, 7\} $,
`level' is its depth encoded as an integer in $ \{0, \ldots, \text{tree-depth}\} $,
and `parent' is its parent's 8-bit occupancy encoded as an integer in $ \{0, \ldots, 255\} $.
We extract an independent deep feature for each node through a multi-layer perceptron (MLP) with the context feature $\bc_i$ as input:

\begin{equation}
\mathbf{h}^{(0)}_{i} = \mathrm{MLP}^{(0)}(\mathbf{c}_i)
\end{equation}
Then, starting with the feature $\mathbf{h}^{(0)}_{i}$ for each node, we perform $K$ aggregations between the current node feature and the feature of its parent. At iteration $k$, the aggregation can also be modeled as an MLP:
\begin{equation}
\mathbf{h}_i^{(k)} = \mathrm{MLP}^{(k)}([\mathbf{h}_i^{(k-1)},  \mathbf{h}_{\pa(i)}^{(k-1)} ])
\end{equation}
where $\mathbf{h}_{\pa(i)}^{(k-1)}$ is the hidden feature of node $i$'s parent. For the root node, we consider its parent feature as all zero features for model consistency.
The final output of our model is a linear layer on top of the $K$-th aggregated feature $\mathbf{h}_i^{(k)}$, producing a 256-dimensional softmax of probabilities for the 8-bit occupancy symbol of the given node:
\begin{equation}
q_i( \cdot \mid \bx_{\an(i)}, \bc_i; \bw) = g(\mathbf{h}_i^{(k)})
\end{equation}
Note that these aggregations only aggregate the node feature with that of its parent, never its child; the child input context is not available during sequential decoding. Moreover, each additional aggregation increases the receptive field of ancestral features by 1, and so the $k$-th aggregation has a receptive field of $k$ ancestors. Fig.~\ref{fig:overall} depicts our proposed stacked entropy model with $K=3$. In this figure, a model with $K$ levels of aggregations predicts the probability of the current node $x_i$ by considering the node feature itself as well as $K-1$ generations of the ancestor's feature.

In this sense, we can view our aggregation as conceptually similar to other autoregressive models, such as the ``masked convolution'' used PixelCNN \cite{oord_pixelcnn} and ``causal convolution'' proposed in Wavenet \cite{oord_wavenet}. Unlike previous work either on 2D grids or 1D sequences, our autoregressive model is applied along the octree traversal path from the root to each node.

\paragraph{Detailed Architecture:} Here we discuss the detailed architecture of the each submodule of our stacked entropy model. The first MLP is a 5-layer MLP with 128 dimensional hidden features. All subsequent MLPs are 3-layer MLPs (with residual layers) with 128 dimensional hidden features. A final linear layer followed by a softmax is used to make the 256-way prediction.  Every MLP is Linear + ReLU without normalization layers.

\vspace{-2mm}
\paragraph{Learning:}

At  training time, the full entropy model is trained end-to-end with the cross-entropy loss on each node:
\begin{equation}
\ell = - \sum_i \sum_j y_{i, j} \log q_{i, j}
\end{equation}
where $\mathbf{y}_i$ is the  one-hot encoding of the ground-truth symbol at node $i$, and $q_{i, j}$ is the predicted probability of symbol $j$'s occurrence at node $i$.
\subsection{Entropy Coder} \label{sec:entropy_coder}

\paragraph{Encoding:}
At the encoding stage, we apply our model sequentially across different levels, from the root to leaves. Our proposed entropy model does not propagate information between nodes at the same level. Therefore, within each level, we are able to parallelize the computation for probability estimation.
Afterwards, we losslessly compress the octree raw bit-stream using an entropy coding algorithm such as arithmetic coding. Our network determines the arithmetic coder's entropy model by predicting the categorical distribution (0 to 255) for each byte $x_i$ in the sequence.

\paragraph{Decoding:}

To decode, the same entropy model is used in the arithmetic coder's decoding algorithm. An octree is then built from the decompressed bitstream and used to reconstruct the point cloud. Due to the auto-regressive fasion of the entropy model, each node probability estimation is only dependent on itself and decoded node features at higher level of the octree.  In addition, the octree is serialized in a breadth-first search fashion.
As a result, given a node $x_i$, its ancestors in the octree $ \bx_{\an(i)} $ are decoded before $ x_i $, making it feasible for the decoder to also decode $ x_i $.

%% file: sections/exp.tex
\newcommand{\bp}{\mathbf{p}}
\newcommand{\bn}{\mathbf{n}}
\newcommand{\cP}{\mathcal{P}}
\newcommand{\hcP}{\hat{\mathcal{P}}}

\section{Experiments}

\begin{figure*}
\vspace{-0.15in}
\begin{center}
\begin{overpic}[width=0.245\textwidth]{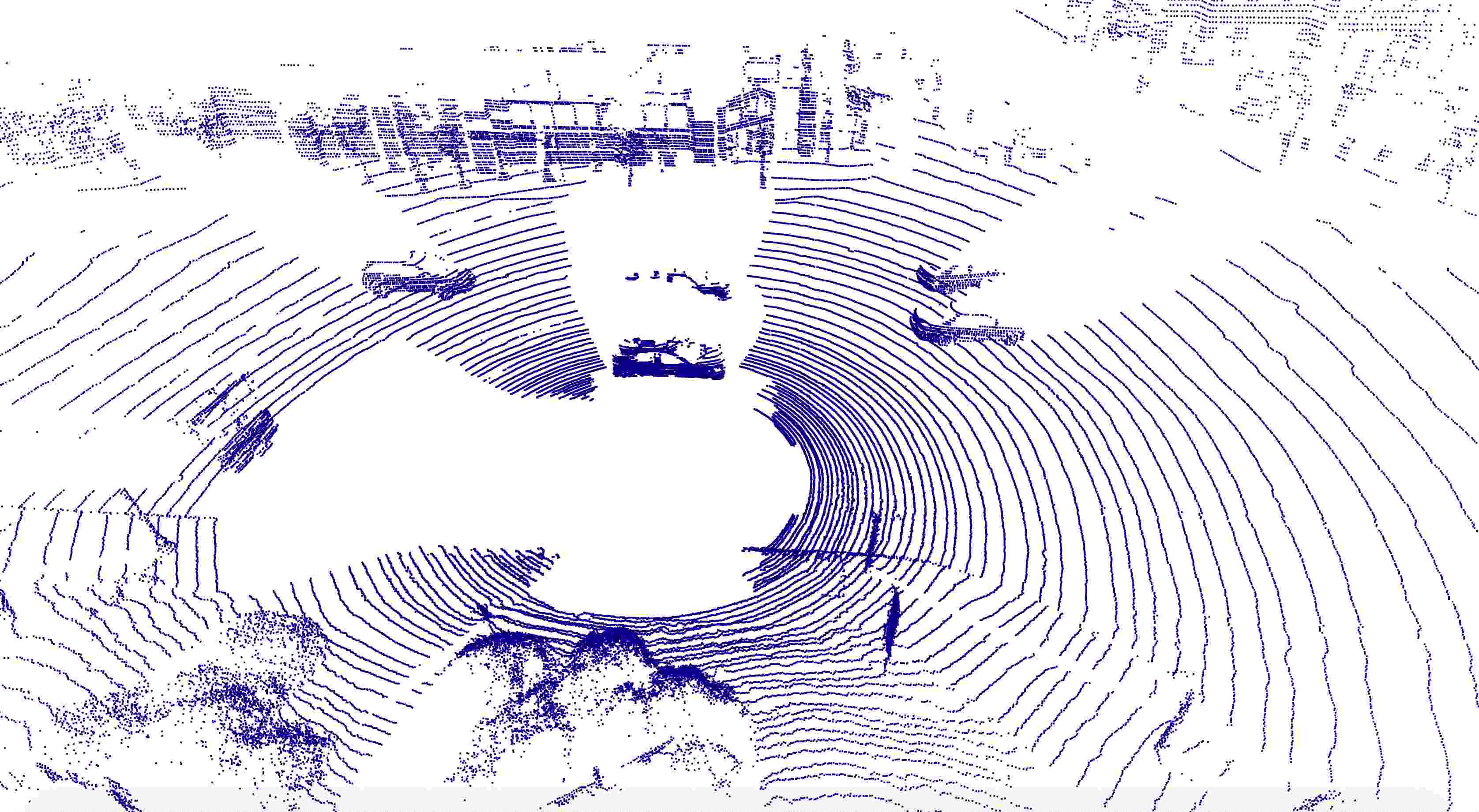}
 \put (0,49) {\colorbox{gray!30}{\scriptsize GT (NorthAmerica)}}
\end{overpic}
\begin{overpic}[width=0.245\textwidth]{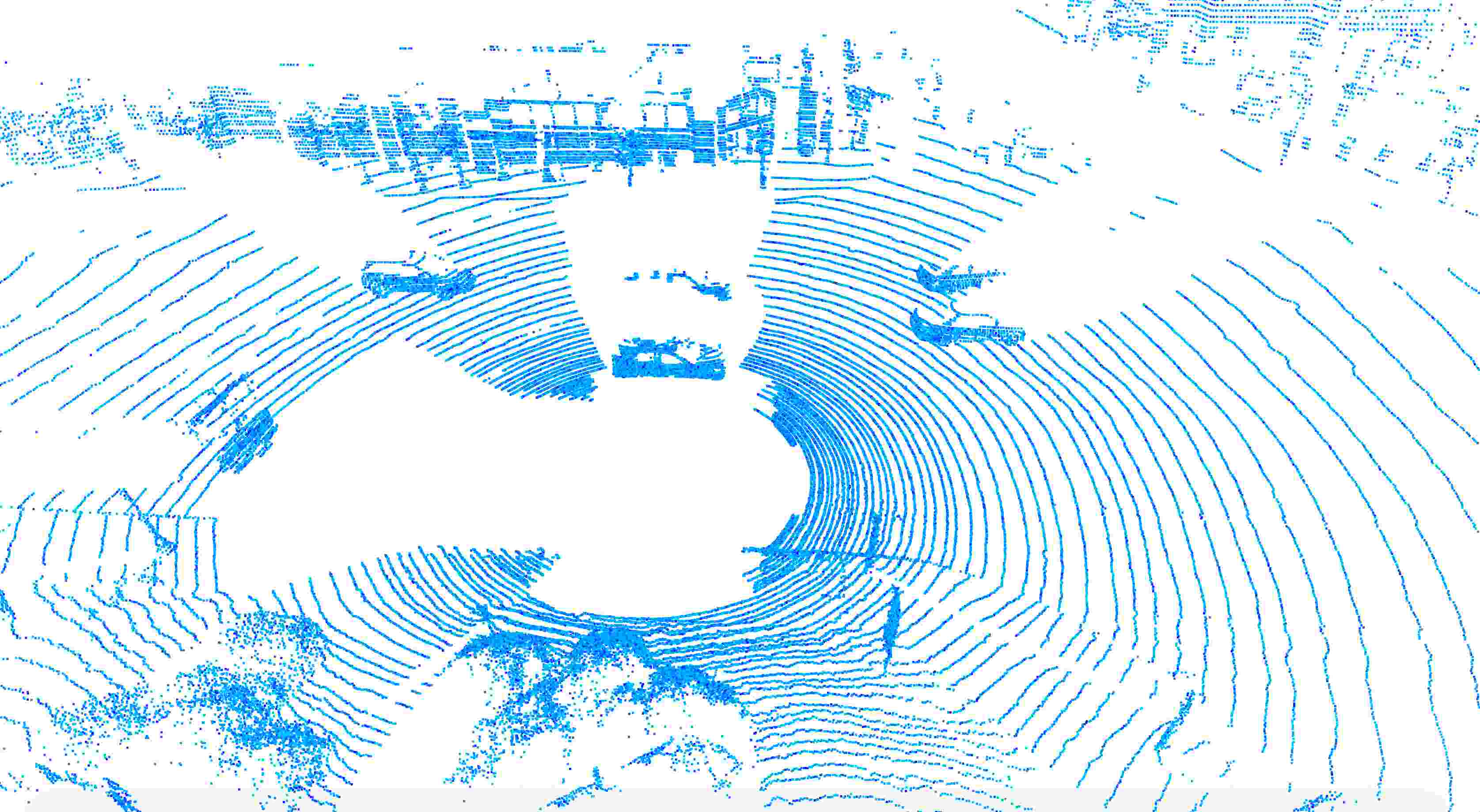}
 \put (0,49) {\colorbox{gray!30}{\scriptsize Ours: PSNR 80.06, Bitrate 11.36}}
\end{overpic}
\begin{overpic}[width=0.245\textwidth]{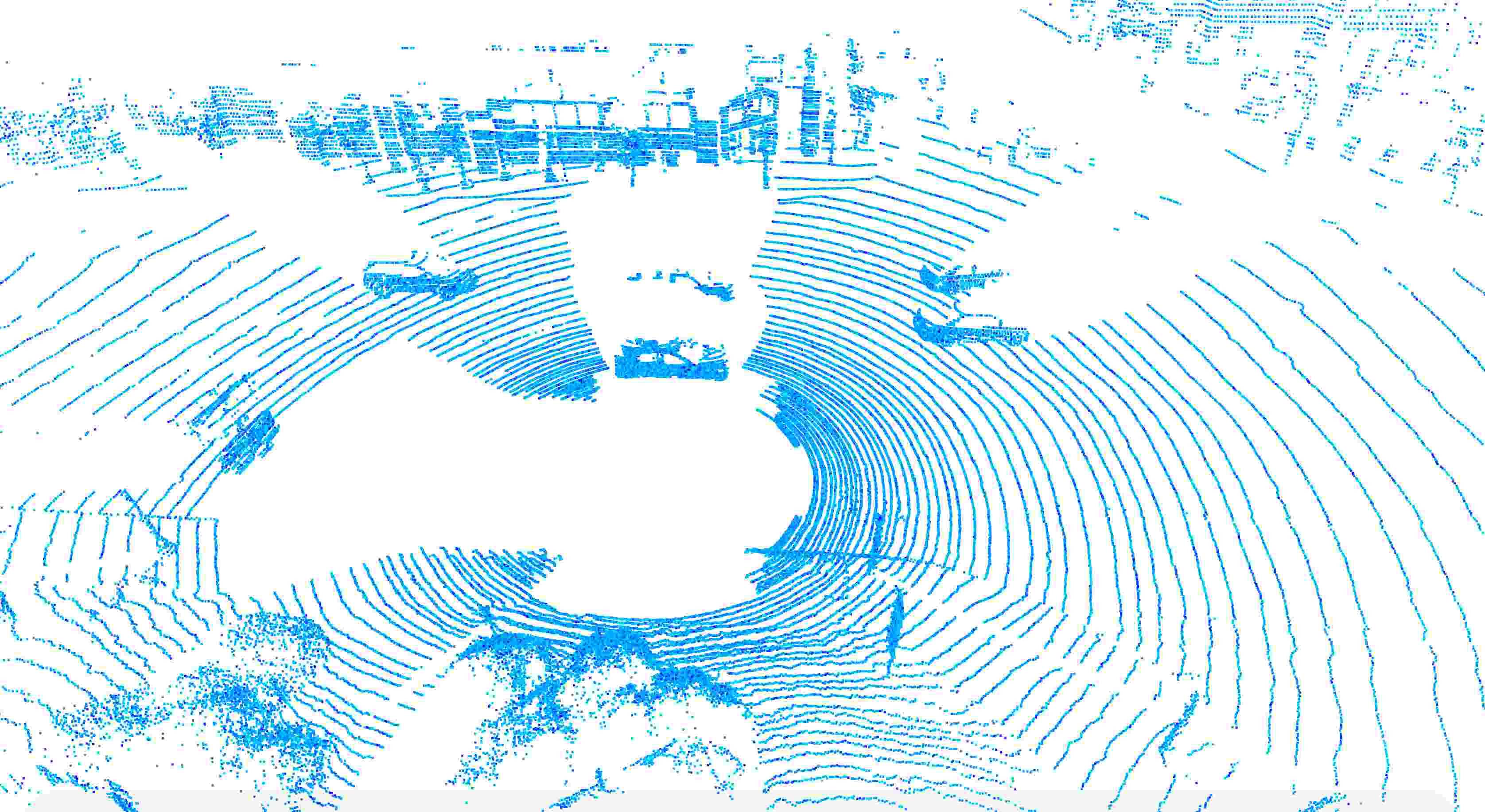}
 \put (0,49) {\colorbox{gray!30}{\scriptsize Draco: PSNR 79.38, Bitrate 12.53}}
\end{overpic}
\begin{overpic}[width=0.245\textwidth]{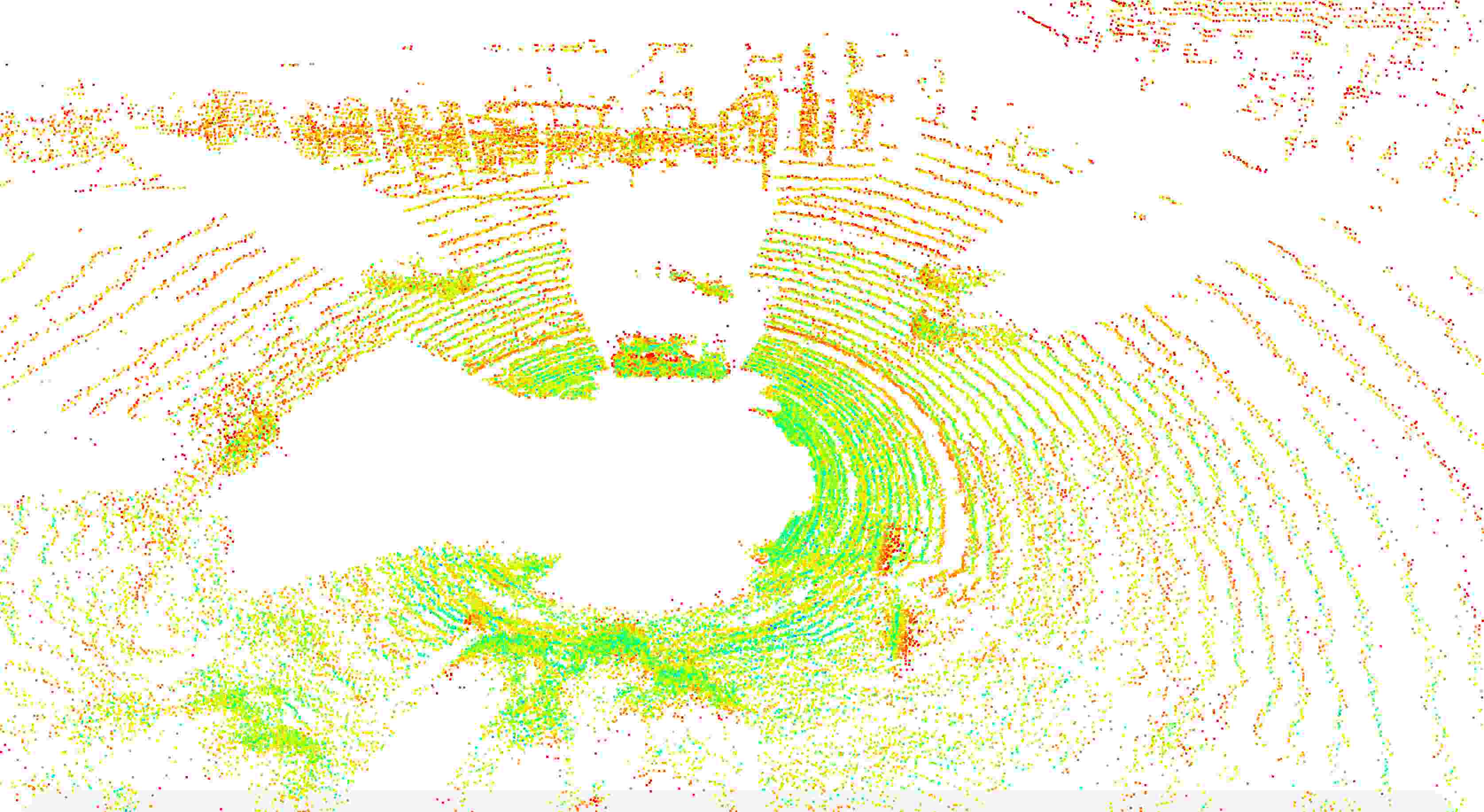}
 \put (0,49) {\colorbox{gray!30}{\scriptsize Range: PSNR 50.35, Bitrate 13.99}}
\end{overpic}

\begin{overpic}[width=0.245\textwidth]{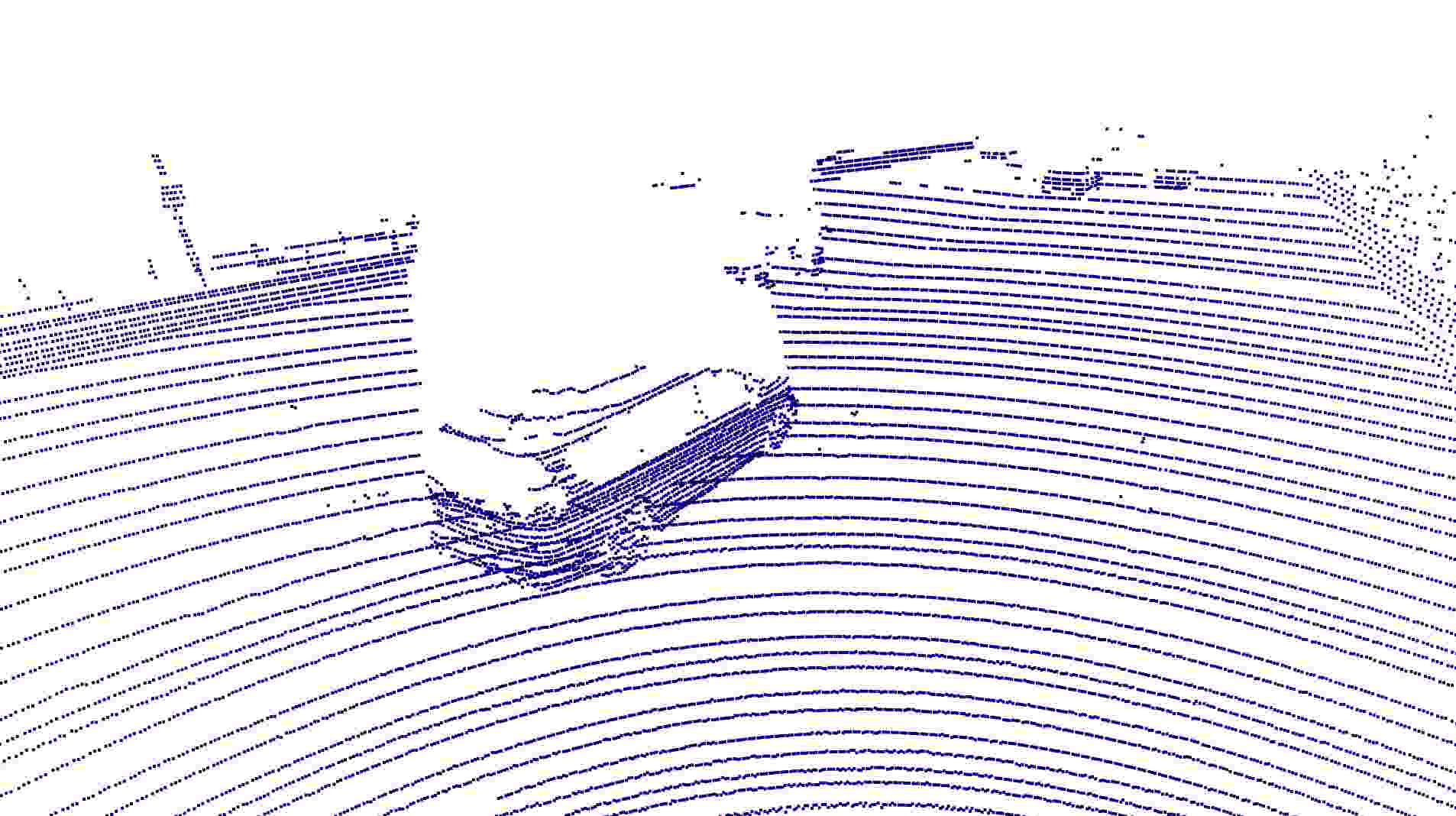}
 \put (0,50) {\colorbox{gray!30}{\scriptsize GT (NorthAmerica)}}
\end{overpic}
\begin{overpic}[width=0.245\textwidth]{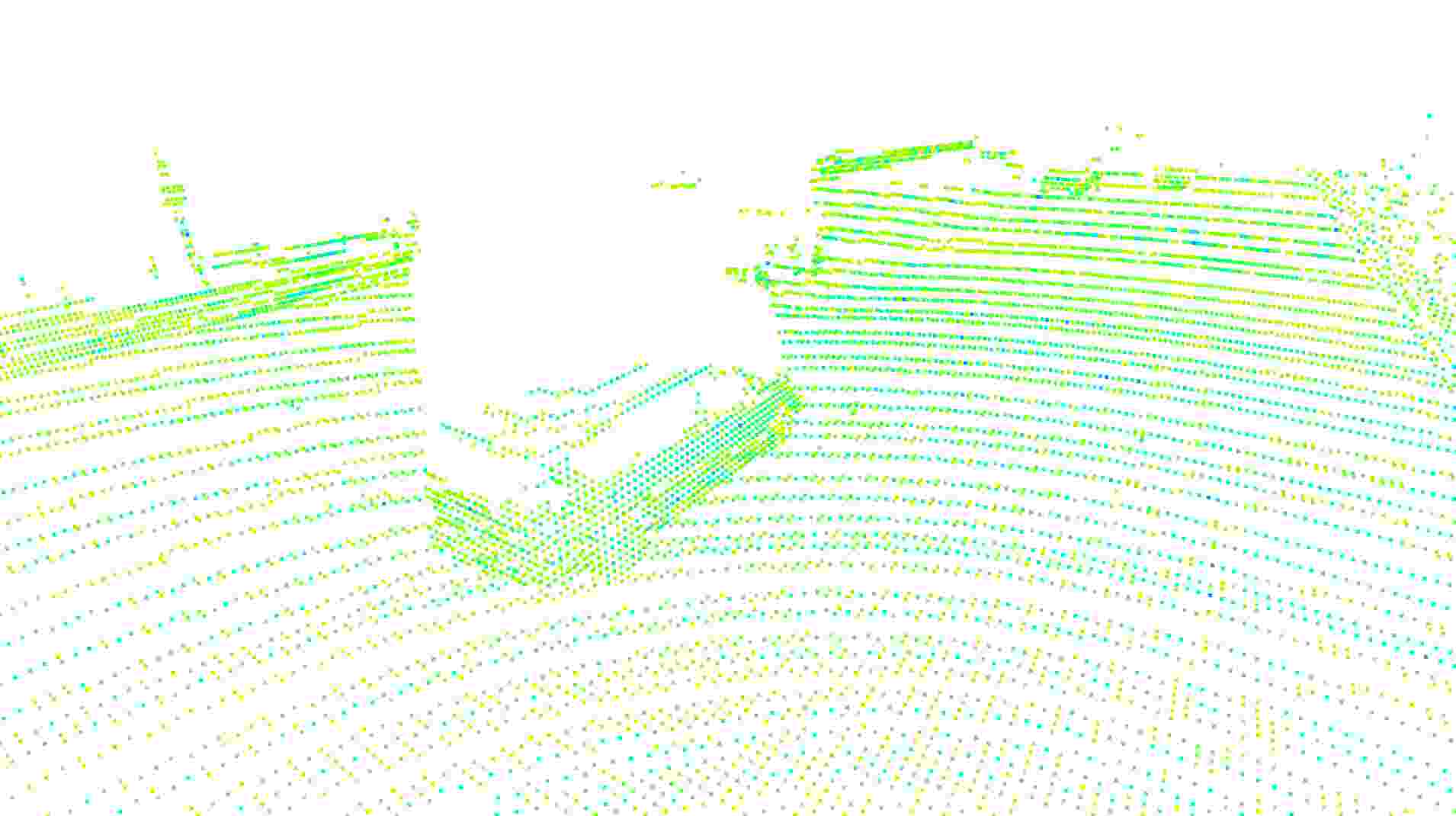}
 \put (0,50) {\colorbox{gray!30}{\scriptsize Ours: PSNR 58.54, Bitrate 2.06}}
\end{overpic}
\begin{overpic}[width=0.245\textwidth]{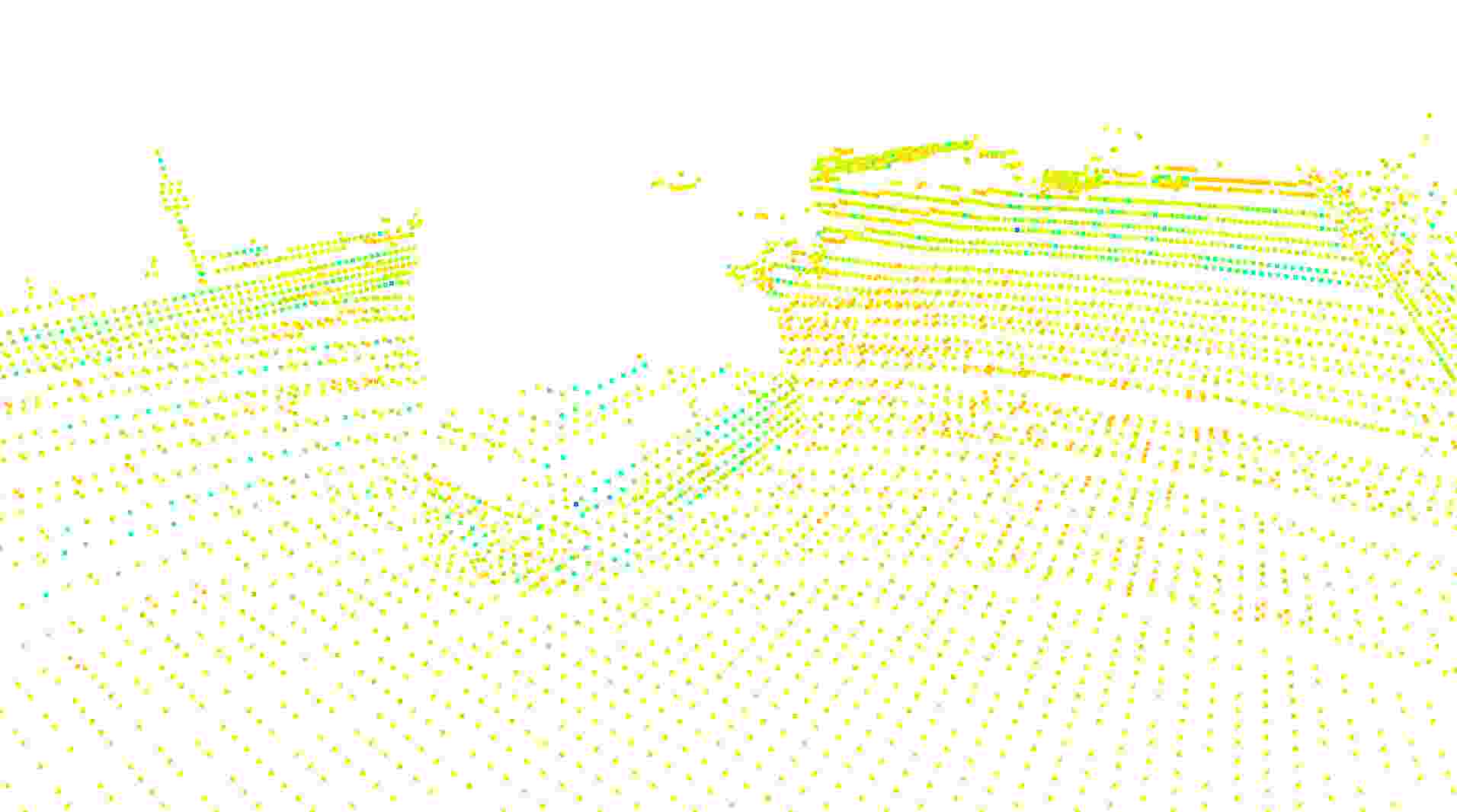}
 \put (0,50) {\colorbox{gray!30}{\scriptsize Draco: PSNR 51.52, Bitrate 2.17}}
\end{overpic}
\begin{overpic}[width=0.245\textwidth]{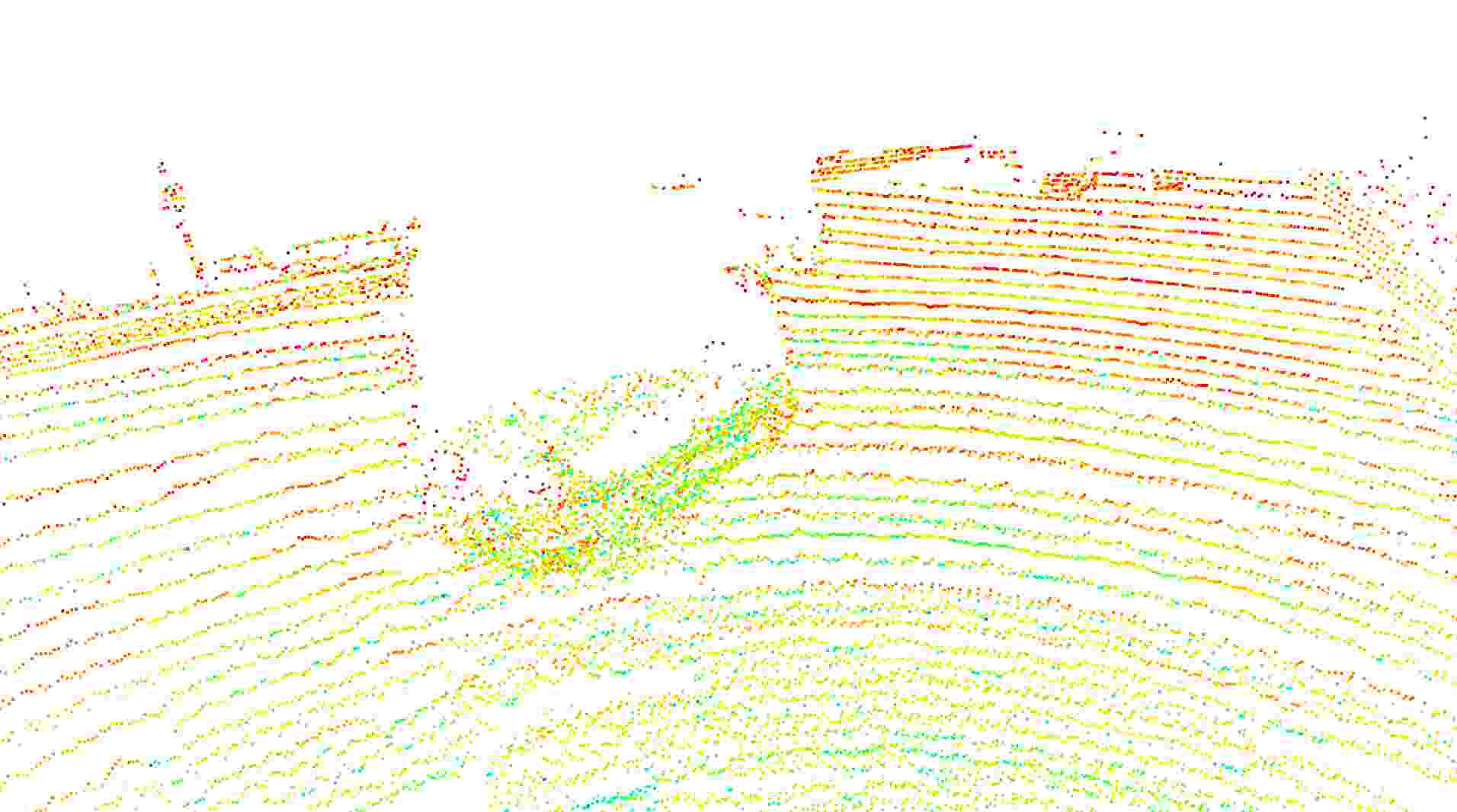}
 \put (0,50) {\colorbox{gray!30}{\scriptsize Range: PSNR 46.50, Bitrate 5.58}}
\end{overpic}

\begin{overpic}[width=0.245\textwidth]{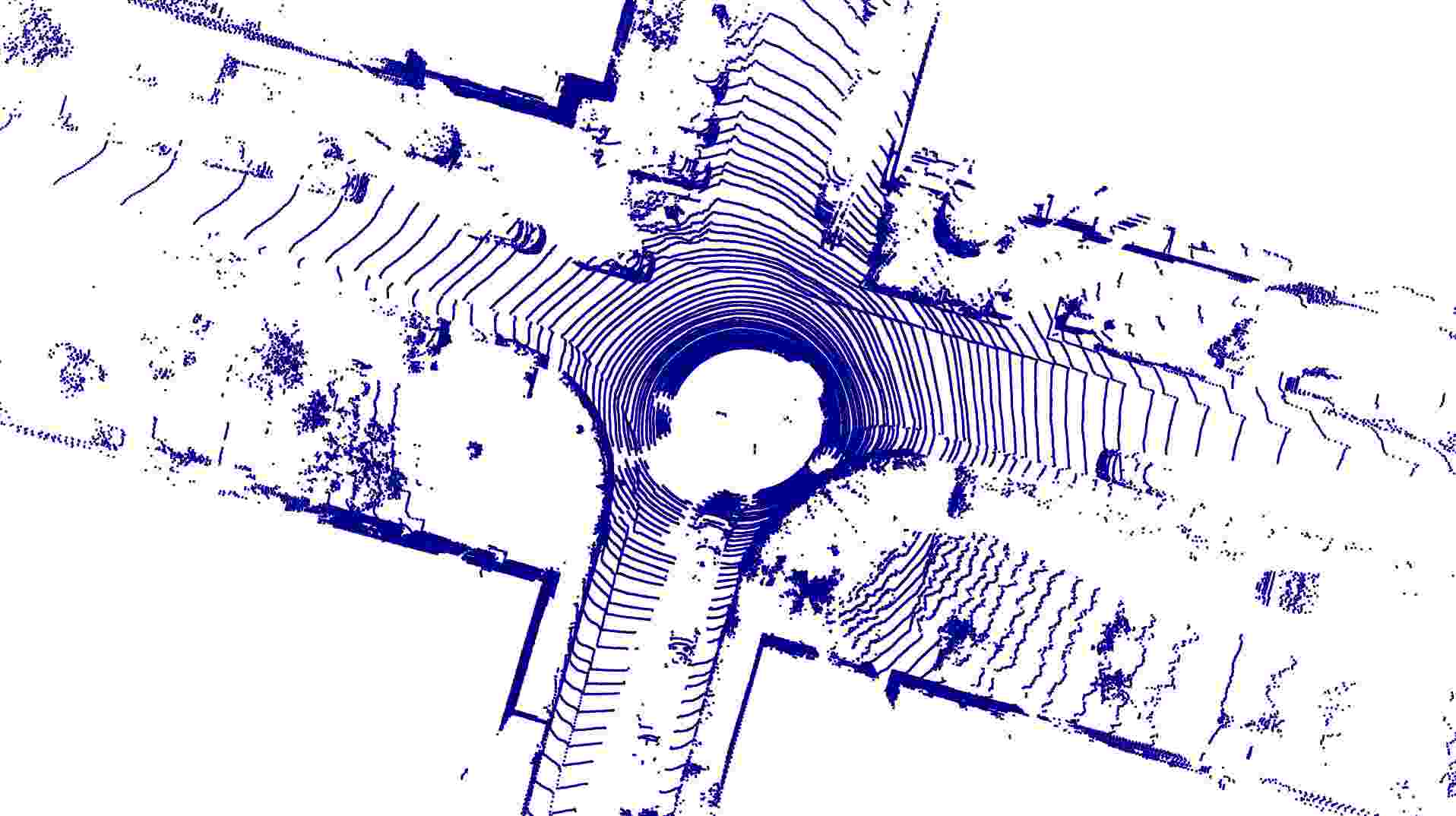}
 \put (0,50) {\colorbox{gray!30}{\scriptsize GT (KITTI)}}
\end{overpic}
\begin{overpic}[width=0.245\textwidth]{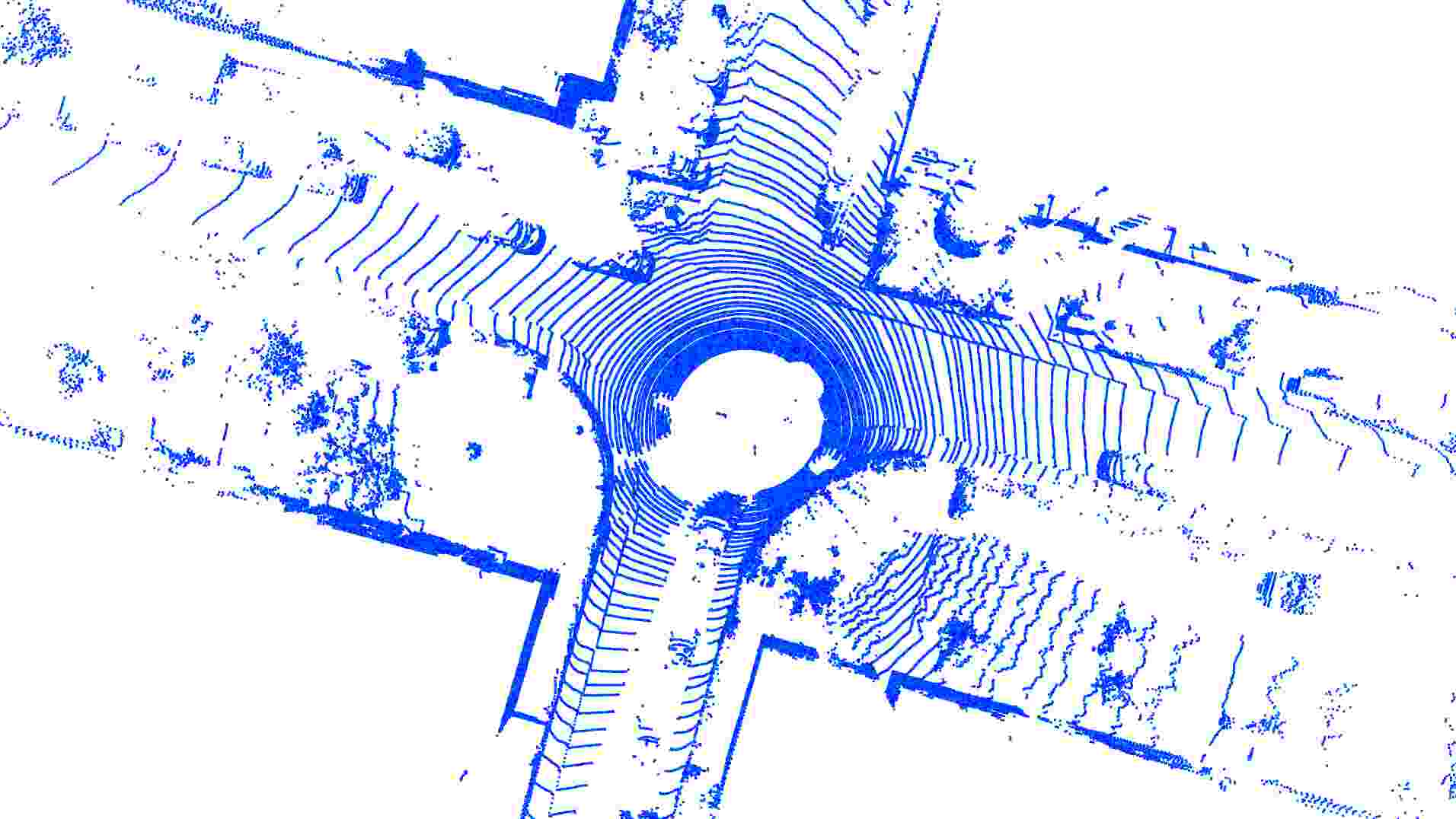}
 \put (0,50) {\colorbox{gray!30}{\scriptsize Ours: PSNR 71.59, Bitrate 13.59}}
\end{overpic}
\begin{overpic}[width=0.245\textwidth]{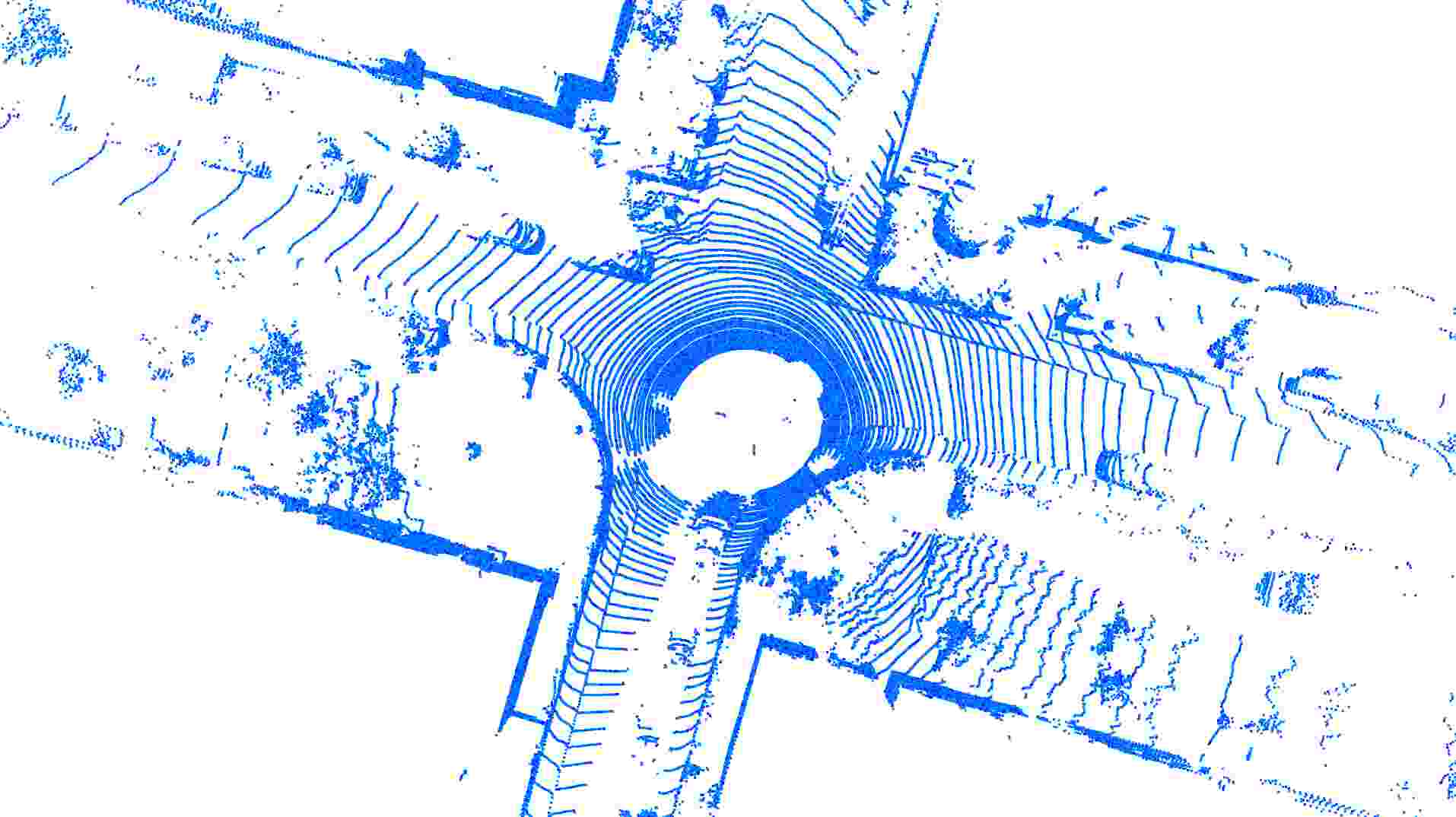}
 \put (0,50) {\colorbox{gray!30}{\scriptsize Draco: PSNR 68.85, Bitrate 13.65}}
\end{overpic}
\begin{overpic}[width=0.245\textwidth]{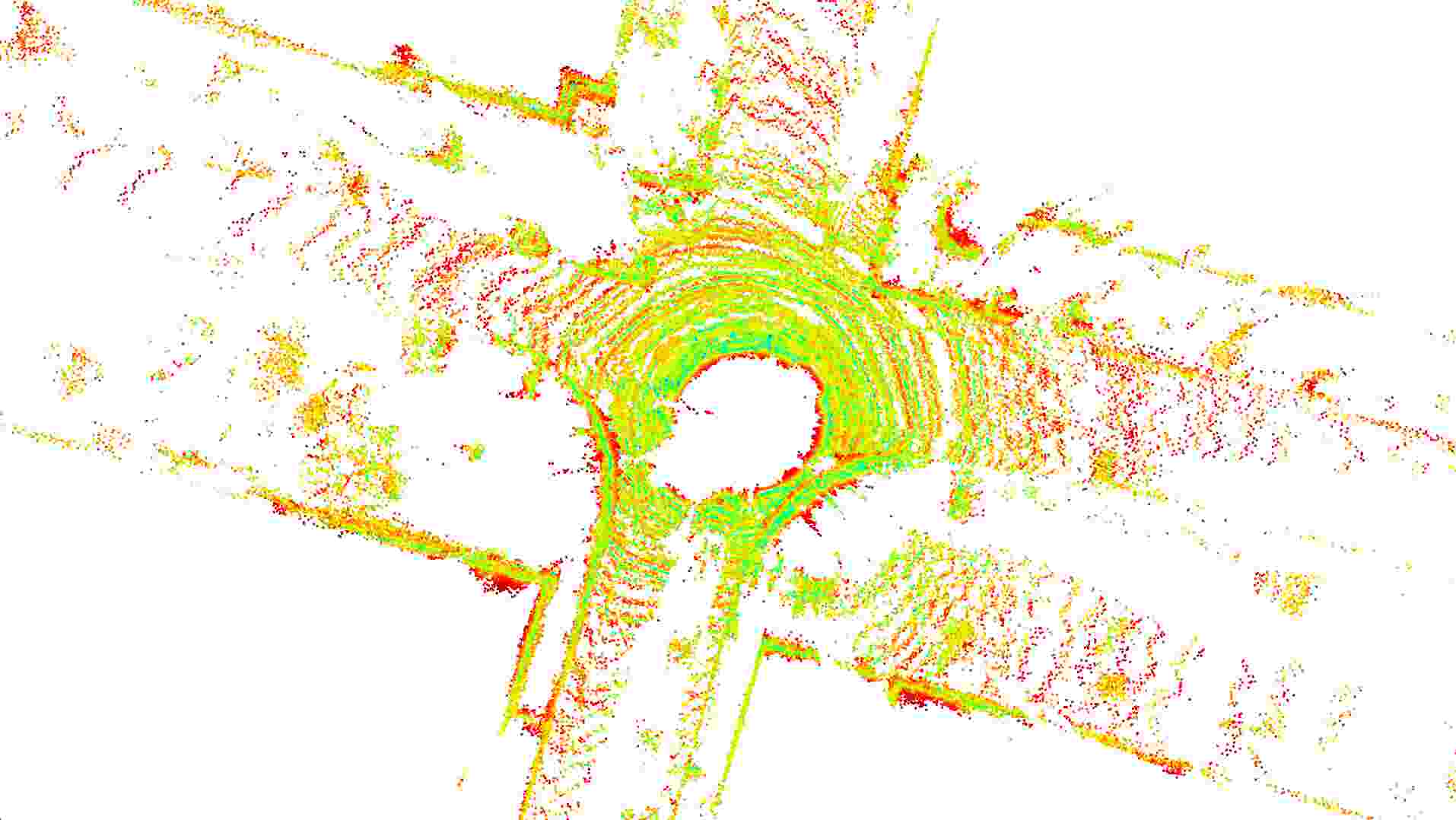}
 \put (0,50) {\colorbox{gray!30}{\scriptsize Range: PSNR 34.43, Bitrate 13.27}}
\end{overpic}

\begin{overpic}[width=0.245\textwidth]{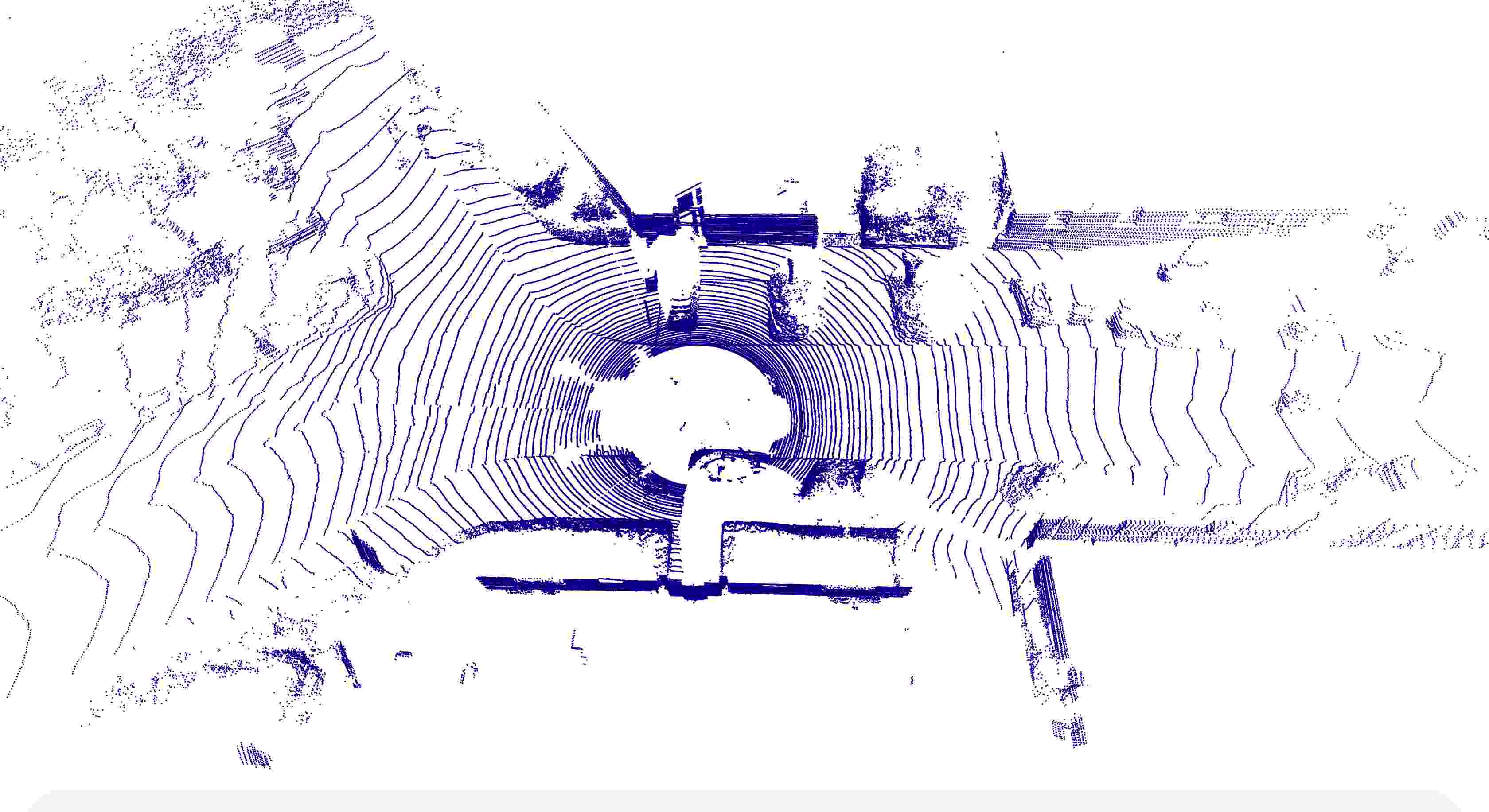}
 \put (0,48) {\colorbox{gray!30}{\scriptsize GT (KITTI)}}
\end{overpic}
\begin{overpic}[width=0.245\textwidth]{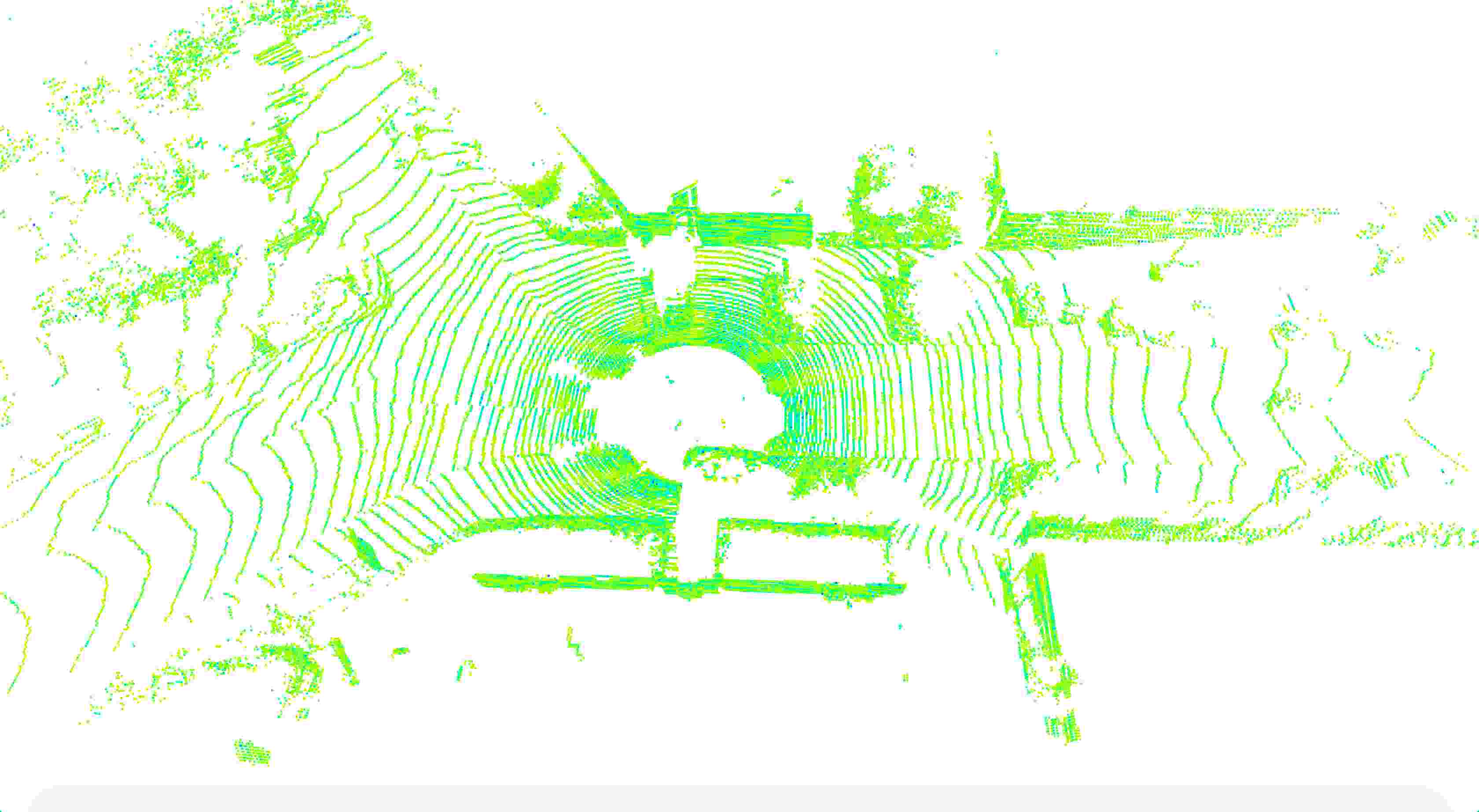}
 \put (0,48) {\colorbox{gray!30}{\scriptsize Ours: PSNR 54.81, Bitrate 2.02}}
\end{overpic}
\begin{overpic}[width=0.245\textwidth]{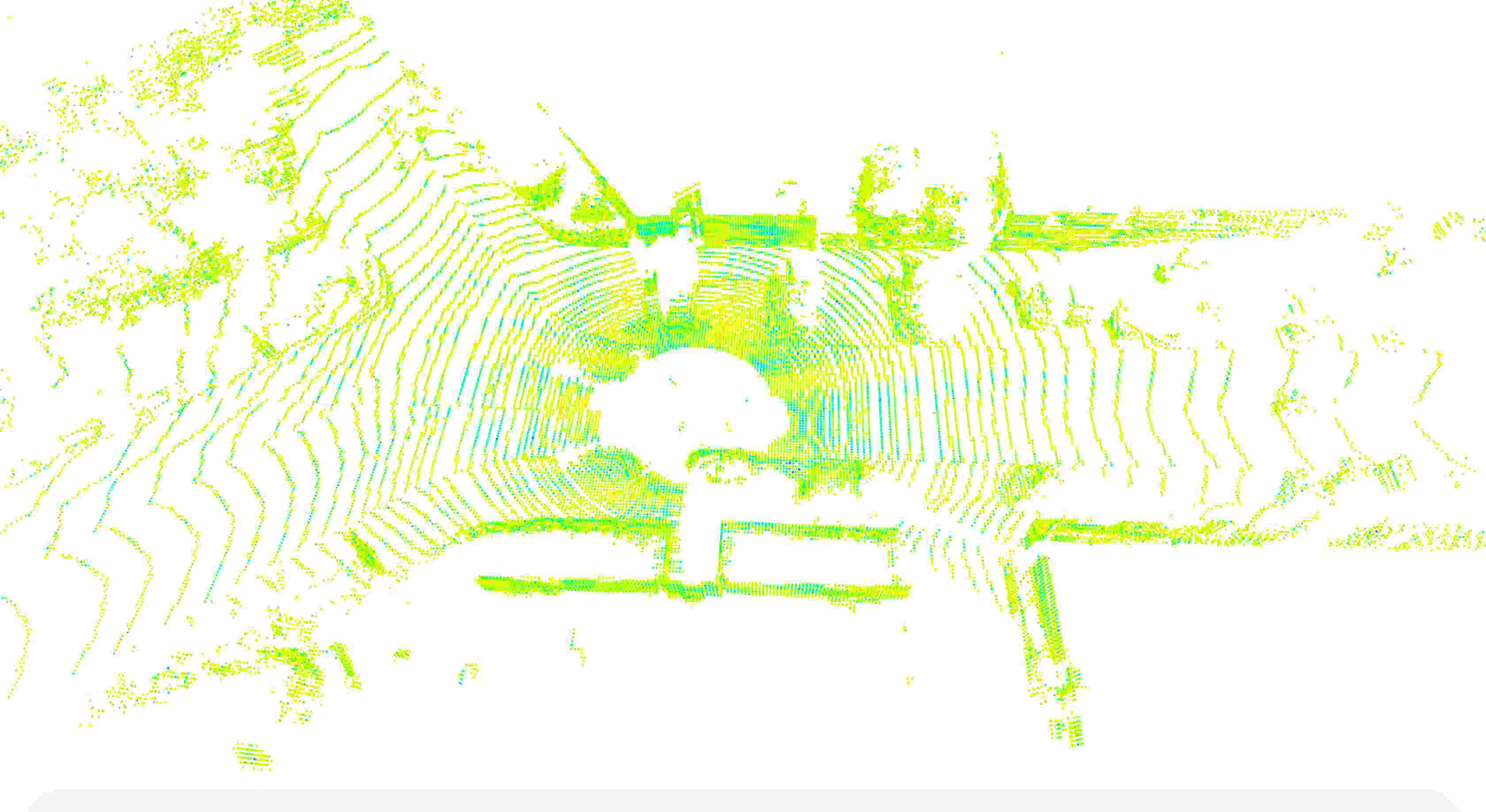}
 \put (0,48) {\colorbox{gray!30}{\scriptsize Draco: PSNR 51.16,  Bitrate 2.35}}
\end{overpic}
\begin{overpic}[width=0.245\textwidth]{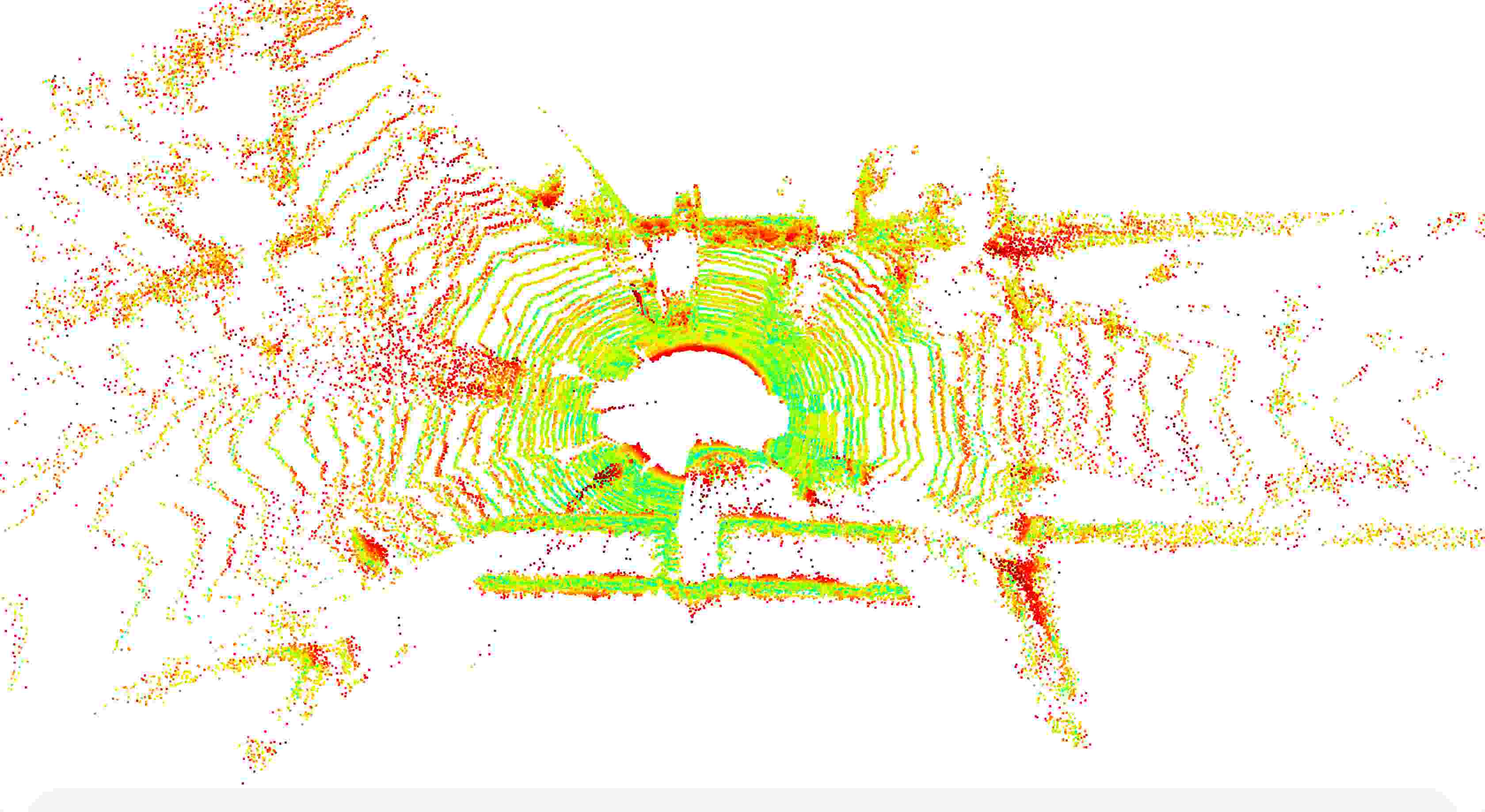}
 \put (0,48) {\colorbox{gray!30}{\scriptsize Range: PSNR 33.30, Bitrate 3.61}}
\end{overpic}

\includegraphics[width=0.5\textwidth]{./figures/colormap.png}
\end{center}
\vspace{-0.2in}
\caption{
    Qualitative results on NorthAmerica and KITTI.
    From left to right: Ground Truth, Ours, Draco, and Deep Range.
}
\label{fig:qual_kitti}
\vspace{-0.15in}
\end{figure*}

In this section we validate the effectiveness of our proposed approach on
two challenging real-world LiDAR datasets with drastically varying scenes.
We compare our method against several state-of-the-art point cloud compression
algorithms in terms of both reconstruction quality and their effects on downstream
perception tasks.

\subsection{Datasets}

\begin{table}[]
\vspace{-0.15in}
\centering
\resizebox{\linewidth}{!}{%
\begin{tabular}{cccc|c|c|c}
\toprule
           &            &            &            & \multicolumn{3}{c}{Bitrate} \\
L          & P          & O          & LL         & Depth = 12                      & Depth = 14                      & Depth = 16        \\
\midrule
           &            &            &            & 3.91                            & 9.99                            & 16.21             \\
\checkmark &            &            &            & 3.86                            & 9.79                            & 15.91             \\
\checkmark & \checkmark &            &            & 3.62                            & 9.33                            & 15.41             \\
\checkmark & \checkmark & \checkmark &            & 3.59                            & 9.27                            & 15.35             \\
\checkmark & \checkmark & \checkmark & \checkmark & \textbf{3.48}                   & \textbf{8.91}                   & \textbf{14.97}    \\
\bottomrule
\end{tabular}%
}
\caption{
    Ablation study on input context features.
    L, P, O, and LL stand for the node's octree level,
    its parent occupancy symbol,
    its octant index,
    and its spatial location respectively.
}
\label{tab:ablation}
\end{table}

\begin{table}[]
\vspace{-0.1in}
\centering
\resizebox{\linewidth}{!}{%
\begin{tabular}{c|c|c|c}
\toprule
                & \multicolumn{3}{c}{Bitrate} \\
\# Aggregations & Depth = 12     & Depth = 14    & Depth = 16    \\
\midrule
0               & 3.48           & 8.91          & 14.97          \\
1               & 3.39           & 8.78          & 14.84          \\
2               & 3.31           & 8.59          & 14.64          \\
3               & 3.25           & 8.47          & 14.51          \\
4               & \textbf{3.17}  & \textbf{8.32} & \textbf{14.33} \\
\bottomrule
\end{tabular}%
}
\caption{Ablation study on the number of aggregations.}
\label{tab:stacks}
\vspace{-0.15in}
\end{table}

\paragraph{NorthAmerica:}
We collected a new internal dataset comprising of driving scenes from a
wide variety of urban and highway environments in multiple cities/states across North America.
From this dataset, we sampled 500K raw LiDAR scans collected by a Velodyne HDL-64
sensor to train our entropy model.
No additional filtering or processing is applied to these LiDAR point clouds.
For evaluation of reconstruction quality, we collected 472 snippets each containing 250 LiDAR scans.
In addition, we also annotate these frames with 2D bird's eye view bounding boxes
for the vehicle, pedestrian, and motorbike classes, as well as per-point semantic
labels for the vehicle, pedestrian, motorbike, road, and background classes.
We use these labels for evaluation on downstream perception tasks.

\paragraph{KITTI:}
To evaluate our method's domain transfer capability, we show results on
SemanticKITTI~\cite{behley_semantickitti}---a public self-driving dataset
containing 21351 scans with 4.5 billion points collected from a Velodyne HDL-64 sensor.
As SemanticKITTI also contains dense point-wise labels from 25 classes, we also
evaluate downstream task performance on this dataset.
Note that there is a significant domain shift from our internal data to KITTI in
terms of the scene layout as well as sensor configuration, such as sensor
height, ego-occlusion, ray angles, \etc.

\subsection{Experimental Details}

\paragraph{Baselines:}
Our baselines include two of the best off-the-shelf point cloud compression approaches, namely Google's \textrm{Draco} encoder (`Draco') \cite{draco} and Mekuria \etal 's octree-based algorithm \cite{mekuria2017} which serves as the MPEG anchor (`MPEG anchor'). In addition, we compare our method against a deep baseline model using a range image representation for the point cloud (`Deep Range').
For the range image representation, we utilize the rolling shutter characteristics to convert each LiDAR scan from Euclidean coordinates to polar coordinates, and store it as a 2.5D range image. We then train the Ball\'e hyperprior model \cite{balle_varhyperprior}, a state-of-the-art image compression model, on these images. During decoding we reconstruct the 2.5D range image and convert it back to Euclidean point cloud.

\paragraph{Implementation Details:}
We train our entropy models on full 16-level octrees.
Training a single model on the full 16-level octree allows for variable rate compression within the same model, since
during test time, we can truncate the same octree over different levels to evaluate
our models over different levels of quantization. Specifically, we evaluate our octree models
with depths ranging from 11 to 16 to measure the bitrate-quality tradeoff. The quantization error ranges from 0.3cm to 9.75cm,
and every decrement in tree height doubles this value.

Our entropy model is implemented in PyTorch and trained over 16 GPUs with the Adam optimizer.
We use a learning rate of $ 1\mathrm{e-}4 $ for 500K iterations.

\subsection{Compression Metrics}

\paragraph{Reconstruction Quality:}

To evaluate reconstruction quality, we use two families of metrics: distance and occupancy.  A commonly used distance-based metric to evaluate point cloud similarity is the symmetric point-to-point Chamfer distance $\textrm{CD}_\textit{sym}$. For a given GT point cloud $\mathcal{P} = \{ \mathbf{p}_i \}_{i = 1, ..., N}$ and reconstructed point cloud $\hat{\cP}$:
\begin{equation}
\textrm{CD}(\cP, \hcP) = \frac{1}{\lvert \cP \rvert}\sum_{i}\min_{j}\lVert \bp_i - \hat{\bp}_j\rVert_2
\end{equation}
\begin{equation} \label{eq:pointtopoint}
\textrm{CD}_\textit{sym} (\cP, \hcP) = \textrm{CD}(\cP, \hcP) + \textrm{CD}(\hcP, \cP)
\end{equation}
A second distance-based metric, symmetric point-to-plane $ \textrm{PSNR}_\textit{sym} $, \cite{pointtoplane} accounts for point cloud resolution:
\begin{equation}
\textrm{PSNR}(\cP, \hcP) = 10 \log_{10} \frac{\max_{i}\lVert \bp_i - \hat{\bp}_i\rVert_2^2}{\textrm{MSE}(\cP, \hcP)}
\end{equation}
\begin{equation} \label{eq:psnr}
\textrm{PSNR}_\textit{sym}(\cP, \hcP) = \min \{\textrm{PSNR}(\cP, \hcP), \textrm{PSNR}(\hcP, \cP)\}
\end{equation}
where $\textrm{MSE}(\cP, \hcP) = \frac{1}{\lvert \cP \rvert}\sum_{i} ((\hat{\bp}_i - \bp_i)\cdot \mathbf{n}_i)^2 $
is the point-to-plane distance,
$\hat{\bp}_i  = \arg\min_{\bp \in \hat{\cP}}\lVert \bp_i - \bp \rVert_2^2$ is the closest point in $\hat{\cP}$ for each point $\bp_i$, and $\bn_i$ is the normal at each $\bp_i$.
We estimate the normal $ \mathbf{n}_i $ at each point $ \bp_i \in \cP $
using the Open3D function \verb+estimate_normals+ with $ k = 12 $ nearest
neighbors~\cite{open3d2018}.

\paragraph{Occupancy Quality:}
It is common practice to use LiDAR point clouds in voxelized form for perception tasks \cite{lang2019, zhou2018, yang2018}. To reflect this, we computed occupancy-based metrics.
In particular, we report the intersection-over-union (IOU) using $0.2 \times 0.2 \times 0.1$ meter voxels:
\begin{equation} \label{eq:iou}
\textrm{IOU} = \frac{\textrm{TP}}{\textrm{TP} + \textrm{FP} + \textrm{FN}}
\end{equation}
where \textrm{TP}, \textrm{FP}, \textrm{FN} are the numbers of true positives, false positives, and false negatives in terms of voxel occupancy.

\begin{figure*}[t]
    \begin{center}
    \includegraphics[width=0.19\textwidth]{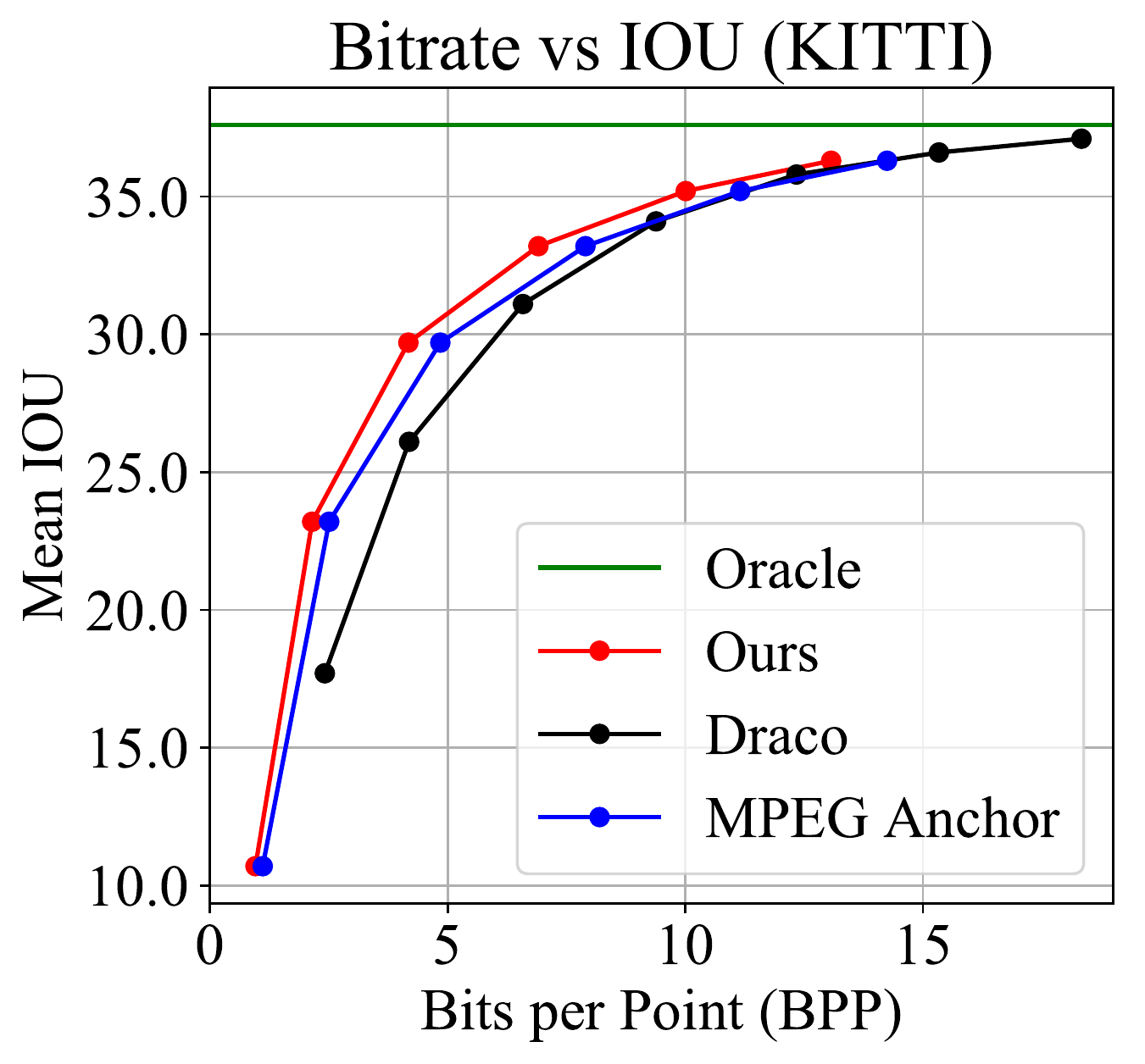}
    \includegraphics[width=0.19\textwidth]{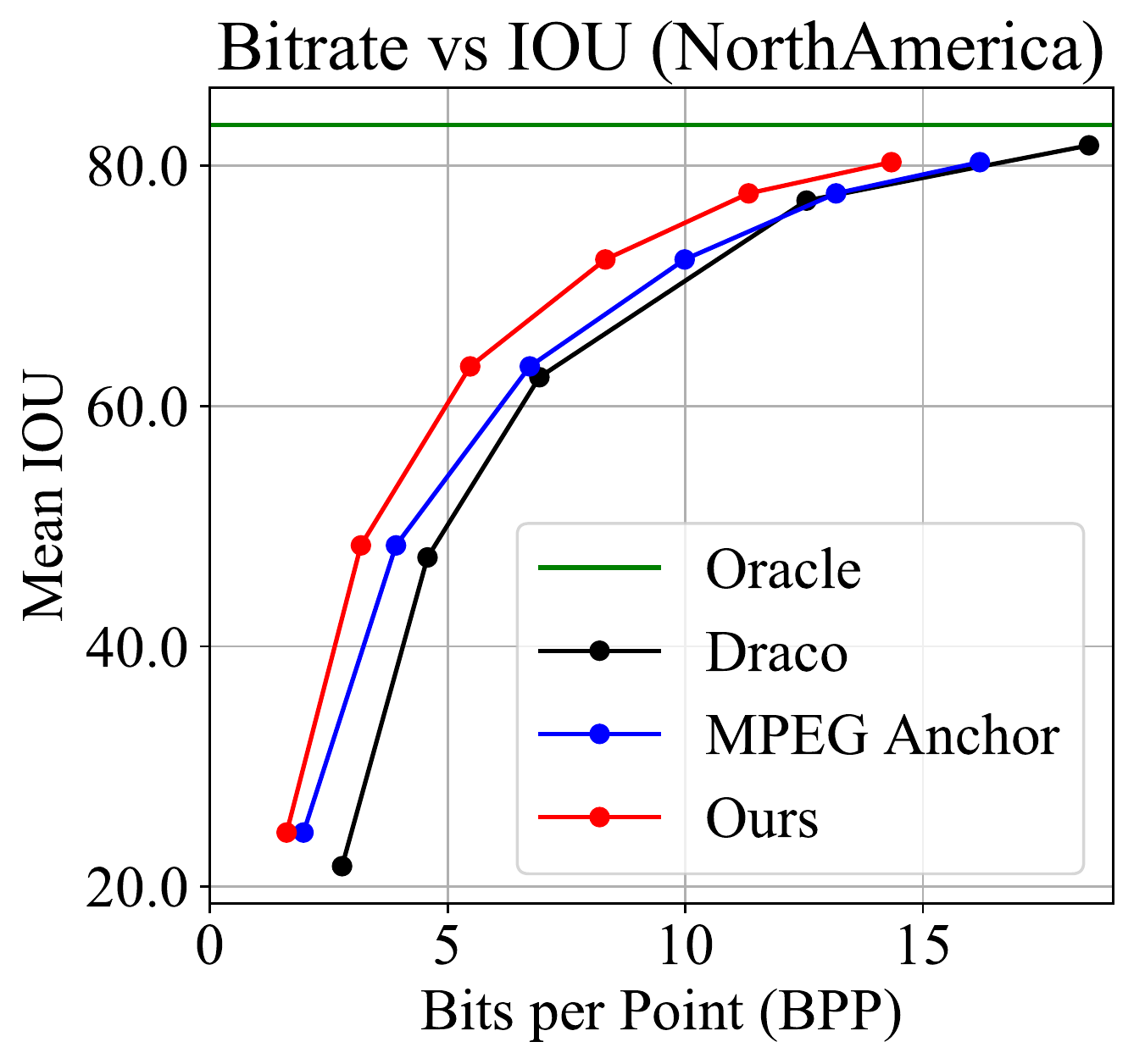}
    \includegraphics[width=0.19\textwidth]{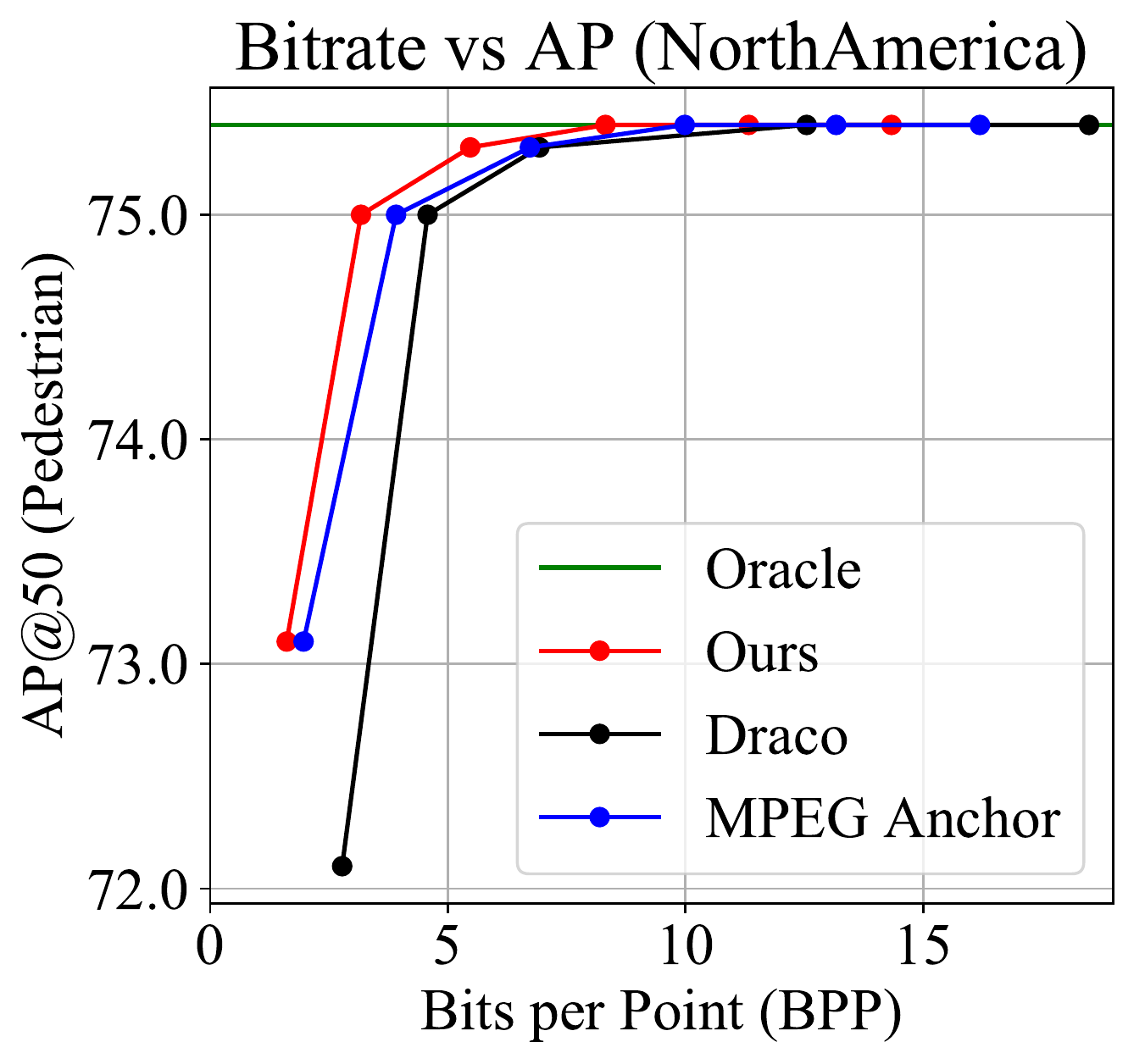}
    \includegraphics[width=0.19\textwidth]{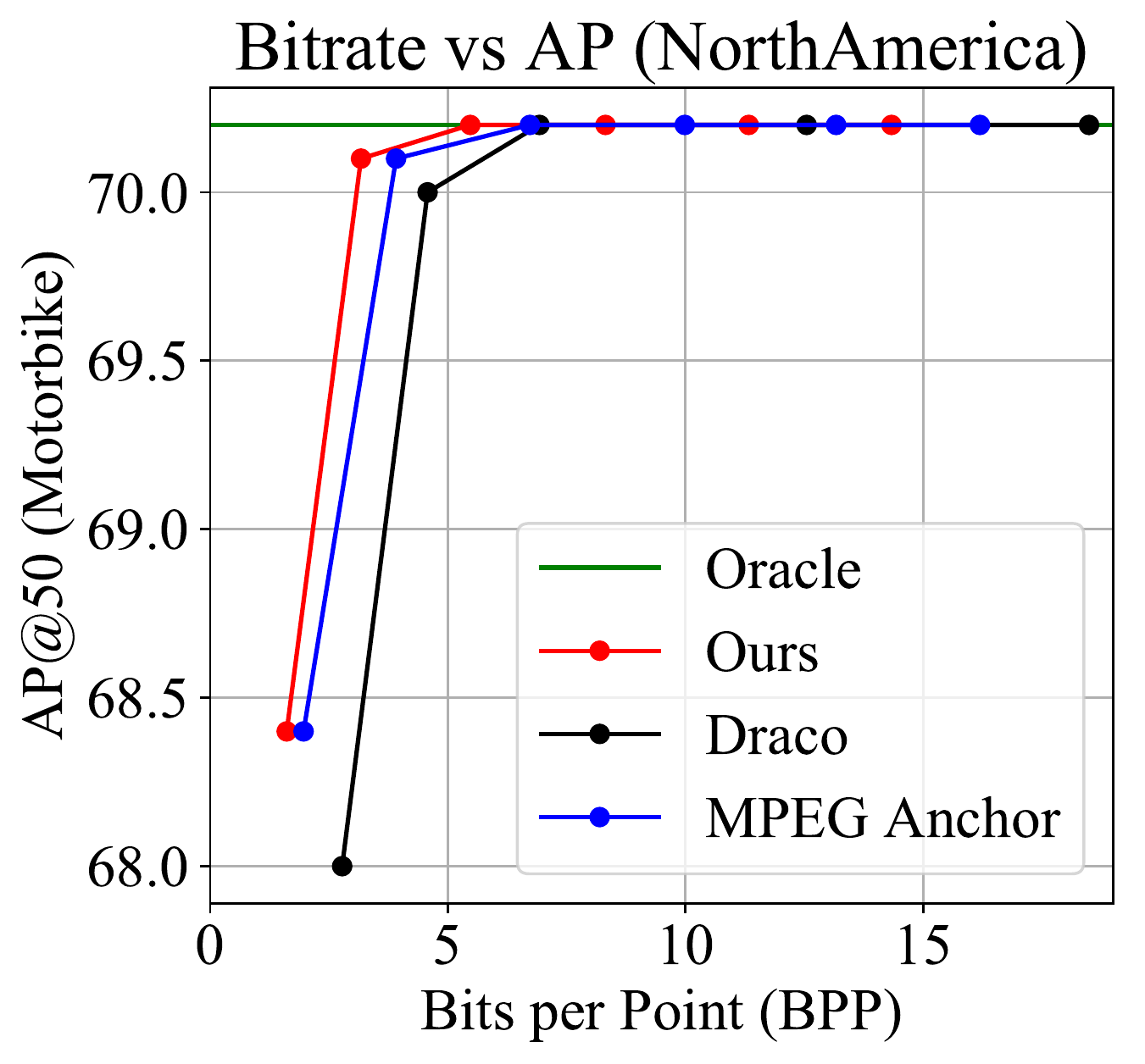}
    \includegraphics[width=0.19\textwidth]{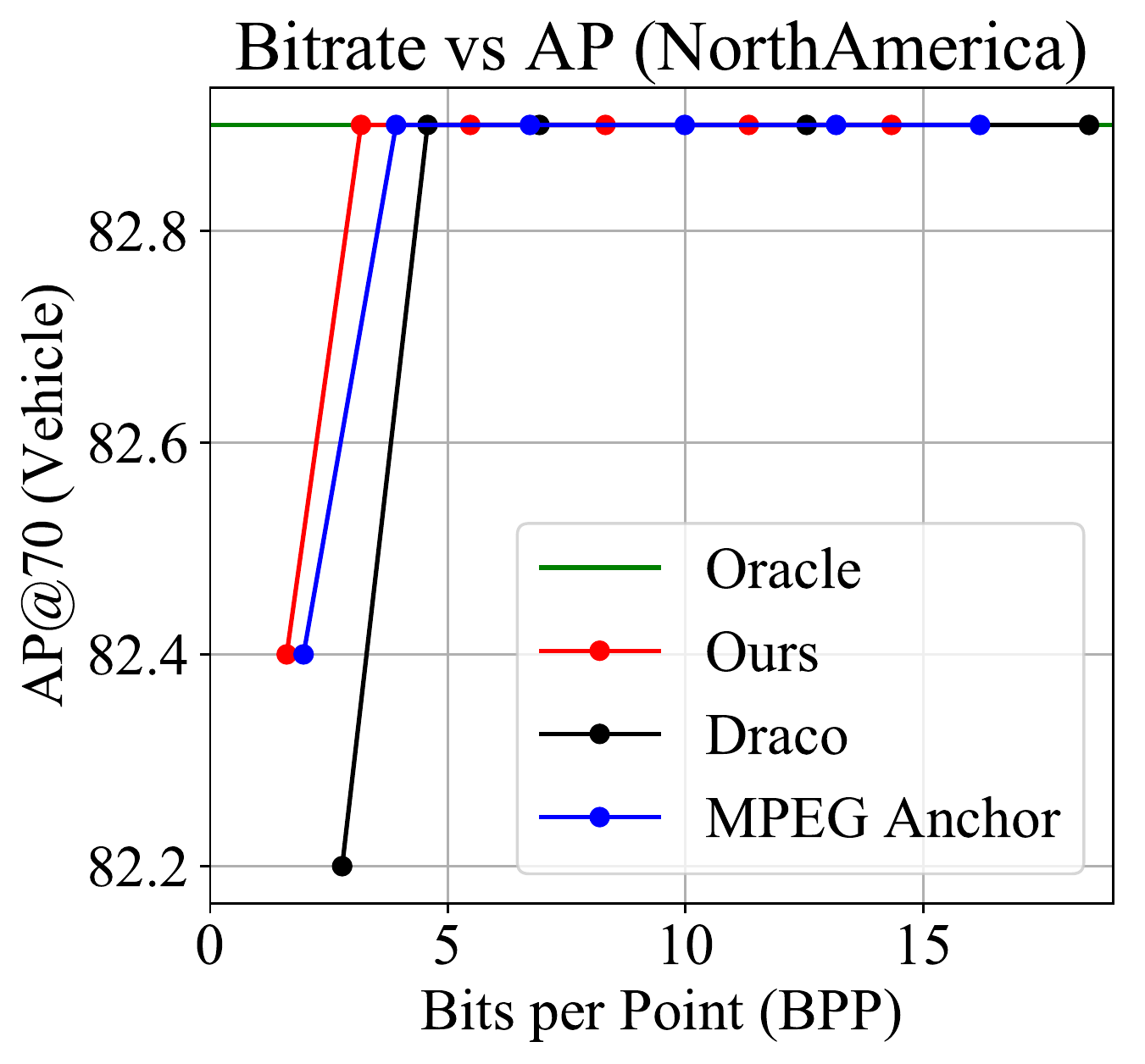}
    \end{center}
    \vspace{-0.15in}
    \caption{
        Quantitative results of downstream perception tasks.
        The leftmost two figures show IOU performance on semantic segmentation for KITTI and NorthAmerica respectively.
        The rightmost three figures show AP performance on object detection for NorthAmerica.}
    \label{fig:downstream_quantitative}
\end{figure*}

\begin{figure*}[t]
\begin{center}
\begin{overpic}[width=0.24\textwidth]{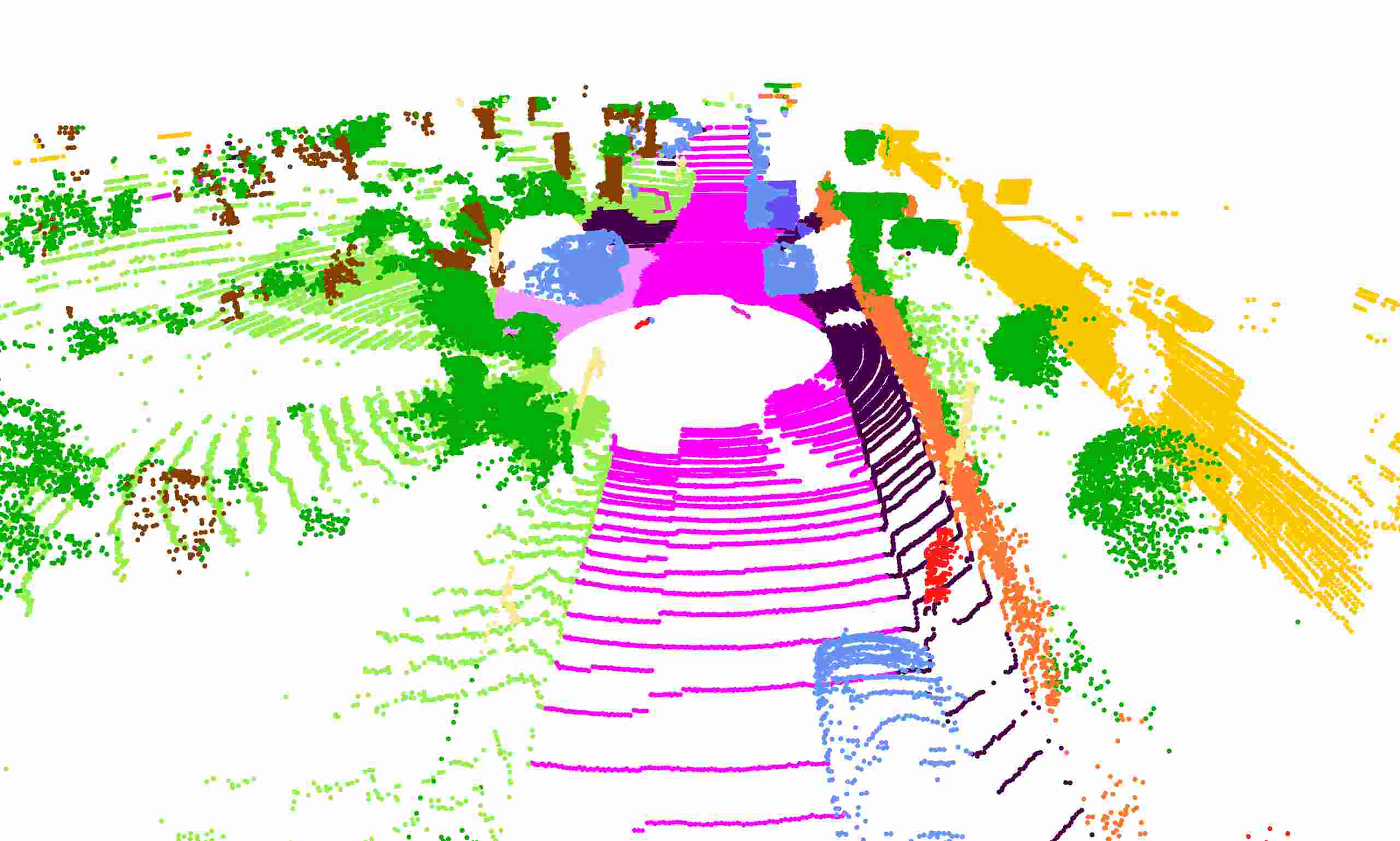}
    \put (0,55) {\colorbox{gray!30}{\scriptsize Oracle: IOU 38.02, Bitrate: 96.00}}
\end{overpic}
\begin{overpic}[width=0.24\textwidth]{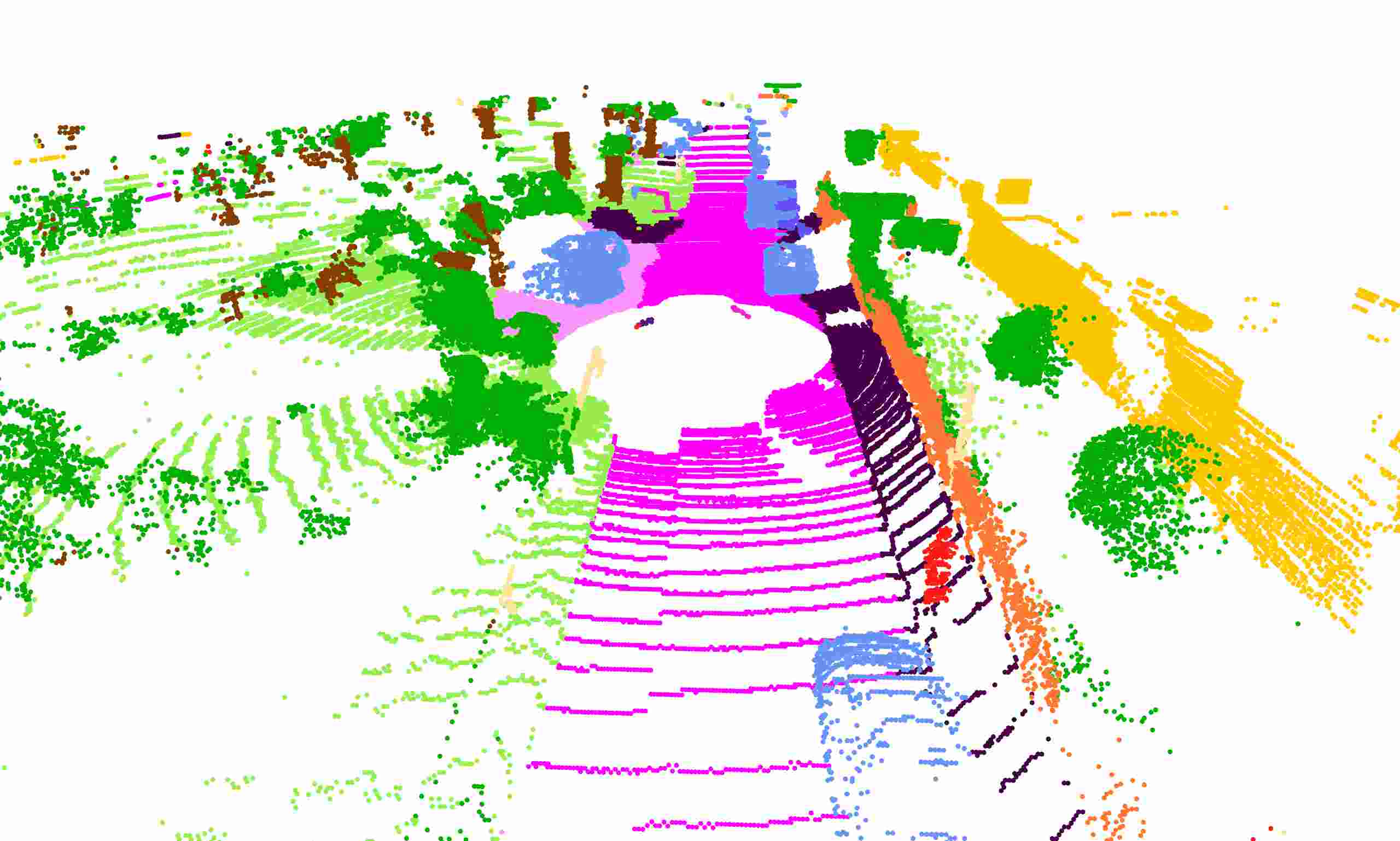}
    \put (0,55) {\colorbox{gray!30}{\scriptsize Ours: IOU 31.94, Bitrate: 4.18}}
\end{overpic}
\begin{overpic}[width=0.24\textwidth]{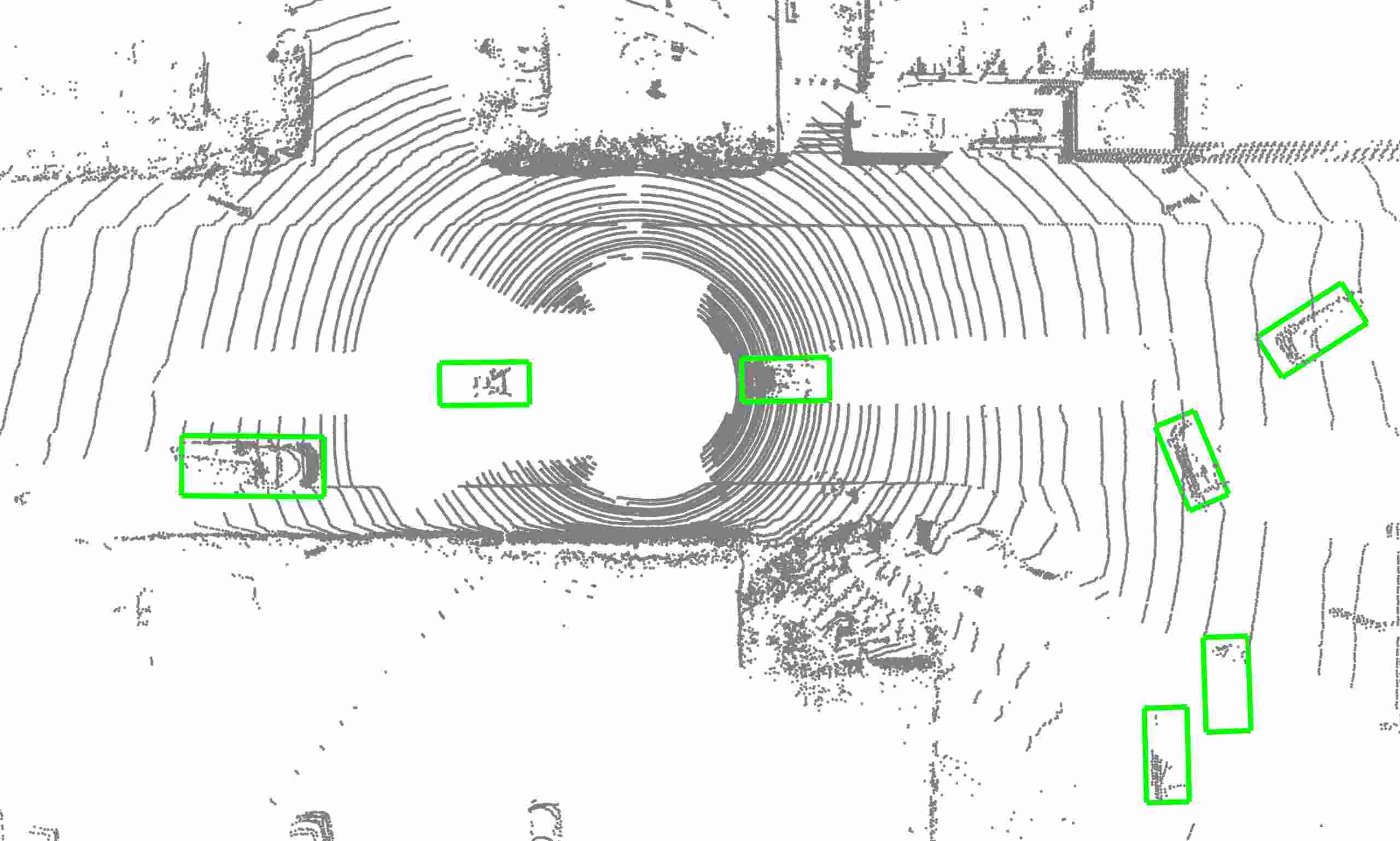}
    \put (0,55) {\colorbox{gray!30}{\scriptsize Oracle: AP@70: 100, Bitrate: 96.00}}
\end{overpic}
\begin{overpic}[width=0.24\textwidth]{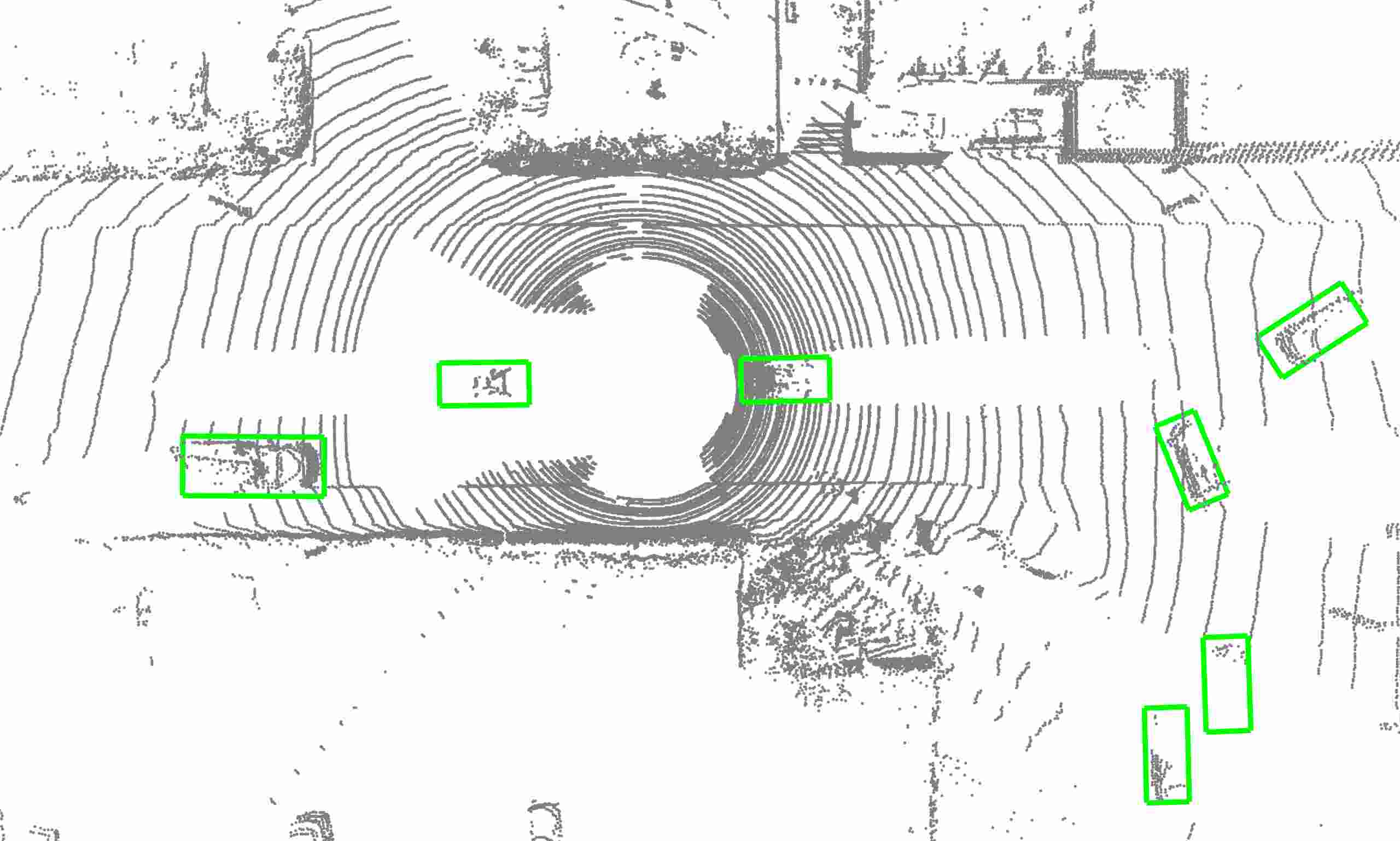}
    \put (0,55) {\colorbox{gray!30}{\scriptsize Ours: AP@70: 100, Bitrate: 6.11}}
\end{overpic}
\end{center}
\vspace{-0.15in}
\caption{Qualitative results of semantic segmentation (right) and object detection (left).}
\label{fig:downstream_qualitative}
\vspace{-0.15in}
\end{figure*}

\subsection{Compression Results}

\paragraph{Quantitative Results on NorthAmerica:}
We report the bitrate versus reconstruction quality metrics (PSNR, IOU, Chamfer)
over all competing algorithms on the NorthAmerica dataset.
As shown in Fig.~\ref{fig:quant_na}, our method outpeforms all previous
state-of-the-art algorithms, with a 10-20\% bitrate reduction over Draco and MPEG Anchor
at the same reconstruction quality.
All three methods significantly outperform the deep range image compression method.
Note that since we use the same octree data structure, our approach has the same reconstruction quality as MPEG Anchor.
However, our bitrate is much lower thanks to the deep entropy model.
These results validate our proposed deep entropy model and our choice of an octree data structure to compress sparse LiDAR point clouds.

\vspace{-2mm}
\paragraph{Quantitative Results on KITTI:}
In Fig.~\ref{fig:quant_na}, we show the bitrate versus reconstruction quality
metrics on KITTI.
Although our model was trained using only data from NorthAmerica, it
can still significantly outperform all competing algorithms, especially at lower bitrates.

\vspace{-2mm}
\paragraph{Qualitative Results:}

Fig.~\ref{fig:qual_kitti} shows point cloud reconstructions on KITTI and NorthAmerica
colored by reconstruction error.
For fair comparison, we choose results from the competing algorithms that have
been compressed at a similar bitrate rate.
All cases show that our method and \textrm{Draco} give more faithful reconstructions than range image compression at comparable bitrates,
as the range image reconstruction suffers from noise and errors at object boundaries as well as lower/upper LiDAR beam.
At the same bitrate, our reconstruction quality is also better than Draco.

\vspace{-2mm}
\paragraph{Ablation Studies:}
We perform ablation studies on the entropy model, both over the context features $\mathbf{c}_i$ as well as over the number of aggregations $K$.
In Tab.~\ref{tab:ablation}, we ablate over context features by progressively incorporating the four features that we use: the node's octree level, its parent occupancy symbol, its octant index, and its spatial location.
Note that these ablations are performed without any aggregations ($K=0$), demonstrating the predictive power of context features alone. As shown in the table, we can see that gradually adding more context information consistently lowers the entropy of our encoding.

Next, we evaluate how the high-order ancestor information helps to predict the probability. We evaluate the proposed entropy model with different levels of aggregation, $K=0,...,4$, incorporating $K$ levels of ``ancestor'' contexts.
Tab.~\ref{tab:stacks} show that in general, conducting more aggregations consistently improves the entropy of our model.

\subsection{Effects on Downstream Perception Tasks}
Another important of metric for compression is its effects on the performance
of relevant downstream tasks.
We quantify these effects for two fundamental perception tasks:
semantic segmentation and object detection.

In our experiments, we evaluate the semantic segmentation and object detection
models described in \cite{wong2019} over point clouds reconstructed from
various compression schemes.
Note that we train these perception models on uncompressed point clouds with detection and segmentation labels---for NorthAmerica, we use the training dataset described in \cite{wong2019}, and for KITTI, we use the official training dataset \cite{behley_semantickitti}.
For semantic segmentation, we report mean intersection-over-union (IOU)
computed using voxelized ground truth labels.
For object detection, we report average precision (AP) at 50\% IOU threshold
for pedestrians and motorbikes, and 70\% for vehicles.

As shown in Fig.~\ref{fig:downstream_quantitative} and Fig.~\ref{fig:downstream_qualitative},
our method outperforms all competing baselines on both NorthAmerica and KITTI.
Our method's strength is particularly highlighted in semantic segmentation where
preserving the fine-grained details of the point cloud is especially important.
For example, at 5 bits-per-point, our method achieves a 5-10\% improvement over
Draco and MPEG for NorthAmerica.
In object detection, our method consistently outperforms the baselines, albeit
more slightly than in segmentation; this is due to the fact that the object
detection model is already robust to a range of bitrates.
Overall, these results attest to the performance of our method and help
illustrate its effects on tasks relevant to many robotics applications.

%% file: sections/conc.tex
\section{Conclusion}

We presented a novel LiDAR point cloud compression algorithm. Our method uses a deep tree-structured entropy model on an octree representation of the points that leverages available context information to reduce the entropy of each intermediate node. This entropy model exploits both the sparsity and structural redundancy between points to reduce the overall bitrate. We validate the effectiveness of our method over two large-scale datasets. The results suggest that our approach significantly reduces the bitrate compared against other competing algorithms at the same reconstruction quality.
In addition, we demonstrate that our compressed representations achieve a lower error on downstream tasks than prior state-of-the-art work.